\documentclass{aa}  

\pdfoutput=1 
\usepackage{graphicx}
\usepackage{booktabs}
\usepackage{lscape}
\usepackage{txfonts}
\usepackage[usenames, dvipsnames]{color}
\usepackage{soul}
\newcommand{\SF}{star-forming }
\newcommand{\HII}{H \textsc{ii} }
\begin{document} 

   \title{The effect of diffuse background on the spatially-resolved Schmidt relation in nearby spiral galaxies}

\titlerunning{Spatially-resolved Schmidt relation}
   \author{Nimisha Kumari,
          \inst{1}
          Mike J. Irwin \inst{1}
          \and
          Bethan L. James \inst{2}
          }
\authorrunning{N. Kumari et al.}
   \institute{$^1$ Institute of Astronomy, University of Cambridge,
              CB3 0HA UK\\
              \email{nkumari@ast.cam.ac.uk}\\
    $^2$ Space Telescope Science Institute, Baltimore MD 21218 US
             }

	\date{Received; accepted}
 
  \abstract
   {The global Schmidt law of star formation provides a power-law relation between the surface densities of star-formation rate (SFR) and gas, and successfully explains plausible scenarios of galaxy formation and evolution. However, star formation being a multi-scale process, requires spatially-resolved analysis for a better understanding of the physics of star formation.}
   {It has been shown that the removal of a diffuse background from SFR tracers, such as H$\alpha$, far-ultraviolet (FUV), infrared, leads to an increase in the slope of the sub-galactic Schmidt relation. We reinvestigate the local Schmidt relations in nine nearby spiral galaxies taking into account the effect of inclusion and removal of diffuse background in SFR tracers as well as in the atomic gas.}
   {We used multiwavelength data obtained as part of the Spitzer Infrared Nearby Galaxies Survey (SINGS), Key Insights on Nearby Galaxies: a Far-Infrared Survey with Herschel (KINGFISH), The H \textsc{i} Nearby Galaxy Survey (THINGS), and HERA CO-Line Extragalactic Survey (HERACLES). Making use of a  novel split of the overall light distribution as a function of spatial scale, we subtracted the diffuse background in the SFR tracers as well as the atomic gas. Using aperture photometry, we study the Schmidt relations on background subtracted and unsubtracted data at physical scales varying between 0.5--2 kpc.}
   {The fraction of diffuse background varies from galaxy to galaxy and accounts to $\sim$34\% in H$\alpha$, $\sim$43\% in FUV, $\sim$37\% in 24 $\mu$m, and $\sim$75\% in H $\textsc{i}$ on average. We find that the inclusion of diffuse background in SFR tracers leads to a linear molecular gas Schmidt relation and a bimodal total gas Schmidt relation. However, the removal of diffuse background in SFR tracers leads to a super-linear molecular gas Schmidt relation. A further removal of the diffuse background from atomic gas results in a
   		slope $\sim$1.4 $\pm$ 0.1, which agrees with dynamical models of
   		star formation accounting for flaring effects in the outer regions of
   		galaxies. 
   	}
   {}


   \maketitle
%

\section{Introduction}

  \label{intro}
  \indent  The physics of star formation is one of the most explored topics of contemporary astrophysics. Though both theorists and observers have worked on different aspects of star formation \citep[see reviews by][]{Shu1987, Kennicutt1998a, Evans1999, Massey2003, Bromm2004, MacLow2004, Glover2005, Zinnecker2007, Mckee2007, KennicuttEvans2012, Krunholz2014}, there are many missing links and we still lack a complete theory. This is mainly because the formation of stars is not a single process but involves a wide range of physical and chemical processes: self-gravity \citep{Goodman2009}; magnetic fields \citep{Li2014}; turbulence \citep{Kritsuk2011}; formation and destruction of molecules; and ionisation from local and background sources \citep{Ward-Thompson2002}. We are familiar with the different conditions involved in the processing of the intergalactic and interstellar medium (ISM), which result in star formation but we still have to understand the primary mechanism that drives and also dominates the formation of stars. 
  
  \indent The widely accepted star-formation law was originally formulated by \citet{Schmidt1959} and relates the volume density of the star-formation rate (SFR) and the gas density as a power-law. Due to the difficulty in measuring the volume densities of the two quantities, generally the Schmidt law of star formation is expressed in terms of surface densities, which are more easily observable: $$\Sigma_{SFR} = A\Sigma_{gas}^{N}$$ where $\Sigma_{SFR}$ and $\Sigma_{gas}$ denote the surface densities of star-formation rate and total gas (atomic and molecular), respectively, A is the average global efficiency of star formation of the system studied (e.g. galaxies, galactic discs, star-forming regions), and N is the power-law index. A value of N = 1.4 $\pm$ 0.1 was found empirically by \citet{Kennicutt1998b}, who derived $\Sigma_{SFR}$  from H$\alpha$ data, and  $\Sigma_{gas}$ by combining CO (molecular gas) and H \textsc{i} (atomic gas) data for normal spirals and starbursts; they established the disc-averaged star-formation law, called the Kennicutt-Schmidt relation. The disc-integrated or  disc-averaged star-formation law implies the averaging out of enormous local variations in the stellar population (age and initial mass function, IMF) in addition to gas and dust geometry \citep{KennicuttEvans2012}, implying that the currently-established law might not follow from a fundamental and causal physical relationship. Moreover, star formation is a multi-scale process. Hence to understand the underlying physical processes and consequently the physics of star formation, a spatially-resolved analysis of star formation is required.  

\indent Thanks to the technical advances resulting in an explosion of spatially-resolved multiwavelength data for nearby galaxies, the Schmidt relation has been studied at sub-galactic scale extensively. Various approaches have been adopted for such analyses, which can be grouped into two main categories: radial profiles (i.e. comparing azimuthally averaged values) of $\Sigma_{SFR}$ and $\Sigma_{gas}$ and  point-by-point analysis (either pixel-by-pixel analysis or aperture photometry). Depending on the approach and data used, different observers have found different values of the power-law index (slope) in the Schmidt relation \citep[see for example,][]{WongBlitz2002, Boissier2003, Heyer2004, Komugi2005, Schuster2007, Kennicutt2007, Bigiel2008, Blanc2009, Liu2011, Rahman2011, Rahman2012, Leroy2013, Momose2013, Shetty2014a, Casasola2015, Roychowdhury2015, Azeez2016, Morokuma-Matsui2017}. 

\indent A common observation in all of these studies is that the molecular gas is more correlated with the SFR than the atomic gas. \citet{WongBlitz2002} analysed radial profiles of seven CO-bright spiral galaxies and found that the $\Sigma_{SFR}$ is directly proportional to the molecular gas density, while \citet{Boissier2003} used the same technique on a sample of 16 galaxies and found a range (0.6 -- 1.3) of power-law slopes. \citet{Kennicutt2007} used aperture photometry of the star-forming regions in M51a and obtained a slope of 1.37$\pm$0.03 between the $\Sigma_{SFR}$ and the molecular gas density at a spatial resolution of 500 pc. In contrast, \citet{Bigiel2008} find a linear molecular law (power-index = 1) from their pixel-by-pixel analysis on seven spiral galaxies. \citet{Bigiel2008} suggest that the slope found by \citet{Kennicutt2007} was steeper because of the subtraction of a local background in the SFR tracers, which was performed as a step in the aperture photometry. \citet{Liu2011} investigate this issue through a pixel-by-pixel analysis on two nearby spiral galaxies where they subtracted a diffuse background using the software $HIIphot$ \citep{Thilker2000}, and reproduced the results of \citet{Kennicutt2007} and \citet{Bigiel2008}. We refer the interested readers to \citet{Liu2011} and \citet{Rahman2011} for a detailed discussion on the astrophysical significance of diffuse background in SFR tracers (H$\alpha$, far-ultraviolet (FUV), 24 $\mu$m hereafter in this paper) described as an emission that is unrelated to the current star formation. 

\indent As discussed in various studies \citep{Lonsdale1987, Leroy2012, Crocker2013, Johnson2013, Boquien2014, Boquien2016}, the observed stellar flux and the dust-emission in a star-forming region not only contains the contribution from the young stars and the related dust component, but also an underlying diffuse component of stellar and dust emission unassociated with the  current star formation. For example, the FUV continuum contains a considerable amount of diffuse emission from evolving or already evolved stars \citep{Tremonti2001} as FUV emission from an instantaneous-burst takes about 100 Myr to diminish by two orders of magnitude \citep{Leitherer1999} compared to the 10 Myr for H$\alpha$ emission. In addition stars migrate away from their birth-site \citep{Chandar2005}. Similarly,  dust in the star-forming regions possibly contains ultra-small grains or large molecules which are unassociated with the current star formation and are likely to substitute for the effects produced by polychromatic hydrocarbon (PAH) e.g single-photon heating \citep{DraineLi2007}. Thus the mid-infrared emission may not only be produced by PAH heated by young stars but also dust grains (or molecules) with vibrational properties of PAH and small (or large) enough to be heated by older stellar population through single-photon heating. Emission at higher infrared wavelength contains contribution from diffuse dust `cirrus' and traces increasingly cooler components.  Thus diffuse component unrelated to the `current star formation' are likely to be present in all SFR tracers. 

\indent In this paper, we also investigate the possibility of diffuse background in the atomic gas. Most of the spatially-resolved observational studies have found no correlation between the surface densities of SFR and atomic gas and hence concluded that molecular gas is the sole driver of star formation. However, some studies \citep{Boissier2003} also suggest that star formation may be considered as a result of dynamical processes where total gas density is important, rather than as purely sequential process where stars form from molecular gas which is initially formed from atomic gas. Some other studies have also pointed out the role of atomic gas in star formation specifically in H \textsc{i}-dominated regions in star-forming galaxies \citep[see][]{Bigiel2010, Roychowdhury2015}. Moreover, some theoretical works also suggest that the stars can form as easily in atomic gas as in molecular gas \citep{Glover2012, Krumholz2012}. The dynamical model of \citet{Elmegreen2015} provides relationships
	between the star-formation rate density and total gas density in the
	central and outer regions of spiral galaxies as well as in dwarf
	irregular galaxies, that is, both molecular and atomic gas. The most commonly used atomic gas tracer, H $\textsc{i}$ 21 cm emission line traces cool H $\textsc{i}$, warm H $\textsc{i}$ and diffuse molecular gas or clouds \citep{Draine2011, Magnani2017}. With regards to the latter component, H I is also detected in and around molecular clouds as demonstrated in previous studies \citep[e.g.][]{Elmegreen1987, GoldsmithLi2005, Fukui2009, Stanimirovic2014, Hayashi2019}. The simulations of \citet{Semenov2017} investigate the
	star-forming and non-star-forming components of the ISM and the role
	of only a small fraction of gas in forming stars in order to explain
	long depletion time scales \citep[see also][]{Semenov2018}. Among these components of the ISM, we try to identify that component of H $\textsc{i}$ 21 cm emission which may contribute to star formation, and study its effect on spatially-resolved Schmidt relation, in case that component is present in the traced atomic gas. 

\indent Hence, the goal of this paper is to extend the previous analyses \citep[e.g.][]{Kennicutt2007, Liu2011} by performing a spatially-resolved study on a larger sample of galaxies, and study the effect of inclusion and removal of a diffuse component not only in the SFR tracers but also in the atomic gas. We find that the observed total gas Schmidt relations after subtraction of diffuse background in SFR tracers and H \textsc{i} gas, are in agreement with dynamical model of star formation. We briefly mention the possibility of a diffuse component in CO gas as well. The paper is organised as follows: Section \ref{obs} describes the galaxies in the sample and the multiwavelength data used in this work. Section \ref{method} describes the essentials of aperture photometry, the SFR and gas density measurements and the novel method we adopted to subtract diffuse background in SFR tracers and atomic gas. Section \ref{results} presents the main results and Section \ref{discussion} presents a comparison with dynamical model of \citet{Elmegreen2015}, a comparison of our results to those in the literature, a comparison with the global Kennicutt-Schmidt relation and  possible implications of our results. Section \ref{section:conclusion} presents the conclusion.
\section{Sample and data }
\label{obs}
\subsection{Sample galaxies}

\begin{table*}
	\centering
	\caption{Properties of sample galaxies}
	\label{sample}
	\resizebox{\textwidth}{!}{\begin{tabular}{@{}lccccccccc@{}}                      
			\hline\hline
			
			\multicolumn{1}{c}{Name} & \begin{tabular}[c]{@{}c@{}}Hubble$^1$\\ Type\end{tabular} & \begin{tabular}[c]{@{}c@{}}Distance$^2$\\ (Mpc)\end{tabular} & \begin{tabular}[c]{@{}c@{}}i$^{3,6}$\\ (deg)\end{tabular} & \begin{tabular}[c]{@{}c@{}}P.A.$^{3,6}$\\ (deg)\end{tabular} & \begin{tabular}[c]{@{}c@{}}r$_{25}^{3,6}$\\ (arcmin)\end{tabular}& E(B--V)$^4$         & \begin{tabular}[c]{@{}c@{}}log(H$\alpha$+NII)$^2$\\ (ergs$^{-1}$cm$^{-2}$)\end{tabular} & {[}NII{]}/H$\alpha$ $^2$\\ \midrule
			NGC 0628                 & Sc                              & 7.3                                                      & 7                &20          & 4.89                                                                                   & 0.0600 $\pm$  0.0012   & -10.84 $\pm$ 0.04                                                           & 0.345 $\pm$ 0.046     \\
			NGC 3184                 & SABc                            & 11.1                                                     & 16               &179         & 3.71                                                                                 & 0.0144 $\pm$ 0.0001 & -11.12 $\pm$ 0.05                                                           & 0.523 $\pm$ 0.052     \\
			NGC 3351  & SBb &9.33 & 41 & 192 &  3.60 &  0.0239 $\pm$ 0.0001 & -11.42 $\pm$ 0.08 & 0.655 $\pm$ 0.027\\
			NGC 3521 & SABb & 8.03 & 73 & 165 & 4.16 &0.0496 $\pm$ 0.0014 & -10.85 $\pm$ 0.04 & 0.558 $\pm$ 0.008 \\
			NGC 4736 & SAab &  5.20 & 41 & 296 &  3.88 & 0.0155 $\pm$ 0.0004 & -10.72 $\pm$ 0.06 & 0.711 $\pm$ 0.006\\
			NGC 5194                 & Sbc                             & 8.2                                                      & 20               &172         & 3.88                                                                                & 0.0350 $\pm$ 0.0017  & -10.45 $\pm$ 0.04                                                           & 0.590 $\pm$ 0.006     \\
			NGC 5055 & Sbc & 7.8 & 59 &102 &  5.93  &0.0153 $\pm$ 0.0003 & -10.80 $\pm$ 0.07 & 0.486$\pm$0.019 \\
			NGC 5457  & Sbc & 6.7 & 18 & 39  & 11.99 & 0.0074 $\pm$ 0.0001 & -10.22$^5$  & 0.54$^5$ \\
			NGC 6946                 & SABc                            & 6.8                                                      & 33               &243         & 5.74                                                                                  & 0.2942 $\pm$ 0.0028 & -10.42 $\pm$ 0.06                                                           & 0.448 $\pm$ 0.087     \\ \bottomrule
			
	\end{tabular}}
	\tablefoot{$^1$Values from NASA/IPAC Extragalactic Database (NED),
		$^2$Values from \citet{Kennicutt2009}, 
		$^3$Values from \citet{Bigiel2008}, 
		$^4$Values from \citet{Schlafly2011}, $^5$Values from \citet{Leroy2013} - no error was given in the paper, $^6$Values from \citet{Schruba2011}, Column 4 (i): inclination angle, Column 5 (P.A.): position angle,
		Other columns are self-explanatory.}

\end{table*}

\begin{table}
	\centering
	\caption{Details and properties of multiwavelength data}
	\label{tab data}
	\begin{tabular}{@{}cclc@{}}
		\hline\hline
		\begin{tabular}[c]{@{}c@{}}Data\\ (1)\end{tabular} & \begin{tabular}[c]{@{}c@{}}Instrument\\ (2)\end{tabular} & \multicolumn{1}{c}{\begin{tabular}[c]{@{}c@{}}Central Wavelength\\ (3)\end{tabular}} & \begin{tabular}[c]{@{}c@{}}PSF\\ (4)\end{tabular} \\ \midrule
		H$\alpha$                                            & CTIO/KPNO                                                     &  \multicolumn{1}{c}{6563 \AA}                                                                                    & $\sim$1.9''                                              \\ 
		FUV                                                & GALEX                                                  &            \multicolumn{1}{c}{1528 \AA}                                                                            & $\sim$5.5''                                             \\
		24 $\mu$m                                             & Spitzer/MIPS                                             & \multicolumn{1}{c}{24 $\mu$m}                                                           & $\sim$6''                                                 \\
		H \textsc{i}                                                & VLA                                                      & \multicolumn{1}{c}{21 cm}                                                            & $\sim$6''                                                  \\
		CO(2-1)                                            & IRAM   & \multicolumn{1}{c}{1.3 mm}                                                                                                                                     & $\sim$11''                                                \\
		
		\bottomrule
	\end{tabular}
	\tablefoot{Column  1:  data  name;  Column  2:  telescope,  instrument,  
		and  filter; Column 3: filters' central wavelength; Column 4: FWHM
		of the PSF}
	
\end{table}

\indent We study a sample of nine nearby spiral galaxies. The criteria for sample selection are as follows: (1) the centres of the galaxies are principally dominated by molecular gas as inferred from CO(2--1) images; (2) the maximum distance of the galaxies in our sample is 11.1 Mpc, beyond which the spatial scale is too compact for detailed spatial studies ; (3) the highest inclination angle of the galaxies studied here is 72.7$^{\circ}$. The criteria (2) and (3) are adopted considering the aperture sizes described in Section \ref{flux extraction}.  

\indent Table \ref{sample} presents the galaxies in our sample along with their relevant properties: galaxy type, distance, inclination angle, position angle, major and minor axes. This table also contains the values of Galactic E(B--V) \citep{Schlafly2011}, which are used to correct foreground galactic extinction for optical (H$\alpha$) and FUV data; log(H$\alpha$ + N \textsc{ii}) and [N \textsc{ii}]/H$\alpha$ to calibrate the H$\alpha$ images. Details related to the multiwavelength data and the respective instruments  are given in Table \ref{tab data}.


\subsection{H$\alpha$ Emission-line Images}
\indent Narrowband images centred at H$\alpha$ and continuum R-band images are  taken either with the 1.5 m telescope at  Cerro Tololo Inter-American Observatory (CTIO) or the 2.1 m telescope at Kitt Peak National Observatory (KPNO). They are either part of the \textit{Spitzer} Infrared Nearby Galaxies Survey \citep[SINGS;][]{Kennicutt2003} or of the `Key Insights on Nearby Galaxies: A Far-Infrared Survey with Herschel' \citep[KINGFISH;][]{Kennicutt2011}. 

\indent To obtain an emission line only image of H$\alpha$ for each galaxy, we rescaled the corresponding R-band image and subtracted it from the narrow-band image. The H$\alpha$ image is  corrected for the foreground galactic extinction using $F(H\alpha)_{corr} = F(H\alpha)_{obs} \times 10^{0.4\times2.535\times E(B-V)}$. A constant ratio of [N \textsc{ii}]/H$\alpha$ for each galaxy is adopted to correct for the contamination from the neighbouring [N \textsc{ii}] $\lambda\lambda$6548, 6584 in the narrowband H$\alpha$ filter. The H$\alpha$ image is finally calibrated using the total flux (H$\alpha$ + N \textsc{ii}). The flux calibration error is $\sim$ 5\% and the point spread function (PSF) is typically $\sim$ 2$\arcsec$. 

\subsection{GALEX FUV Images}
\indent  FUV images are taken from the Galaxy Explorer (GALEX) and form part of KINGFISH except NGC 5194  which is from the GALEX NGS Survey \citep{GildePaz2007}. Technical details are described in \citet{Morrissey2005}. We correct the FUV images for foreground galactic extinction using  $FUV_{corr} = FUV_{obs} \times 10^{0.4\times8.02\times E(B-V)}$, and calibrate them assuming an AB magnitude system. The calibration uncertainty is $\sim$ 0.15 mag. The images obtained from KINGFISH were already corrected for obvious artefacts and the bright foreground galactic stars were entirely removed.

\subsection{Spitzer MIPS 24 $\mu$m Images}
\indent We use the 24 $\mu$m infrared (IR) data to estimate and correct for the H$\alpha$ and FUV radiation obscured by dust (details in Section \ref{conversion}). The 24 $\mu$m images are taken with the Multiband Imaging Photometer (MIPS) installed on the \textit{Spitzer Space Telescope} and are obtained either from KINGFISH or SINGS. The calibration error is 5\%. For details on the observing strategy and data reduction procedures of the MIPS Instrument, see \citet{Kennicutt2003} and \citet{Gordon2005}, respectively. 

\subsection{HERACLES CO (J=2--1) Images} 
\indent We use CO (J= 2--1) images  from `The HERA CO-Line Extragalactic Survey' (HERACLES) for estimating the molecular gas content of the galaxies in our sample. These maps are obtained  using the IRAM 30 m telescope; details of observation and data reduction are given in \citet{Leroy2009}. The angular resolution of these maps is 11$\arcsec$ \citep{Bigiel2008} and the flux calibration error is 20\% \citep{Leroy2009}. For all galaxies, we adopt a constant standard value of X(CO)  = 2$\times$10$^{20}$ cm$^{-2}$ (K km s$^{-1}$)$^{-1}$ as established for the Milky Way \citep{Dame2001} and ratio CO(2-1)/CO(1-0) = 0.8  \citep{Leroy2009}, for converting the CO intensity maps to the mass maps of molecular hydrogen, noting that uncertainties in the X(CO) factor contribute to the overall calibration error. 

\subsection{VLA H \textsc{I} Images}
\indent The neutral atomic hydrogen gas content in the galaxies is estimated from the moment 0 maps (integrated H \textsc{i} map from `robust' weighting) obtained from `The H \textsc{i} Nearby Galaxy Survey' \citep[THINGS;][]{Walter2008}. These maps are obtained by using the National Radio Astronomy Observatory (NRAO) Very Large Array (VLA). The typical beam size for robust weighted maps is $\sim$ 6$\arcsec$. We created the atomic mass maps by using the formulae given in \citet{Walter2008}.The calibration error for the H \textsc{i} is 5\%.

\indent We note here that neither H \textsc{i} nor H$_2$ is corrected for the presence of helium and other metals. Equation (4) from \citet{Leroy2009} used for calculating H$_2$ surface density includes the multiplicative factor of 1.36 to account for helium and other elements in the galaxies, which is dependent on metallicity and other environmental factors. We do not include this factor for comparison and consistency with other published works such as \citet{Bigiel2008} which contains eight out of nine galaxies in our sample. 

\section{Methodology}
\label{method}

\subsection{Aperture photometry: apertures sizes and aperture correction}
\label{flux extraction}
\indent We employ aperture photometry on potential star forming regions as
	well as the regions in between them selected on
	the basis of the H$\alpha$ image of each galaxy (an example in Figure \ref{figure:region selection}). We note that the regions
	in between star-forming regions must have H$\alpha$ detection, and are
	selected in order to sample low-luminosity regions which might have
	diffuse background. Such a selection is essential to study the effect
	of diffuse background on the star-formation scaling relations, between gas and
	recent star formation ($<$ 5 Myr). In what follows, we describe the motivation behind choosing different aperture sizes for different galaxies, flux-extraction from apertures on all images and aperture correction. The galaxies in our sample lie at distances ranging from 3.13--11.10 Mpc, with inclination angles varying between 7--72.7$^{\circ}$. In practice this drives us towards adopting a range of physical aperture sizes to accomodate the different angular resolution available. The SFR calibration combining observed H$\alpha$ and 24 $\mu$m (see equations \ref{Halpha_corr} and \ref{sfr_halpha}) is valid for regions larger than the H \textsc{ii} complexes ($\sim$0.8--5.1 kpc) which produce the ionizing radiation \citep{Calzetti2007}. The proportionality coefficient in equation \ref{Halpha_corr} varies by only $\sim$20\%  (within 1$\sigma$ uncertainty) when derived for NGC 5194 \citep{Kennicutt2007} on regions of physical diameter 520 pc. Hence for our study, we adopt a minimum physical diameter of $>$ 500 pc. However, we can not use this physical diameter for the galaxies with high inclination angle and large distance (NGC 3351, NGC 3521 and NGC 5055) as this physical size corresponds to radii of 1--2 pixels and results in undersampling for the aperture. To determine a viable maximum aperture size, we performed an experiment described in Appendix \ref{appendix: aperture size}, and hence decided to use aperture sizes varying between $\sim$0.5--2.0 kpc depending on the distance and inclination angle of individual galaxy. 

\indent For flux extraction, images in all wavelengths in each galaxy are aligned and registered on the astrometric grid of the CO image which has the coarsest pixel size ($\sim$ 2$\arcsec$). This ensures that flux is extracted in all image apertures over the same region. Potential star-forming regions and diffuse regions are selected by visual inspection of the H$\alpha$ images. For NGC 5194, we used
	the catalogue of regions provided in \citet{Kennicutt2007} for
	comparison. Fluxes are measured in the apertures placed over these
	regions in all images using the `phot' task available in IRAF\footnote{IRAF is distributed by the National Optical Astronomy Observatory, which is operated by the Association of Universities for Research in Astronomy (AURA) under a cooperative agreement with the National Science Foundation.}. The same procedure is applied on images where the diffuse background has been subtracted via a software \textit{Nebulosity Filter (Nebuliser)}, as described later in Section \ref{nebuliser}. As a check we estimate an overall clipped median value for the background in each image (except H \textsc{i}) though we note here that the \textit{Nebulosity Filter} reduces the background in the corrected maps to negligible levels. 

\indent Since the H \textsc{ii} regions in the sample galaxies are extended sources, or are composed of overlapping point sources, an aperture correction is required to obtain total fluxes. We use synthetic PSFs for 24 $\mu$m and FUV, and Gaussian PSFs of 2$\arcsec$, 6$\arcsec$, and 11$\arcsec$ for H$\alpha$, H \textsc{i}, and CO, respectively\footnote{http://www.astro.princeton.edu/~ganiano/Kernels.html}. We then create a curve-of-growth by calculating flux in successive apertures centred on the PSF and hence calculate aperture correction as a function of radius for each wavelength.

\subsection{Uncertainty calculation}
\label{section: error}
\indent Three main sources of uncertainty (variance) are present in the derived fluxes: systematic offsets in estimates of the local background $\sigma^2_{sys}$; {\it rms} errors in the summed flux $\sigma^2_{rms}$; and the systematic uncertainty from physical calibration of the source images during data reduction $\sigma^2_{cal}$. The dominant pixel-level {\it rms} error $\sigma_b$ is estimated by placing several apertures in the outer regions of the galaxies for each image and analysing the pixel-level flux variation. The main random error in each aperture pixel is then given by $\sqrt{n}\,\sigma_b$, where $n$ is the number of pixels in the aperture. The systematic uncertainty due to large scale variations in the background is smaller than the random error because of the large number of effective pixels $m$ used in tracking the background variation. All uncertainties are added in quadrature to calculate the final uncertainties in the fluxes in different wavebands following equation \ref{err equation}, 
\begin{equation}
\sigma^2_{flux} = \sigma^2_{rms} + \sigma^2_{sys} + \sigma^2_{cal} 
\simeq n\sigma^2_{b} + n^2 \sigma^2_{b}/m + \sigma^2_{cal}.
\label{err equation}
\end{equation}
These errors are then propagated to errors in $\Sigma_{SFR}$ and gas surface densities ($\Sigma_{H \textsc{i}}$, $\Sigma_{H_2}$ and $\Sigma_{H  \textsc{i} + H_2}$). 


\begin{figure*}
	\centering
	
	\includegraphics[width=\textwidth, trim={4cm 0 4cm 0}, clip]{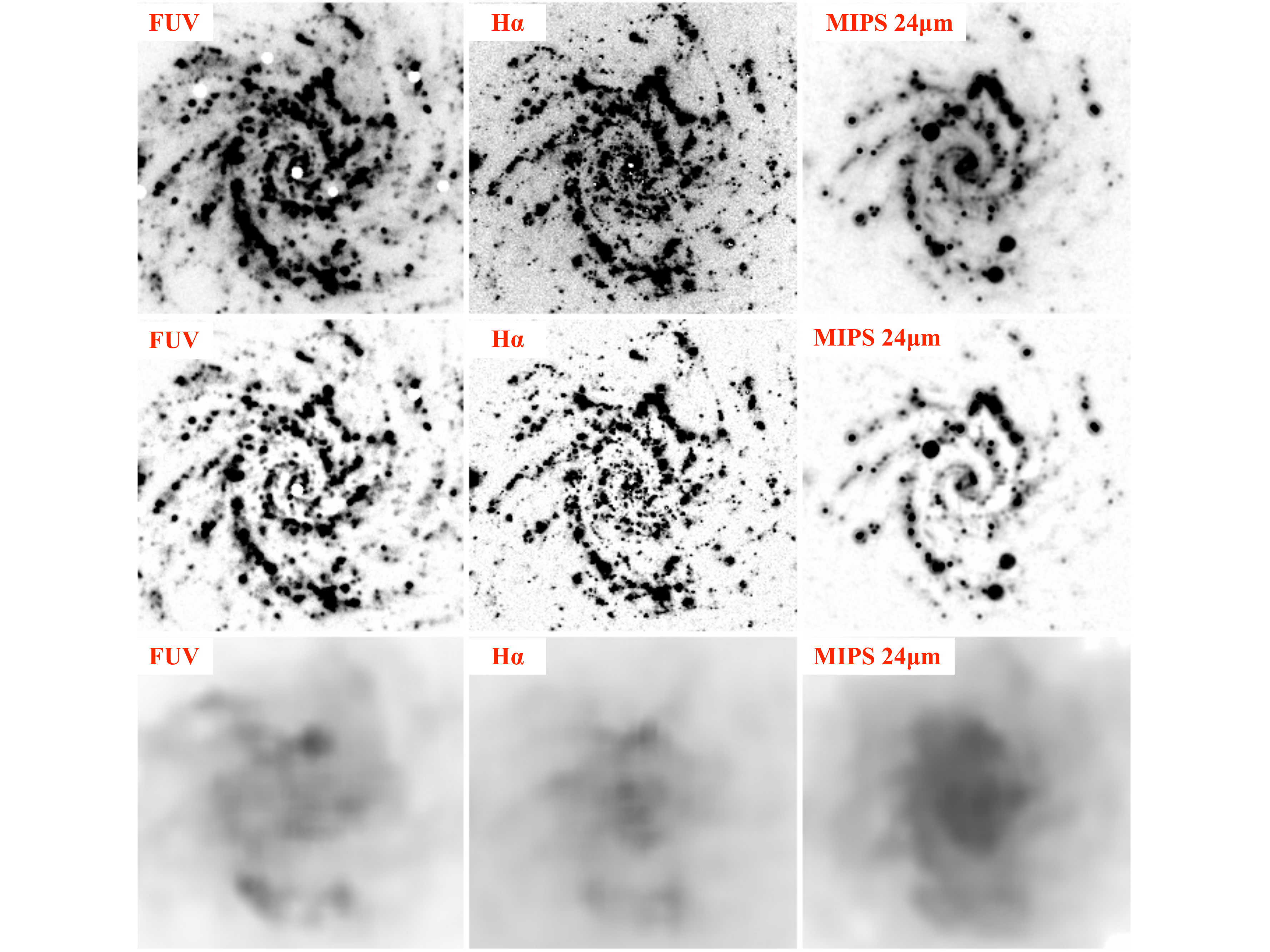}
	\caption{Upper panel: Original images of galaxy NGC 0628 before subtraction of diffuse background. Middle panel: Images after subtraction of diffuse background, which show likely current star-forming regions. Lower panel:  Images of estimated diffuse background, showing its complex spatially varying nature. The images in a particular wavelength have the same flux scale limits. }
	\label{Nebuliser Images}
\end{figure*}

\begin{figure*}
	\centering
	\includegraphics[width=\textwidth, trim={1 1.6cm 0 0}, clip]{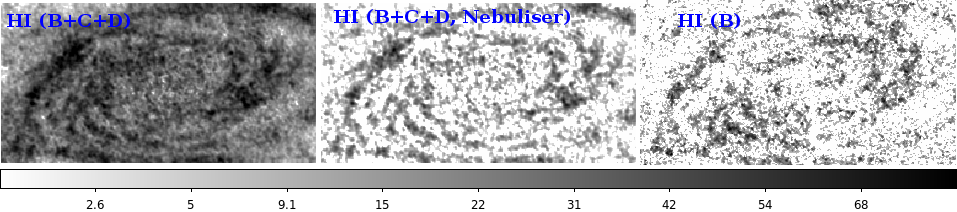}
	\caption{Left panel: The HI map of NGC 5055 obtained by combining data taken in B, C and D configurations of VLA. Middle panel: The HI map of NGC 5055 obtained after processing the HI map in left panel with the \textit{Nebuliser}, i.e. after subtraction of diffuse background. Right panel: The HI map of NGC 5055 obtained by reducing data taken in B configuration of VLA, which should not contain any diffuse background. The images obtained via the two methods (middle panel and right panel) show similar structure, and similar fraction of diffuse background ($\sim$ 76\% from $Nebuliser$ and $\sim$78\% from B-configuration map). All images have the same flux scale limits. The pixel size of the HI (B-configuration) (right panel) is 0.5 arcsec while the other two have pixel size of 2 arcsec, same as CO map.}
	\label{figure: HI nebulosity}
\end{figure*}


\subsection{Conversion of luminosity into SFR}
\label{conversion}
\indent Before converting optical (H$\alpha$) or FUV luminosity to SFR, they need to be corrected for internal dust-attenuation. We use a simple energy-balance argument \citep{Gordon2000, Kennicutt2007,Calzetti2007} involving the use of IR luminosities (here MIPS 24 $\mu$m) to estimate the attenuation of H$\alpha$ and FUV luminosities. This method allows both dust-obscured and dust-unobscured fluxes to be taken into consideration. The reason behind using MIPS 24 $\mu$m is the strong correlation between H$\alpha$ and 24 $\mu$m emission in H \textsc{ii} regions as found by \citet{Calzetti2005}. The attenuation-correction coefficients of H$\alpha$ are taken from \citet{Calzetti2007} and FUV from \citet{Leroy2008}
\begin{equation}
L(H\alpha)_{corr} = L(H\alpha)_{obs} + 0.031L(24),
\label{Halpha_corr}
\end{equation}
\begin{equation}
L(FUV)_{corr} = L (FUV)_{obs}+6L(24),
\label{FUVcorr}
\end{equation}
where the subscripts $obs$ and $corr$ denote the observed and attenuation-corrected luminosities, respectively.

\indent The conversion of attenuation-corrected luminosity into SFR requires assumptions on the IMF related to its form and sampling. Two forms of IMF are widely used - Salpeter IMF \citep{Salpeter1955} and Kroupa IMF \citep{Kroupa2001}. Adopting the more realistic Kroupa IMF and applying the dust-attenuation correction as described above, the following formulae \citep{Liu2011} are used to calculate the SFR:
\begin{equation}
SFR(H\alpha) (M_{\odot} yr^{-1} )= 5.3\times10 ^{-42}L(H\alpha)_{corr}(erg s^{-1}),
\label{sfr_halpha}
\end{equation}
\begin{equation}
SFR(FUV) (M_{\odot} yr^{-1} ) = 3.4\times10^{-44}L(FUV)_{corr}(erg s^{-1}).
\label{sfr_fuv}
\end{equation}

 The two SFR calibrations noted here are generally used in spatially-resolved star-formation studies, though \citet{Kennicutt2007} applied the global conversion (derived in \citealt{Kennicutt1998b}) for the spatially-resolved study of NGC 5194. As pointed by \citet{Kennicutt2007}, using the global conversion to calculate the SFR for an individual H $\textsc{ii}$ region, has limited physical meaning because stars are younger and the region under study go through an instantaneous event compared to any galactic evolutionary or dynamical timescale. The purpose of using the global conversion in \citet{Kennicutt2007} was to make a comparison of the slope and zero-point of the spatially-resolved Schmidt relation with the global Kennicutt-Schmidt relation \citep{Kennicutt1998b}. 
 
 \indent  While comparing our spatially-resolved results with the global relation in Section \ref{section: comparison global}, we take into account the change of IMF. A conversion from a Kroupa IMF to a Salpeter IMF can be easily done by multiplying the calibration constant in equations \ref{sfr_halpha} and \ref{sfr_fuv} by 1.6. As \citet{Calzetti2007} notes, the choice of IMF contributes more significantly (59\%) to the $\sim$50\% difference between the global SFR recipe \citep{Kennicutt1998b} and the spatially-resolved calibration used here (equation \ref{sfr_halpha}), derived by \citet{Calzetti2007}. Other factors such as assumptions on the stellar populations (100 Myr in \citet{Calzetti2007} versus infinite age in \citet{Kennicutt1998b}) gives only a 6\% decrease in the discrepancy given by the different IMFs. The contribution of the above-mentioned factors should be taken into account while making comparisons between different star-formation studies.

\indent Since our eventual aim is to study the spatially-resolved Schmidt relation, the SFRs are converted to SFR surface densities ($\Sigma_{SFR}$) normalising the SFRs by the area of the aperture and dividing by an additional factor of $\sim$ 1/cos\,i to correct for the inclination of the galaxies given in Table \ref{sample}.


\begin{figure}
	\centering
	\includegraphics[width = 0.40\textwidth ]{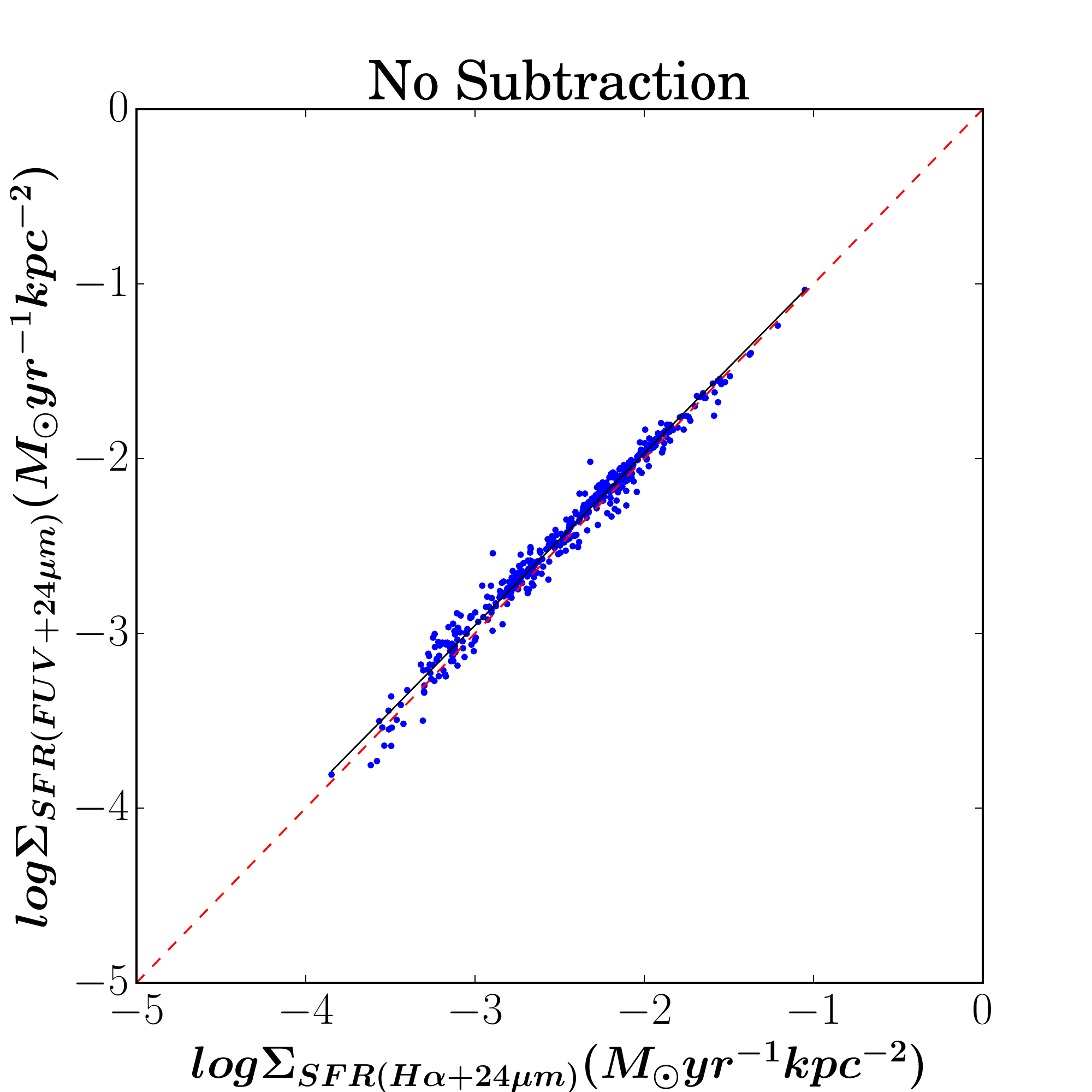}
	\includegraphics[width = 0.40\textwidth]{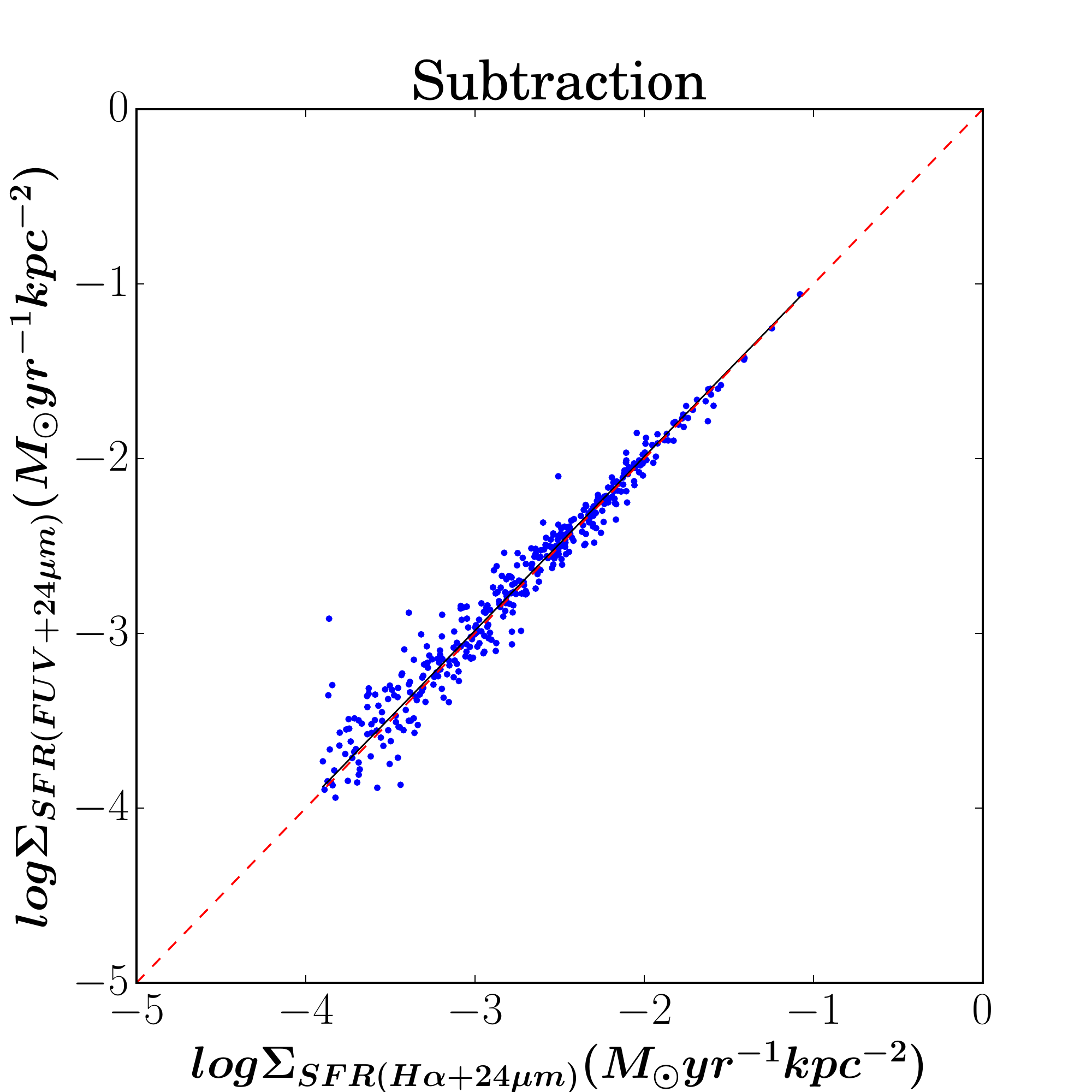}
	\caption{Comparison of attenuation-corrected $\Sigma_{SFR}$ calculated from H$\alpha$ and FUV before (upper panel) and after subtraction (lower panel) of diffuse background in SFR tracers for NGC 0628. The solid black line shows the best-fit to the data and the dashed red line is the one-to-one relation between the two recipes of $\Sigma_{SFR}$. The scaling relation does not change significantly which shows that the subtraction of diffuse background is done consistently in all SFR tracers.}
	\label{Figure: scaling relation}
\end{figure}


 \subsection{Subtraction of diffuse background}
 \label{nebuliser}


\begin{table}
	\centering
	\caption{Percentage of diffuse background in each of the SFR tracers and atomic gas in the \SF regions in the sample galaxies, along with the filtering scale used in Nebulosity Filter}
	\label{diffuse}
	\begin{tabular}{@{}lllllc@{}}
		\hline\hline
		\multicolumn{1}{c}{Galaxy} &  \multicolumn{4}{c}{\begin{tabular}[c]{@{}c@{}}Diffuse\\ Background\end{tabular}} & \multicolumn{1}{c}{\begin{tabular}[c]{@{}c@{}}Filtering\\ Scale$^a$\end{tabular}}\\ \midrule
		& H$\alpha$                   & FUV                    & 24 $\mu$m         & H \textsc{i}      & (kpc)   \\
		
		NGC 0628 & 25 & 28 & 38 & 67 & 1.8\\
		NGC 3184 & 25 & 20 & 32 & 63 & 2.7\\
		NGC 3351 & 45 & 33 & 25 & 76 & 2.3\\
		NGC 3521 &22 & 45 & 17 & 80  & 1.9\\
		NGC 4736 & 26 & 15 & 16 & 76 & 1.3\\
		NGC 5055 & 26 & 38 & 43 & 76 & 1.9\\
		NGC 5194 & 2 & 25 & 20 & 58  & 2.0\\
		NGC 5457 & 5 & 14 & 23 & 37  & 2.9\\
		NGC 6946 & 19 & 28 & 17 & 63 & 3.0\\ \bottomrule
		
	\end{tabular}
\\
Notes: $^a$: Scale lengths are not corrected for inclination of galaxies.
\end{table} 
\begin{figure*}
	\centering
	\includegraphics[width = 0.345\textwidth,trim={0.2cm 0.5cm 1.6cm 0.2cm},clip]{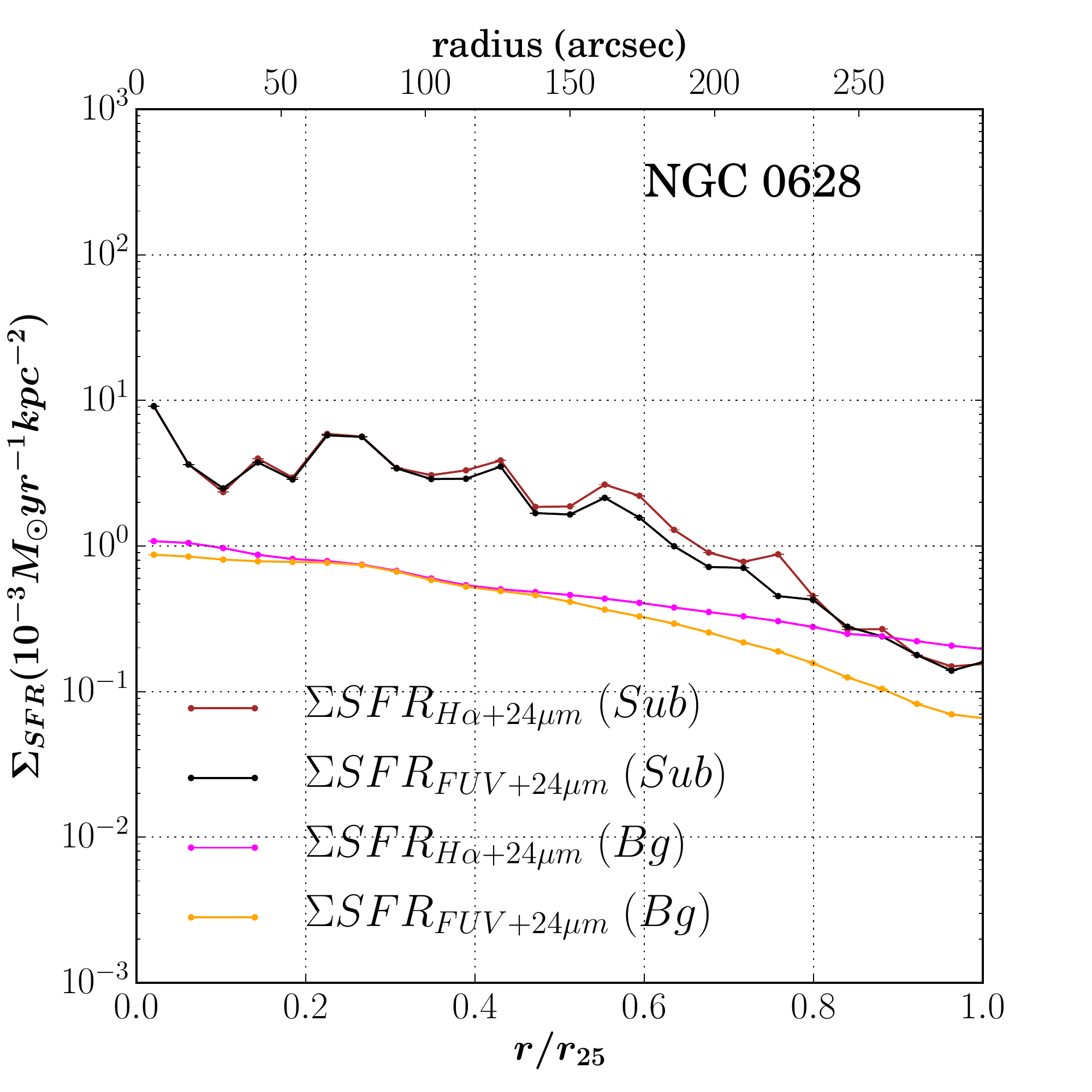}
	\includegraphics[width = 0.32\textwidth,trim={1.2cm 0.5cm 1.6cm 0.2cm},clip]{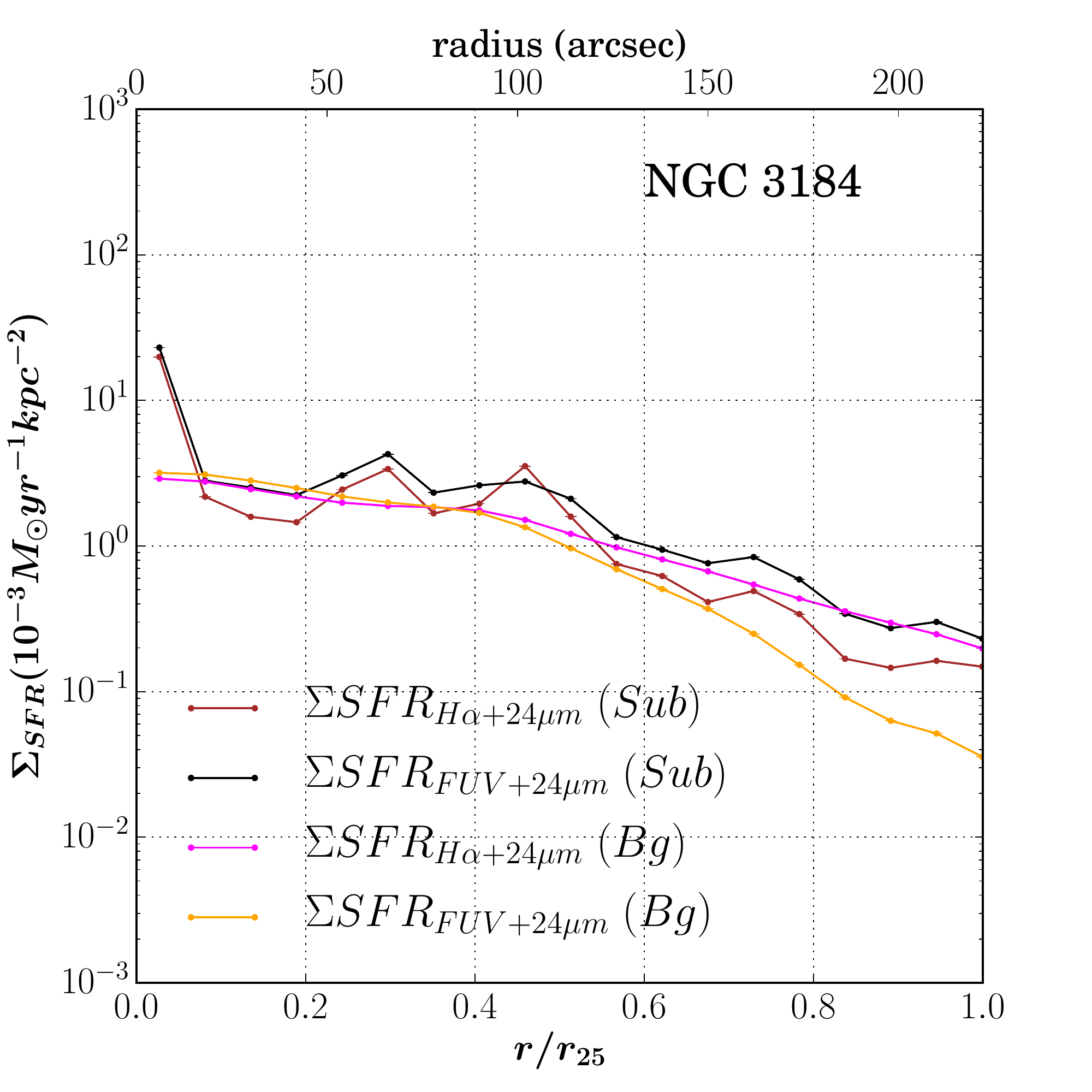}
	\includegraphics[width = 0.32\textwidth,trim={1.2cm 0.5cm 1.6cm 0.2cm},clip]{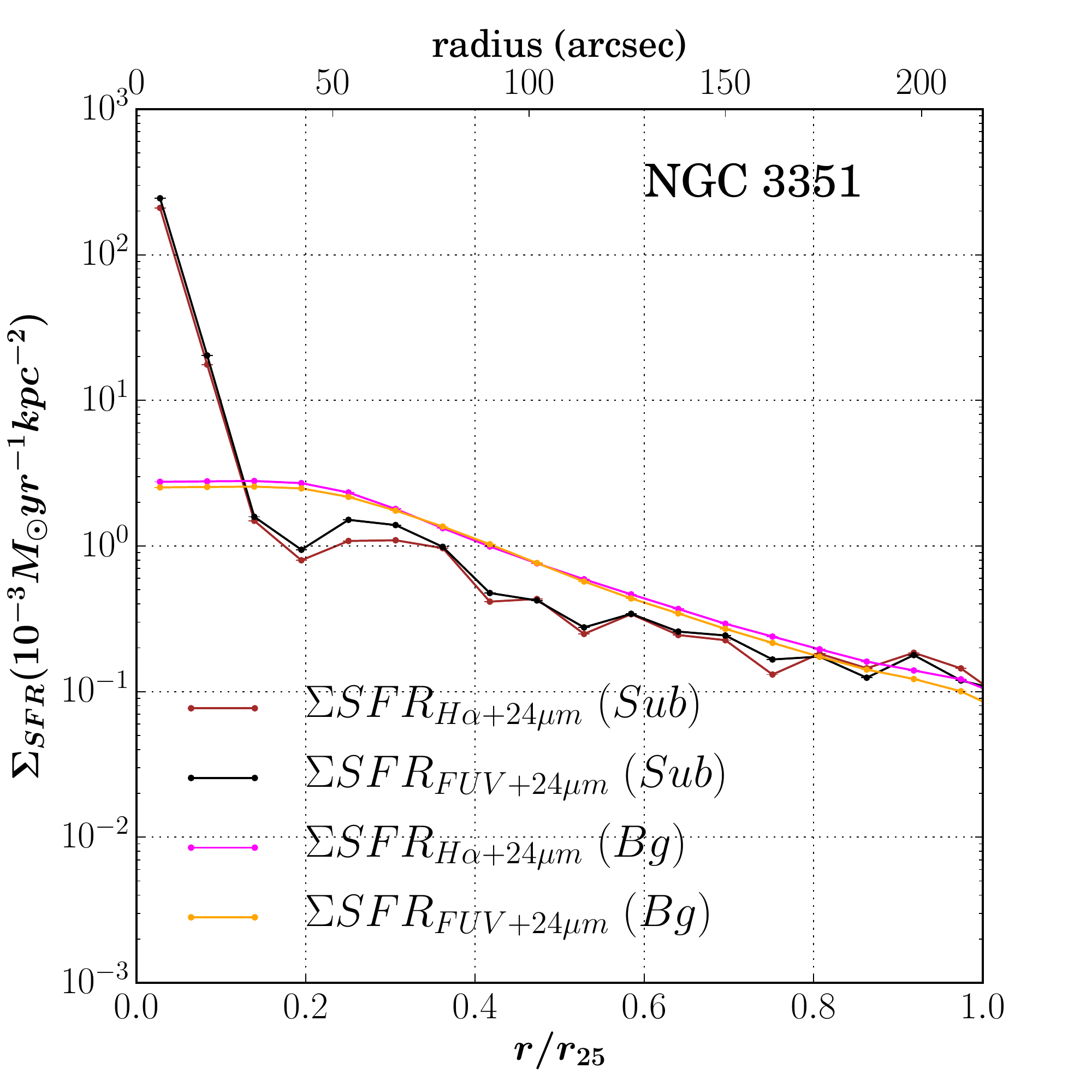}
	\includegraphics[width = 0.345\textwidth,trim={0.2cm 0.5cm 1.6cm 0.2cm},clip]{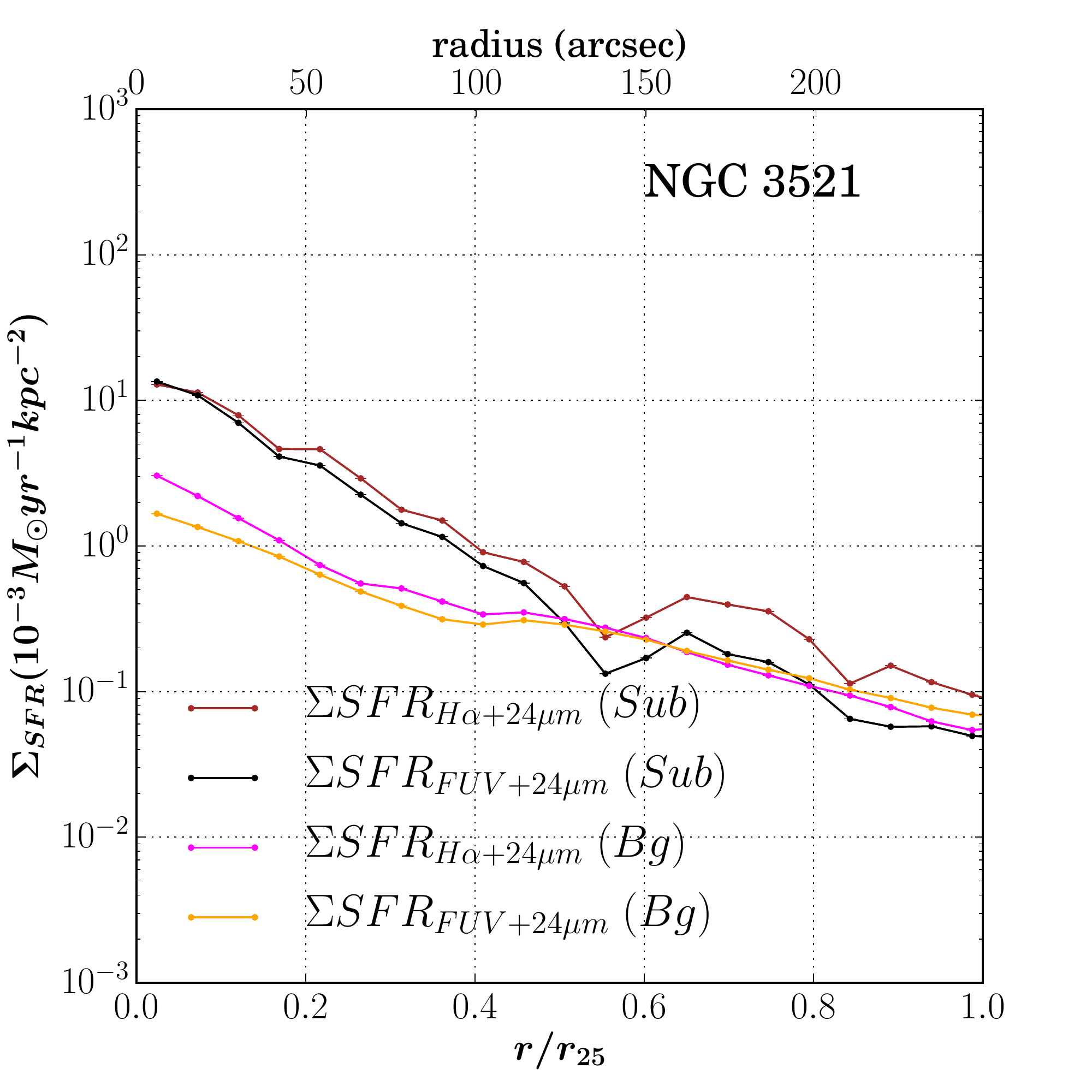}
	\includegraphics[width = 0.32\textwidth,trim={1.2cm 0.5cm 1.6cm 0.2cm},clip]{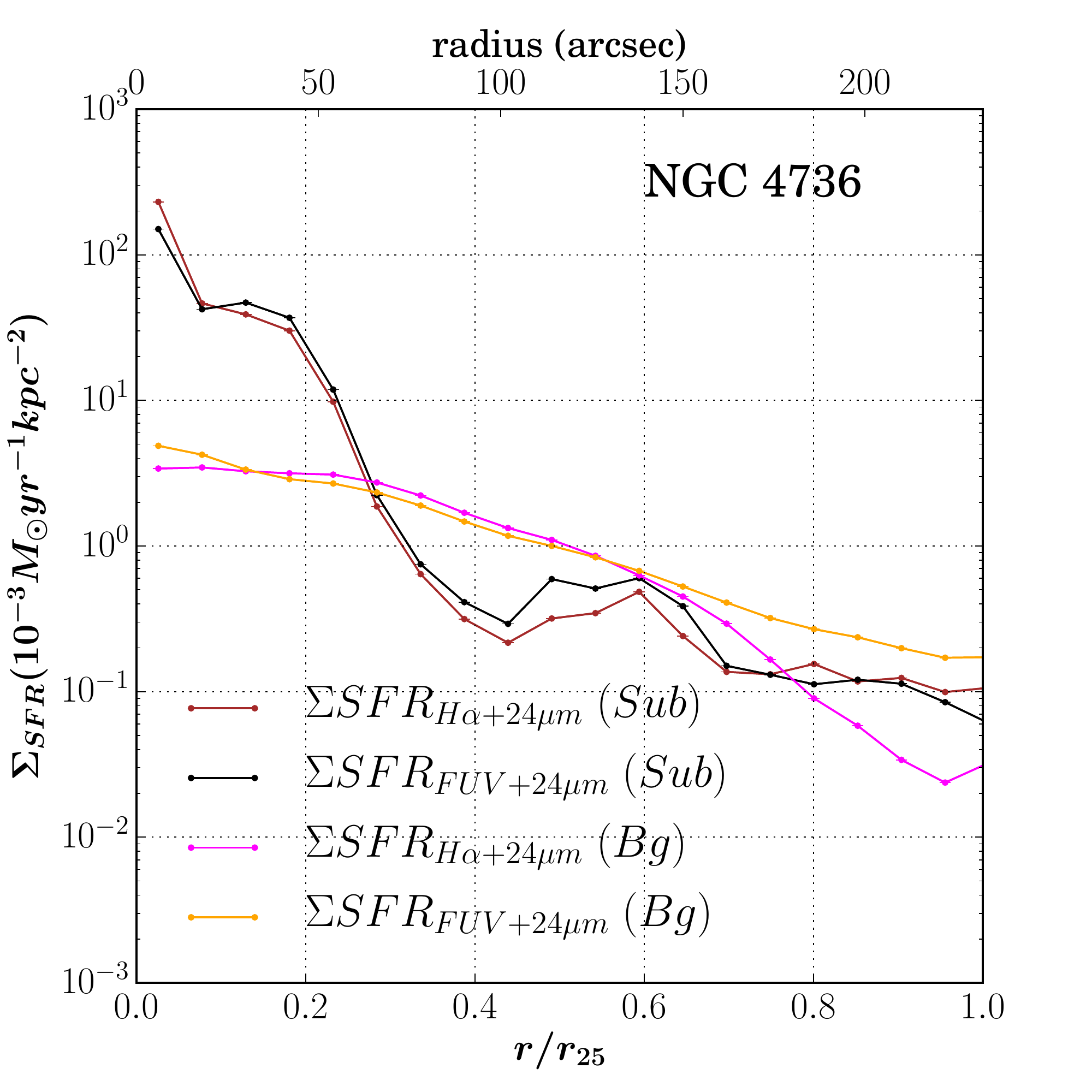}
	\includegraphics[width = 0.32\textwidth,trim={1.2cm 0.5cm 1.6cm 0.2cm},clip]{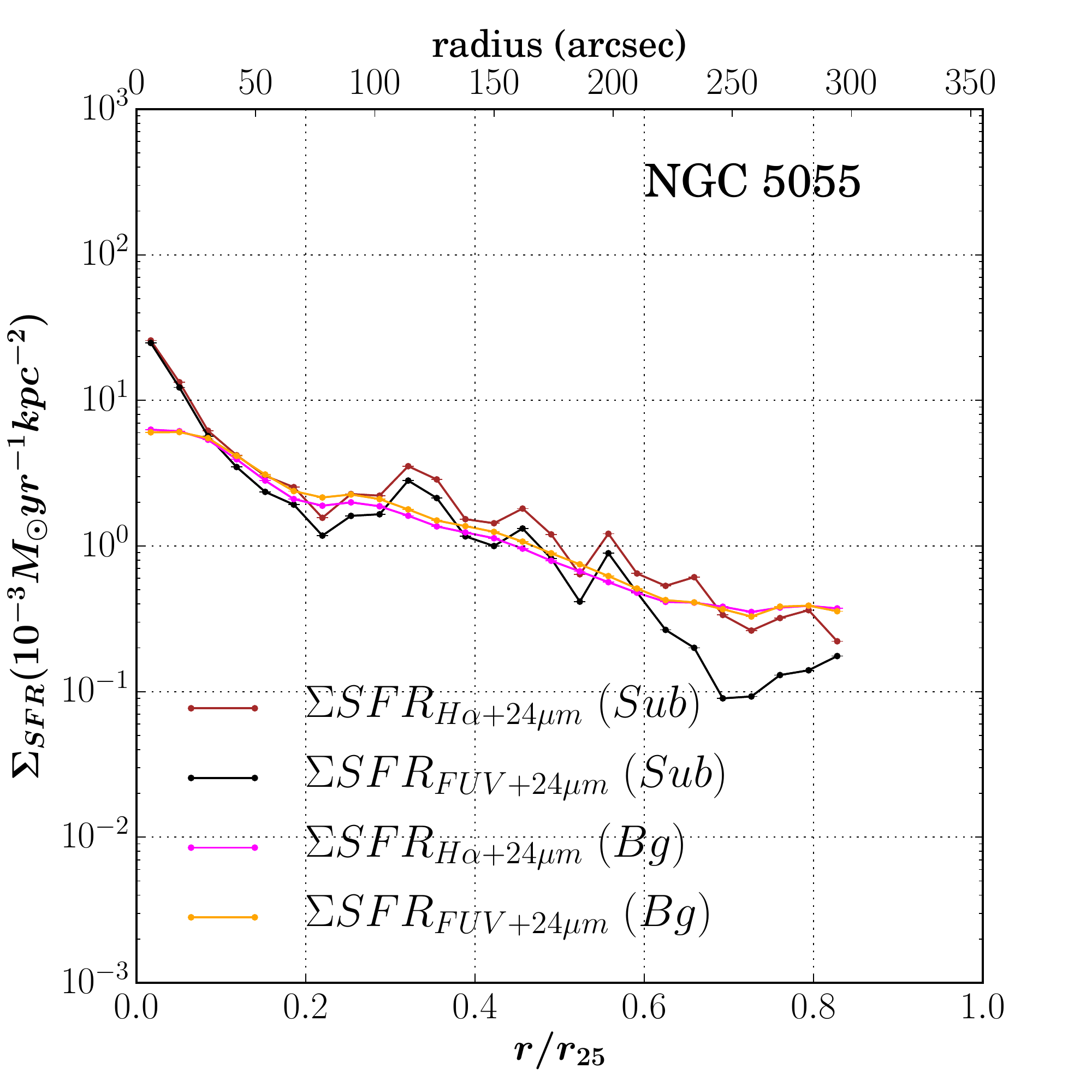}
	\includegraphics[width = 0.345\textwidth,trim={0.2cm 0.5cm 1.6cm 0.2cm},clip]{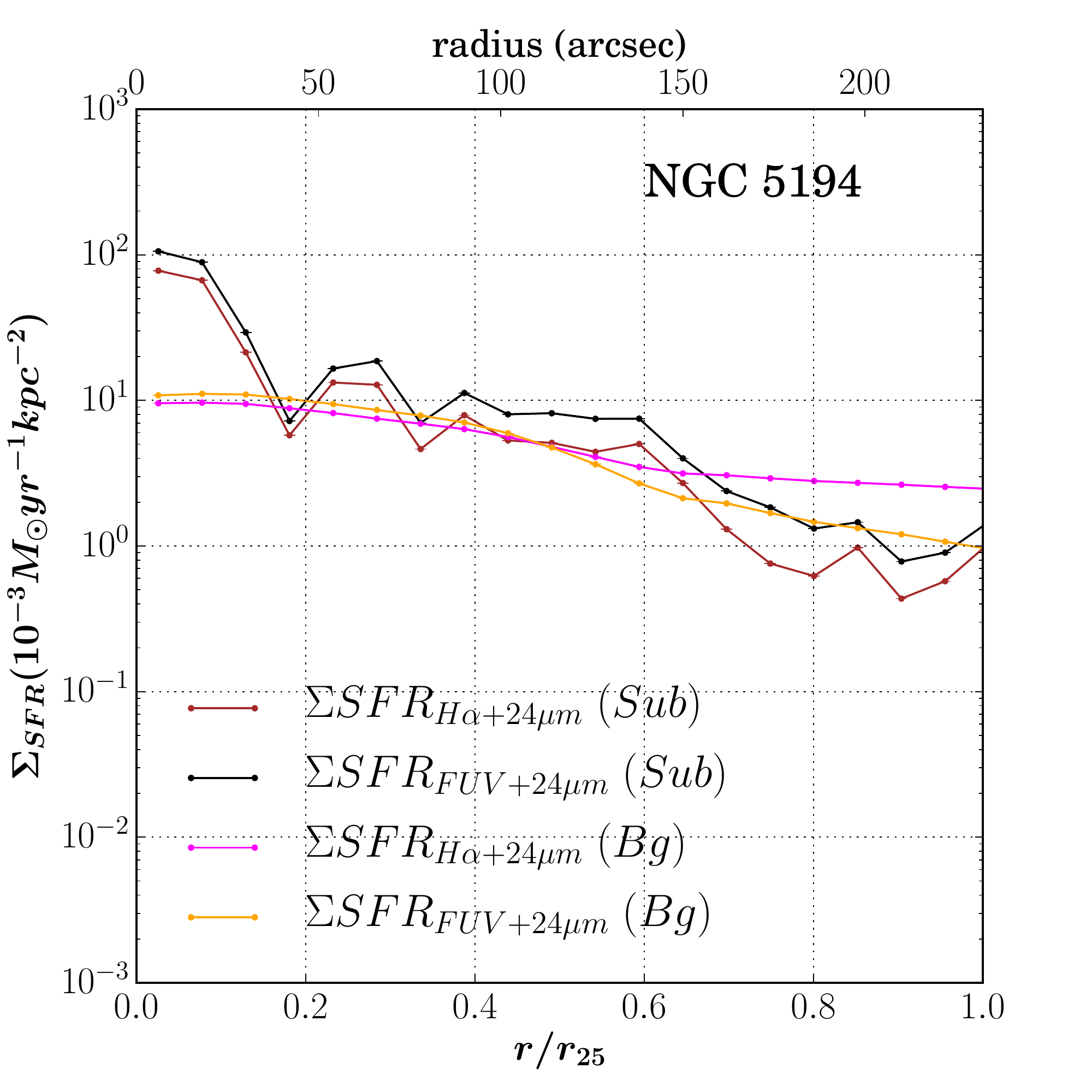}
	\includegraphics[width = 0.32\textwidth,trim={1.2cm 0.5cm 1.6cm 0.2cm},clip]{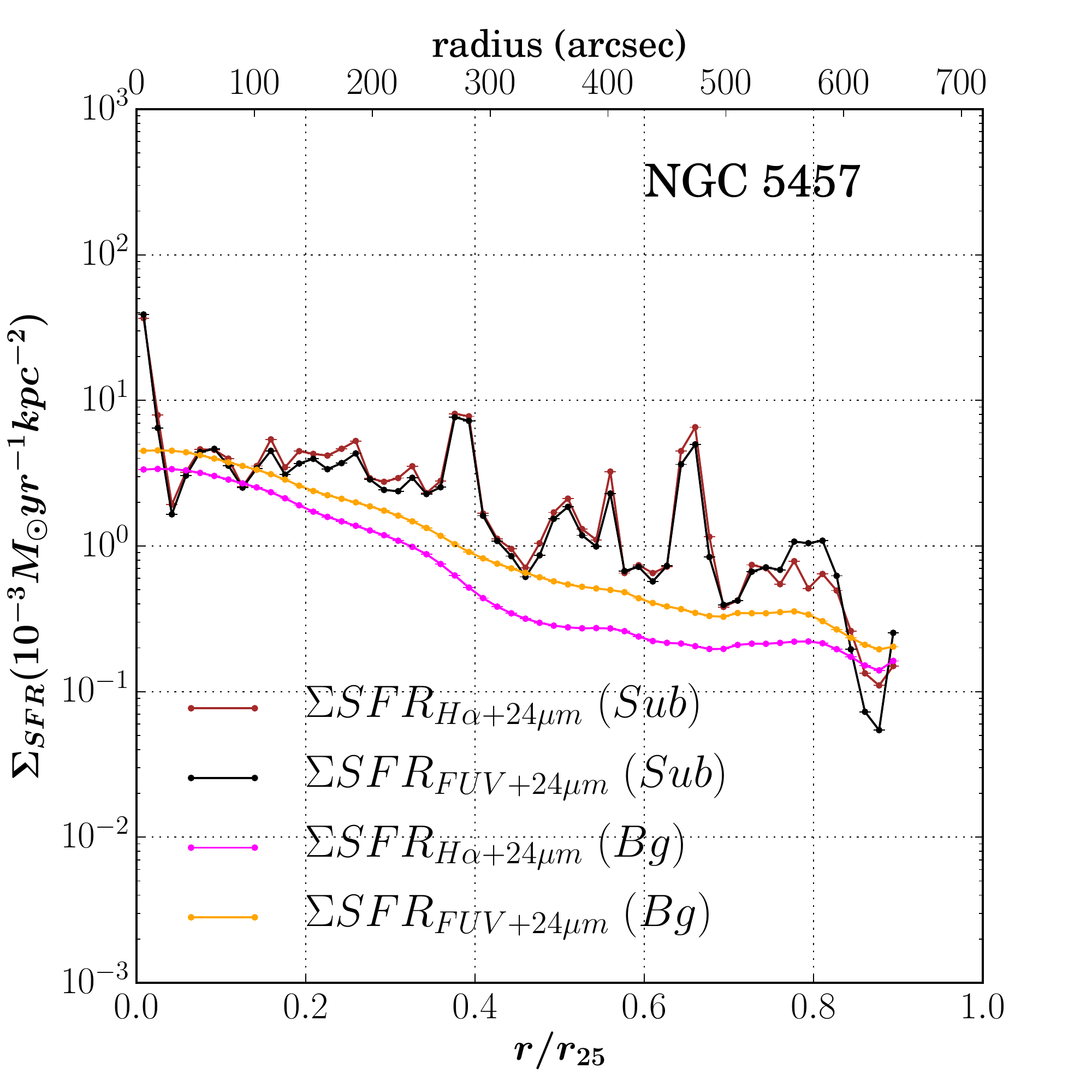}
	\includegraphics[width = 0.32\textwidth,trim={1.2cm 0.5cm 1.6cm 0.2cm},clip]{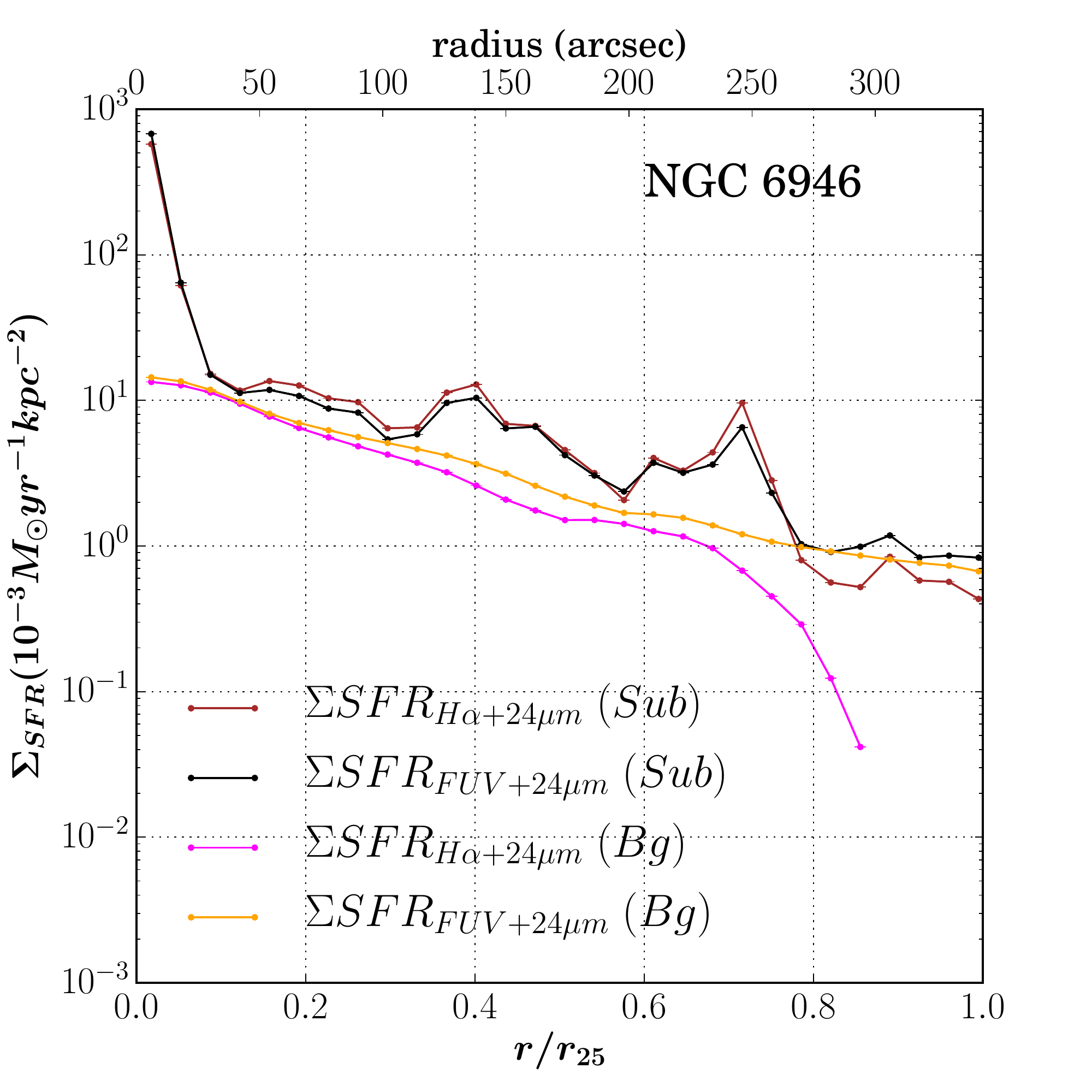}
	\caption{Radial profiles of $\Sigma_{SFR}$  for all nine galaxies in sample after removal of diffuse background. The x-axis presents the galactocentric radius normalised by r$_{25}$ (bottom) and in arcsecs (top). The y-axis presents $\Sigma_{SFR}$ in units of 10$^{-3}$ M$_{\odot}$ yr$^{-1}$ kpc$^{-2}$ (the scaling is performed for comparison with radial plots from original unsubtracted data in Fig. \ref{Figures: app original radial}). The brown and black curves denote $\Sigma_{SFR} (H\alpha + 24 \mu m)$ and $\Sigma_{SFR} (FUV + 24 \mu m)$ respectively  obtained from the data where diffuse background is subtracted from the SFR tracers. The magenta and orange curves denote  $\Sigma_{SFR} (H\alpha + 24 \mu m)$ and $\Sigma_{SFR} (FUV + 24 \mu m)$, respectively obtained from the diffuse background maps of the SFR tracers. }
	\label{Figure: sub radial}
\end{figure*}



\begin{figure*}
	\centering
	
	\includegraphics[width = 0.33\textwidth,trim={0.2cm 0cm 1.5cm 0cm},clip]{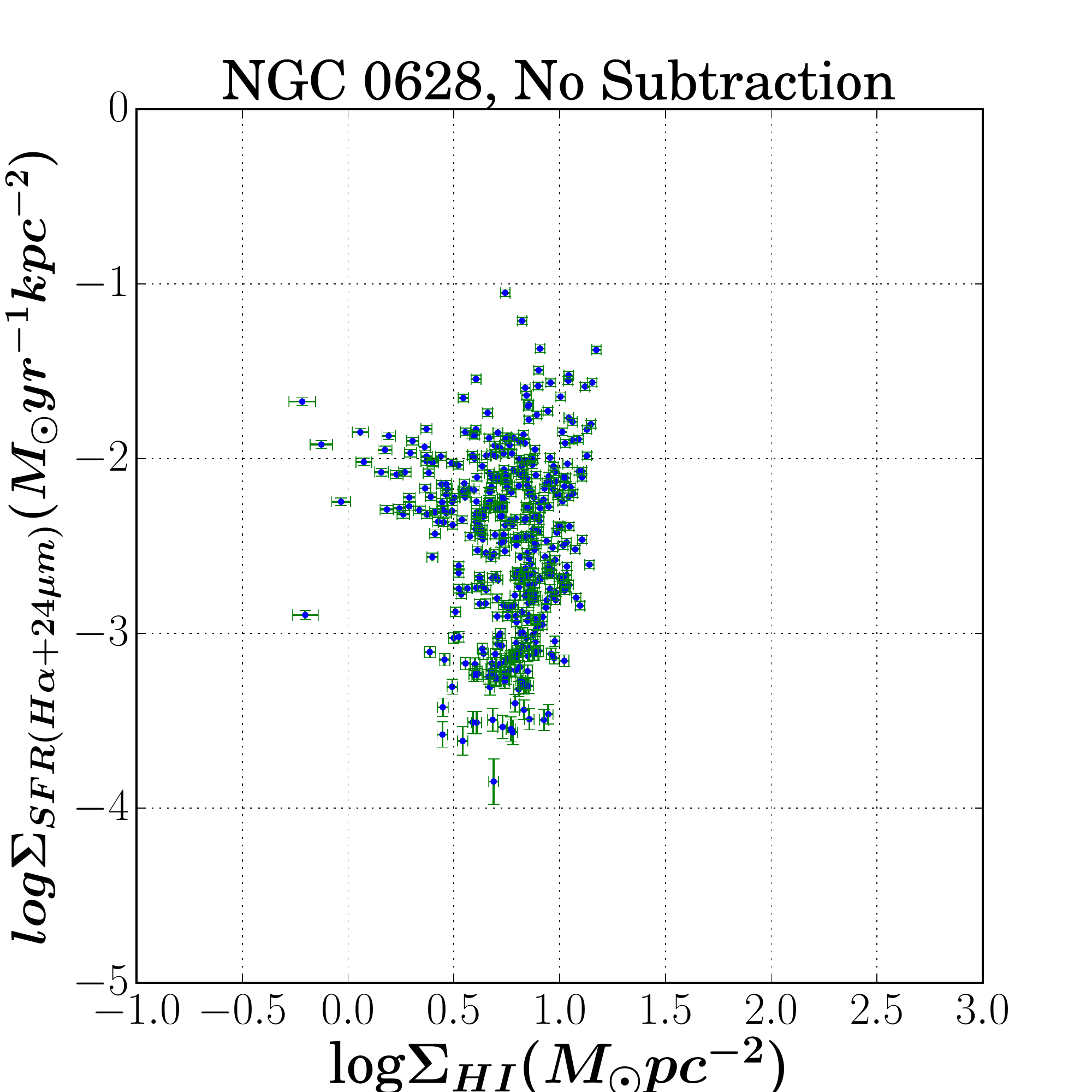}
	\includegraphics[width = 0.33\textwidth,trim={0.2cm 0cm 1.5cm 0cm},clip]{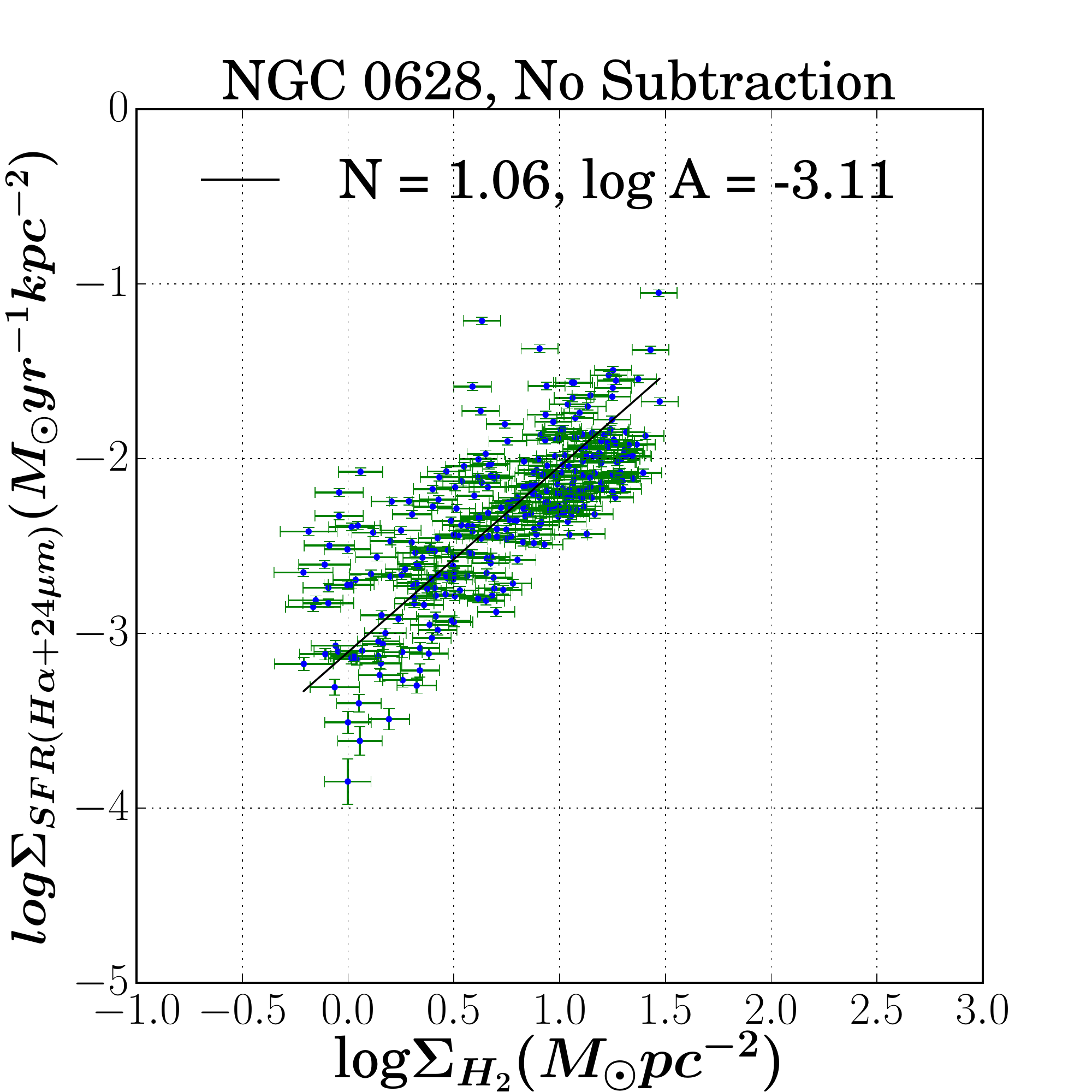}
	\includegraphics[width = 0.33\textwidth,trim={0.2cm 0cm 1.5cm 0cm},clip]{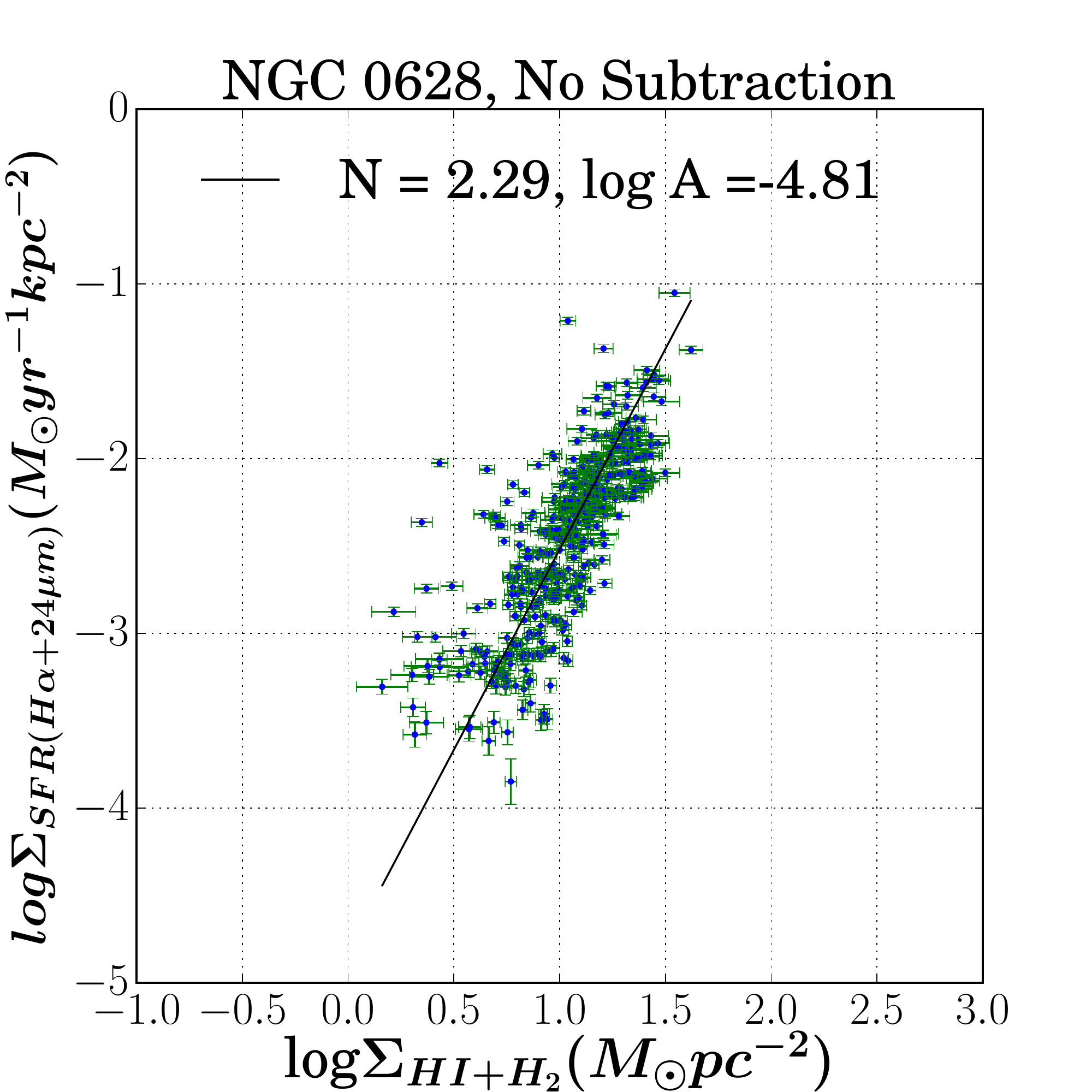}
	
	\includegraphics[width = 0.33\textwidth,trim={0.2cm 0cm 1.5cm 0cm},clip]{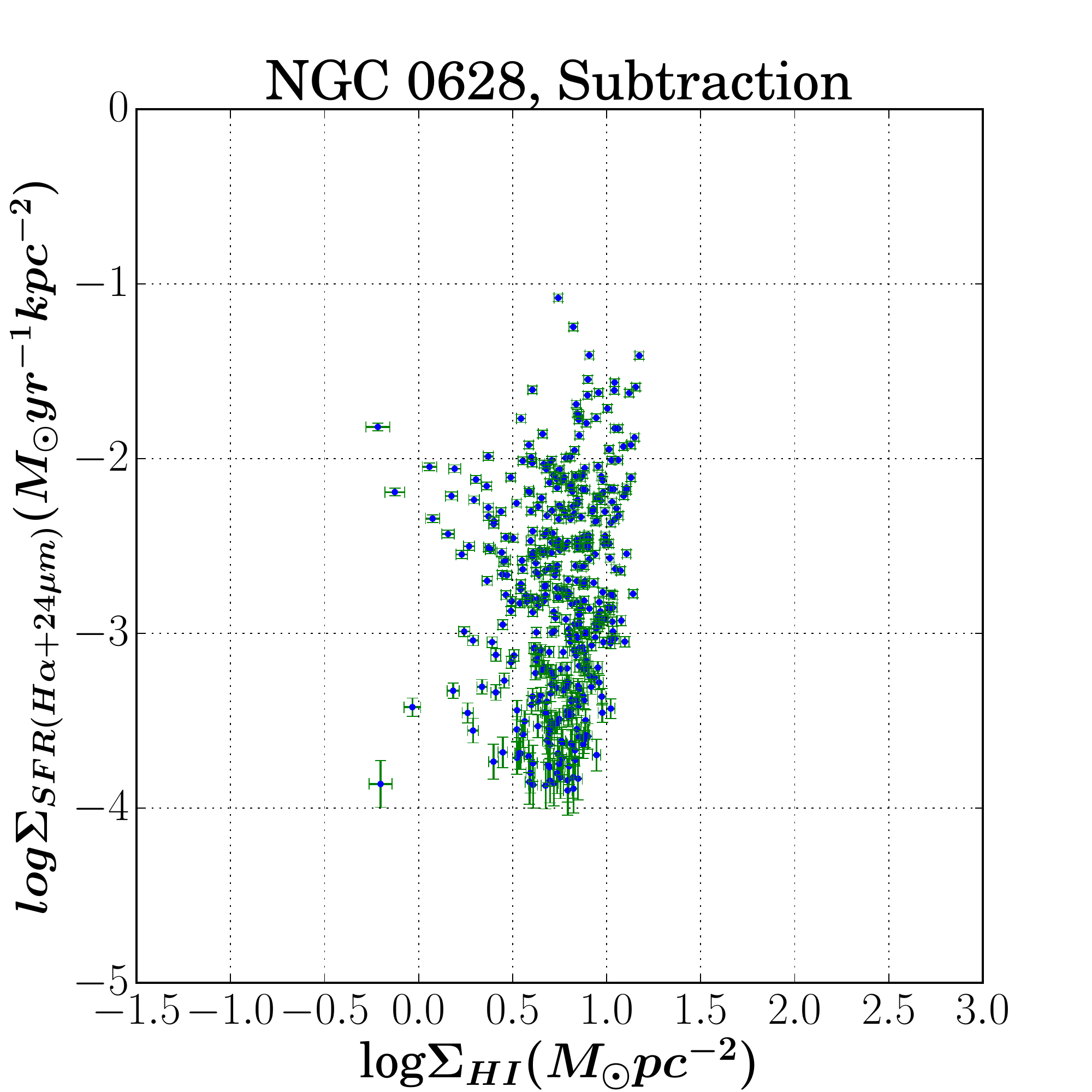}
	\includegraphics[width = 0.33\textwidth,trim={0.2cm 0cm 1.5cm 0cm},clip]{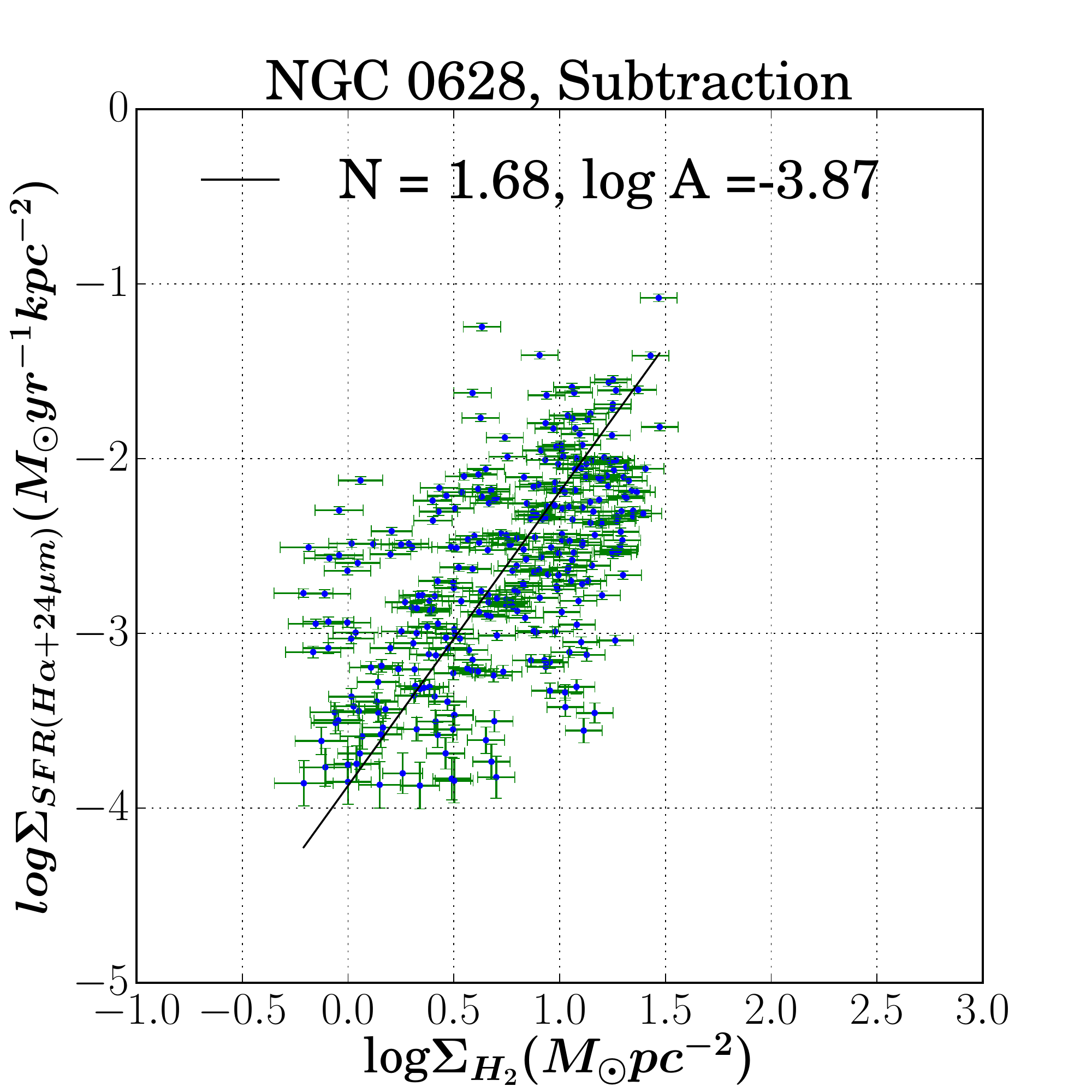}
	\includegraphics[width = 0.33\textwidth,trim={0.2cm 0cm 1.5cm 0cm},clip]{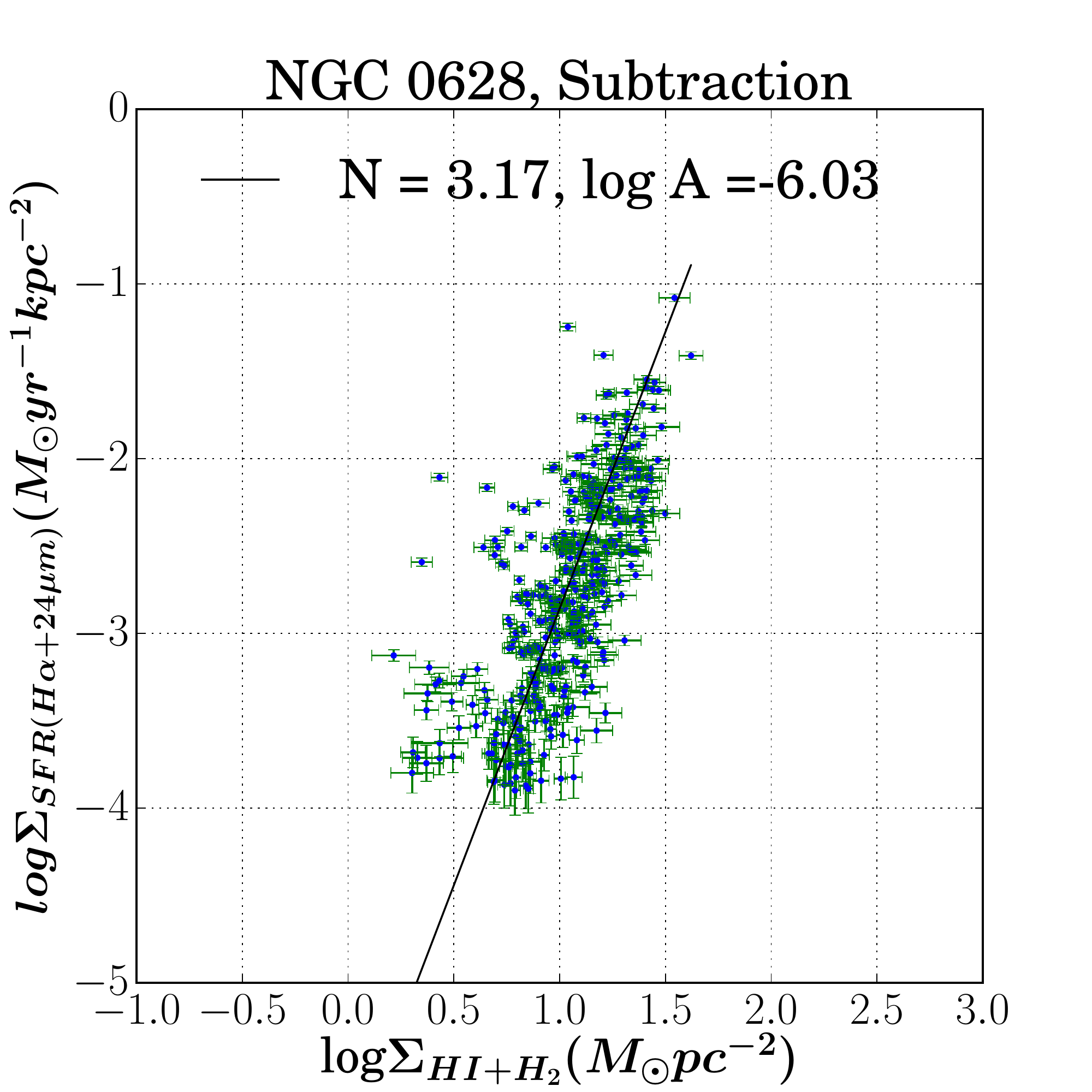}
	\caption{Effect of subtraction of diffuse background in SFR tracers on spatially-resolved ($\sim$560 pc) Schmidt relations in NGC 0628. Upper panel: analysis on the images with diffuse background (no subtraction). Lower panel:  analysis on the images where diffuse background is subtracted. Left panel: $\Sigma_{SFR}$ versus $\Sigma_{H \textsc{i}}$, middle panel: $\Sigma_{SFR}$ versus $\Sigma_{H_2}$. Right panel: $\Sigma_{SFR}$ versus $\Sigma_{H \textsc{i} + H_2}$. The error bars include the random error and systematic error on the flux measurements. The solid black line in each plot denotes the best-fit line to the data where equal weight is given to each data point on both axes. N and  A  denote the power-law index and the  star-formation efficiency. The slope of the $\Sigma_{SFR}-vs-\Sigma_{H_2}$ and $\Sigma_{SFR}-vs-\Sigma_{H \textsc{i} + H_2}$ increases after subtraction of diffuse background.  We note the saturation of H \textsc{i} at $\sim$10 M$_{\odot}$pc$^{-2}$. Here, we assumed a Kroupa IMF and adopted a constant X(CO) factor = 2.0$\times$10$^{20}$ cm$^{-2}$ (K km s$^{-1}$)$^{-1}$.}
	\label{Figure: NGC 0628}
\end{figure*}


\indent Various methods have been devised and employed to subtract the underlying
diffuse component of stellar and dust emission \citep{Calzetti2005, Kennicutt2007, Prescott2007, Blanc2009, Liu2011}. The methods based on statistics of the regions in galaxies seem to work better than a method with astrophyscial approach \citep{Liu2011}, in terms of the estimated fraction of diffuse component. This might be because the nature of diffuse background is not fully understood in all wavelengths. In this work, we adopt a novel statistical approach, making use of the \textit{Nebulosity~Filter}\footnote{ http://casu.ast.cam.ac.uk/publications/nebulosity-filter} software to subtract the diffuse background. This nebulosity filtering algorithm uses an iteratively clipped non-linear filter to separate different spatial scales in an image, in this case the diffuse background from the more compact H \textsc{ii} regions. The algorithm works by forming successive estimates of the diffuse background using clipped two-dimensional median filters. The difference between the smoothed image and the original is used to define regions for masking for subsequent application of the median filter. The size of the median filter defines the scale. For the analysed galaxies, we have adopted this scale as 50 arcsec or more, which is larger than the angular resolution of any dataset on which background subtraction is performed. After the final iteration the diffuse background estimate is further smoothed using a conventional boxcar two-dimension linear filter 1/3 of the scale of the median filter. This final step ensures the diffuse background image is free of the discrete steps that median filters are prone to delivering while maintaining the desired overall scale. We tested the results of using this software on a modelled galaxy (see Appendix \ref{appendix: background} for further details) to assess possible systematic biases.

\indent  Fig. \ref{Nebuliser Images} shows an example of this process for the two split components of the  FUV (left panel),  H$\alpha$ (middle panel) and 24 $\mu$m (right panel) images of NGC~0628. As found in Fig. 3 of  \citet{Liu2011}, our background images also show higher diffuse background in star-forming regions than elsewhere; however, the relative effect is stronger in fainter regions (Section \ref{section: radial}). These images highlight the complex spatially-varying nature of the diffuse background and the variation of this component in the different wavelength regions. To investigate the efficacy of background subtraction, we check the scaling relations between attenuation-corrected $\Sigma_{SFR}$ estimated from H$\alpha$ and FUV before and after subtraction of the diffuse background (Fig. \ref{Figure: scaling relation}). We find that the slope of the scaling relation changes by typically $\pm 0.01$  after the subtraction of the diffuse background, which is an insignificant change and indicates that the background subtraction has been done consistently in all the SFR tracers. This test has been done for each galaxy in the sample. A slight increase in scatter is observed because the background subtraction decreases the signal to noise (S/N) due to loss of signal from the removal of the diffuse background, whereas the noise remains essentially the same. The scale-free scatter in the log-log domain when applied to the linear signal, decreases by more than a factor of three on average, hence decreasing the scatter in the linear difference in the derived SFRs. 

\indent Adopting the same filtering scale length in the \textit{Nebulosity Filter} as found for the SFR tracers, we also experimented with removing a diffuse background component from the H\,\textsc{i} map. To check the efficacy of Nebulosity Filter on HI maps, we reduced VLA B-configuration data for one of the typical galaxies (NGC5055)
	in the sample, and created a moment 0 map with $\sim$6 arcsec
	resolution, which by construction should not contain much of the
	diffuse emission. Figure \ref{figure: HI nebulosity} shows that this image (right panel) is similar to the image obtained after
	processing the THINGS HI map of this galaxy with the \textit{Nebulosity
		Filter} (middle panel). The fraction of diffuse background estimated in the two cases
	are in good agreement with each other, further showing the
	robustness of the \textit{Nebulosity Filter} in separating diffuse background
	from the maps.

\indent The aperture-corrected fluxes (section \ref{flux extraction}) from the original and subtracted images are used to calculate the fraction of diffuse background in the H \textsc{ii} regions in H$\alpha$, FUV, MIPS 24 $\mu$m and H \textsc{i}.  Since our aim is to remove the diffuse background in order to trace `current' star formation, the fraction of diffuse background in the selected regions is a more relevant quantity than the diffuse background in the entire galaxy. The combined results for each galaxy in the sample are tabulated in Table \ref{diffuse}. On the entire images of the galaxies, we find a mean diffuse fraction of $\sim$34\% in H$\alpha$, $\sim$43\% in FUV, $\sim$37\% in 24$\mu$m, and $\sim$75\% in atomic gas. The diffuse fraction in SFR tracers is in agreement with previous studies ($\sim$30--50\% diffuse ionised gas (DIG) related to H$\alpha$ \citep{Ferguson1996}; 30--40\% related to 24$\mu$m in galactic centres and 20\% in  discs \citep{Verley2009}; at least a 40\% diffuse UV \citep{Liu2011}). The diffuse FUV fraction of 46\% for NGC 5194 and 58\% for NGC 3521 found in this work, are comparable to that (44\% for NGC 5194 and 56\% for NGC 3521) found by \citet{Liu2011}. We find a fraction of 36\% in 24$\mu$m image of NGC 5194, which is close to the range (15--34\%) given in \citet{Kennicutt2007}. We find a  H$\alpha$ diffuse fraction of $\sim$17\% in the central region of NGC 5194, which is mid-way between the values calculated by \citet[][11\%]{Blanc2009} and \citet[][32\%]{Liu2011}. We note here that the diffuse fraction of H$\alpha$ is more difficult to estimate accurately because of the necessity of scaling and subtracting the appropriate R-band continuum image from the H$\alpha$ image. In all cases the fraction of diffuse background in the SFR tracers is significantly less than that in the atomic gas.

\section{Results}
\label{results}

\subsection{Radial profiles}
\label{section: radial}
\indent We study the spatial variation of diffuse background in SFR tracers in individual galaxies through their radial profiles, which are created by averaging over the pixels of the  $\Sigma_{SFR}$ maps in elliptical annuli of constant width 12\arcsec centred on the galaxies. $\Sigma_{SFR}$ maps are created by combining the optical/FUV and IR using formulae \ref{sfr_halpha} \& \ref{sfr_fuv} and normalised by the de-projected area of each pixel. Fig. \ref{Figure: sub radial}  presents radial profiles of $\Sigma_{SFR}$ of the diffuse-background-subtracted data  and $\Sigma_{SFR}$ of diffuse-background, showing the relative contribution of the two components (potential currently star-forming regions and diffuse background) in the star-forming galaxies. We find that the effect of the diffuse background in the faint outer star-forming regions is more significant than in the bright central star-forming region. For comparison, we created radial profiles for the original unsubtracted SFR and gas data presented in Appendix \ref{appendix: radial}. This comparison shows that $\Sigma_{SFR}$ even after subtraction of the diffuse background, follows a similar pattern to the molecular gas profile, highlighting the importance of molecular gas in star formation in the central regions of galaxies.

 \subsection{Effect of subtraction of diffuse background in spatially-resolved Schmidt relations}
 \label{section effect}


 \begin{figure}
 	\centering
 	\includegraphics[width=0.33\textwidth,trim={0.2cm 0cm 1.5cm 0cm},clip]{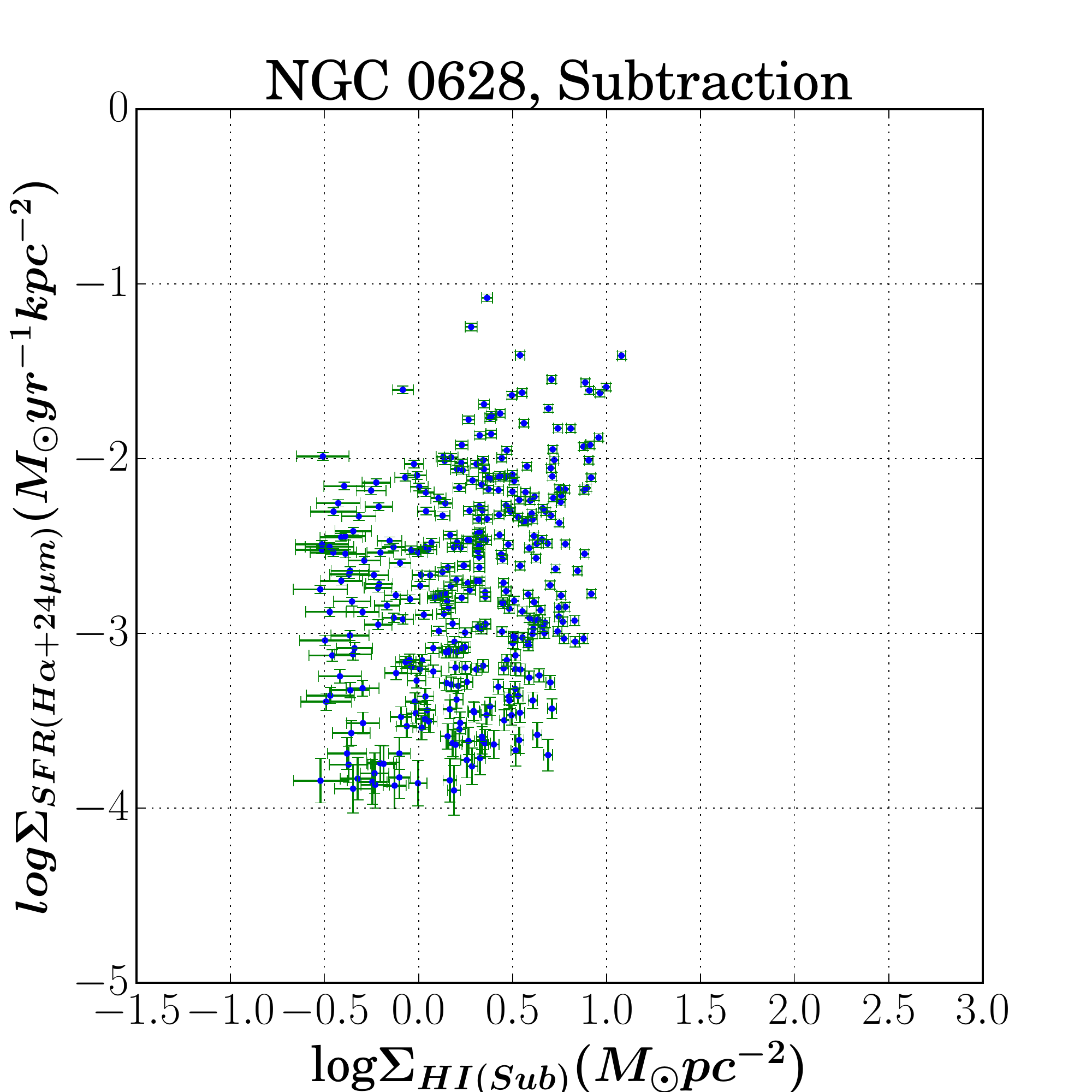}
 	
 	\includegraphics[width=0.33\textwidth,trim={0.2cm 0cm 1.5cm 0cm},clip]{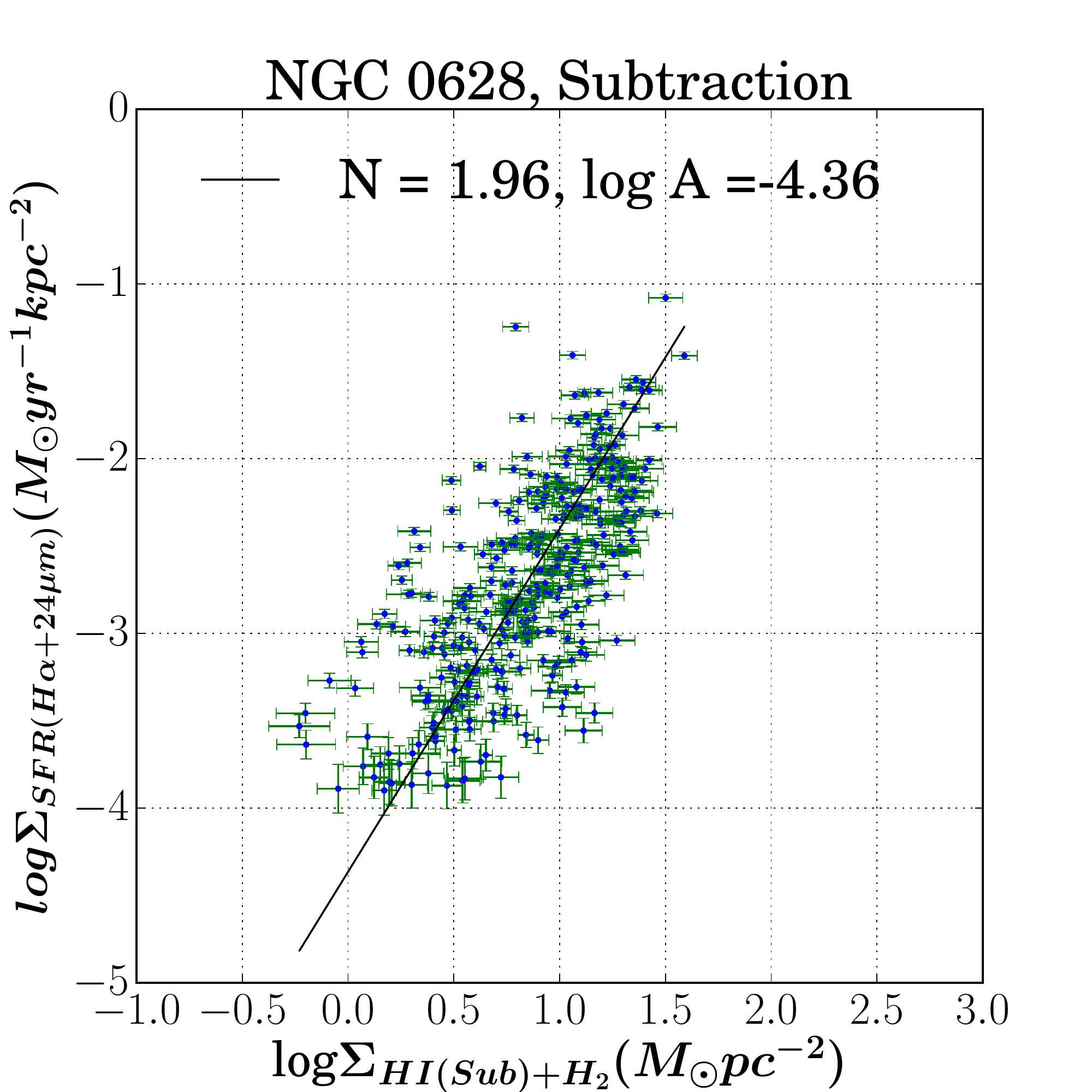}
 	\caption{Effect of subtraction of diffuse background in atomic gas and SFR tracers on spatially-resolved Schmidt relations in NGC 0628. Upper panel: $\Sigma_{SFR}$ versus $\Sigma_{H \textsc{i}}$. Lower panel: $\Sigma_{SFR}$ versus $\Sigma_{H \textsc{i} + H_2}$.  See caption of Fig. \ref{Figure: NGC 0628} for details on legends.}
 	\label{Figure: diffuse HI}
 \end{figure}

 \begin{figure}
	\centering
	\includegraphics[width = 0.45\textwidth]{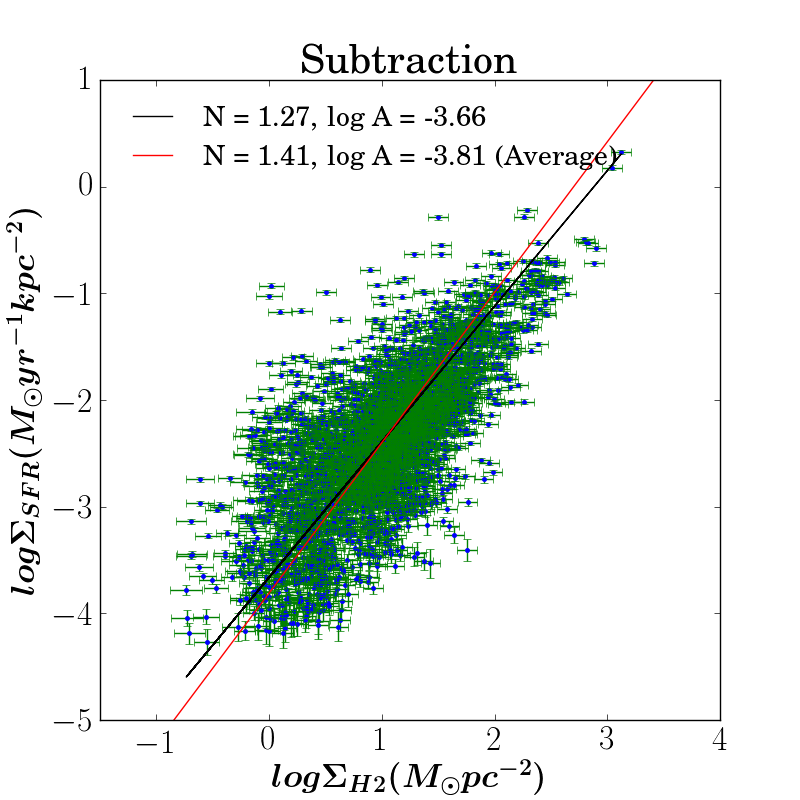}
	\caption{Spatially-resolved molecular gas Schmidt relation ($\Sigma_{SFR}$ vs $\Sigma_{H_2}$) for all sample galaxies, where the diffuse background has been subtracted from the SFR tracers. The red line is the fit obtained by averaging N  (power-law indices) and log A (star-formation efficiencies) from individual galaxies (Table \ref{table:KS Ha}). The solid black line shows the best-fit to the spatially-resolved data for all galaxies. A Kroupa IMF and a constant X(CO) factor = 2.0$\times$10$^{20}$ cm$^{-2}$ (K km s$^{-1}$)$^{-1}$ have been assumed.}
	\label{Figure: Schmidt H2}
\end{figure}


 \indent We fitted $\Sigma_{SFR}$ versus $\Sigma_{gas}$ in logarithmic space: $log(\Sigma_{SFR}) = N log(\Sigma_{gas}) + log  A$, using the orthogonal distance regression (ODR) algorithm \citep[see reference for ODR in][]{Virtanen2019}, where maximum likelihood estimation of parameters is performed assuming that the distribution of errors on both axes is Gaussian. A S/N-cut of 3 was performed on all fits. In what follows we make use of unweighted fits for analysing the Schmidt relation variation. Using unweighted fits is standard practice in Schmidt relation studies \citep[see for example][]{Bigiel2008, Liu2011}. Systematic uncertainties as large as 30-50\% are present on both axes related to the SFR calibration and the CO-to-$H_2$ conversion \citep{Leroy2013}, which explains the adoption of unweighted fits giving equal weight to each point on both axes. Problems with fitting  power-laws in star-formation studies is a well-known issue \citep{Blanc2009, Leroy2013, Casasola2015}. In Appendix \ref{appendix: line-fitting}, we revisit this problem by comparing different fits including unweighted fits and maximum likelihood fits weighted by uncertainties calculated as in section \ref{section: error}. Including systematic calibrations errors of 20\% on both axes is necessary to yield data models with reduced $\chi^2$ values of order unity. With this additional calibration error the difference between weighted fit and unweighted fit parameters lies within the measurement error.
 
 \indent To study the effect of inclusion and removal of diffuse background, we explored the spatially-resolved Schmidt relation on the original and diffuse-background-subtracted data for each individual galaxy. Fig. \ref{Figure: NGC 0628} and Fig. \ref{Figure: diffuse HI} show this analysis for NGC 0628. The study on rest of the galaxies are shown in Appendix \ref{appendix: individual galaxies}. We have performed the entire analysis using both SFR(H$\alpha$+24$\mu$m) and SFR(FUV + 24$\mu$m), though for space considerations we only show the Schmidt relations where SFR is determined using  H$\alpha$. The use of H$\alpha$ is justified by its ability to trace the current SFR due to its short age sensitivity ($\sim$ 5 Myr), which is thus more relevant for the present analysis. However, the SFR estimated from H$\alpha$ has been shown to deviate significantly from the true SFR accompanied by large scatter at lower fluxes/SFRs due to stochastic fluctuations. Given that the FUV is less sensitive to such effects, its use may be preferable as an SFR tracer \citep{daSilva2014}. To investigate this we analysed the sample using FUV as a SFR tracer and present the best-fit parameters for all Schmidt relations derived using H$\alpha$ as well as FUV as SFR tracers. Tables \ref{table:KS Ha} and \ref{table:KS FUV} present the best-fit parameters (N, log A) for the original and subtracted data individually for each galaxy obtained from H$\alpha$ and FUV, respectively. Table \ref{table:KS} compares the parameters of Schmidt relation obtained using H$\alpha$ and FUV from the combined data of all galaxies. The uncertainty on N is $\sim$ 0.1 and log A is $\sim$ 0.1 dex for both H$\alpha$ and FUV, and are obtained from the dispersion of values of these parameters from galaxy to galaxy. Thus, the parameters obtained from the two tracers agree with each other within the uncertainties, which adds confidence to our analysis. 

 \indent We note here that the value of log A in Tables \ref{table:KS Ha}, \ref{table:KS FUV} and \ref{table:KS} has no practical meaning as it is representative of data points lying at the far lower-end of SFR and gas density. We have presented log A for the sake of comparison with global relation presented in Section \ref{section: comparison global}. It would be more meaningful to compare log A in different galaxies at gas surface density of $\sim$ 10 M$_{\odot}$pc$^{-2}$, where the majority of the data points lie.

\begin{table*}[]
	\centering
	\caption{Summary of best-fit parameters of Schmidt relation for individual galaxies. Here SFRs are calculated from a combination of H$\alpha$ and 24$\rm\mu$m (Equation \ref{sfr_halpha}).}
	\label{table:KS Ha}
	\resizebox{\textwidth}{!}{%
	\begin{tabular}{ccccccccclccccc}
		\toprule
		Galaxy   & \multicolumn{5}{c}{No subtraction}                     &  & \multicolumn{5}{c}{Subtraction (SFR only)}     &  & \multicolumn{2}{c}{Subtraction (SFR \& H \textsc{i})} \\
		\cmidrule(r){2-6} \cmidrule(lr){8-12} \cmidrule(l){14-15}
		& \multicolumn{2}{c}{H2} &   & \multicolumn{2}{c}{HI+H2} &  & \multicolumn{2}{c}{H2} &    & HI+H2   &        &  & \multicolumn{2}{c}{HI+H2}                  \\
		\cmidrule(r){2-3} \cmidrule(lr){5-6} \cmidrule(lr){8-9} \cmidrule(lr){11-12} \cmidrule(l){14-15}
		& log A      & N         &   & log A        & N          &  & log A      & N         &    & log A   & N      &  & log A                & N                   \\
		NGC 0628 & -3.11      & 1.06      &   & -4.81        & 2.29       &  & -3.87      & 1.68      &    & -6.03   & 3.17   &  & -4.36                & 1.96                \\
		NGC 3184 & -3.18      & 0.85      &   & -4.29        & 1.55       &  & -3.93      & 1.32      &    & -5.49   & 2.31   &  & -4.33                & 1.56                \\
		NGC 3351 & -3.36      & 1.29      &   & -4.22        & 2.02       &  & -4.13      & 1.8       &    & -5.29   & 2.77   &  & -4.39                & 2.03                \\
		NGC 3521 & -2.89      & 0.84      &   & -5.25        & 2.54       &  & -3.22      & 1.1       &    & -6.14   & 3.18   &  & -3.9                 & 1.66                \\
		NGC 4736 & -2.99      & 0.99      &   & -3.43        & 1.19       &  & -4.13      & 1.7       &    & -5.02   & 2.15   &  & -4.26                & 1.75                \\
		NGC 5055 & -3.10      & 0.85      &   & -4.01        & 1.38       &  & -3.9       & 1.25      &    & -5.12   & 1.97   &  & -4.15                & 1.4                \\
		NGC 5194 & -3.18      & 0.97      &   & -3.70        & 1.21       &  & -4.32      & 1.52      &    & -5.17   & 1.92   &  & -4.63                & 1.66                \\
		NGC 5457 & -2.64      & 0.66      &   & -3.98        & 1.63       &  & -3.24      & 1.16      &    & -4.78   & 2.21   &  & -4.03                & 1.70                \\
		NGC 6946 & -2.95      & 0.86      &   & -3.66        & 1.23       &  & -3.55      & 1.15      &    & -4.51   & 1.65   &  & -3.94                & 1.35        \\
		   \bottomrule    
	\end{tabular}%
	}
\end{table*}

\begin{table*}[]
		\centering
	\caption{Summary of Schmidt relation results for individual galaxies. Here SFRs are calculated from a combination of FUV and 24$\rm\mu$m (Equation \ref{sfr_fuv}).}
	\label{table:KS FUV}
	\resizebox{\textwidth}{!}{%
	\begin{tabular}{ccccccccccccccc}
		\toprule
		Galaxy & \multicolumn{5}{c}{No subtraction}                     &  & \multicolumn{5}{c}{Subtraction (SFR only)}              &  & \multicolumn{2}{c}{Subtraction (SFR \& H \textsc{i})} \\
		\cmidrule(r){2-6} \cmidrule(lr){8-12} \cmidrule(l){14-15}
		& \multicolumn{2}{c}{H2} &   & \multicolumn{2}{c}{HI+H2} &  & \multicolumn{2}{c}{H2} &    & \multicolumn{2}{c}{HI+H2} &  & \multicolumn{2}{c}{HI+H2}                  \\
		\cmidrule(r){2-3} \cmidrule(lr){5-6} \cmidrule(lr){8-9} \cmidrule(lr){11-12} \cmidrule(l){14-15}
		& log A      & N         &   & log A        & N          &  & log A      & N         &    & log A        & N          &  & log A                & N                   \\
		NGC 0628        & -3.06      & 1.04      &   & -4.78        & 2.3        &  & -3.86      & 1.67      &    & -5.91        & 3.08       &  & -4.32                & 1.93                \\
		NGC 3184        & -3.05      & 0.82      &   & -4.14        & 1.52       &  & -3.63      & 1.17      &    & -5.08        & 2.09       &  & -4.03                & 1.42                \\
		NGC 3351        & -3.29      & 1.27      &   & -4.11        & 1.93       &  & -4.07      & 1.82      &    & -5.07        & 2.59       &  & -4.25                & 1.93                \\
		NGC 3521        & -2.93      & 0.85      &   & -5.33        & 2.57       &  & -3.41      & 1.25      &    & -6.76        & 3.63       &  & -4.23                & 1.95                \\
		NGC4736         & -3.03      & 1.05      &   & -3.51        & 1.27       &  & -4.00      & 1.67      &    & -4.81        & 2.06       &  & -4.1                 & 1.69                \\
		NGC 5055        & -3.13      & 0.88      &   & -4.11        & 1.46       &  & -4.11      & 1.38      &    & -5.47        & 2.18       &  & -4.38                & 1.54                \\
		NGC 5194        & -2.92      & 0.85      &   & -3.39        & 1.07       &  & -4.24      & 1.48      &    & -5.05        & 1.85       &  & -4.56                & 1.62                \\
		NGC 5457        & -2.51      & 0.56      &   & -3.77        & 1.48       &  & -3.07      & 0.97      &    & -4.71        & 2.15       &  & -3.93                & 1.61                \\
		NGC 6946        & -2.92      & 0.86      &   & -3.64        & 1.22       &  & -3.13      & 0.91      &    & -3.95        & 1.34       &  & -3.51                & 1.11    \\
		\bottomrule           
	\end{tabular}%
}
\end{table*}

\begin{table*}[]
	\caption{Comparison of parameters of Schmidt relation obtained by using different SFR tracers (H$\alpha$ and FUV) by fitting data points from all galaxies}
	\label{table:KS}
	\centering
	\begin{tabular}{@{}ccccccccc@{}}
		\toprule
		Tracer & \multicolumn{5}{c}{Subtraction (SFR only)}       &  & \multicolumn{2}{c}{Subtraction (SFR \& H \textsc{i})} \\ \cmidrule(r){2-6} \cmidrule(r){8-9}
		& \multicolumn{2}{c}{H$_2$} && \multicolumn{2}{c}{H\textsc{i}+H$_2$} && \multicolumn{2}{c}{H \textsc{i}+H$_2$}                   \\ \cmidrule(r){2-3} \cmidrule(r){5-6} \cmidrule(r){8-9}
		& log A      & N         && log A        & N          && log A                 & N                   \\ 
		Ha + a24$\mu$m    & -3.66      & 1.27    & & -4.82        & 1.93       && -4.01                 & 1.46                \\ 
		FUV + b24$\mu$m   & -3.63      & 1.26     & & -4.78        & 1.92       && -3.99                 & 1.46                \\ \bottomrule
	\end{tabular}
\tablefoot{a and b indicate the attenuation-correction coefficients for H$\alpha$ and FUV tracer given in Equations \ref{Halpha_corr} and \ref{FUVcorr}. The typical uncertainties on derived parameter $\sim$ 0.1.}
\end{table*}


 \subsubsection{Molecular gas and star-formation}
 \label{molecular SF}
 \indent We find that the scaling relation between $\Sigma_{SFR}$ and $\Sigma_{H_2}$ is approximately linear where no subtraction of diffuse background is done whereas the relation is super-linear when diffuse background is subtracted from the SFR tracers. Considering the individual galaxies, NGC 5457 is worth mentioning. For this galaxy, we find the slope of the molecular gas Schmidt relation before subtraction of the diffuse background to be very low at 0.66 (Fig. \ref{NGC 5457}, top-middle panel), which falls in the sub-linear regime \citep{Shetty2013} and has been explained by bright diffuse CO emission \citep{Shetty2014b, Shetty2014a}. However, after subtraction of the diffuse background from the SFR tracers the Schmidt relation slope lies within the range found for the rest of the sample.
 
\indent Taking the average of slopes found for  each of the galaxies in the sample (Table \ref{table:KS Ha}), we find the slope to be 0.93 $\pm$ 0.06 before subtraction and 1.41 $\pm$ 0.09\footnote{error = standard deviation/$\sqrt{N-1}$ where N is the total number of galaxies.} after subtraction of diffuse background. We note that $\Sigma_{SFR}$ is estimated using H$\alpha$ and 24 $\mu$m (equation \ref{sfr_halpha}) in the above analysis. If we use instead the combination of FUV and 24 $\mu$m (equation \ref{sfr_fuv}) to estimate $\Sigma_{SFR}$, we find the slope to be  0.91 $\pm$ 0.07 before subtraction and 1.37 $\pm$ 0.11 after subtraction of diffuse background. 
 
 \indent Fig. \ref{Figure: Schmidt H2} shows the spatially-resolved Schmidt relation between $\Sigma_{SFR (H \alpha + 24\mu m)}$ and $\Sigma_{H_2}$ for all galaxies in the sample on a single plot. The red line denotes the fit obtained above by averaging the slopes and intercepts from individual galaxies (Table \ref{table:KS Ha}) which gives equal weight to each galaxy. This method was adopted for a comparison with the pixel-by-pixel analysis of \citet{Bigiel2008}. Each galaxy in principle should  be weighted by the number of star-forming regions used in each galaxy. This can also be achieved by deriving parameters directly from the best-fit line to the spatially-resolved data of all galaxies (solid black line in Fig. \ref{Figure: Schmidt H2}). 
 The best-fit slope is now 1.27 $\pm$ 0.1 (compared to average slope of 1.41 reported earlier) with a large {\it rms} scatter (0.34 dex). The formal fitting error in the slope is however small due to the large number of points (n$\sim$3000) used in the fit. We note here that owing to the large systematic uncertainties on the determination of quantities on each axes, the formal fitting error can not reflect the true error on the estimated best-fit parameters. Hence, we estimate the uncertainties from  the dispersion in the best-fit parameters (N and log A) obtained for each galaxy.

 \subsubsection{H\,\textsc{i} Saturation in all galaxies}
 \indent We do not find any correlation between $\Sigma_{SFR}$ and $\Sigma_{H\,\textsc{i}}$ before or after subtraction of the diffuse background in the SFR tracers. Prior to subtraction of the diffuse H \textsc{i} background, saturation of H\,\textsc{i} is observed around $\sim$ 10$^{1.5}$ M$_{\odot}$ pc$^{-2}$ (leftmost panels in Figs. \ref{Figure: NGC 0628} and \ref{NGC 3184}--\ref{NGC 6946}). After subtraction, the saturation of $\Sigma_{H\,\textsc{i}}$ is still observed but at a lower value of $\sim$10 M$_{\odot}$ pc$^{-2}$ (Fig \ref{Figure: diffuse HI}, upper panel and Fig. \ref{HI sub}).

 \subsubsection{Total gas and star-formation}

 \indent The relation between $\Sigma_{SFR}$ and $\Sigma_{H \textsc{i} + H_2}$ is always super-linear irrespective of the background subtraction in SFR tracers (rightmost panels in Figs. \ref{Figure: NGC 0628}, \ref{NGC 3184}--\ref{NGC 6946}). When diffuse background is removed from the SFR tracers , the slope steepens and varies from $\sim$1.65--3.18 depending on the relative quantity of $\Sigma_{H \textsc{i}}$ and $\Sigma_{H_2}$ (bottom-right panels in Figs. \ref{Figure: NGC 0628}, \ref{NGC 3184}--\ref{NGC 6946}). In the majority of galaxies in the sample (e.g. NGC3184 \& NGC 6946),
 	molecular gas surface density extends over two orders of magnitude
 	whereas the atomic gas surface density saturates around 3--10 M$_{\odot}$pc$^{-2}$. The
 	dominance of molecular gas typically at the higher end of total gas
 	surface density leads to a flatter slope in most galaxies compared to
 	galaxies like NGC 0628 and NGC 3521 where neither gas component
 	significantly dominates the higher end.

 \indent When diffuse background is removed from SFR tracers as well as atomic gas data, the slope of  $\Sigma_{SFR}$ and $\Sigma_{H \textsc{i} + H_2}$ varies from $\sim$1.3--2.0 depending on the galaxy (Figs. \ref{Figure: diffuse HI}, lower panel and \ref{total sub}). The range of slope is significantly lower than the case with no
 background subtraction ($\sim$1.19--2.54) or the case where the background is subtracted only from SFR tracers ($\sim$1.65--3.18).

\begin{figure*}
 	\centering
 	\includegraphics[width=0.45\textwidth]{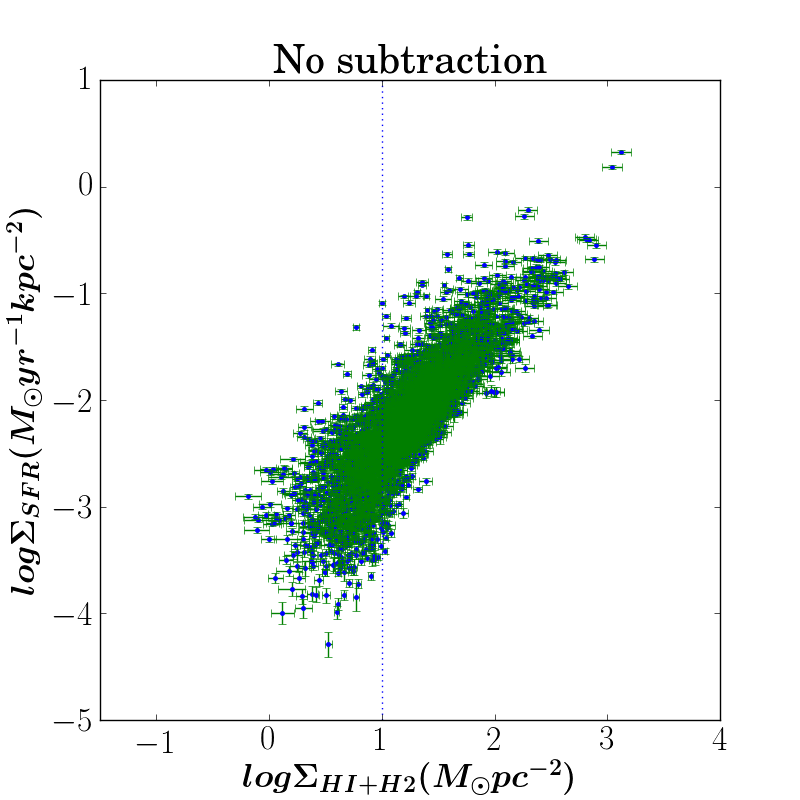}
 	\includegraphics[width=0.45\textwidth]{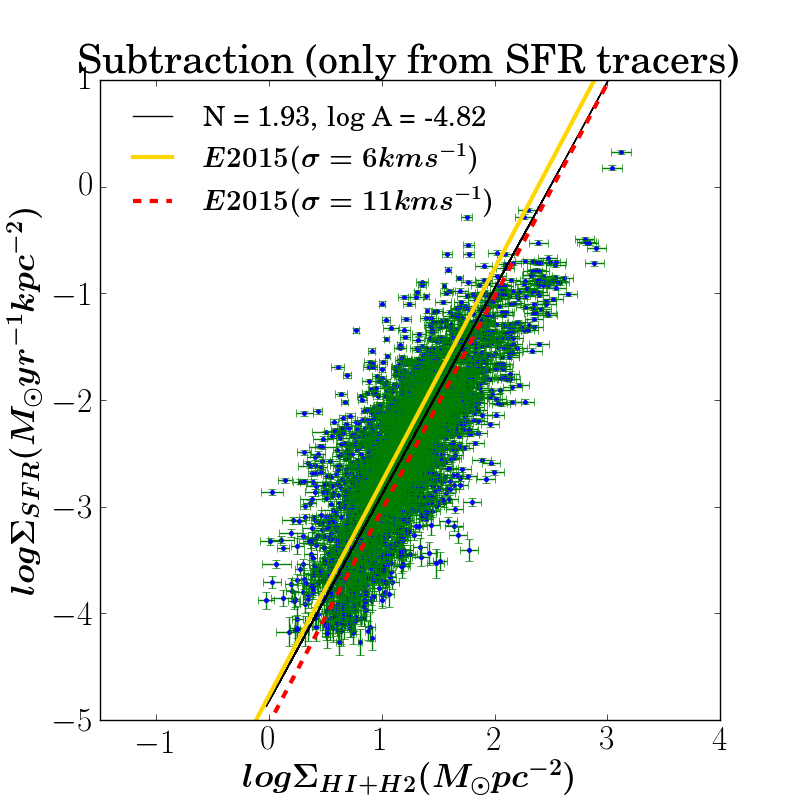}
 	\includegraphics[width=0.45\textwidth]{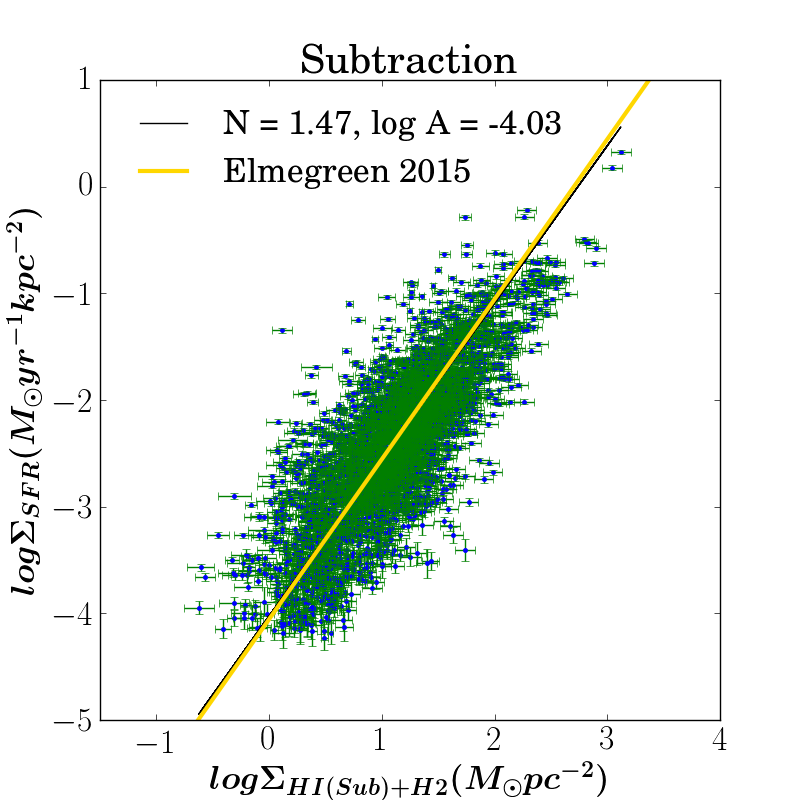}
 	\includegraphics[width=0.45\textwidth]{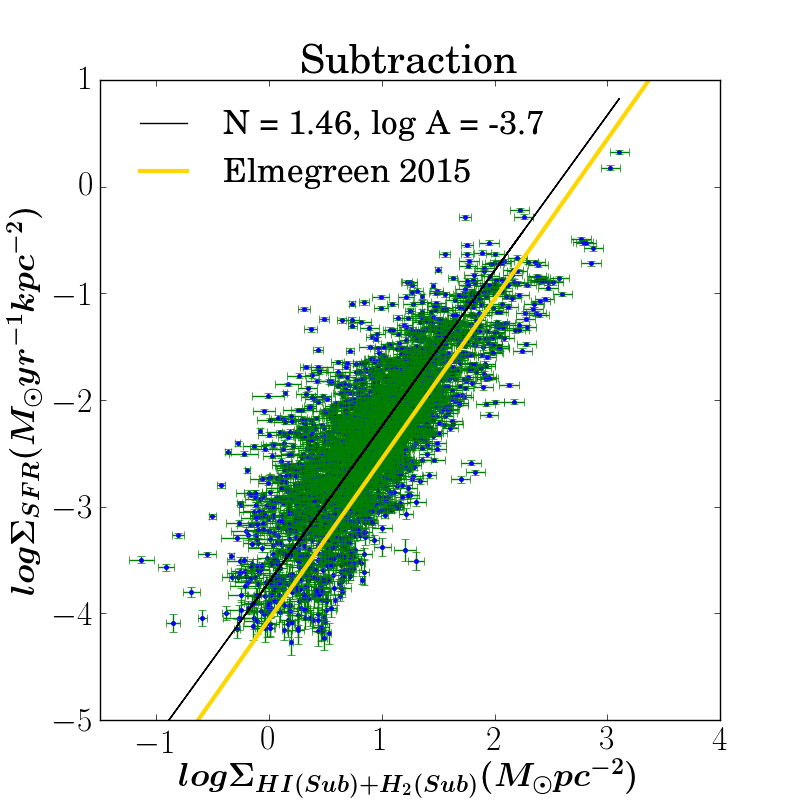}
 	\caption{Spatially-resolved total gas Schmidt relation ($\Sigma_{SFR}$ and $\Sigma_{ H \textsc{i} + H_2}$).  In each panel, blue dots with green error bars are the spatially-resolved data, and the solid black line shows the best-fit to the spatially-resolved data for all galaxies. We have assumed a Kroupa IMF and adopted a constant X(CO) factor = 2.0$\times$10$^{20}$ cm$^{-2}$ (K km s$^{-1}$)$^{-1}$. Upper-left panel: No subtraction of diffuse background is done. The vertical dotted line corresponds to 10 M$_{\odot}$ pc$^{-2}$ around which atomic gas surface density saturates. Upper-right panel: The diffuse background is subtracted from the SFR tracers. The solid yellow line and the dashed red line denote star-formation relation from \citealt{Elmegreen2015} (E2015) and correspond to velocity dispersion ($\sigma$) of 6 km s$^{-1}$ and 11 km s$^{-1}$ respectively. Lower-left panel: The diffuse background is subtracted from the SFR tracers as well as from H I. Lower-right panel: The diffuse background is subtracted from the SFR tracers, HI and CO. The solid yellow solid line on lower panels corresponds to the star-formation relation based on a dynamical model from \citet{Elmegreen2015} obtained at a constant scale height of $H$ = 100 pc. See text in Section \ref{section:comparison models} for further details on models.}
 	\label{Schmidt HI}
 \end{figure*}

 \indent Fig. \ref{Schmidt HI} presents the spatially-resolved total gas Schmidt relation for all galaxies in the sample. Upper-left panel shows the original data where diffuse background is not subtracted, where we observe a bimodal relation. Upper-right panel shows the Schmidt relation where diffuse background is subtracted from SFR tracers only, and the slope is super-linear.  Lower-left panel shows data where diffuse background is subtracted from SFR tracers as well as atomic gas data. Here, the best-fit line (solid black) to the spatially-resolved data (blue dots with green error bars) for all galaxies  results in a slope of $\sim$1.47 $\pm$ 0.08 with a scatter of 0.27. The reported error on the slope is the dispersion in slopes obtained for each galaxy in the sample. Thus, comparing the upper-left panel with other panels in Figure \ref{Schmidt HI}, we find that the removal of the diffuse background from the SFR tracers results in a continuous relation. In Figure \ref{Schmidt HI FUV}, we present the equivalent figures of Fig. \ref{Schmidt HI}, where SFR is obtained using FUV instead of H$\alpha$. As mentioned beforehand, change of SFR tracer do not result in any significant changes in the best-fit parameters or in general trend of spatially-resolved total gas Schmidt relation (see Table \ref{table:KS} for a comparison).
 
 
\section{Discussion}
\label{discussion}
\subsection{Comparison with models}
\label{section:comparison models}

\indent In Figure \ref{Schmidt HI}, we compare
	spatially-resolved total gas Schmidt relations with dynamical
	models of star formation where we take into account  disc flaring
	as discussed in \citet{Elmegreen2015}. The flaring of the gas  disc is
	parametrised by the scale height $H$ of the gas, which varies with
	surface gas density such that $ H \propto \sigma^2/\Sigma_{gas}$. where
	$\sigma$ is the gas velocity dispersion. For a uniform spherical gas
	cloud of density $\rho \propto \Sigma_{gas}/H$, star formation is
	related to the available gas reservoir by $\Sigma_{SFR} =
	\epsilon_{ff}\Sigma_{gas}/t_{ff}$ where the free-fall time for cloud
	collapse $t_{ff} = \sqrt{3\pi/32 G\rho}$ and $\epsilon_{ff}$ denotes
	the efficiency per unit free-fall time. For $\sigma = 6$ km s$^{-1}$
	\citep{Kennicutt1989} and $\epsilon_{ff} = 0.01$ \citep{Krumholz2007},
	star formation is then expected to follow:

\begin{equation} \rm \frac{\Sigma_{SFR}}{M_{\odot} kpc^{-2} yr^{-1}} =
1.7 \times 10^{-5} \Bigg( \frac{\Sigma_{gas}}{M_{\odot}
	pc^{-2}}\Bigg)^2, \label{Elmegreen2} \end{equation}

\noindent as derived by \citet{Elmegreen2015}. We
	overlay this relation (\ref{Elmegreen2}) as the solid yellow line on
	Figure \ref{Schmidt HI} (upper-right panel), which shows the total gas
	spatially-resolved Schmidt relation where the diffuse background has been
	subtracted from the SFR tracers only. The dashed red line corresponds to
	a star-formation relation where we assume $\sigma = $ 11 km s$^{-1}$
	based on more recent THINGS galaxies (\citealt{Leroy2009} and also see
	\citealt{Tamburro2009}). We find that the model with both values of
	$\sigma$ is in reasonable agreement with our data.

\indent For the total gas Schmidt relation where
	the diffuse background is subtracted from SFR tracers as well as atomic
	gas, we use the dynamical model with a constant scale height, meaning that,  there 
	is no flaring. In this case for a fixed scale height of $H$ = 100 pc and
	$\epsilon_{ff} = 0.01$, \citep{Elmegreen2015} derived the following
	star-formation relation:

\begin{equation} \rm \frac{\Sigma_{SFR}}{M_{\odot} kpc^{-2} yr^{-1}} =
8.8 \times 10^{-5} \Bigg( \frac{\Sigma_{gas}}{M_{\odot}
	pc^{-2}}\Bigg)^{1.5}.  \label{Elmegreen15} \end{equation}

\noindent The above equation (\ref{Elmegreen15})
	corresponds to the yellow line in Figure \ref{Schmidt HI} (lower-left
	panel) showing a very good agreement with the best-fit line to the
	spatially-resolved data. The agreement of the model with the
	background subtracted atomic data probably indicates that removing the
	diffuse background component using the $Nebulosity ~Filter$ mainly removes 
	the atomic gas from the vertically-extended regions in
	the flared outskirts of galaxies, effectively rendering the scale
	height to be constant throughout the sample of galaxies. While it is expected
	that the star-forming atomic gas is most likely the cold neutral medium 
	component of the ISM, it is however difficult to infer the nature of this
	component with the current data.

\indent We also tested the potential impact of diffuse CO emission by applying the
	$Nebulosity ~Filter$ on CO maps adopting the same parameters used
	for removing diffuse background in the SFR tracers. We then revisited the
	total gas Schmidt relation using the background subtracted SFR,
	atomic and molecular gas data and compared with predictions from the
	dynamical models discussed earlier (Figures \ref{Schmidt HI} and \ref{Schmidt HI FUV}, lower-right panels). It is perhaps not surprising that models
	incorporating flaring do not agree with these Schmidt relation parameters
	because the scale height of the molecular gas disc is significantly less
	than the neutral gas and also varies less with radius as found for the
	edge-on galaxy NGC 891 (\citealt{Yim2011} and also see \citealt{Barnes2012}). Hence, we do not
	expect any significant large scale diffuse CO emission that could be
	reliably picked up by our current analysis method.

\indent We note that we found a non-linear molecular
	gas Schmidt relation (Figure \ref{Figure: Schmidt H2}), where the diffuse
	background is subtracted only from SFR tracers. This result is also in
	agreement with \citet{Elmegreen2015}, who mentions that the molecular gas
	relation should not be exactly linear because of the change of
	$\epsilon_{ff}$ for different molecules considering the effects of
	molecule depletion on grains or change in gas sub-structure due to
	turbulence.

\indent In summary, a comparison with dynamical
	models incorporating flaring effects indicates that a diffuse
	component is present in SFR tracers as well as atomic gas, which does
	not contribute to current star formation. Though this model was
	originally meant to explain the origin of a power-law index of 1.5 for
	the inner regions of spiral galaxies and power-law index of 2 for the
	outskirts and for dwarf irregular galaxies, we find that accounting
	for vertically-extended diffuse background in atomic gas leads to a
	single power-law index of 1.5 for the inner as well as outer regions
	of spiral galaxies. Further works need to be done to test if
	accounting for diffuse background could lead to a power-law index of
	1.5 in dwarf galaxies as well.

\subsection{Comparison with Literature}

\indent The current work is a spatially-resolved star-formation study where diffuse background has been taken into account in both the SFR tracers as well as the atomic gas while studying the  Schmidt relations. Earlier spatially-resolved star-formation studies have either been through radial-profile analysis \citep{WongBlitz2002} or through point-by-point analysis \citep[e.g.][]{ Bigiel2008, Kennicutt2007, Liu2011}, where either diffuse background is not considered at all \citep[e.g.][]{Bigiel2008, WongBlitz2002} or it is accounted for in the SFR tracers \citep[e.g.][]{Kennicutt2007, Liu2011, Momose2013, Morokuma-Matsui2017} or in the molecular gas \citep{Rahman2011} but not in the atomic gas.

\indent The saturation of H\,\textsc{i} is a common result found in all works. From the aperture photometry analysis, we find the mean saturation value to be 16 $\pm$ 8 M$_{\odot}$ pc$^{-2}$ when diffuse background is not subtracted, which is consistent with the value of $\sim$ 25 M$_{\odot}$ pc$^{-2}$ derived by aperture photometry analysis of \citet{Kennicutt2007} on NGC 5194. Our radial profile analysis (Figure \ref{Figures: app original radial}) also shows a saturation below 10 M$_{\odot}$ pc$^{-2}$ in agreement with $\sim$ 9--10 M$_{\odot}$ pc$^{-2}$ derived by earlier radial profile analyses \citep{WongBlitz2002, Bigiel2008}. The difference in results can be attributed to the different approaches adopted in these  studies. Unlike aperture photometry,  pixel-by-pixel analysis or radial profile analysis does not take into account the flux lost in the adjacent pixels, which might be the cause of the lower level of saturation for H\,$\textsc{i}$ in these analyses. \citet{Roychowdhury2015} finds a power-law slope of 1.5 between $\Sigma_{SFR}$ and
	$\Sigma_{HI}$, which appears to be in contrast with this and previous studies mentioned above and
	may be simply because \citet{Roychowdhury2015} exclude regions detected in CO to concentrate mainly on regions dominated by HI, while other studies
	(including this one) do not make such selections. We discuss the agreement of results of \citet{Roychowdhury2015} and the current work later in Section \ref{section: comparison global}.

\indent Comparison of the power-law index of the molecular gas Schmidt relation found by pixel-by-pixel analysis \citep{Bigiel2008} and aperture photometry (upper panels in Figs. \ref{Figure: NGC 0628} and \ref{NGC 3184}--\ref{NGC 6946}) for individual galaxies in the sample shows a large variation in the case where diffuse background is not subtracted. However, results from the two works agree with each other statistically as found in Section \ref{molecular SF}. The slope of the molecular gas Schmidt relation reported by \citet{Bigiel2008} (0.96 $\pm$ 0.07) agrees within error with the slope we found here  by averaging the slopes found for individual galaxies (i.e. 0.93 $\pm$ 0.06 using H$\alpha$ and 0.91 $\pm$ 0.07 derived using FUV). Hence, both of these works show that before background subtraction, $\Sigma_{SFR}$ scales linearly with molecular gas surface density statistically, in agreement with various other studies \citep[e.g.][]{Liu2011, Schruba2011, Leroy2013}. 
The total gas Schmidt relation before subtraction of diffuse background shows a knee at 10 M$_{\odot}$pc$^{-2}$, which is in agreement with the results of \citet{Bigiel2008}. Compared to the bimodal relation of \citet{Bigiel2008}, we find a higher
dispersion at the lower end of the total gas Schmidt relation. This is
simply because \citet{Bigiel2008} only include the molecular gas data in the regime where $\Sigma_{H_2}$ $>$ 3 $M_{\odot}$ pc$^{-2}$ for HERACLES data and $\Sigma_{H_2}$ $>$ 10 $M_{\odot}$ pc$^{-2}$ for BIMA SONG data. Since H \textsc{i} saturates around this cut-off value for all the galaxies, the total gas Schmidt relation at the lower end is basically the atomic gas Schmidt relation. In our work, by using forced aperture photometry based on positions derived from the H$\alpha$ map, we can legitimately include regions with minimal ab-initio detection of molecular gas, particularly in the outer regions of galaxies. Our results are supported by those of \citet{Schruba2011}, whose stacking analysis allow them to trace the molecular gas surface density even in the outskirts of the galaxy. They report marginal CO detections which are as low as 0.1 M$_{\odot}$ pc$^{-2}$ equivalent to our S/N $>$ 3 cut. Hence our analysis shows that the Schmidt relations on sub-galactic scales do not depend on the method adopted - aperture photometry or pixel-by-pixel analysis.

\indent Some studies \citep{Heyer2004, Komugi2005, Momose2013} have found a super-linear molecular Schmidt relation even using data where 
diffuse background was not subtracted from the SFR tracers. However, all of these studies have used CO(1--0) data to trace the molecular gas, instead of CO(2--1) used in this work. The excitation for CO(2--1) is significantly affected by a slight change in the kinetic temperature and volume density of molecular gas, which in turn affects the power-law index of the Schmidt relation \citep{Momose2013}. The change of the power-law index resulting from using a different molecular gas tracer (for example, CO(1--0), CO(3--2), HCN(1--0)) has been observed before by other studies as well \citep{Narayanan2008, Bayet2009, Iono2009}.

\indent Subtraction of the diffuse background in SFR tracers naturally leads to an increase in the slope of the molecular gas Schmidt relation in all galaxies in the sample. This result is in agreement with \citet{Liu2011}, who reports that the subtraction of diffuse background in SFR tracers leads to a super-linear slope of the molecular gas Schmidt relation from their analysis on two galaxies (NGC 5194 and NGC 3521). Adopting the same method of subtraction as \citet{Liu2011}, \citet{Momose2013} and \citet{Morokuma-Matsui2017}  found the steepening of slope in agreement with our study. \citet{Rahman2011} also shows that the spatially-resolved molecular Schmidt relation can be non-linear if SFR tracers contain a significant fraction of diffuse emission. \citet{Leroy2013} subtracted diffuse background only in the 24 $\mu m$ data before combining it to H$\alpha$ and found subtle effects in the molecular gas Schmidt relation. This is likely due to the presence of diffuse background in H$\alpha$ as well, which should also be taken into account while studying the current star-formation rate. However, their results are consistent with ours in the sense of increase in scatter of the relation after removal of diffuse background. After subtraction of the diffuse background in SFR tracers, the best-fit line to our spatially-resolved data gives a slope of 1.26 $\pm$ 0.10 for the molecular gas Schmidt relation. This result is in agreement with previous works \citep{Kennicutt2007, Liu2011}. Using integral field spectroscopic data (IFS) for NGC 5194, \citet{Blanc2009} determined the diffuse emission in H$\alpha$ using the [S \textsc{ii}]/H$\alpha$ line ratio and found the slope of the molecular gas Schmidt relation to be 0.84 in this galaxy. However, a direct comparison between their work and ours is complicated by the large differences in the methodology adopted, particularly the difference in data (photometric versus IFS), physical scale of study (0.5--2 kpc versus $\sim$170 pc) and  diffuse background estimation. 

\indent For both molecular and total gas Schmidt relations of individual galaxies, we find that the {\it rms} scatter increases after subtraction of diffuse background (Figs. \ref{Figure: NGC 0628}, \ref{NGC 3184}-\ref{NGC 6946}). This result is consistent with previous works \citep{Liu2011, Leroy2013, Morokuma-Matsui2017}, which have employed different techniques for diffuse background subtraction. This consistency in the increase in scatter in the Schmidt relation in a log-log domain may point towards an astrophysical origin but is quite likely a result of removal of diffuse signal whilst minimally changing the underlying noise.

\subsection{Comparison with the global Kennicutt-Schmidt relation}
\label{section: comparison global}


\begin{figure}
	\includegraphics[width = 0.55\textwidth]{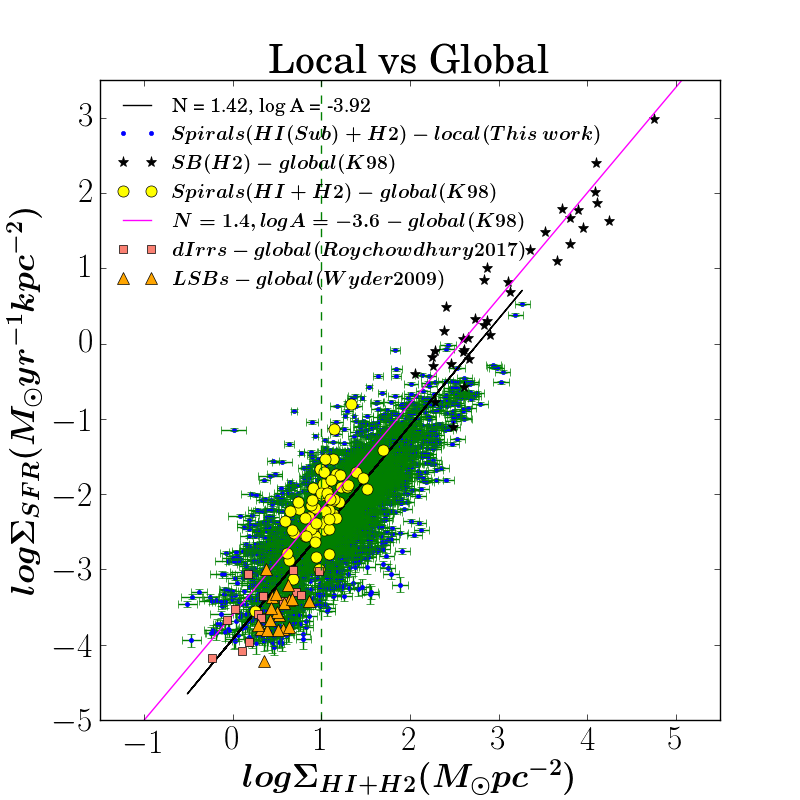}
	\caption{Comparison of the  spatially-resolved total gas Schmidt relation for nine spiral galaxies in this work with the global Kennicutt-Schmidt relation of the starbursts (black stars) and spirals (yellow dots) taken from \citet{Kennicutt1998b}, low surface brightness galaxies (LSBs, orange triangles) from \citet{Wyder2009} and dwarf irregular galaxies (dIrrs, pink squares) from \citet{Roychowdhury2017}. The diffuse background has been subtracted from the SFR tracers as well as from the atomic gas for the spatially-resolved data (blue dots with green error bars). These local measurements assume a Salpeter IMF and an X(CO) factor of 2.8$\times$10$^{20}$ cm$^{-2}$ (K km s$^{-1}$)$^{-1}$ to match the parameters adopted in \citet{Kennicutt1998b}. The pink line represents the global Kennicutt-Schmidt relation from \citet{Kennicutt1998b} while the solid black line is the best-fit line to the spatially-resolved data from all galaxies in this work.}
	\label{global local}
\end{figure}

\indent Fig. \ref{global local} presents a summary of the Schmidt relation with all the star-forming regions from our galaxy sample together with the integrated measurements of starburst and spiral galaxies taken from \citet{Kennicutt1998b}, low surface brightness galaxies (LSBs, orange triangles) from \citet{Wyder2009}, dwarf irregulars (dIrrs, pink squares) from \citet{Roychowdhury2017}, which is a sub-sample of \citet{Roychowdhury2014}. For the spatially-resolved data, the diffuse background has been subtracted from the SFR tracers and the atomic gas, and we adjusted the IMF and X(CO) factor of our spatially-resolved data to those of integrated data (i.e. a Salpeter IMF and an X(CO) factor of 2.8$\times$10$^{20}$ cm$^{-2}$ (K km s$^{-1}$)$^{-1}$). Hence, we find the following best-fit line:
\begin{equation}
\Sigma_{SFR} = 10^{-3.9\pm0.1}\Sigma_{gas}^{1.4\pm0.1}.
\label{best fit}
\end{equation}

\indent Interestingly, the above best-fit line passes through integrated data
corresponding to dIrrs \citep{Roychowdhury2017} and LSBs \citep{Wyder2009},
which only has atomic gas observations, and is in agreement with the
conclusions of \citet{Roychowdhury2015}, who suggest a power law slope of
$\sim$1.5 in the HI-dominated regions.

\indent The slope of the global and local best-fit lines is identical within the errors which we reason as follows. The radial profiles of SFR (Figure \ref{Figure: sub radial}) highlight that the diffuse
	background affects the low surface brightness outer regions of galaxies more
	significantly than the central regions of galaxies. Since the global
	Kennicutt-Schmidt relation targets entire star-forming  discs of
	galaxies, or bright star-forming regions, the effect of any diffuse
	background either averages out or is negligible. However, at local
	scales, the effect on the Schmidt relation becomes prominent when
	star-forming peaks as well as regions dominated by diffuse background
	are selected. That is possibly a reason why the global
	Kennicutt-Schmidt relation and local Schmidt relation have similar
	slopes. We note here that the similarity of slope in local and global results must be interpreted with caution, given that there has been much debate about the relative importance of molecular and atomic gas in the formation of stars, and the presence of diffuse H \textsc{i} component is even more uncertain. 

\indent Furthermore, there is an offset in the zeropoints, that is, $-$3.9 in this study versus $-$3.6 in the global relation. A Kolmogorov-Smirnov test on the global and spatially-resolved data  shows that the observed offset in the zeropoints of the global and local results is significant enough to affect the overall interpretation of the star-formation relation. There are several possible reasons for this difference in zero-points. In local sample selection, the chosen sub-regions of star formation in any spatially-resolved study will be biased by the selection criteria. For example, in this study we used H$\alpha$ luminous regions but given our understanding of the  formation of stars from molecular gas clouds, we could also have selected star-forming regions on the basis of molecular gas peaks. There is no such sampling effect in global studies. In addition, \citet{Kennicutt2007} describes the filling factor of star-forming regions in the disc of galaxies as a potential cause of the observed offset. Other reasons which can contribute to the observed offset are the different SFR calibration recipes at local and global scale. We note here that the non-negligible uncertainties due to age and the IMF are present in the conversion factor of luminosity to SFR. While comparing the spatially-resolved Schmidt relation with the global Kennicutt-Schmidt Law, we took into account the IMF which contributes significantly to the differences in SFR calibration at global and local scale (in section \ref{conversion}) but we did not take into account for example, the age of the H $\textsc{ii}$ regions. Hence, it is not surprising to see differences in the global-vs-local total gas Schmidt relations. 

\indent  Though the slope of the combined total gas spatially-resolved Schmidt relation matches the global
Schmidt-Kennicutt relation, there is considerable variation in slopes from galaxy to galaxy. Such variation has been observed before in previous studies, even without the subtraction of diffuse background \citep[see e.g.][]{WongBlitz2002, Boissier2003, Bigiel2008} though note here the subtraction of diffuse background from H \textsc{i} gas has led to a decrease in the variation of slopes. To explore the variation in the background subtracted data, we examined the variation of slope with respect to the ratio of the molecular-to-atomic gas surface density (within the analysis apertures) for all galaxies, however we did not find any trend. The large variation in slopes from galaxy to galaxy may also result from the higher uncertainties at lower SFRs \citep{daSilva2014}.

\indent The variation of X(CO) factor within galaxies may be another cause of the observed variation of slopes, for example, by using a radially-varying metallicity-dependent X(CO) factor, \citet{Boissier2003} derived a steeper slope ($\sim$2). \citet{Sandstrom2013} reports a radial variation of X(CO) in eight out of nine galaxies in our sample, with an overall decrease of 0.3 dex in the centre of galaxies. Since in our work, we have assumed a constant X(CO) factor (for comparison with other works), it is likely that the reported molecular gas surface densities are underestimated in the centres. It is possible that the variation of slope from galaxy-to-galaxy will further reduce if we adopt a radially-varying X(CO) factor in conjunction with diffuse background subtraction in SFR tracers and atomic gas. In our analysis, we did not take into account a local diffuse CO emission, mainly because of the poor resolution of CO map. However, various studies suggest the presence of a diffuse molecular gas forming at least 30\% of the total CO intensity  \citep[see][and references therein]{Shetty2014b}. As in the case of SFR tracers and atomic gas, a diffuse CO component will affect the lower intensity regions more than the higher intensity regions. This will result in the flattening of the slope both for the molecular gas and total gas Schmidt relations. Like SFR tracers and atomic gas, the fraction of CO bright diffuse gas will also vary from galaxy to galaxy. Correcting for the diffuse CO gas may result in a reduced overall variation of slopes. 

\indent Another possible source of systematics affecting the derived slopes are selection effects and sample biases from using the H$\alpha$ maps for picking out the star-forming regions. Ideally, to check this we should use the peaks of molecular gas (from CO maps), as indicators of star-forming regions. However, to check the effect of such systematics, we would need high resolution molecular gas maps, where the peaks can be readily resolved. \citet{Kruijssen2014} discusses how centring an aperture on gas or star-forming peaks might introduce a spatial-scale-dependent bias of the gas depletion time-scale and allows us to quantify this bias via an uncertainty principle for star formation. However, we could not use this principle to quantify the potential bias because it is only applicable to the case where the depletion time-scale is constant. In our study, the subtraction of diffuse background in SFR tracers has resulted in super-linear molecular gas Schmidt relation. The total gas Schmidt relation is super-linear irrespective of background subtraction in SFR tracers or atomic gas. The super-linearity implies that the depletion time-scale is not constant. Further theoretical and observation work needs to be done for quantification of such biases.

\section{Conclusion}
\label{section:conclusion}

\indent In summary, we have studied the spatially-resolved molecular and total gas Schmidt relation individually for nine galaxies, and also the overall combined fit for all galaxies for both molecular and total gas (Figures \ref{Figure: Schmidt H2}  and \ref{Schmidt HI}), by performing aperture photometry on the regions selected from H$\alpha$ maps. Though we find the slopes of the global and local total gas Schmidt relation to be similar, it is difficult to draw conclusions from this similarity especially when a large variation in slopes has been observed from galaxy to galaxy. The smaller scatter in log-log space in the total gas Schmidt relation compared to the molecular gas Schmidt relation may simply be the result of having more signal with respect to the noise contributions from the combination of H \textsc{i} and H$_ 2$. However, it may also indicate that atomic gas may be contributing to the star formation. In the spatially-resolved total gas Schmidt relation, we obtain a slope of 1.4 when diffuse background is subtracted from SFR tracers and the atomic gas. We find that the fraction of diffuse atomic gas in all galaxies is higher than the fraction of diffuse background in the SFR tracers (Table \ref{diffuse}) at the same spatial scale. It constitutes $\sim$ 37--80\% depending on the galaxy.  Excluding NGC 5457, diffuse atomic gas is $>$ 50\% in all galaxies. This might indicate that at the local scale in projection, a vast amount of atomic gas (i.e. diffuse background) in the ISM is not relevant for star formation. However, about 20--50\% of this atomic gas is available locally for star formation and hence its role can not be neglected. It is probably the cold component of the atomic gas, that is,  the cold
	neutral medium rather than the warm neutral medium, which contributes
	to star formation, though it is difficult to infer the nature of the
	diffuse background seen in atomic gas on the basis of the current data.

\indent A comparison with the dynamical model of Elmegreen 2015 indicates that the
	diffuse background in HI gas is likely the vertically-extended atomic
	gas which does not contribute to star formation. It is probably both
	atomic and molecular gas in the  disc (rather than out of  disc) of
	galaxies which contribute to star formation. Diffuse CO emission is
	not compatible with the dynamical model considered here, indicating that
	molecular gas is still the prime driver of star formation. However, the role of atomic gas in star formation especially in the outskirts of galaxies should not be neglected, where there is an abundance of atomic gas with no detection of molecular gas. Moreover, in the nearby Universe, atomic gas forms a large fraction of cold gas reservoir in galaxies. The molecular gas which directly feeds star formation, constitutes only $\sim$ 30\% of the cold gas \citep{Catinella2010, Saintonge2011, Boselli2014}. A recent study by \citet{Cortese2017} also shows that the gas reservoirs of star-forming  discs at z $\sim$ 0.2 are not predominantly molecular. So in such systems, atomic gas seems to be the source of star formation. Though from the current study it is difficult to infer the relative importance of atomic and molecular gas in star formation, we may say that  both of the components (atomic and molecular) of the ISM may be collectively important for forming stars.

\indent In this study, we find that the dynamical models of star formation with
	a constant scale height throughout galaxy  discs explains very well the
	spatially-resolved total gas Schmidt relation when diffuse background
	from SFR tracers as well as H $\textsc{i}$ are removed. The consideration of
	this diffuse background results in a single power-law relation between
	$\rm{\Sigma_{SFR}}$ and $\rm{\Sigma_{H \textsc{i} + H_2}}$ with
	power-law index of 1.5 in both inner and outer parts of spiral
	galaxies. Furthermore, a super-linear Schmidt relation at local and
	global scales would imply the dominant role of non-linear processes in
	driving star formation at spatially-resolved as well as global scales,
	and a non-constant star formation efficiency, or time scale, for
	different gas surface densities at different spatial scales.

\indent

\indent To explore the above hypotheses, we need to study the distribution of different phases of the ISM at the scales of H $\textsc{ii}$ regions, their spatial correlation, and also study these star-forming regions in different evolutionary stages. This will allow us to understand the relative importance of the different components of the ISM responsible for star formation and the nature of diffuse emission in SFR tracers, for example warm ionised medium (DIG) traced by H$\alpha$. It is now possible to study in great detail the characteristic properties of DIG in the nearby star-forming regions, with the advent of IFS facilities like  the Multi-Unit Spectroscopic Explorer (MUSE) and the Keck Cosmic Web Imager (KCWI). Moreover, the IFS technique allows us to correct the internal dust-attenuation in the galaxies directly from the Balmer-decrement (H$\alpha$:H$\beta$:H$\gamma$), bypassing the use of the infrared (24$\mu$m) images, which means that we can reduce the uncertainties related to the diffuse component in infrared images. Similarly, observational facilities like the Atacama Large Millimeter/submillimeter Array (ALMA), the Submillimeter Array (SMA), the Robert C. Byrd Green Bank Telescope (GBT) are apt to characterise the neutral gas (both atomic and molecular) in the nearby star-forming regions. Thus these combined analyses will help us better understand the intricacies of star formation.

\begin{acknowledgements}
		\indent We sincerely thank the reviewer of the paper whose helpful comments and suggestions greatly improved the paper. It is a pleasure to thank Rob Kennicutt for providing us with archival KINGFISH data, reading manuscripts of this paper and giving invaluable comments which greatly improved the scientific presentation of this analysis. NK also thanks Gareth Jones for helping with CASA. NK acknowledges the support from Institute of Astronomy, Cambridge and the Nehru Trust for Cambridge University during PhD, and the Schlumberger Foundation during the post-doc. This research made use of the NASA/IPAC Extragalactic Database (NED) which is operated by the Jet Propulsion Laboratory, California Institute of Technology, under contract with the National Aeronautics and Space Administration; SAOImage DS9, developed by Smithsonian Astrophysical Observatorys;  Astropy, a community-developed core Python package for Astronomy.
      
\end{acknowledgements}

%
  \bibliographystyle{bibtex/aa} 
  \bibliography{Bibliography1} 
%
\begin{appendix}

\section{Region selection}
\label{appendix:region selection}

\begin{figure}
	\centering
	\includegraphics[width=0.48\textwidth, trim={0 1.5cm 0 0},clip]{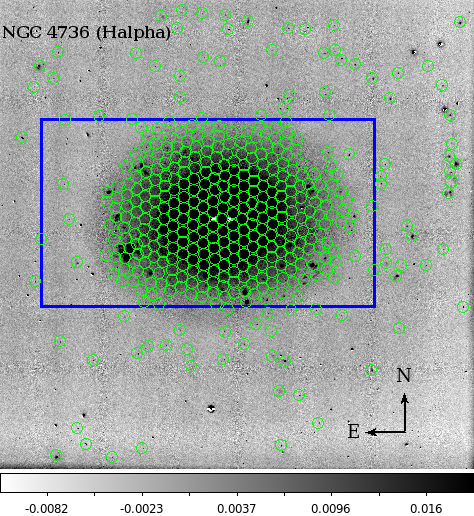}
	\caption{H$\alpha$ image of NGC 4736 where green circles denote the apertures covering star-forming regions as well as the regions in between them which have H$\alpha$ detection. The blue rectangular box shows the region covered by the CO map. The apertures within the blue rectangular box are used for Schmidt relation analysis.}
	\label{figure:region selection}
\end{figure}

\indent Figure \ref{figure:region selection} shows an example of region identification using the H$\alpha$ image of a sample galaxy NGC 4736.

\section{Maximum aperture-size}
\label{appendix: aperture size}
\begin{figure*}
	\centering
	\includegraphics[width = 0.48\textwidth]{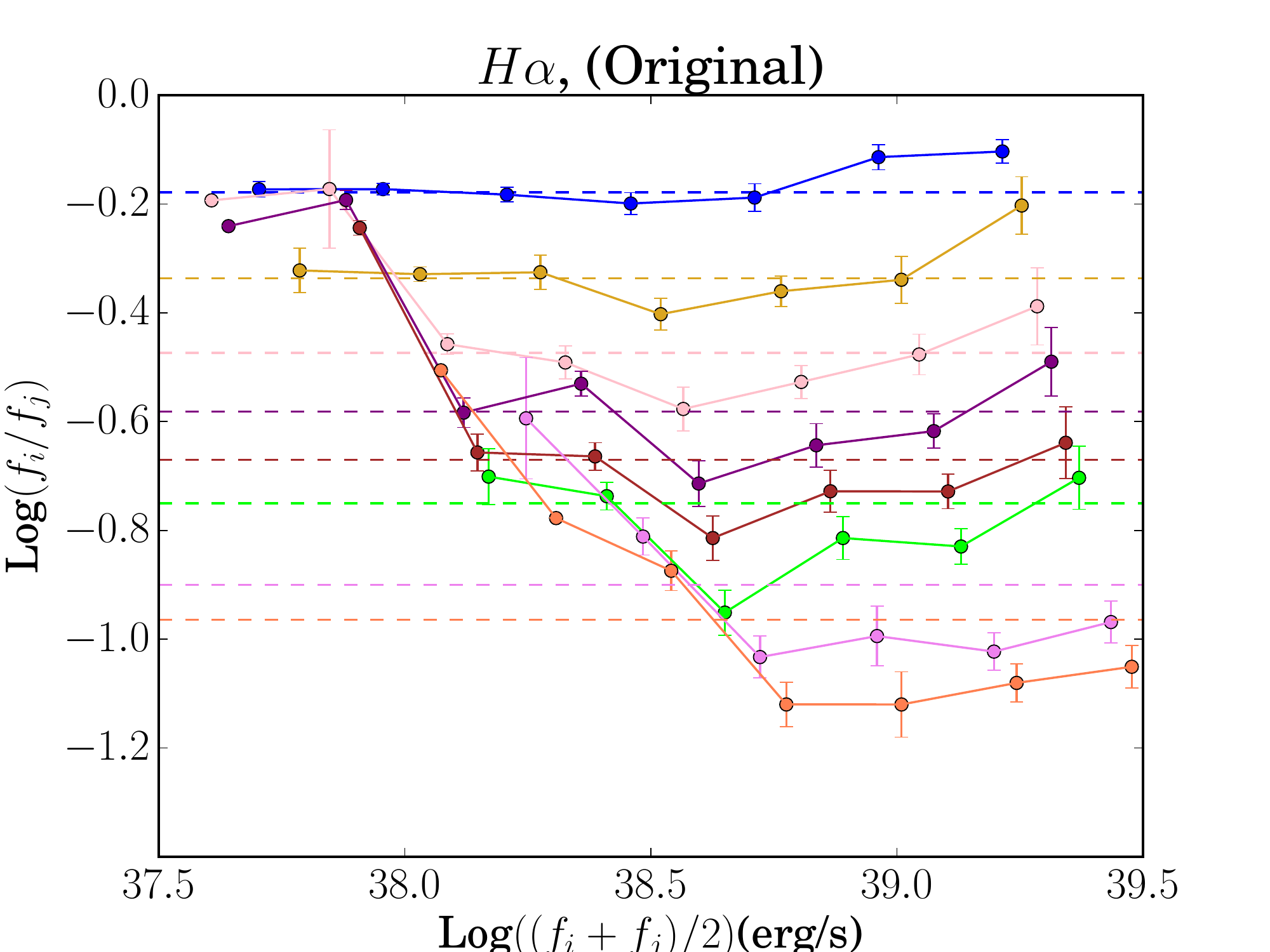}
	\includegraphics[width = 0.48\textwidth]{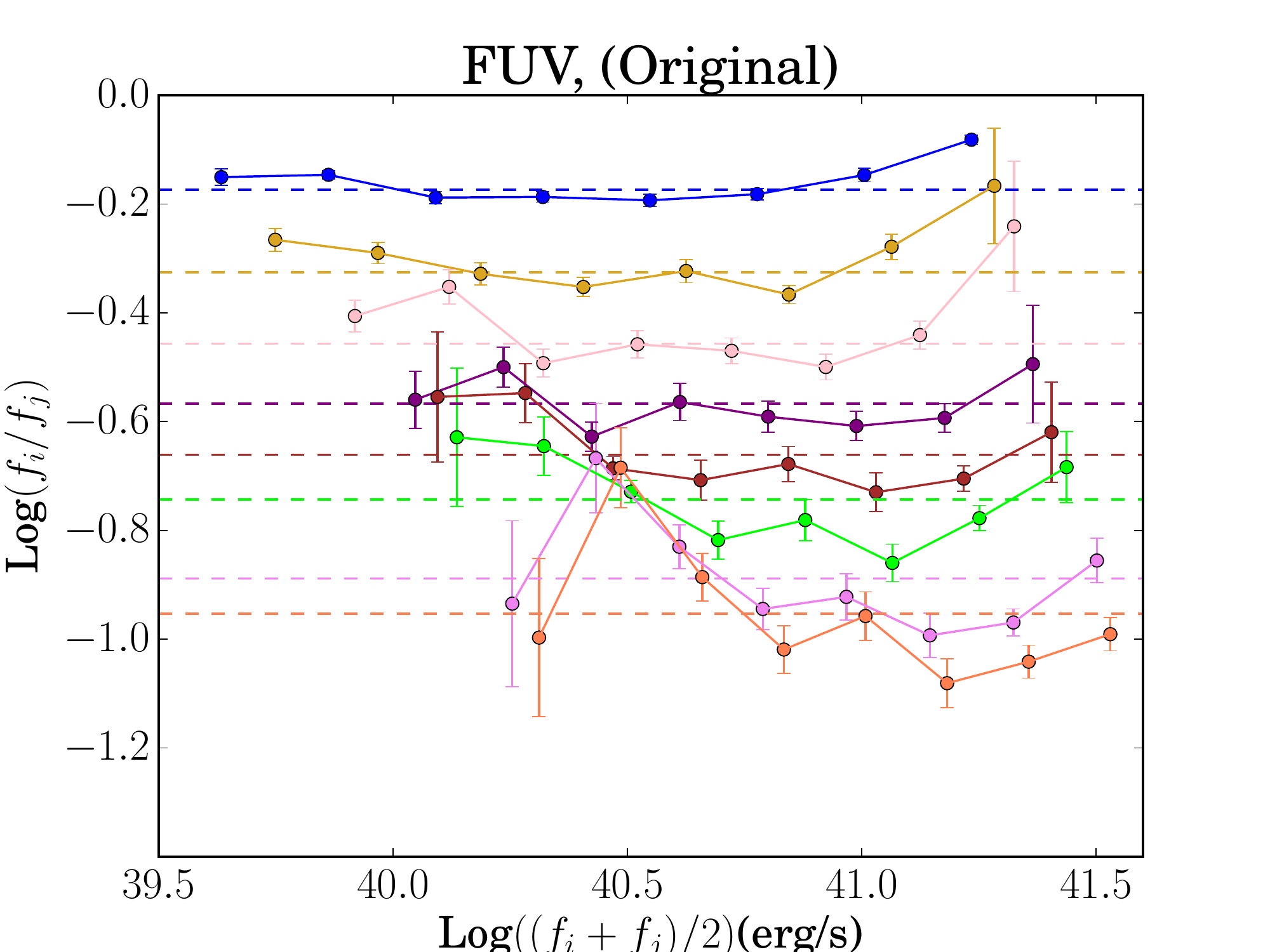}
	\includegraphics[width = 0.48\textwidth]{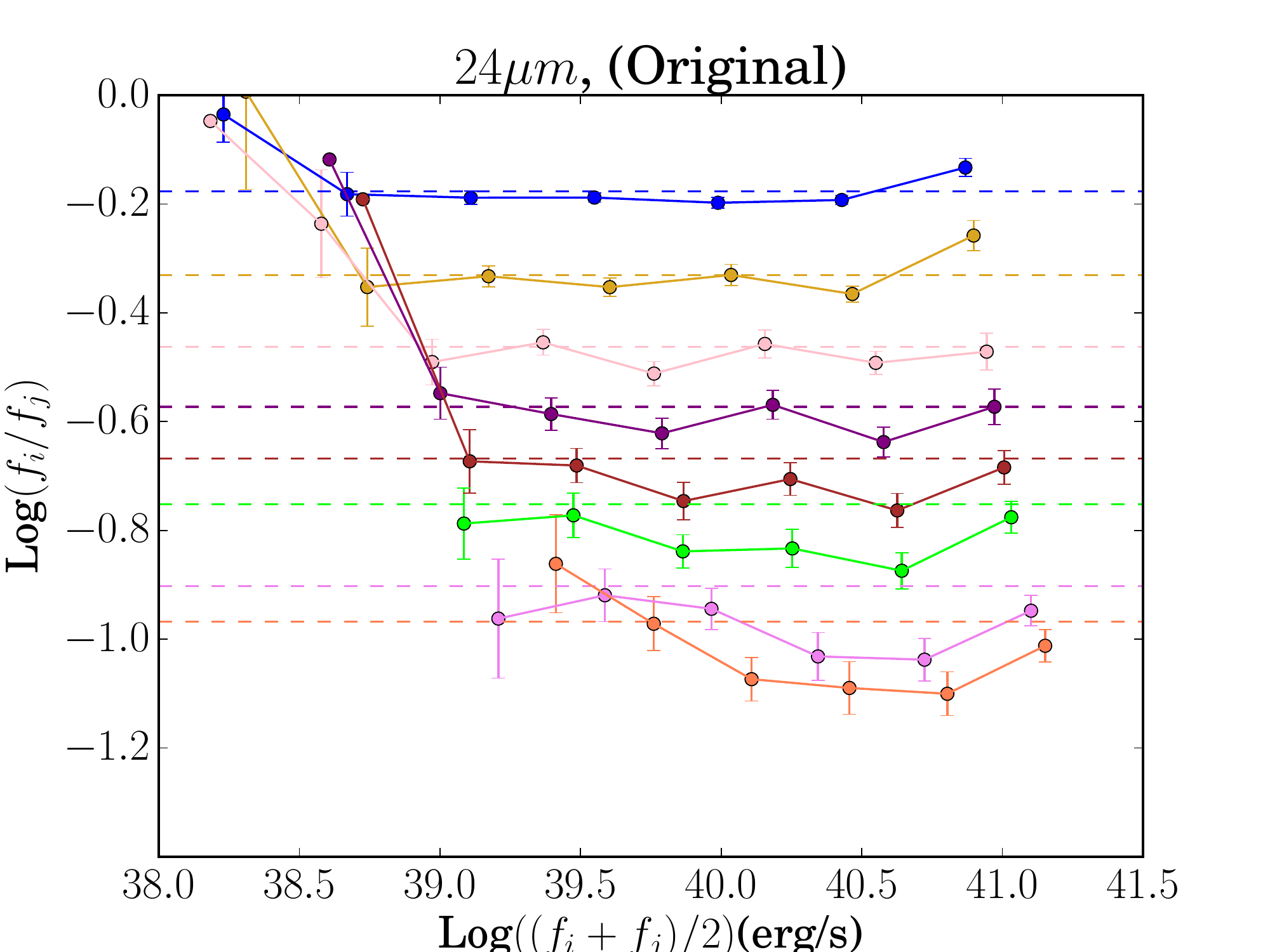}
	\includegraphics[width = 0.48\textwidth]{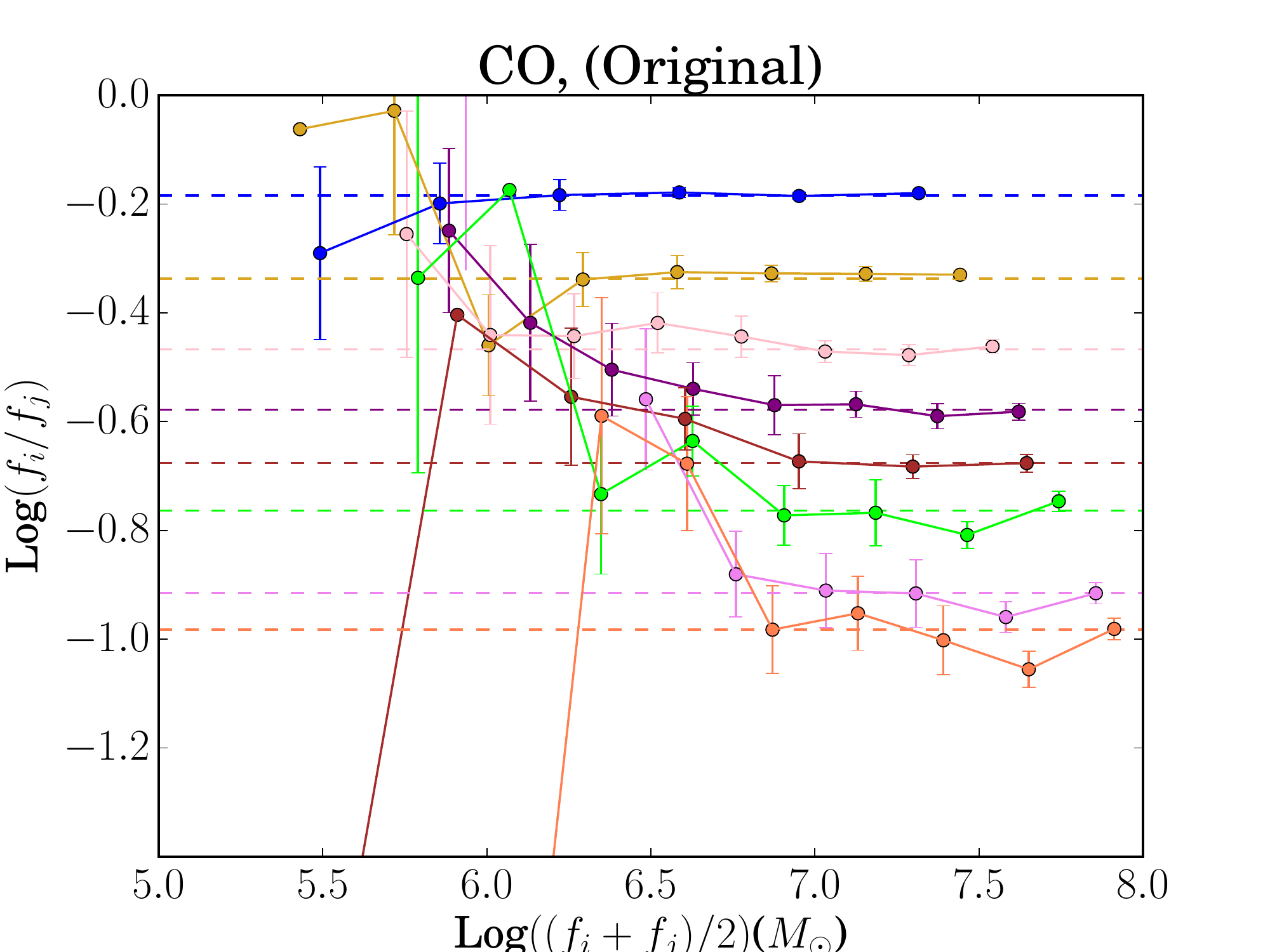}
	\includegraphics[width = 0.75\textwidth]{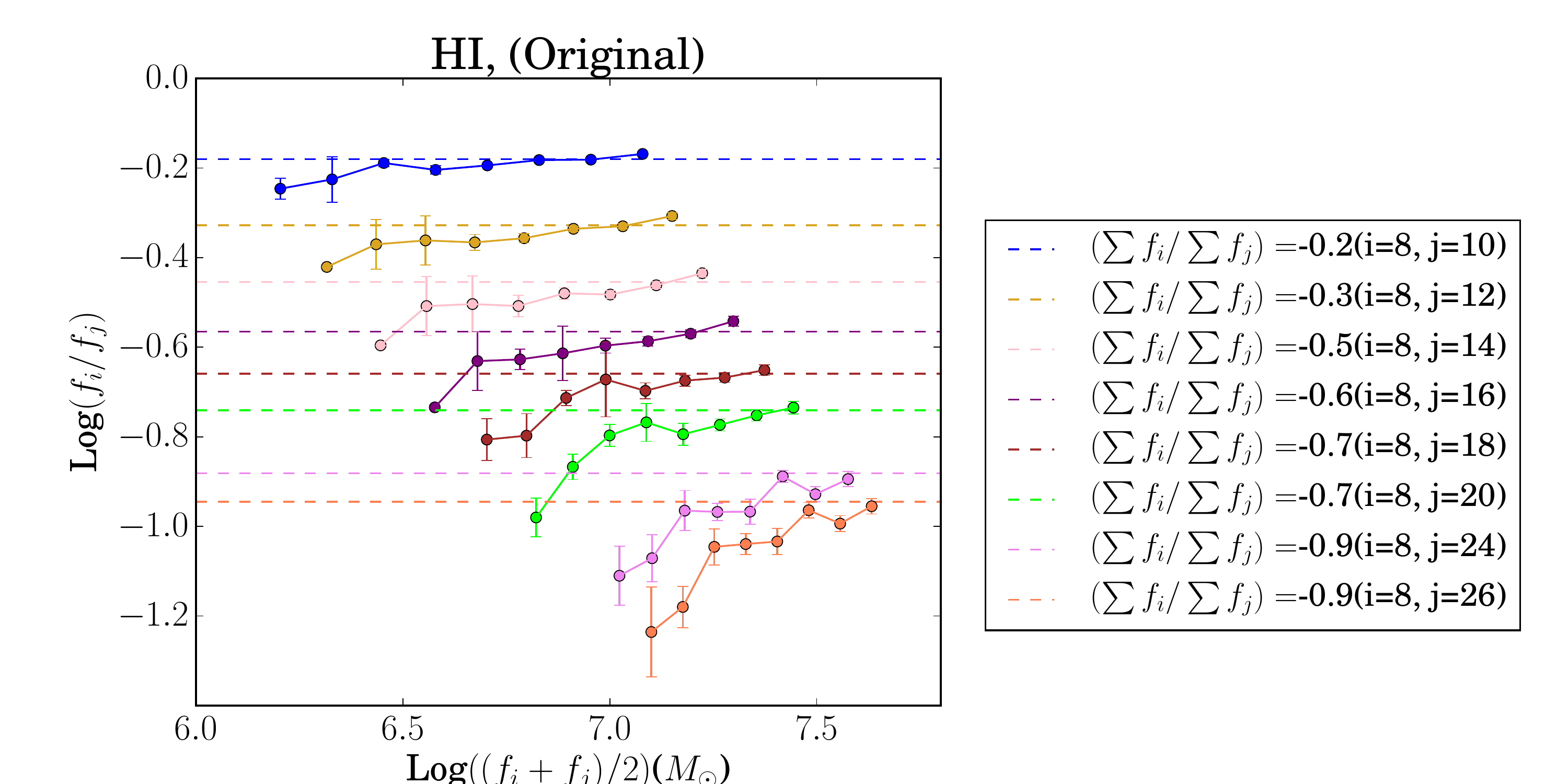}
	\caption{The flux (mass) in apertures of size varying from 0.6--2.0 kpc normalised by the flux (mass) in an aperture of size $\sim$500 pc compared to the average flux (mass) in the corresponding apertures. The solid colour lines show the normalised flux (mass) while the dashed colour lines show the ratio of the total flux in the apertures under study, where i \& j denote the aperture radii in arcsec. These figures correspond to analysis of the galaxy NGC 0628, for which the minimum radius (8$\arcsec$) and maximum radius (26$\arcsec$) correspond to physical sizes of $\sim$ 600 pc and 2 kpc, respectively.}
	\label{Figures: app aperture choice}
	
\end{figure*}

\indent To determine a viable maximum aperture size  we experimented on NGC 0628 which is at an intermediate distance ($\sim$ 7.3 Mpc) and is almost-face-on (inclination angle $\sim$ 7$^{\circ}$), mitigating any effect due to inclination angle.  Figure \ref{Figures: app aperture choice} shows the results of this experiment where in each panel, the y-axis shows the flux (for SFR tracers), or mass for gas data, in successive apertures of size varying from 0.6--2.0 kpc, normalised by the flux, or mass, in the smallest aperture ($\sim$ 500 pc). The x-axes shows the average flux/mass in the corresponding apertures. In the first four panels, we find that for each tracer (H$\alpha$, FUV, 24 $\mu$m and CO) the normalised flux/mass (solid colour line) shows the same trend, following closely the ratio of the total flux or mass (dashed colour line) for all apertures varying from 0.6--2.0 kpc. However, the corresponding plot for H \textsc{i}  (bottom panel), shows a deviation in the trend at an aperture radius of $\sim$14$\arcsec$ (solid pink line) which corresponds to physical size of $\sim$1 kpc. This deviation becomes significant at physical sizes greater than $\sim$2 kpc. A physical size of 1.0 kpc was found suitable for two galaxies  (NGC 3351 and NGC 5055) in the sample owing to their distances and inclination angles. Due to the high inclination angle of NGC 3521, a physical size of 1.0 kpc still leads to problems of under-sampling the enclosed pixels in apertures. Hence for a comparison with \citet{Liu2011} who studied Schmidt relation for this galaxy at 2.0 kpc scale, we set the aperture size for NGC 3521 to correspond to a physical size of $\sim$2.0 kpc. The individual results for this galaxy should be interpreted with caution due to the potential deviation in trend of H \textsc{i} beyond 1 kpc.	
	
\section{Background subtraction on a mock galaxy}
\label{appendix: background}

\begin{figure}
	\centering
	\includegraphics[width=0.45\textwidth]{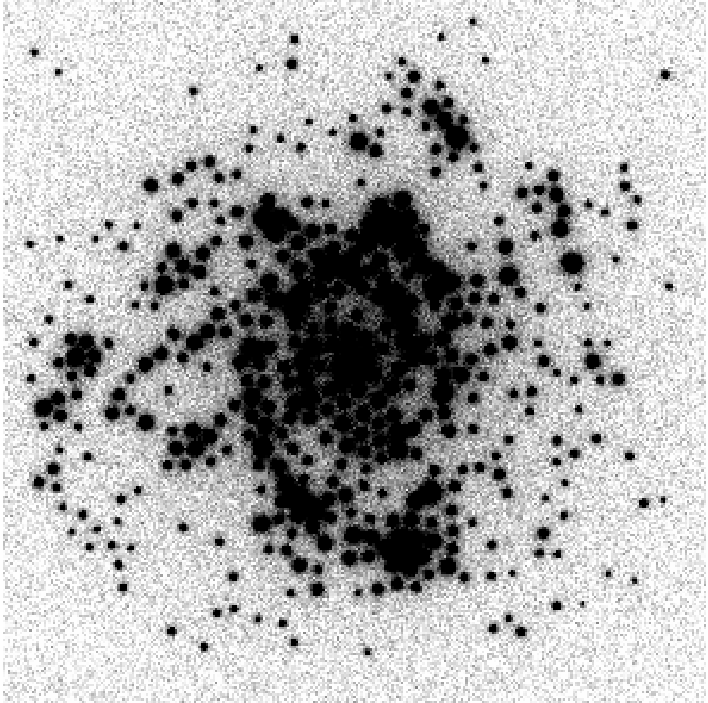}
	\caption{Mock galaxy where H $\textsc{ii}$ regions are modelled as Moffat profiles with $\beta$ = 1.5. For more details on the modelling see the text in  Appendix \ref{appendix: background}.}
	\label{Figure: model}
\end{figure} 
\begin{figure}
	\centering
	\includegraphics[width=0.45\textwidth]{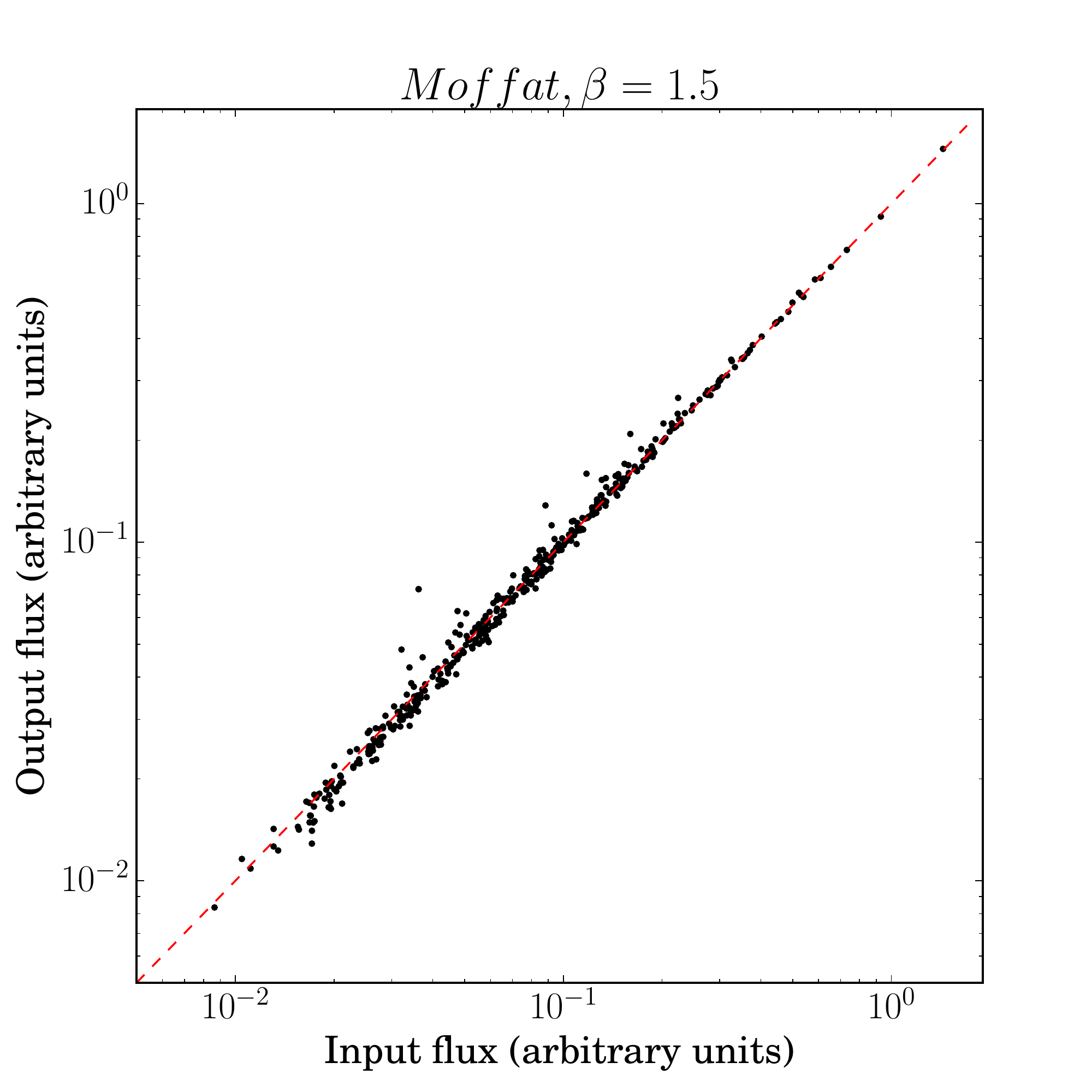}
	\includegraphics[width=0.45\textwidth]{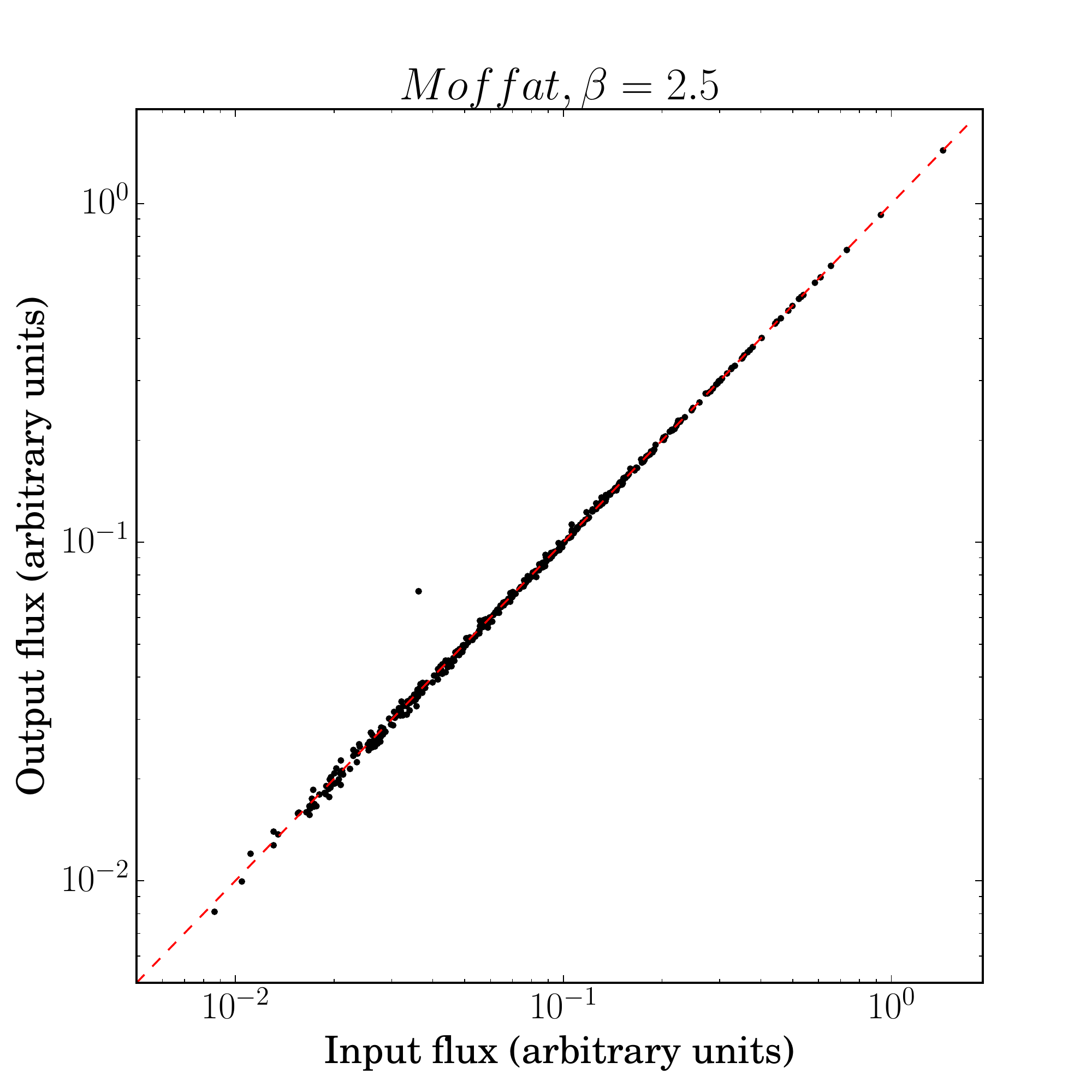}
	\caption{In both panels, x-axis shows the input fluxes to model H \textsc{ii} regions in a mock galaxy with an underlying diffuse component and y-axis shows the fluxes in H  \textsc{ii} regions after diffuse background subtraction. In both panels, the dashed red line is the one-to-one line. H $\textsc{ii}$ regions are modelled as the Moffat profiles with $\beta$ = 1.5 (upper panel) and $\beta$ = 2.5 (lower panel).}
	\label{Figure: app bg}
\end{figure}

We modelled the mock galaxy (Fig. \ref{Figure: model}) as a combination of \SF regions, a  discy exponential profile and used similar noise model properties to the real data. For simulating \HII regions, we used Moffat profiles defined by $I(r) = I_o[1 + (r/\alpha)^2]^{-\beta}$ where I$_o$ is the peak intensity at the centre, and  $\alpha$ and $\beta$ are related to FWHM as: $FWHM = 2\alpha\sqrt{2^{1/\beta} - 1}$. The FWHM was set to be 2\arcsec similar to the real data. The intensity enclosed in an aperture of radius R is given by $I(<R) = \frac{\pi\alpha^2}{\beta - 1}I_o[1 - [1+(R/\alpha)^2]^{-\beta + 1}]$ from which aperture corrections can be readily computed. In our modelling, we experimented with two values of $\beta$ = 1.5, 2.5. A Moffat profile with $\beta$=2.5 produces a more Gaussian-like profile, while a Moffat profile with $\beta$=1.5 produces extended outer wings, providing a better representation of the \HII regions. Model flux values for the \HII regions were taken from the original unsubtracted H$\alpha$ image of the galaxy NGC 0628. The diffuse background exponential disc was modelled based on the effective radius of the galaxy and scaled to yield a diffuse background component of 25\%. On the modelled galaxy images ($\beta$ = 1.5, 2.5), as for the real data, we used the Nebulosity Filter software to estimate ab-initio the diffuse background component. We then performed aperture photometry with apertures of radii of 8\arcsec (the same as adopted for NGC 0628) to extract fluxes in the background subtracted modelled \HII regions. Aperture corrections corresponding to the two Moffat profiles were applied on the extracted fluxes, and aperture-corrected fluxes were compared with the input fluxes used to model \HII regions. Figure \ref{Figure: app bg} shows comparisons of the input and output fluxes for the modelled galaxies with $\beta = 1.5$ (left panel) and $\beta$ = 2.5 (right panel). The input and output fluxes lie on the expected one-to-one line (red-dashed line) expanding over two order of magnitudes.

\section{Radial profiles on unsubtracted data}
\label{appendix: radial}

\begin{figure*}
	\centering
	\includegraphics[width = 0.345\textwidth,trim={0.2cm 0cm 1.6cm 0cm},clip]{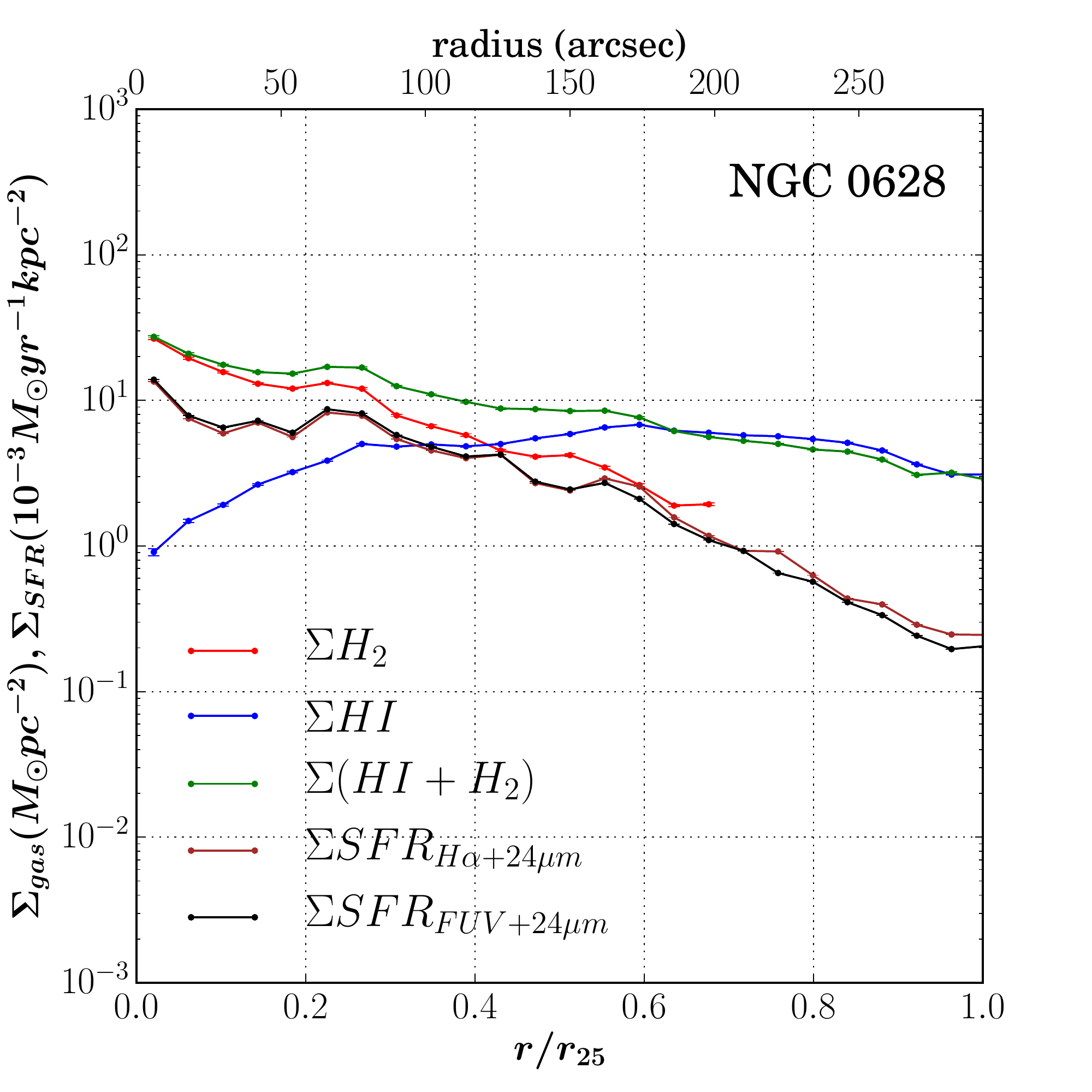}
	\includegraphics[width = 0.32\textwidth,trim={1.2cm 0cm 1.6cm 0cm},clip]{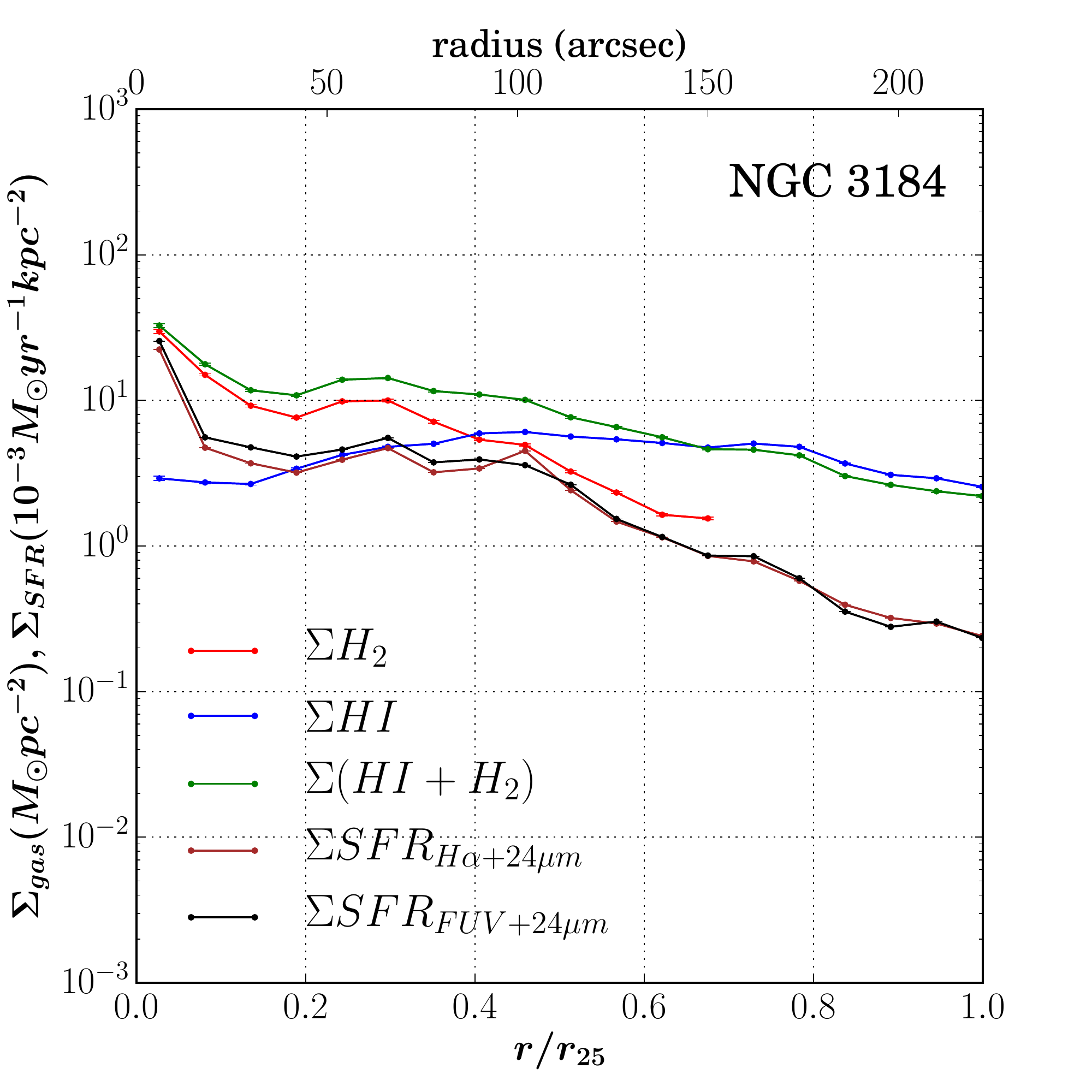}
	\includegraphics[width = 0.32\textwidth,trim={1.2cm 0cm 1.6cm 0cm},clip]{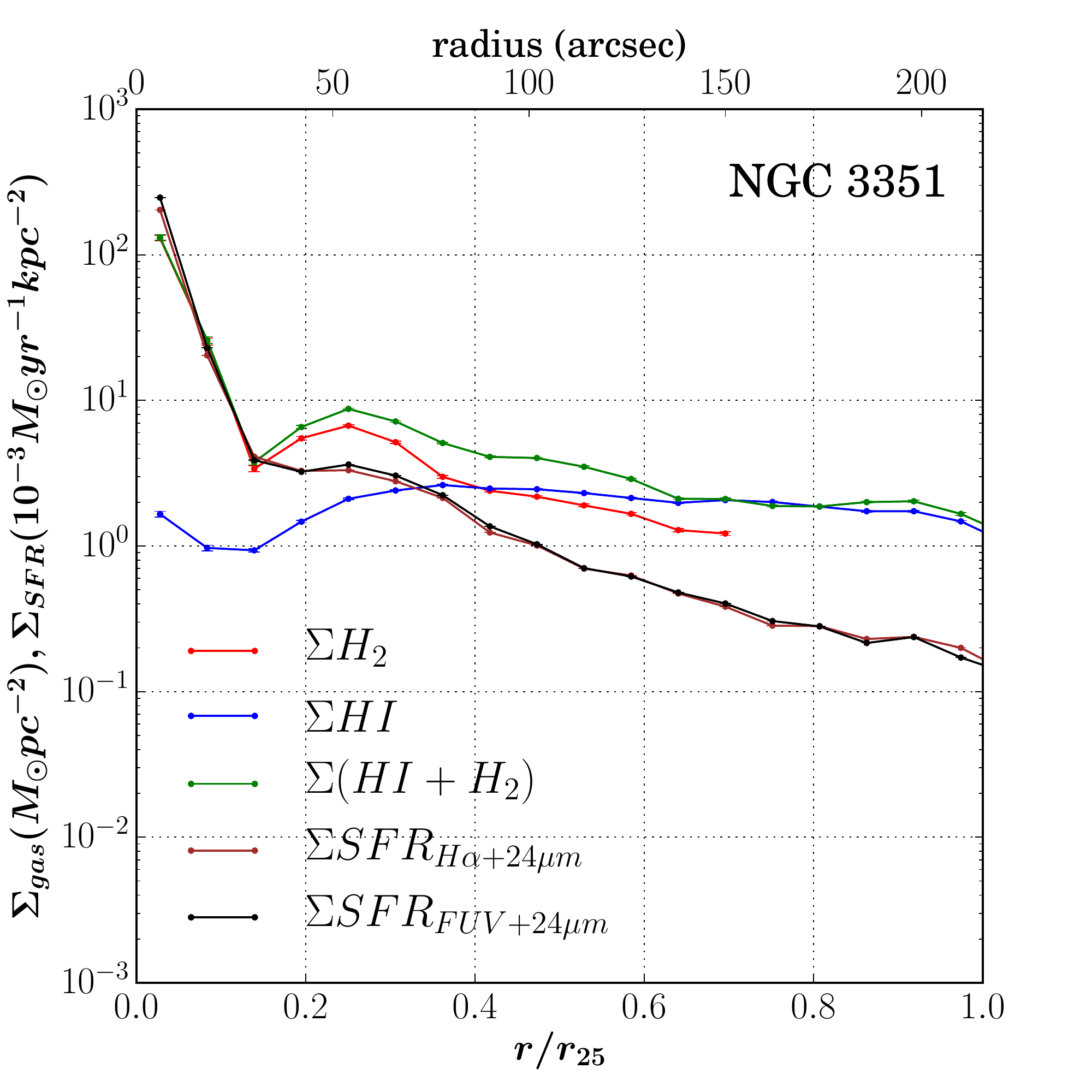}
	\includegraphics[width = 0.345\textwidth,trim={0.2cm 0cm 1.6cm 0cm},clip]{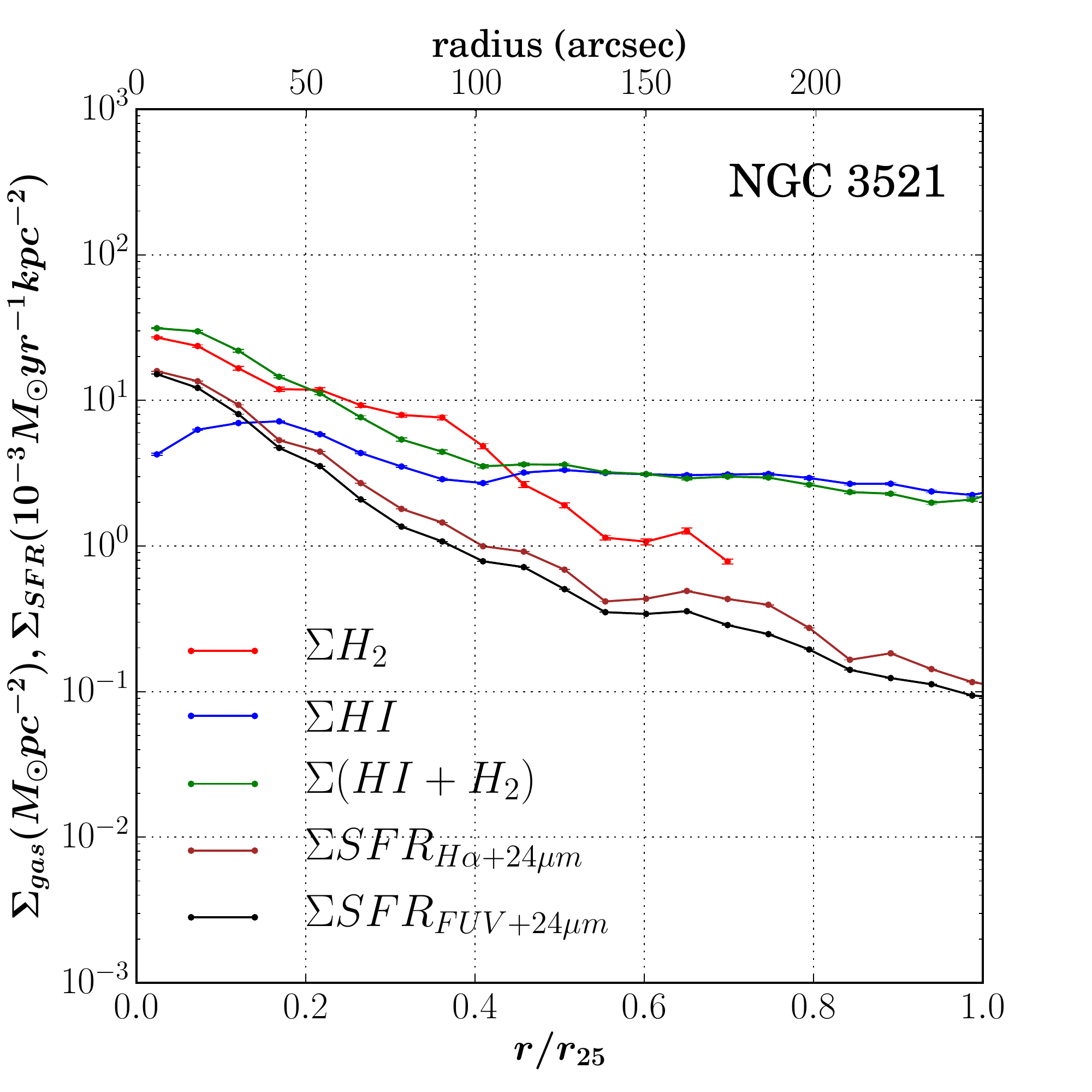}
	\includegraphics[width = 0.32\textwidth,trim={1.2cm 0cm 1.6cm 0cm},clip]{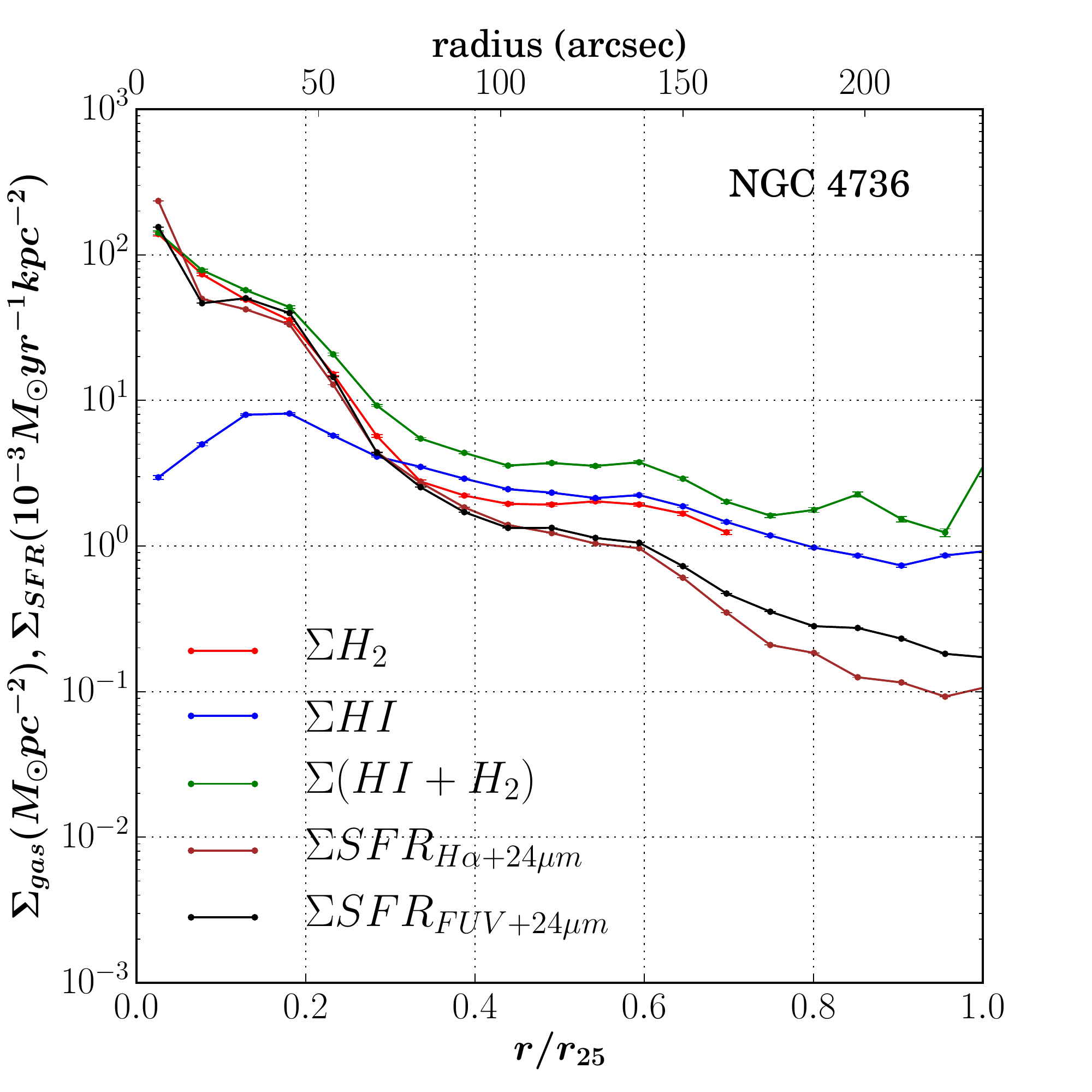}
	\includegraphics[width = 0.32\textwidth,trim={1.2cm 0cm 1.6cm 0cm},clip]{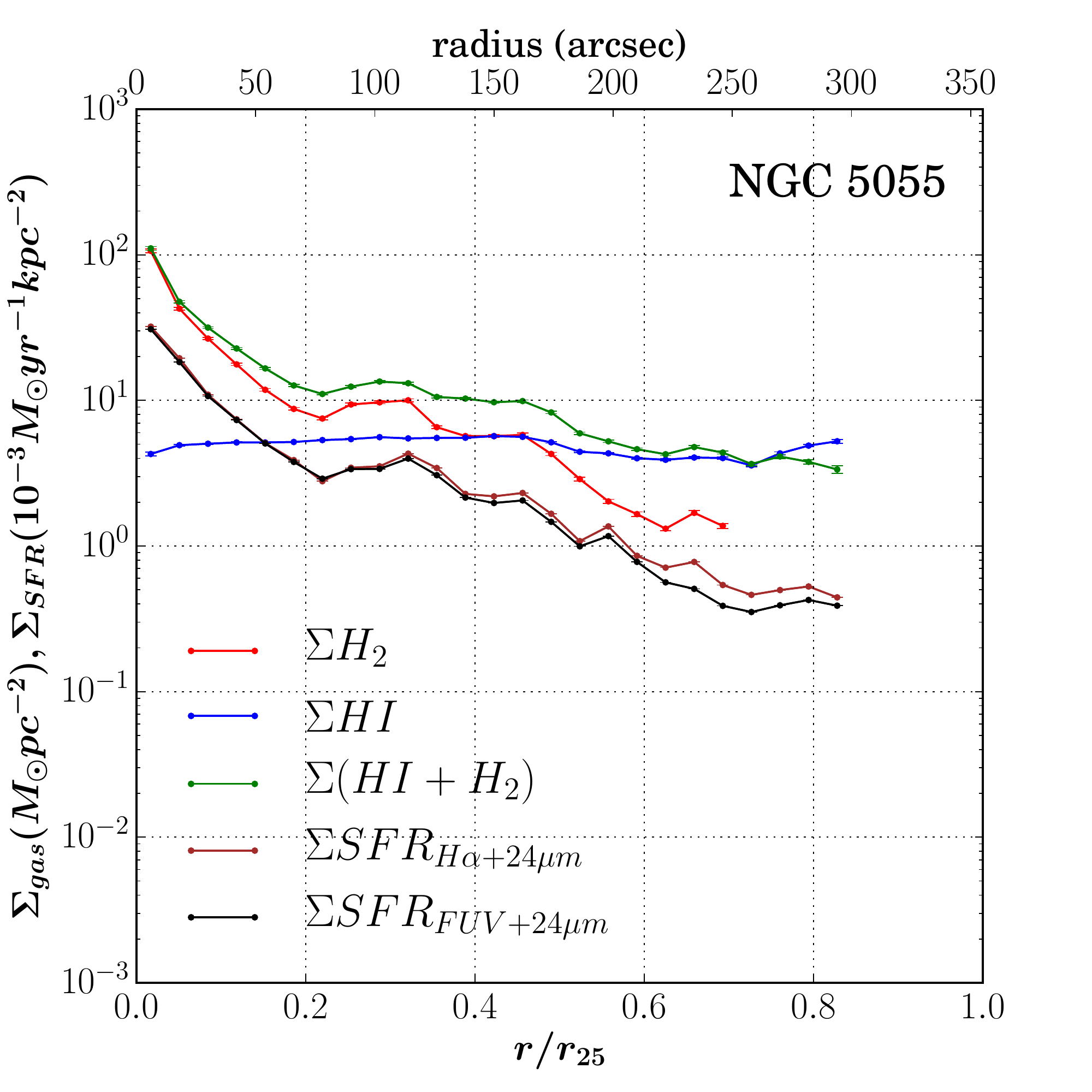}
	\includegraphics[width = 0.345\textwidth,trim={0.2cm 0cm 1.6cm 0cm},clip]{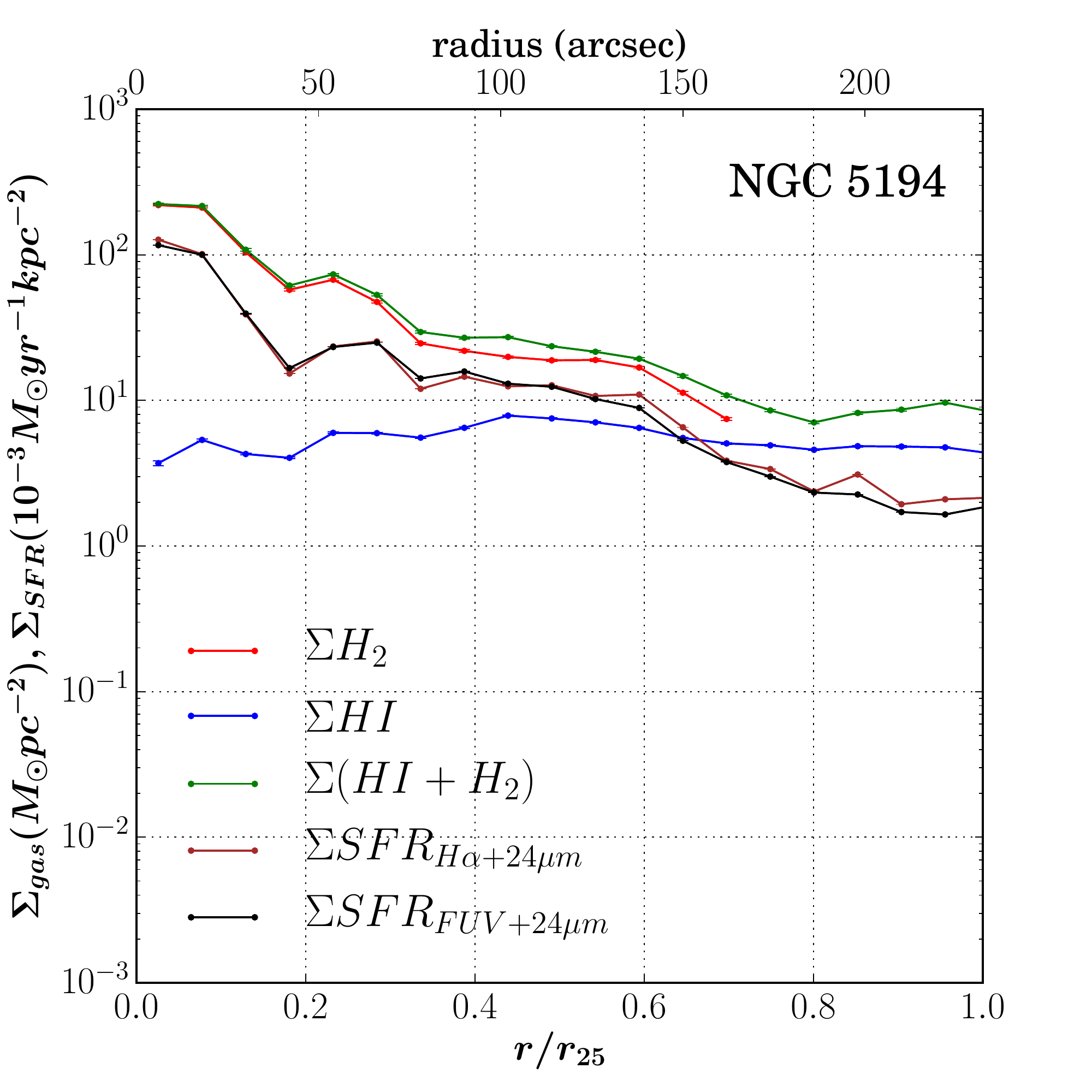}
	\includegraphics[width = 0.32\textwidth,trim={1.2cm 0cm 1.6cm 0cm},clip]{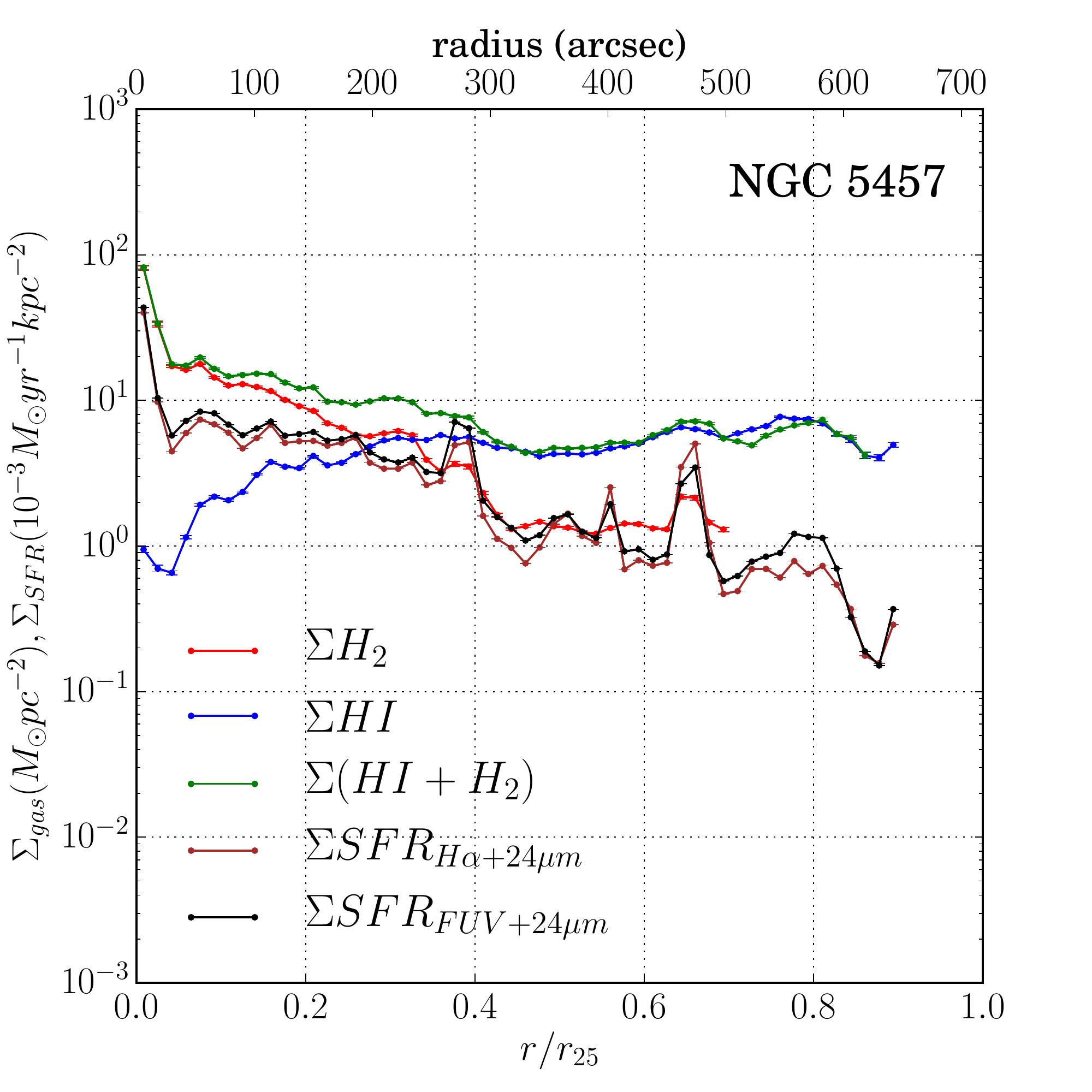}
	\includegraphics[width = 0.32\textwidth,trim={1.2cm 0cm 1.6cm 0cm},clip]{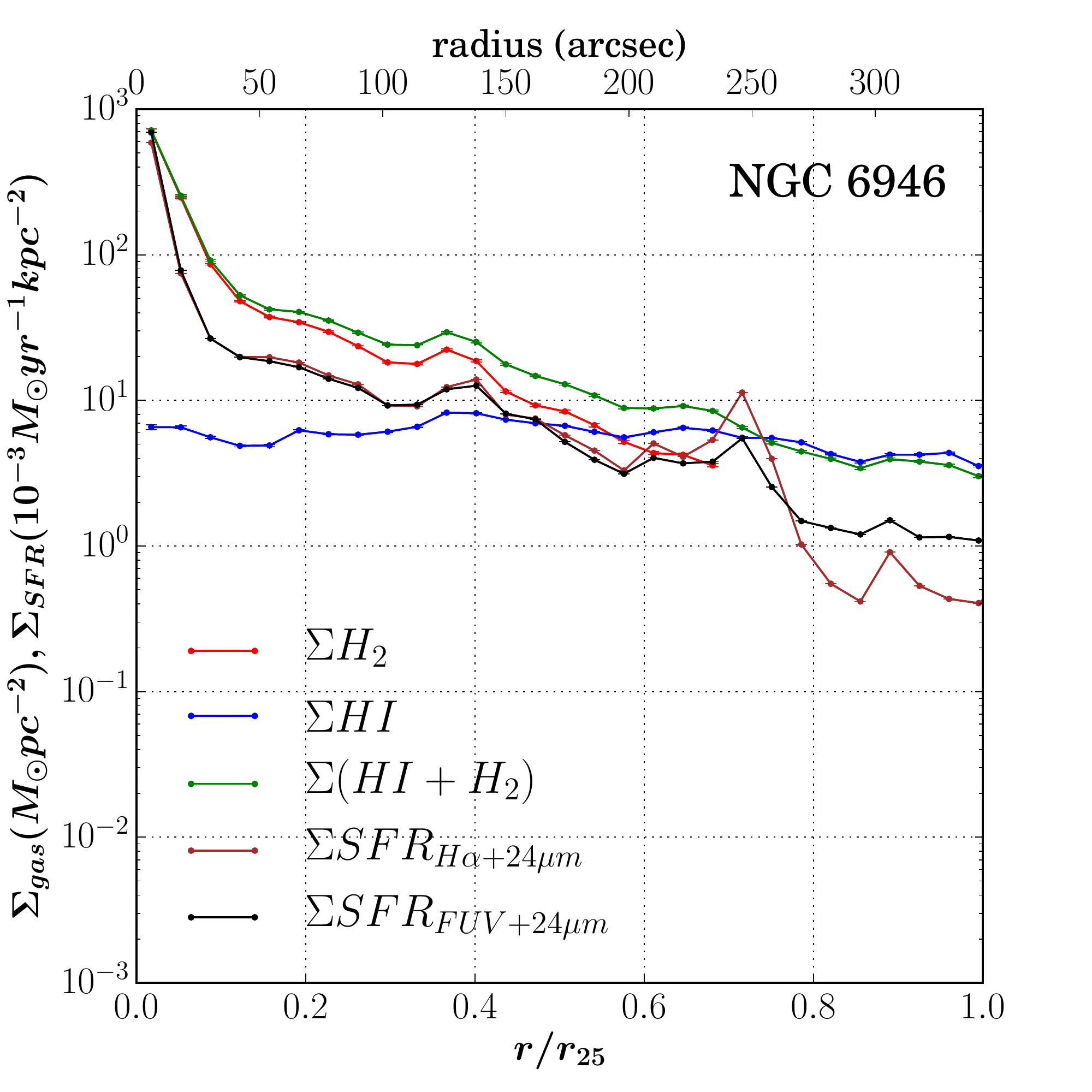}
	\caption{Radial profiles of $\Sigma_{SFR}$ (original unsubtracted data), $\Sigma_{H \textsc{i}}$, $\Sigma_{H_2}$ and $\Sigma_{H \textsc{i} + H_2}$ for all nine galaxies in the sample. The x-axis presents the galactocentric radius normalised by r$_{25}$ (bottom) and in arcsecs (top). The y-axis presents $\Sigma_{gas}$ in units of M$_{\odot}$ pc$^{-2}$ and $\Sigma_{SFR}$ in units of 10$^{-3}$ M$_{\odot}$ yr$^{-1}$ kpc$^{-2}$ (the scale range is the same for all galaxies). $\Sigma_{H_2}$ (red curves) and $\Sigma_{SFR}$ (brown and black curves) show a similar radial falloff for all galaxies. $\Sigma_{H \textsc{i}}$ (blue curves) have a relatively flat radial profile, usually a deficit towards the centre and show an upper limit of $\sim$10 M$_{\odot}$pc$^{-2}$}
	\label{Figures: app original radial}
\end{figure*}

\indent Fig. \ref{Figures: app original radial} presents radial profiles of  $\Sigma_{SFR}$ (brown curve: H$\alpha$ and black curve: FUV), $\Sigma_{H \textsc{i}}$ (blue curve),  $\Sigma_{H_2}$ (red curve),  $\Sigma_{H \textsc{i} + H_2}$ (green curve) for all nine galaxies in the sample before subtraction of diffuse background. We do not show radial profiles of $\Sigma_{H_2}$ for r $>$ $0.7$ r$_{25}$ because the radial profiles of CO intensity are found to fall below 3$\sigma$ approximately after this radius \citep{Leroy2009}. The radial profiles $\Sigma_{H_2}$ (red curves) are decreasing strongly with radius in all galaxies and are closely correlated with the $\Sigma_{SFR}$ radial profiles (brown and black curves). 

\indent The radial profiles of $\Sigma_{H \textsc{i}}$ (blue curves) on the other hand show a well-defined upper limit of $\sim$10 M$_{\odot}$pc$^{-2}$ and in general a strong depletion towards the galaxy centres. There is a relatively high surface density of molecular gas compared to atomic gas in the centre of all galaxies (r $<$ 0.2 r$_{25}$), while in the outer parts $\Sigma_{H_2}$ decreases in the regions where H \textsc{i }saturates. Clearly the formation and destruction of H\,\textsc{i} and H$_2$ are interdependent and these trends presumably follow naturally from the interplay of H\,\textsc{i}, H$_2$ and star formation. In the centres of these galaxies most of the H\,\textsc{i} is presumably in the form of H$_2$ which is itself being used to form stars. In the outer regions, either molecular gas dissociates back to form atomic gas which suppresses star formation, or formation of molecular gas is itself low because of the lower amount of dust\footnote{Infrared maps of galaxies are brighter in the inner regions of galaxies compared to the outer regions} needed for molecule formation, and sufficient provision of energetic FUV photons in the extended arms of galaxies leading to further dissociation of molecular gas into atomic gas.

We note that the radial profiles of the original unsubtracted  data for these galaxies have been published before by \citet{Bigiel2008}, \citet{Leroy2008} and \citet{Schruba2011}, though the methods adopted are slightly different in each analysis. We present them here to compare with the profiles obtained from the subtracted data presented in Fig. \ref{Figure: sub radial}.

\section{Power-law fit: weighted versus unweighted}
\label{appendix: line-fitting}
\begin{figure*}
	\centering
	\includegraphics[width=0.6\textwidth]{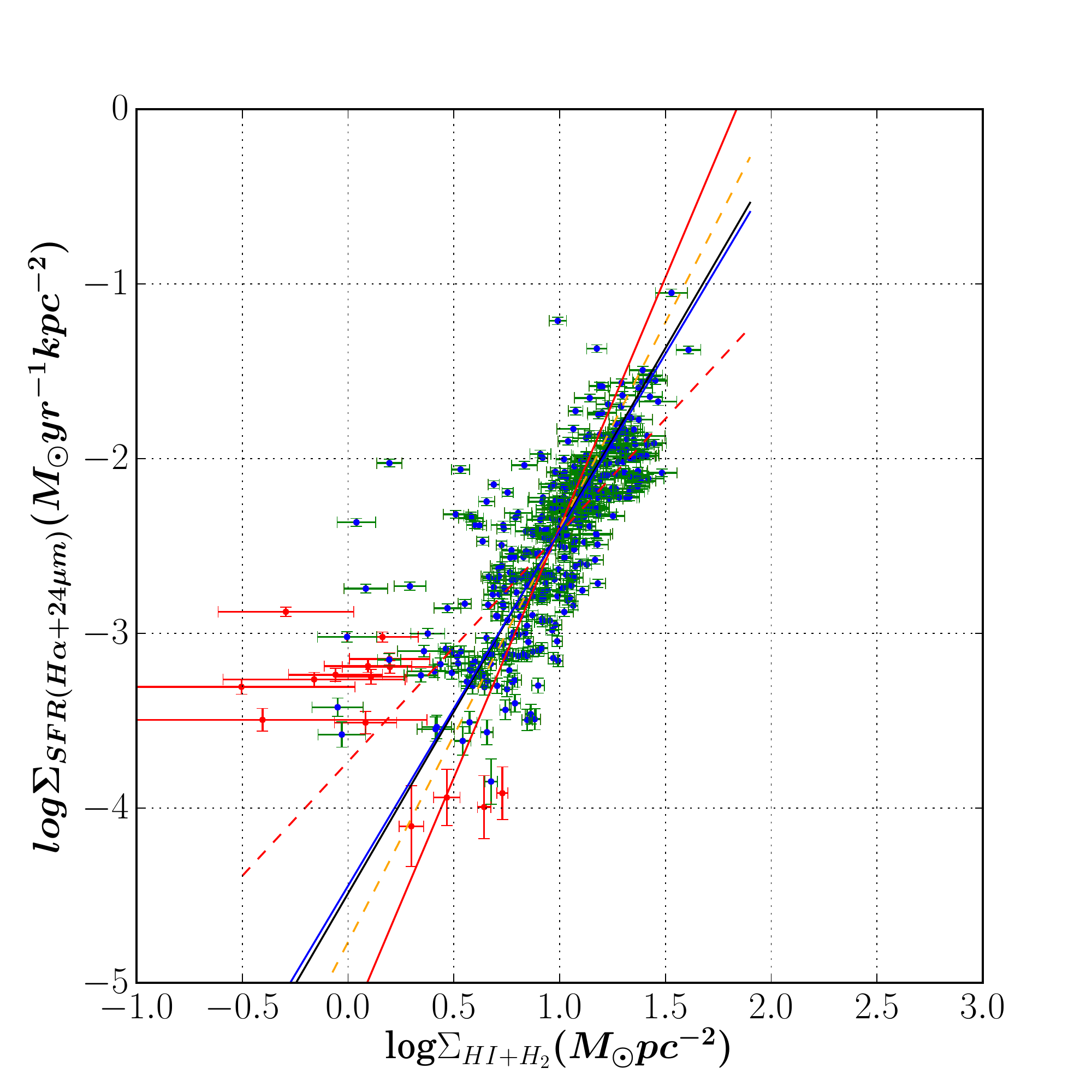}
	\caption{Different fits to the spatially-resolved total Schmidt relation for NGC 0628. Only blue points with green error bars are used for fitting. The red data points with errorbars correspond to points with S/N$<$3, and have not been included in the fits. The error bars shown here include the random error and systematic error on flux measurements as explained in Section \ref{section: error}, but not the additional calibration error. Dashed red line: standard unweighted regression,  dashed orange line: inverse weighted regression. solid black line: maximum likelihood estimate where each point on both axes is given equal weight, solid blue line: maximum likelihood estimate where 20\% calibration error has been included on both axes in addition to the uncertainties estimated in Section \ref{section: error}, solid red line: maximum likelihood estimate where points have been weighted solely with respect to the uncertainties estimated in Section \ref{section: error}. }
	\label{weighted fit}
\end{figure*}

We experimented with different methods for fitting the total gas Schmidt relation to the spatially-resolved data of NGC 0628, where no diffuse background was subtracted. These fitting methods include a standard unweighted regression, inverse unweighted regression, and maximum likelihood fits (both weighted and unweighted).  The results of all the fits are shown in Fig. \ref{weighted fit}. Previous studies have shown that calibration errors as large as 30--50\%  may exist on both axes owing to uncertainties on the SFR calibration and the CO-to-H$_2$ conversion. Hence in the weighted maximum likelihood fit (solid blue line), we include a 20\% calibration error on both axes in addition to the uncertainties estimated in section \ref{section: error}.  We test the goodness of fit from the value of the reduced-$\chi^2$ obtained. We find that the reduced-$\chi^2$ is close to unity (0.98 in this case) when additional calibration uncertainties are included in the maximum likelihood fit.  Without this additional uncertainty the reduced-$\chi^2$ is much larger than one indicative of an incorrect data model. The parameters obtained from the weighted fit are similar to those derived from a maximum likelihood fit where equal weight is given to each point on both axes (solid black line). Hence based on these experiments, we decided to estimate the best-fit parameters using an equal weight maximum likelihood fit in all the analysis.   

\section{Schmidt Law fits to individual galaxies}
\label{appendix: individual galaxies}
\begin{figure*}
		\centering
		\includegraphics[width = 0.33\textwidth,trim={0.2cm 0cm 1.5cm 0cm},clip]{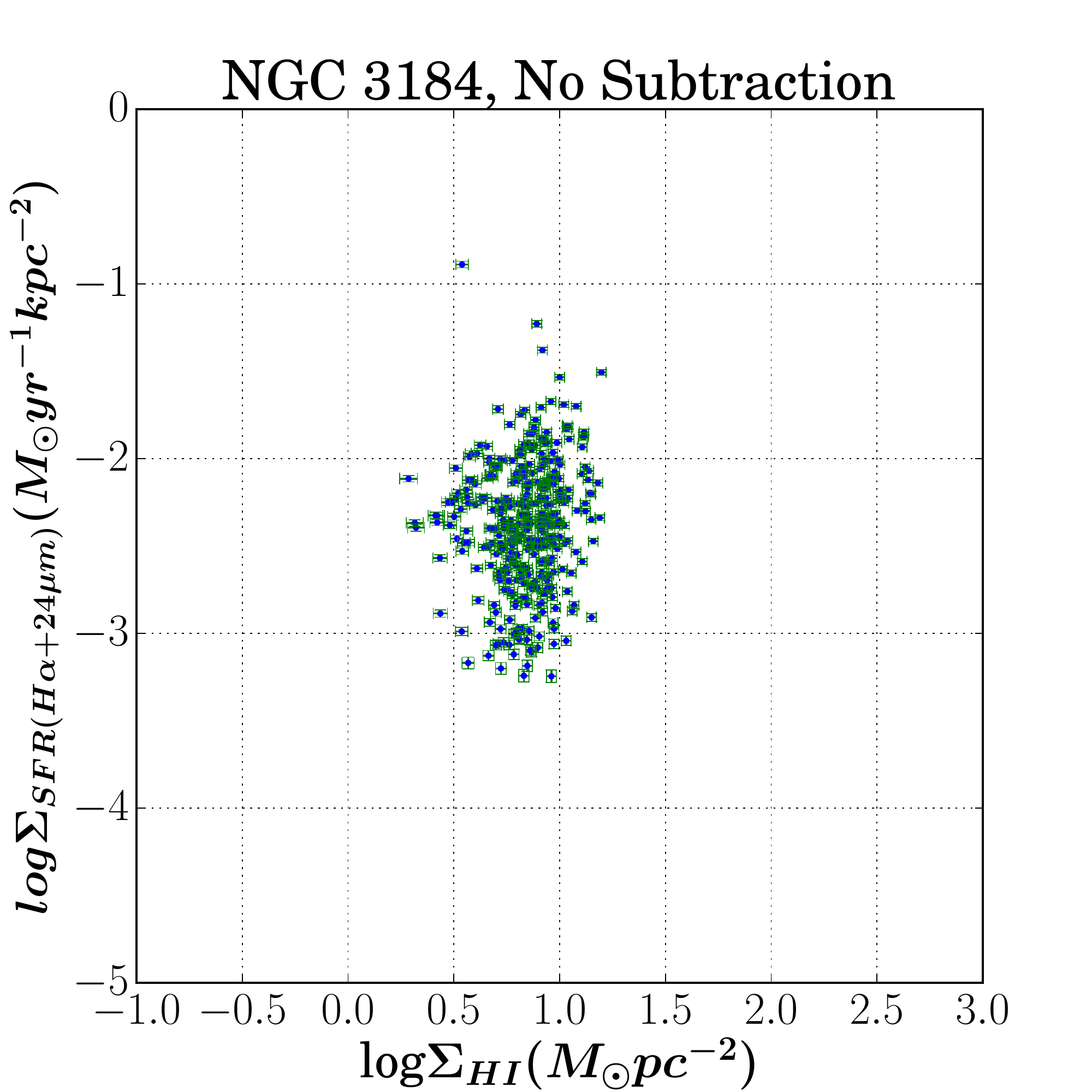}
		\includegraphics[width = 0.33\textwidth,trim={0.2cm 0cm 1.5cm 0cm},clip]{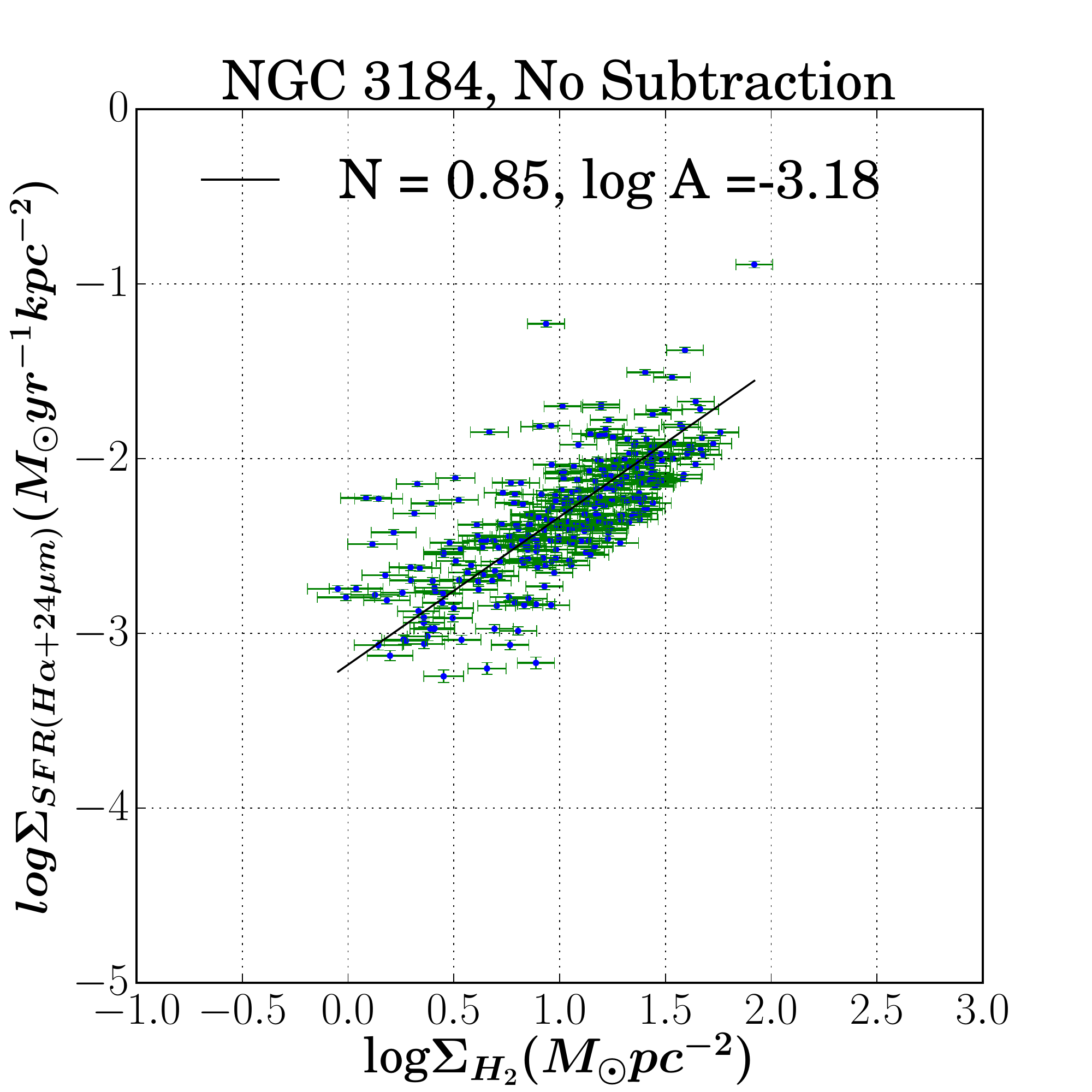}
		\includegraphics[width = 0.33\textwidth,trim={0.2cm 0cm 1.5cm 0cm},clip]{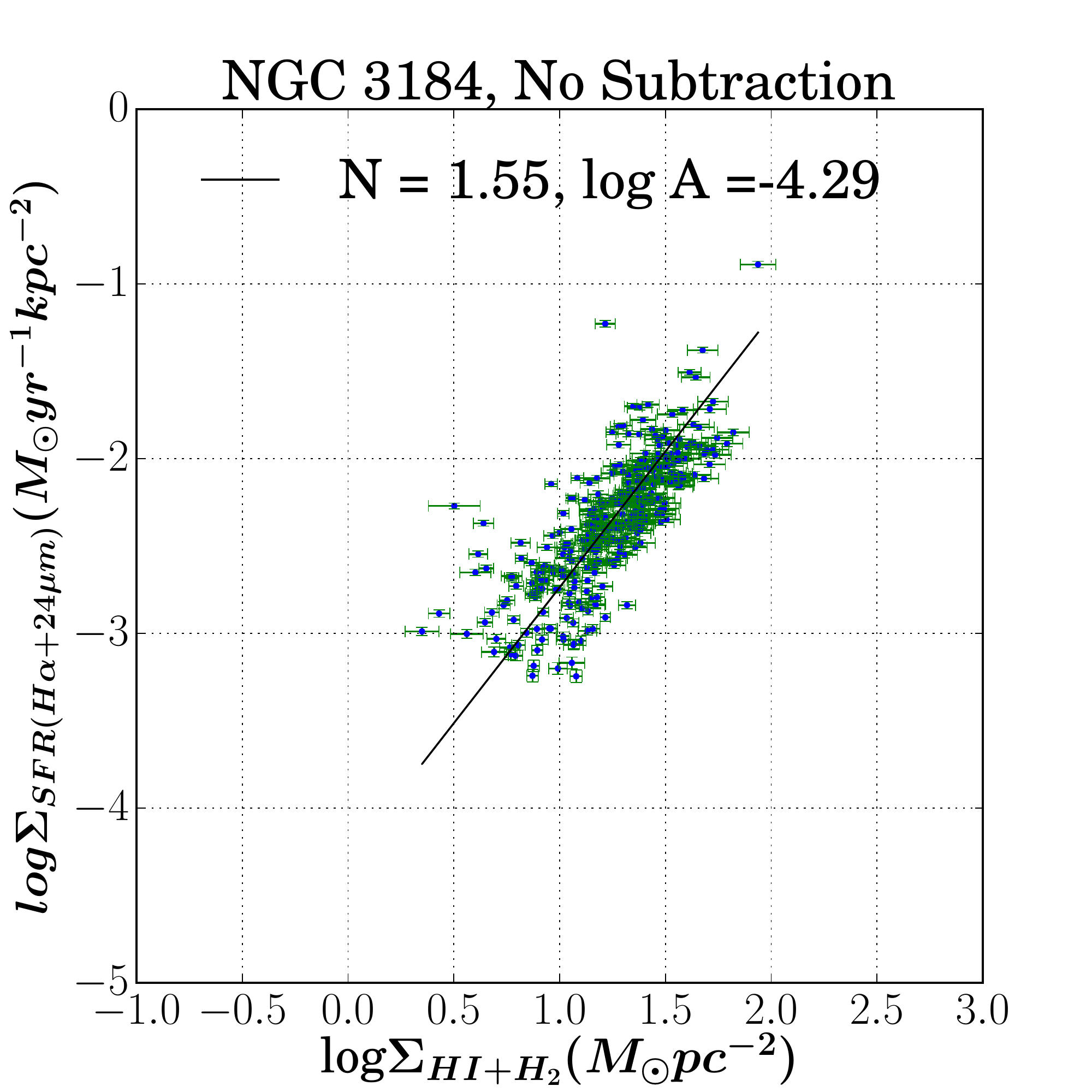}

		\includegraphics[width = 0.33\textwidth,trim={0.2cm 0cm 1.5cm 0cm},clip]{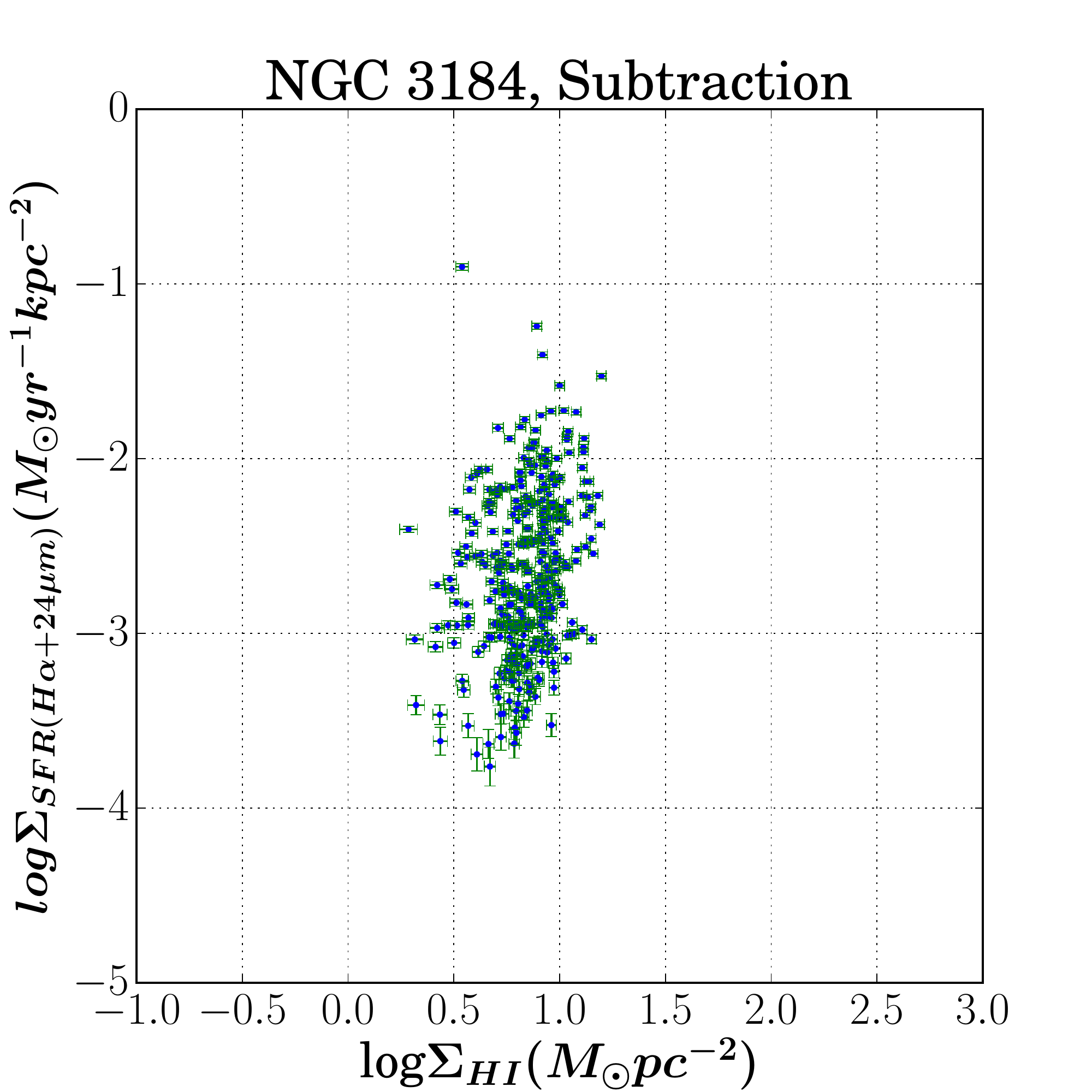}
		\includegraphics[width = 0.33\textwidth,trim={0.2cm 0cm 1.5cm 0cm},clip]{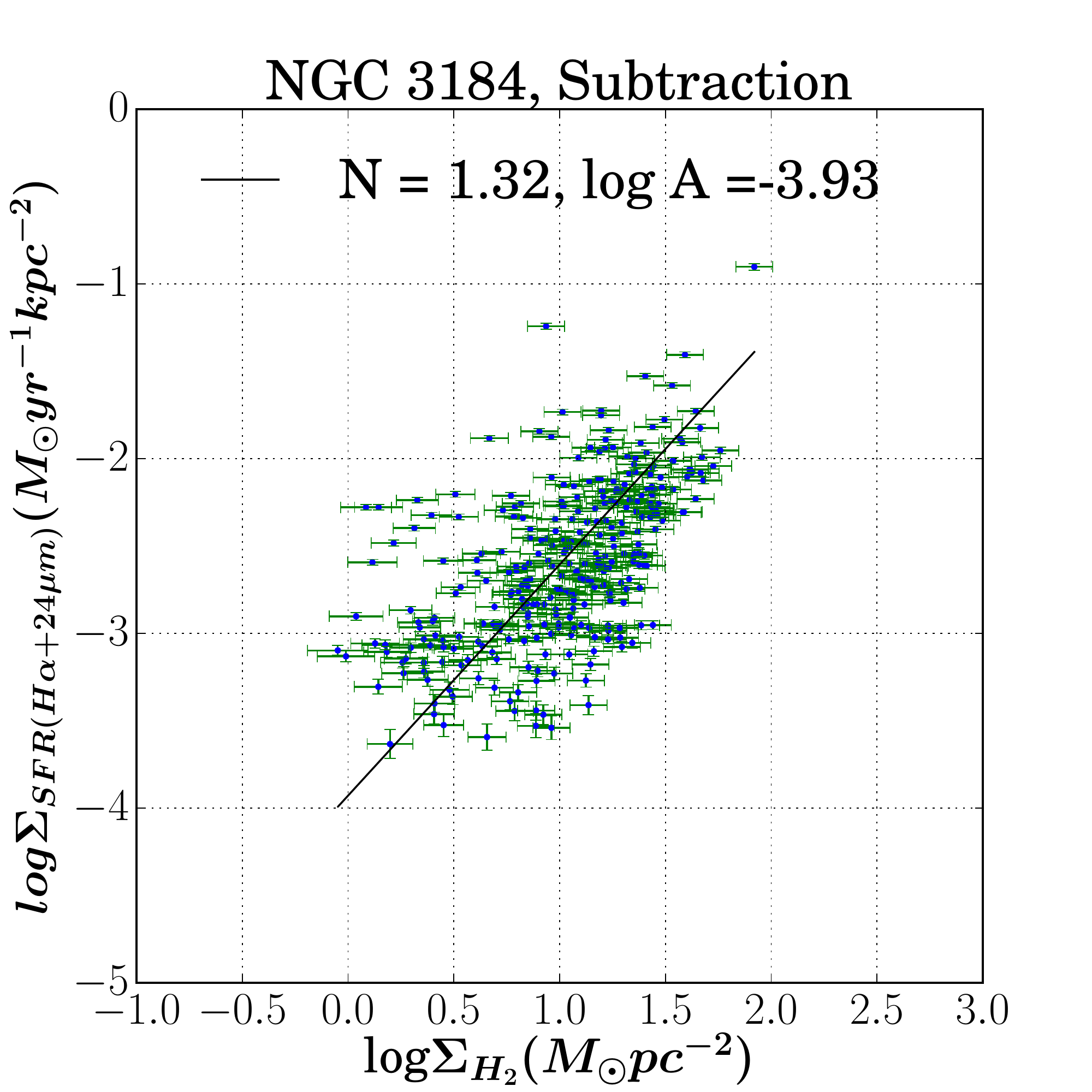}
		\includegraphics[width = 0.33\textwidth,trim={0.2cm 0cm 1.5cm 0cm},clip]{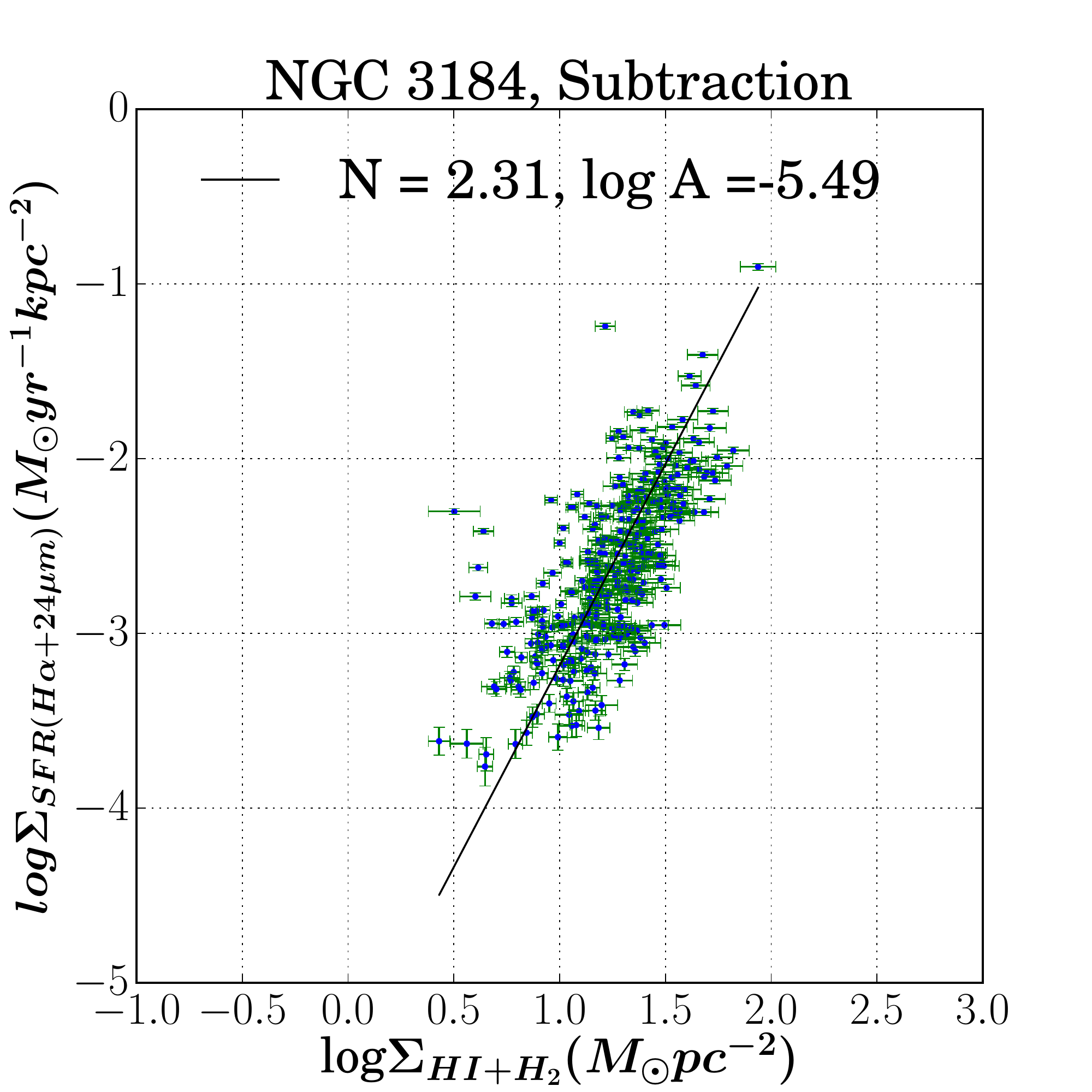}
		\caption{NGC 3184: an aperture size of 10\arcsec is adopted, which corresponds to a physical diameter of $\sim$ 520 pc for NGC 3184, at a distance of 11.1 Mpc and inclination angle of  18$^{\circ}$. See caption of Fig. \ref{Figure: NGC 0628} for details.}
		\label{NGC 3184}
	\end{figure*}

\begin{figure*}
		\centering
		\includegraphics[width = 0.33\textwidth,trim={0.2cm 0cm 1.5cm 0cm},clip]{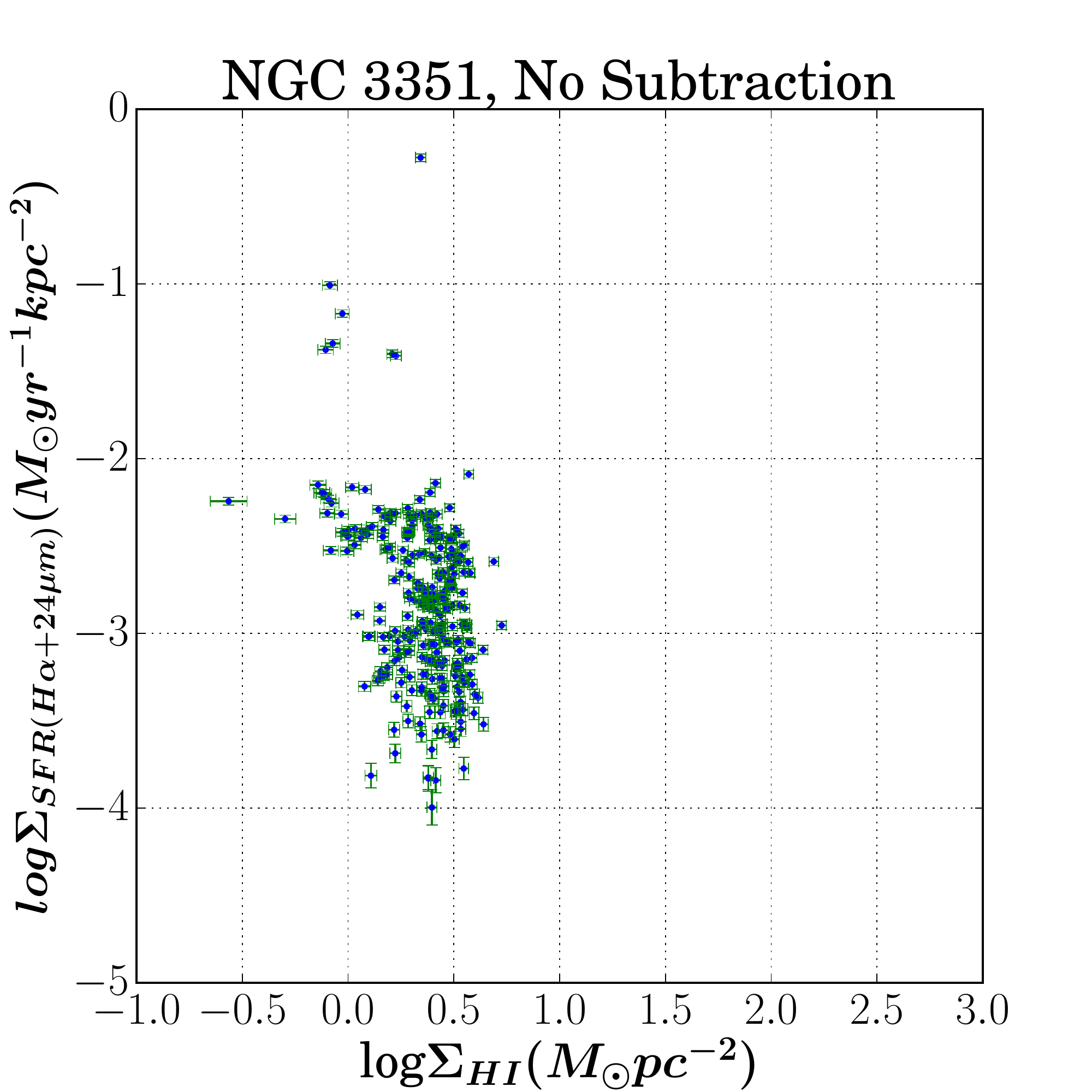}
		\includegraphics[width = 0.33\textwidth,trim={0.2cm 0cm 1.5cm 0cm},clip]{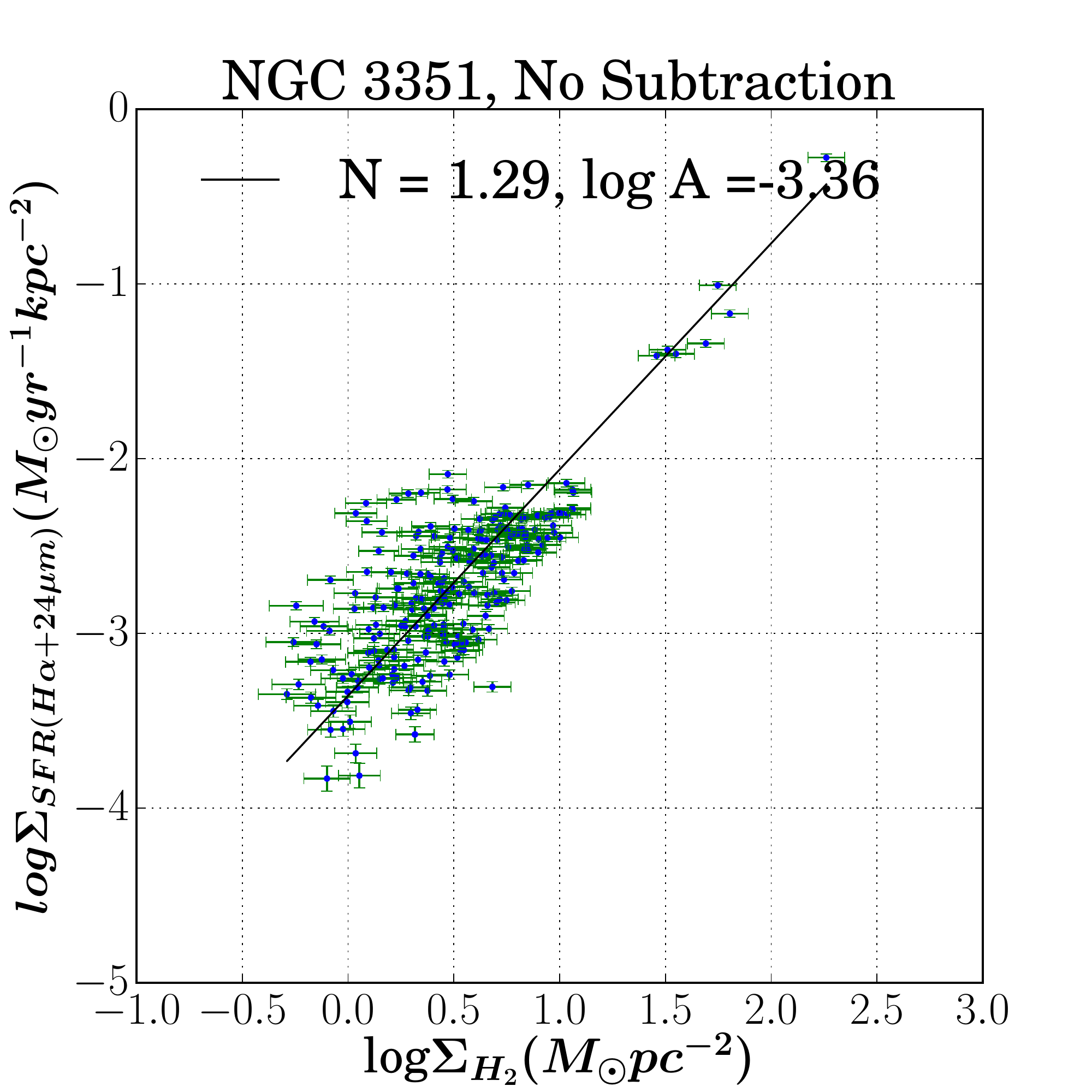}
		\includegraphics[width = 0.33\textwidth,trim={0.2cm 0cm 1.5cm 0cm},clip]{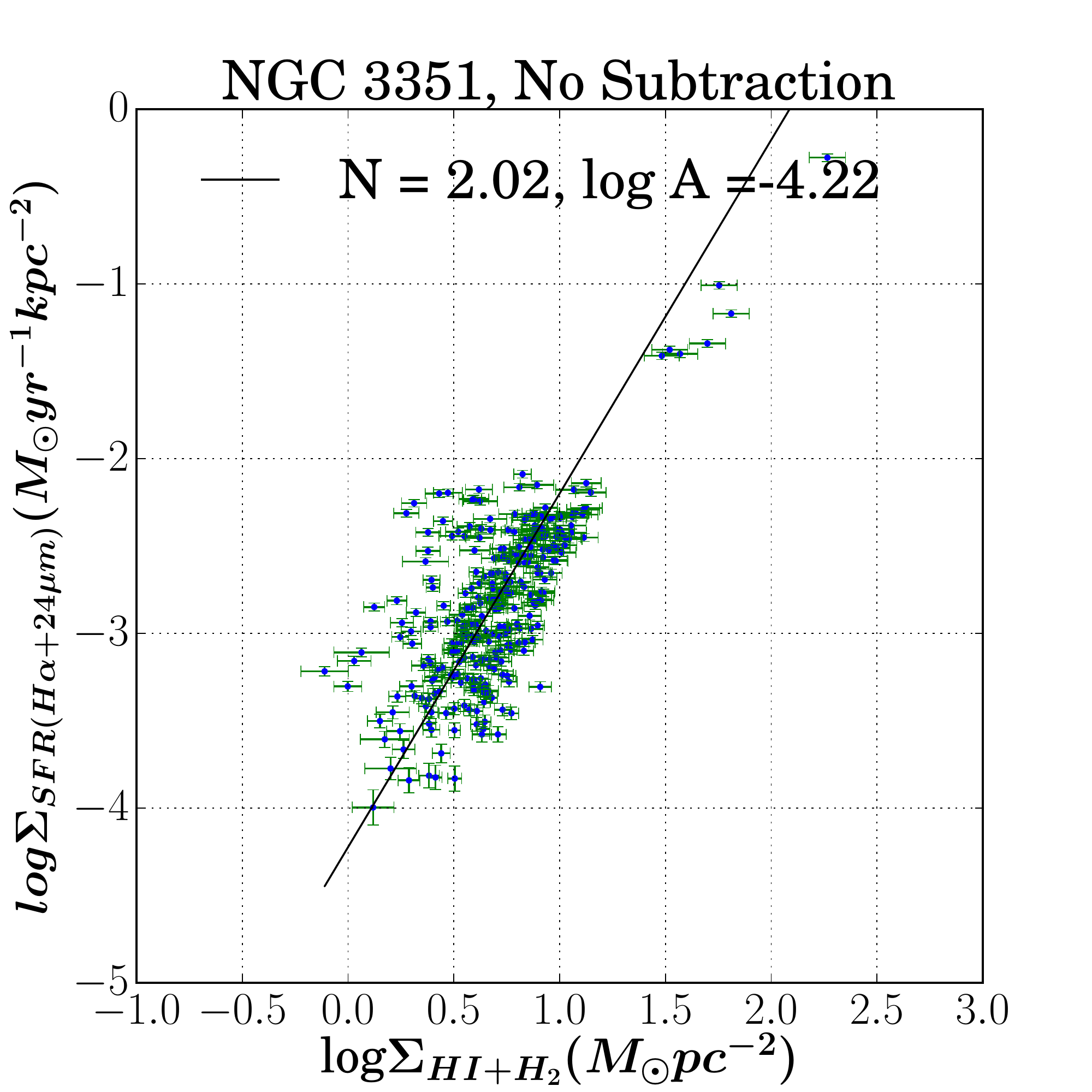}
		\includegraphics[width = 0.33\textwidth,trim={0.2cm 0cm 1.5cm 0cm},clip]{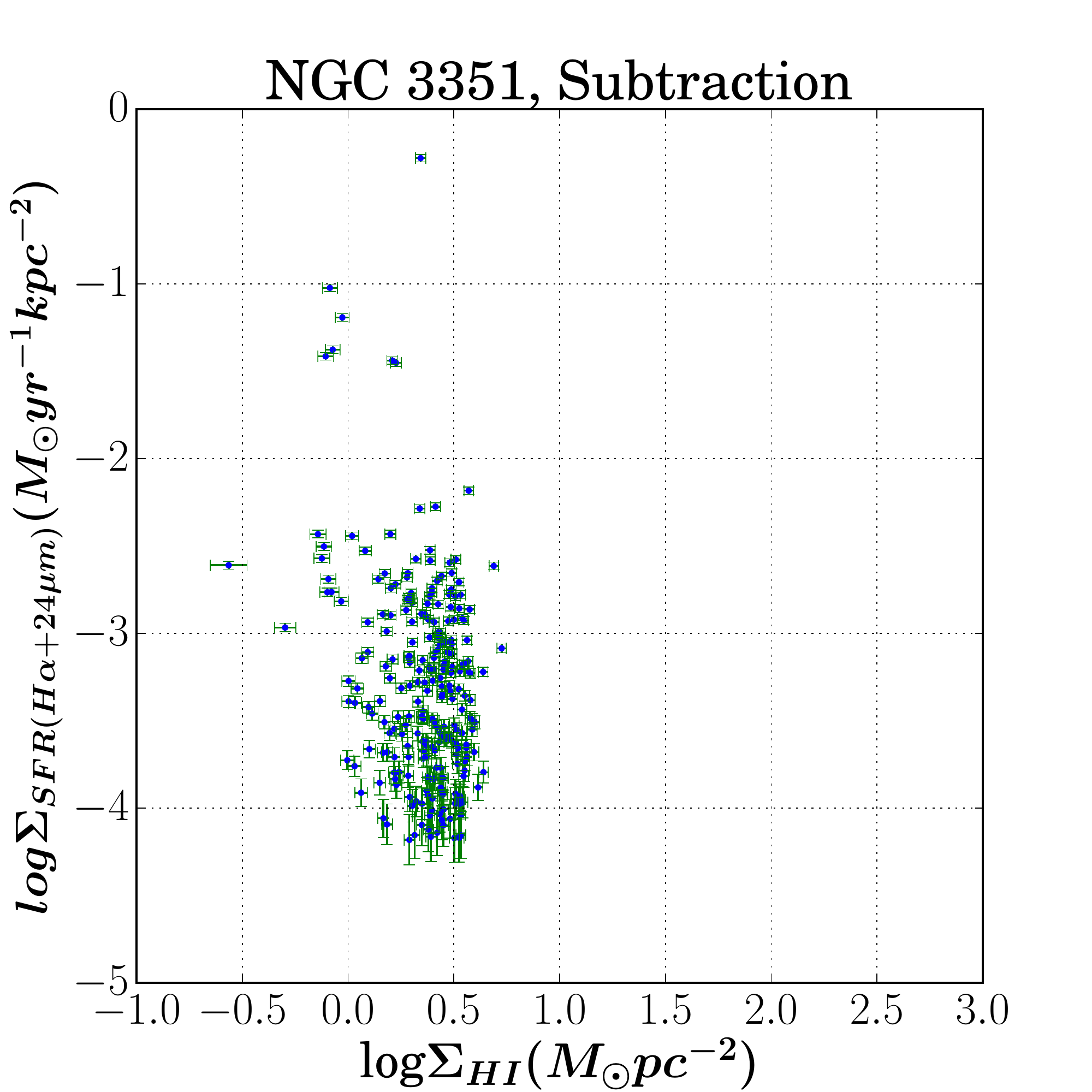}
		\includegraphics[width = 0.33\textwidth,trim={0.2cm 0cm 1.5cm 0cm},clip]{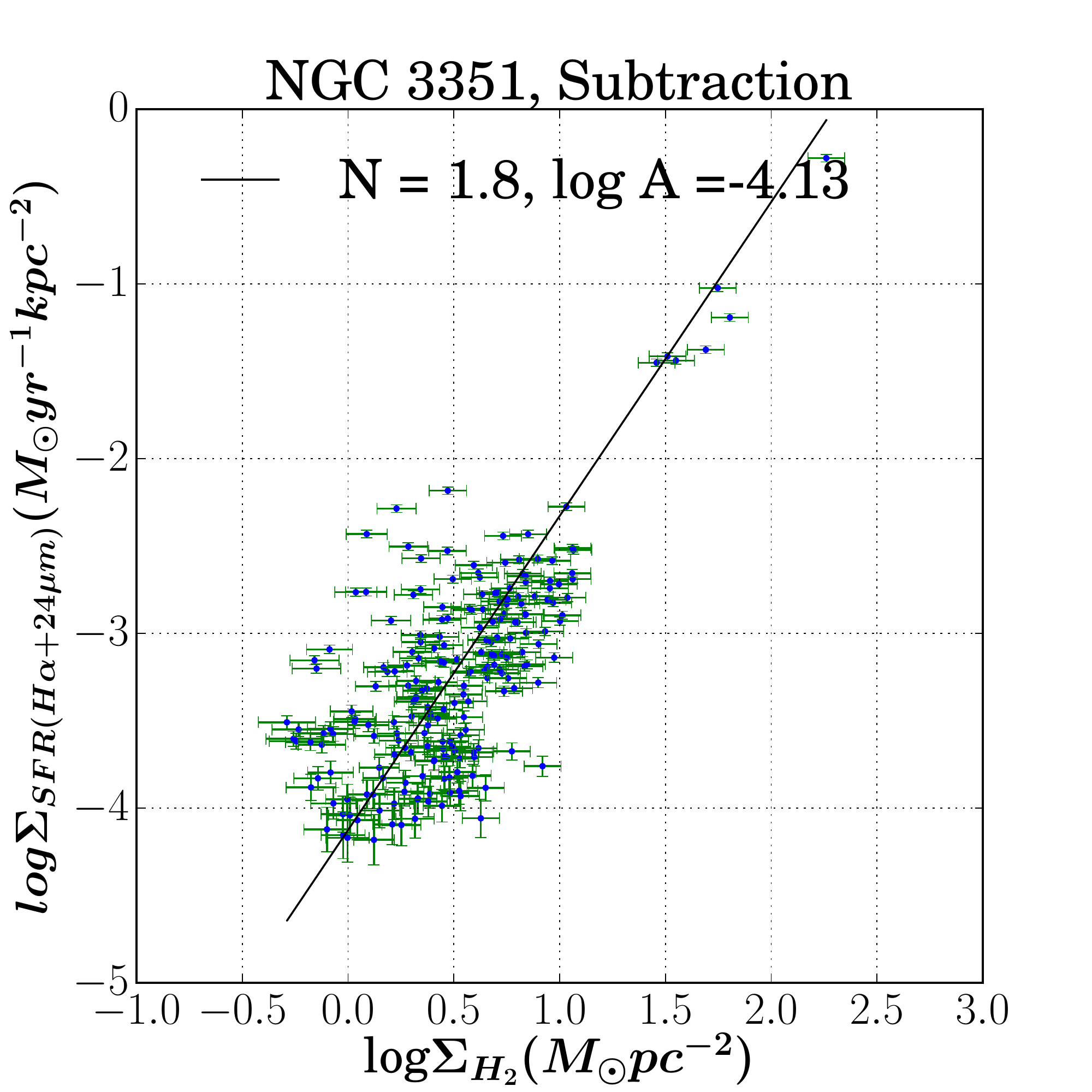}
		\includegraphics[width = 0.33\textwidth,trim={0.2cm 0cm 1.5cm 0cm},clip]{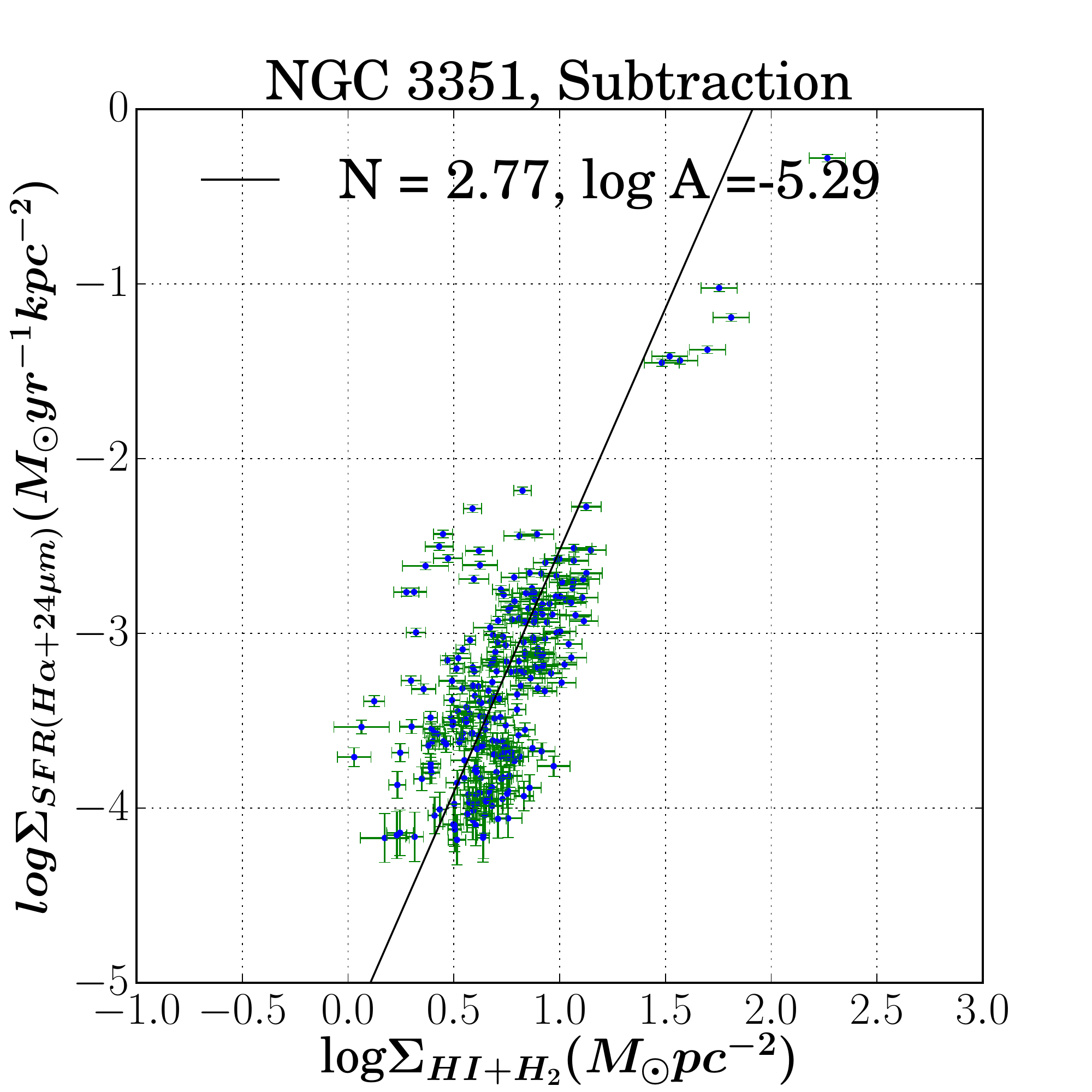}
		\caption{NGC 3351: an aperture size of 16\arcsec is adopted, which corresponds to the physical diameter of $\sim$ 1.0 kpc at a distance of  9.33 Mpc and inclination angle of  41$^{\circ}$.  See caption of Fig. \ref{Figure: NGC 0628} for details.}
		\label{NGC 3351}
	\end{figure*}

\begin{figure*}
		\centering
		\includegraphics[width = 0.33\textwidth,trim={0.2cm 0cm 1.5cm 0cm},clip]{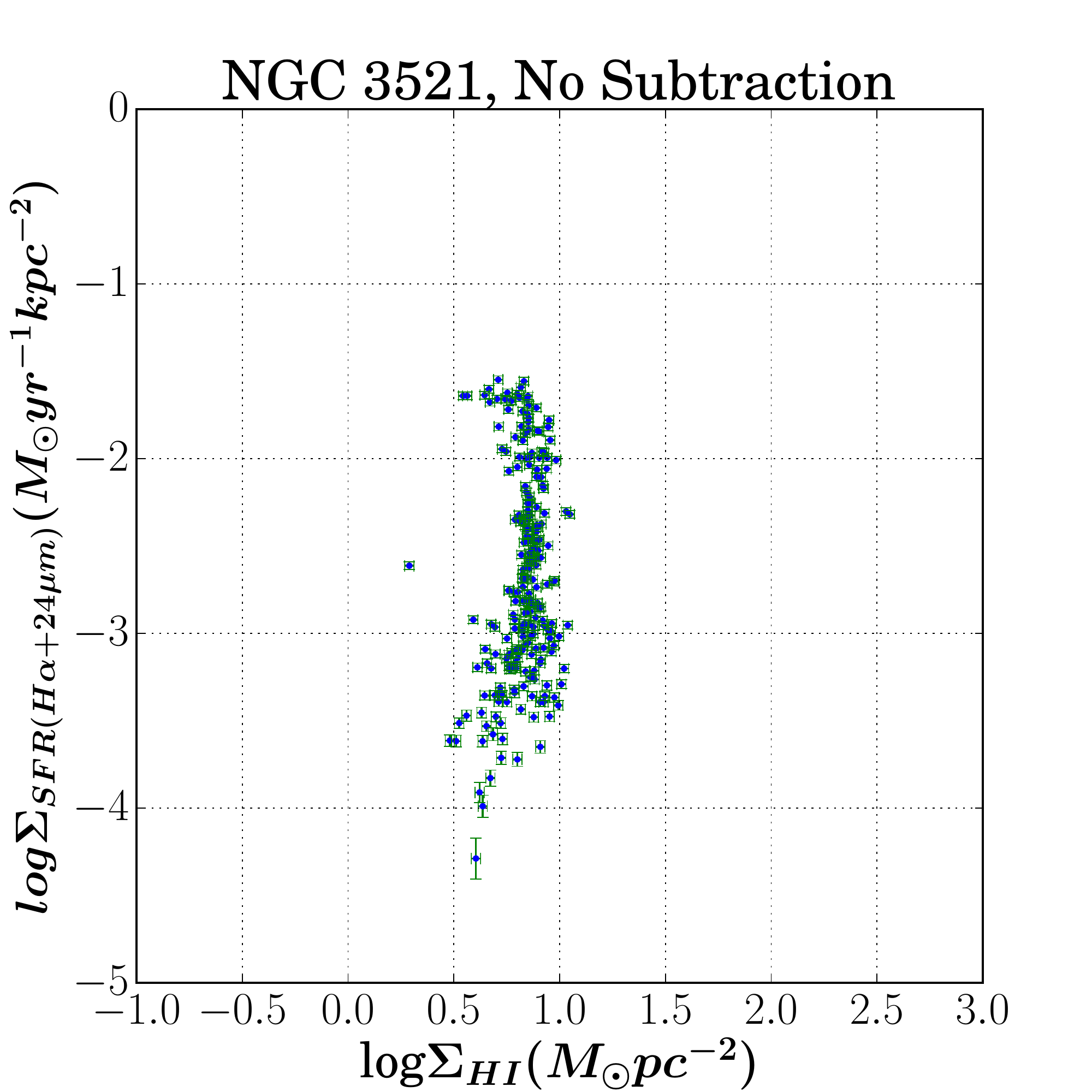}
		\includegraphics[width = 0.33\textwidth,trim={0.2cm 0cm 1.5cm 0cm},clip]{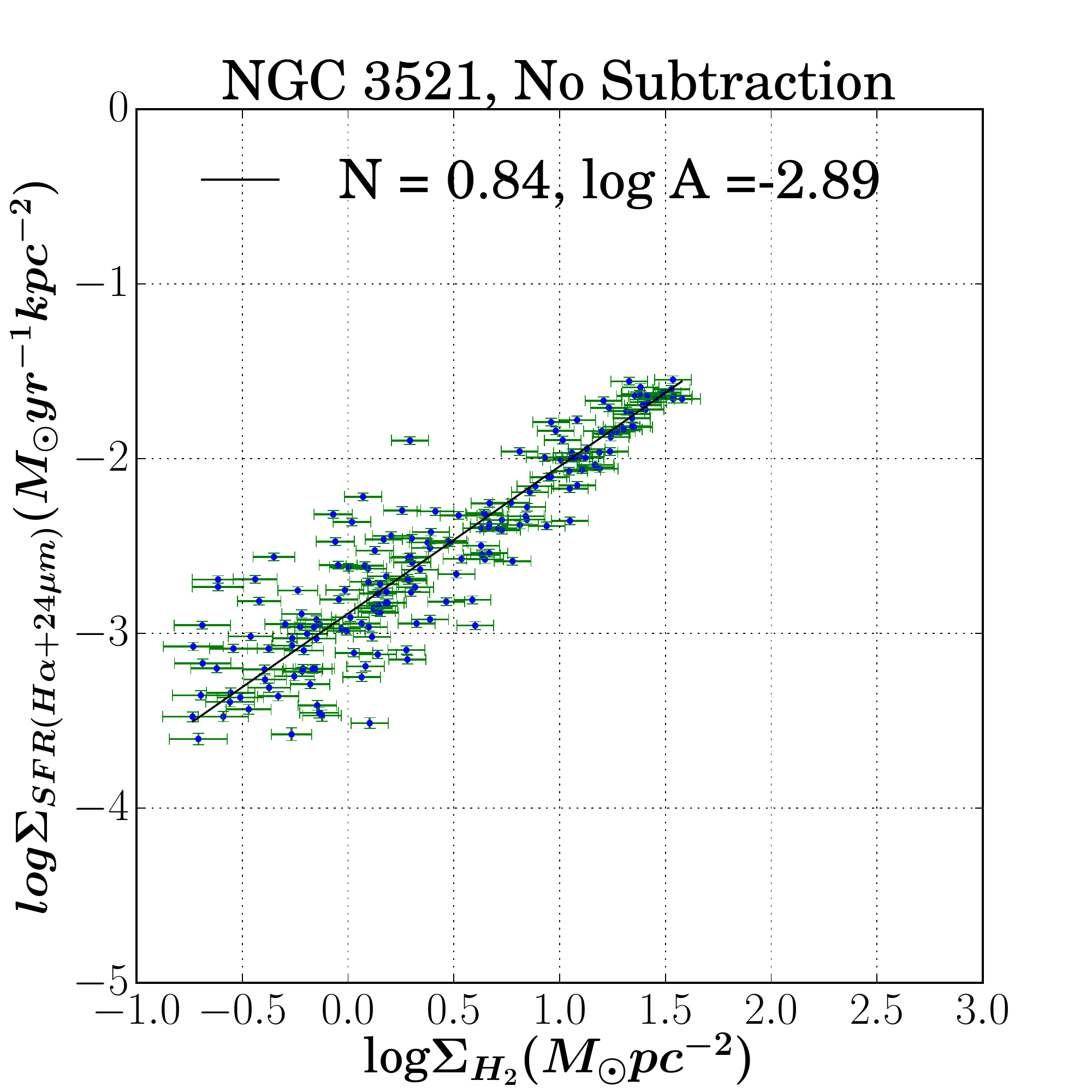}
		\includegraphics[width = 0.33\textwidth,trim={0.2cm 0cm 1.5cm 0cm},clip]{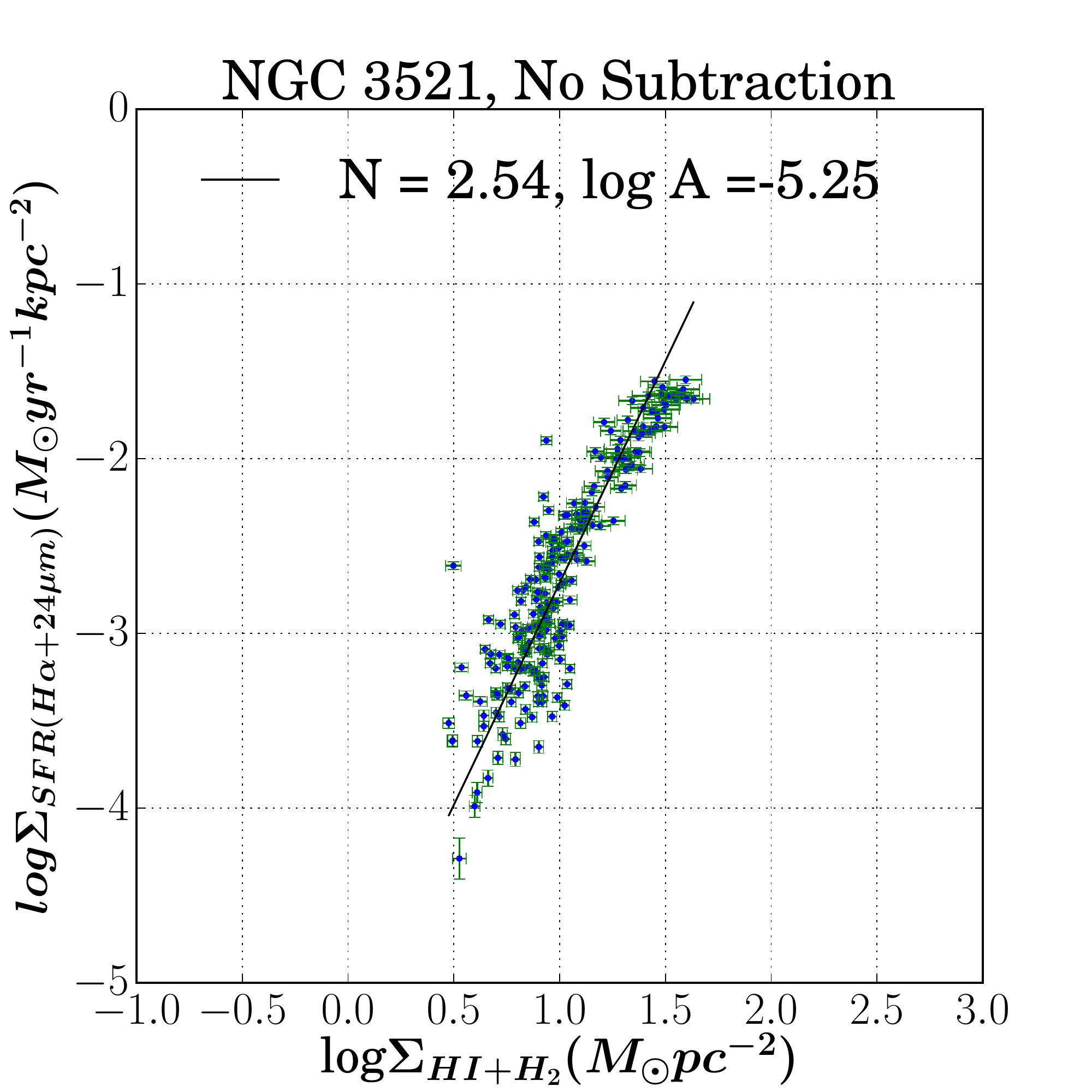}
		\includegraphics[width = 0.33\textwidth,trim={0.2cm 0cm 1.5cm 0cm},clip]{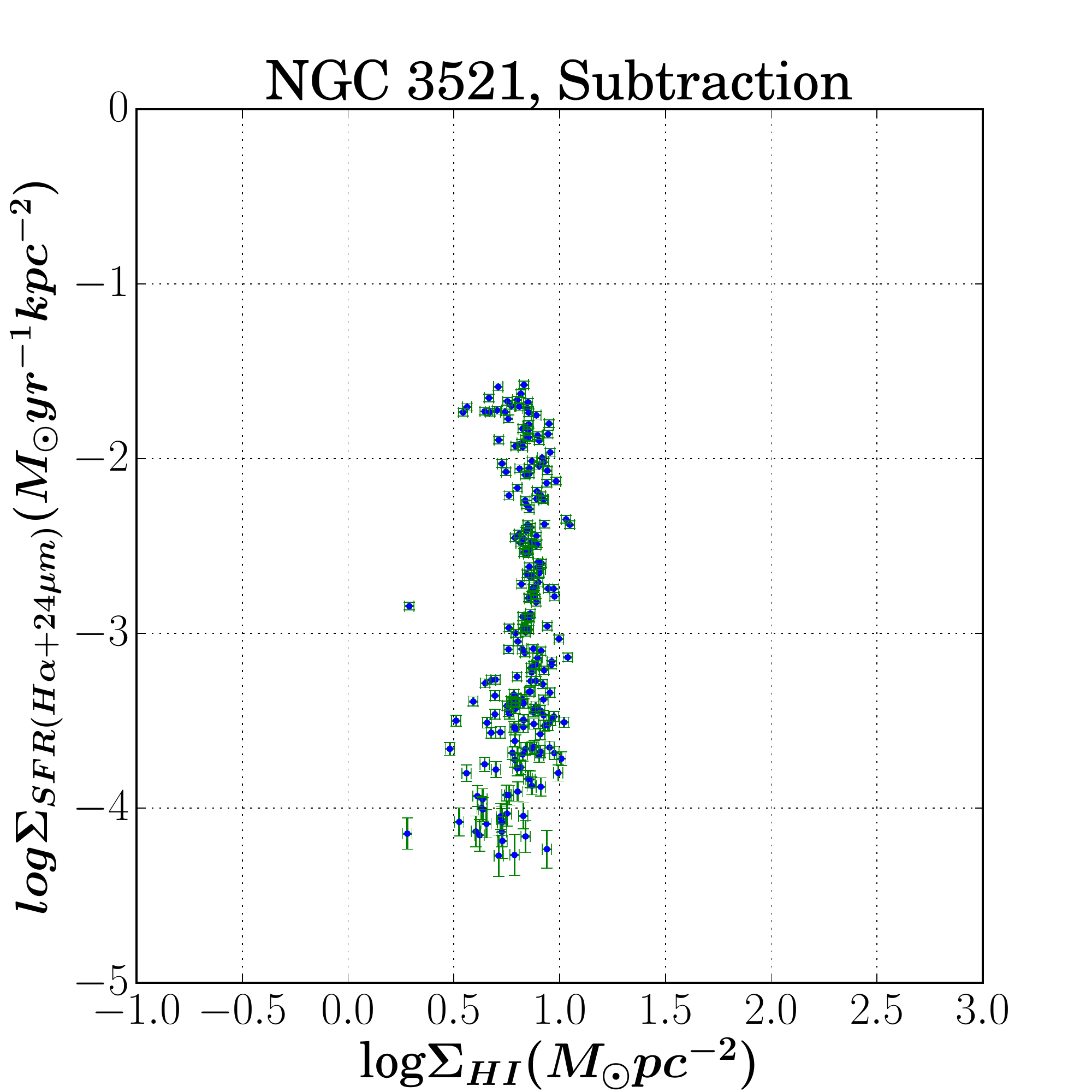}
		\includegraphics[width = 0.33\textwidth,trim={0.2cm 0cm 1.5cm 0cm},clip]{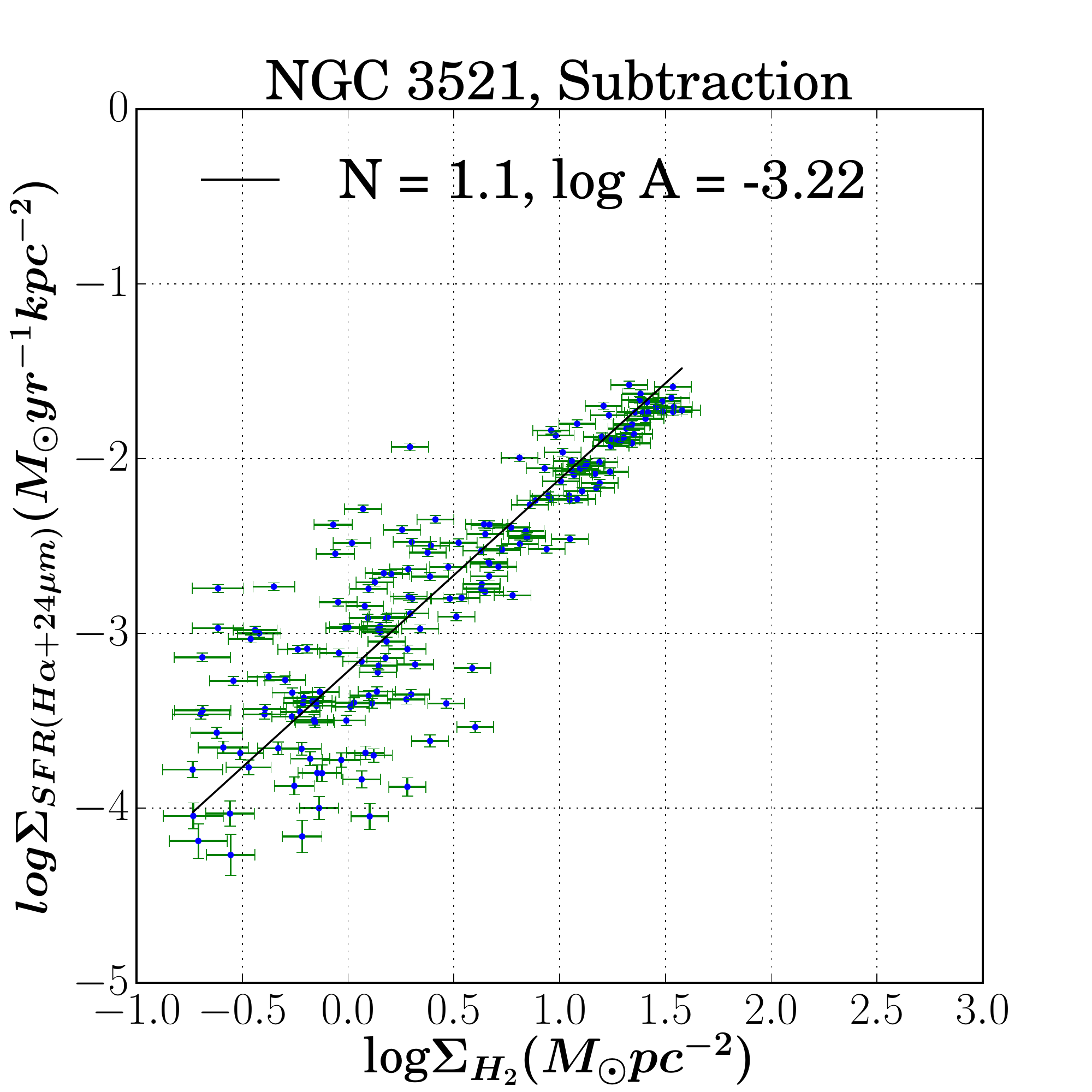}
		\includegraphics[width = 0.33\textwidth,trim={0.2cm 0cm 1.5cm 0cm},clip]{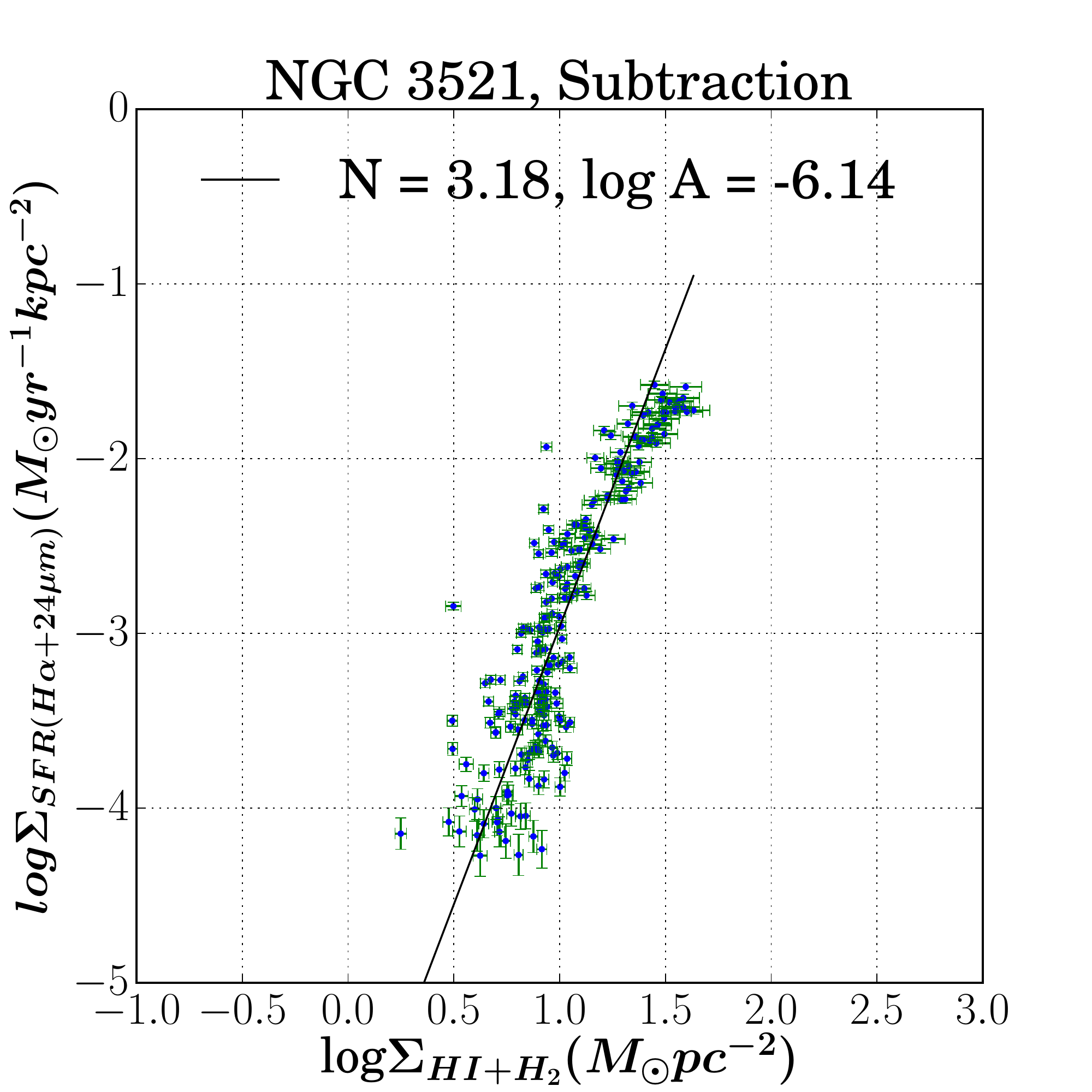}
		\caption{NGC 3521: an aperture size of 18\arcsec is adopted, which corresponds to the physical diameter of $\sim$ 2.0 kpc at a distance of 8.03 Mpc and inclination angle of  73$^{\circ}$. See caption of Fig. \ref{Figure: NGC 0628} for details.}
		\label{NGC 3521}
	\end{figure*}

\begin{figure*}
		\centering
		\includegraphics[width = 0.33\textwidth,trim={0.2cm 0cm 1.5cm 0cm},clip]{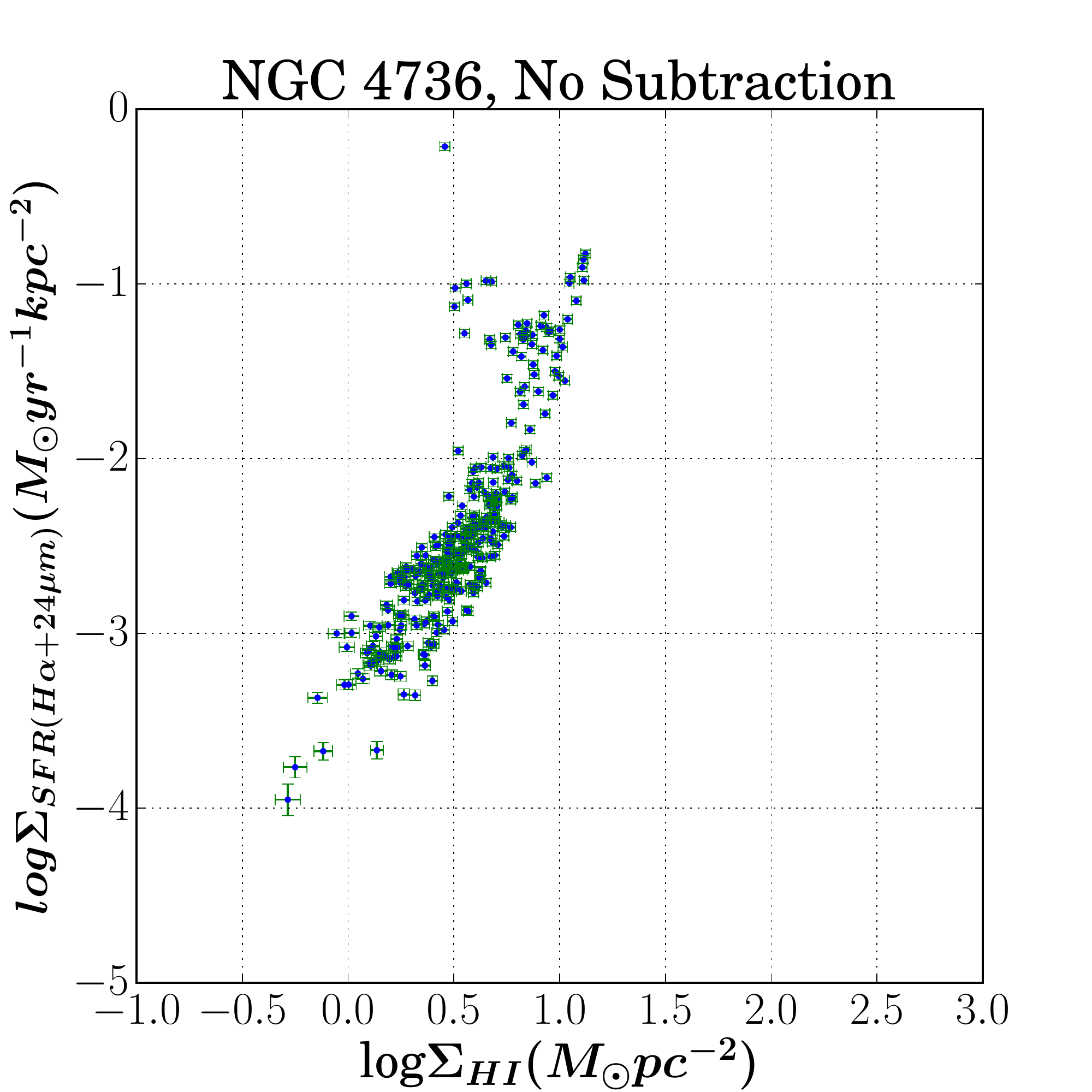}
		\includegraphics[width = 0.33\textwidth,trim={0.2cm 0cm 1.5cm 0cm},clip]{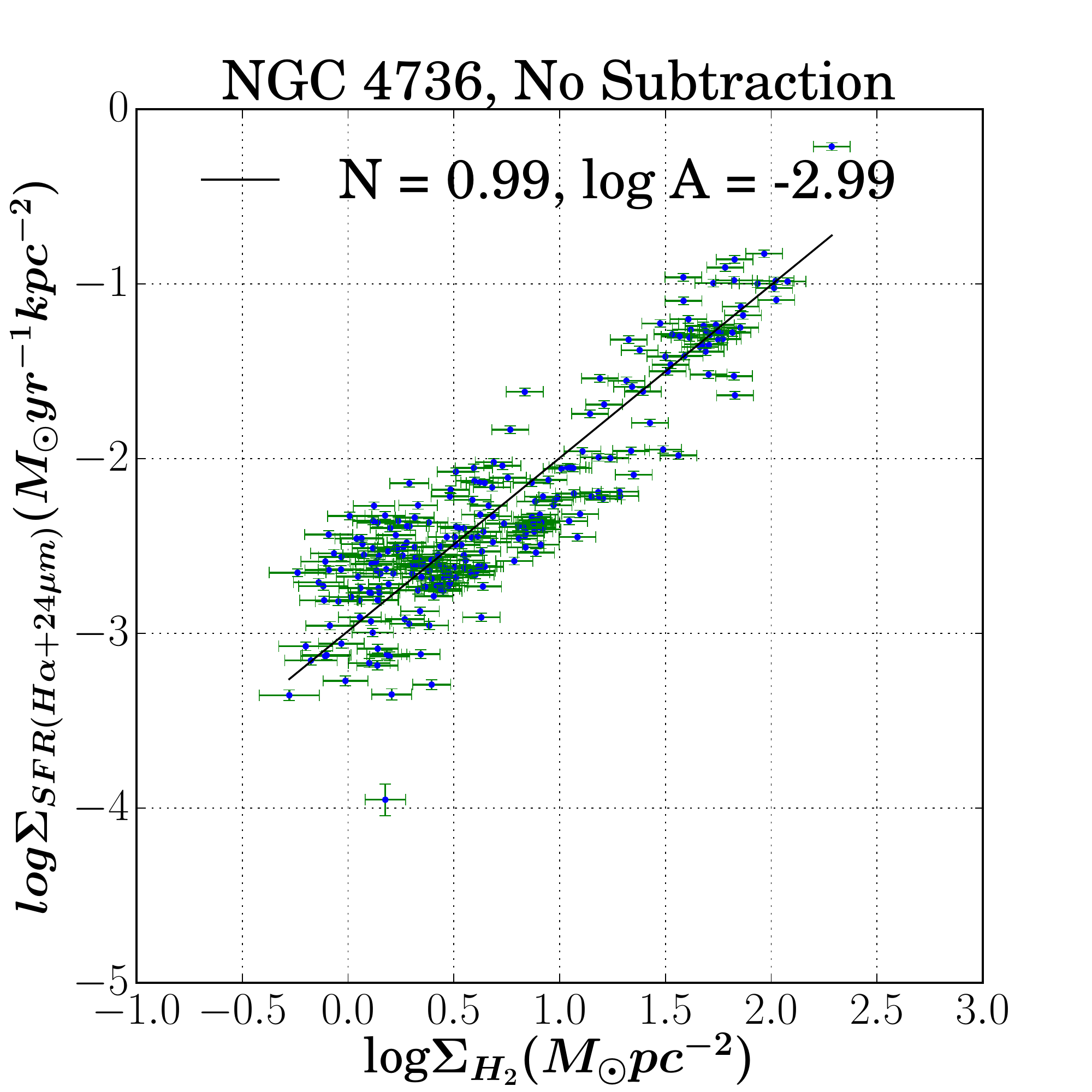}
		\includegraphics[width = 0.33\textwidth,trim={0.2cm 0cm 1.5cm 0cm},clip]{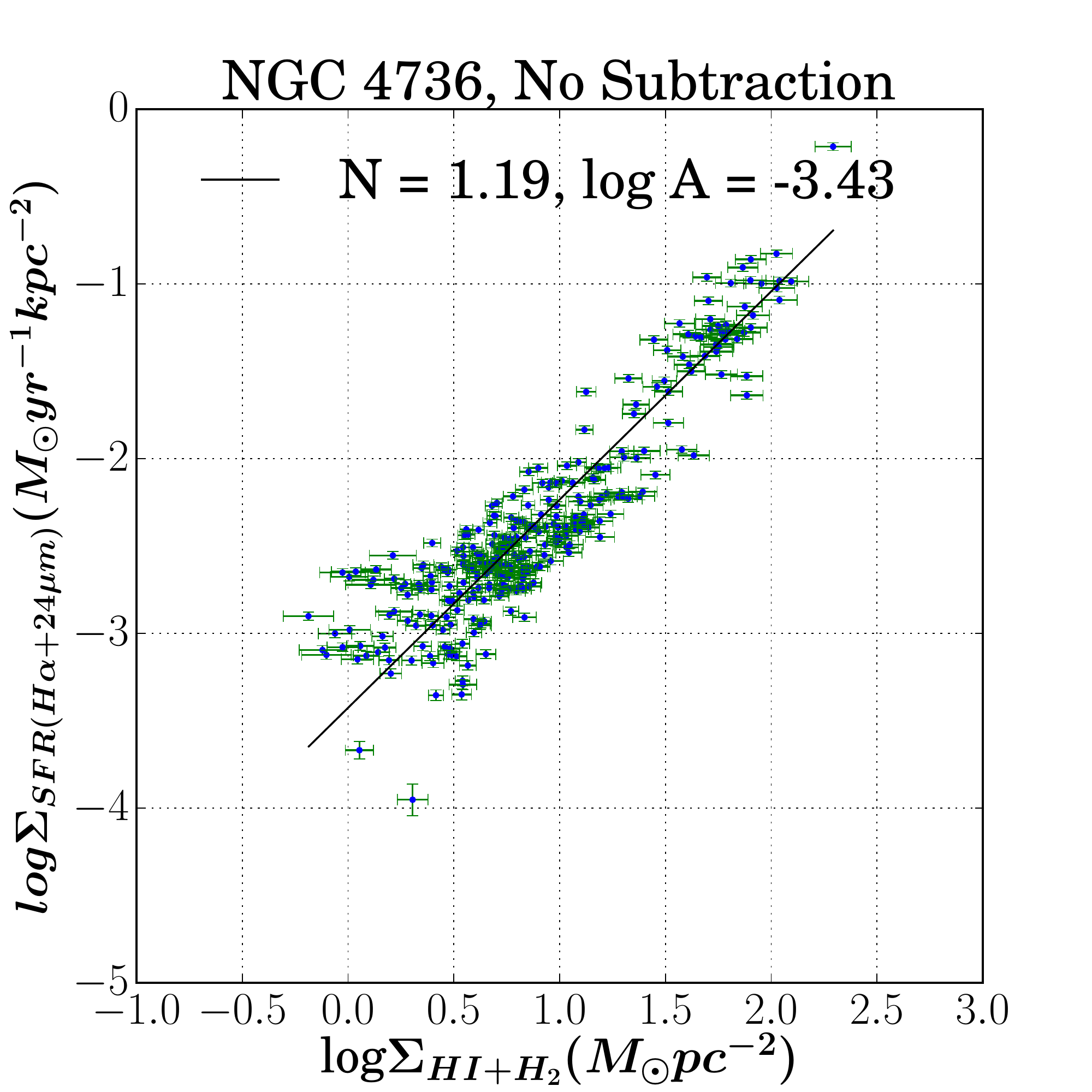}
		\includegraphics[width = 0.33\textwidth,trim={0.2cm 0cm 1.5cm 0cm},clip]{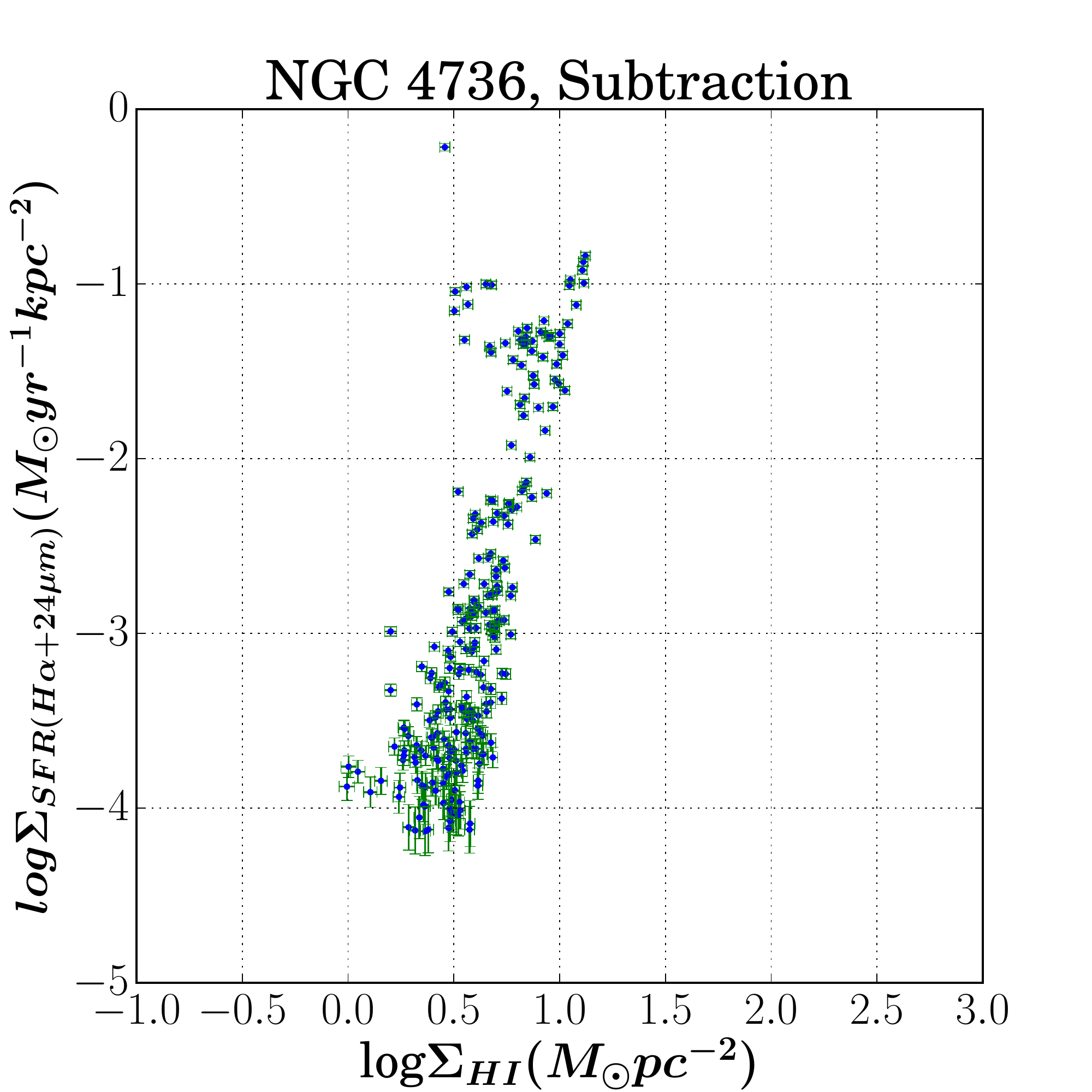}
		\includegraphics[width = 0.33\textwidth,trim={0.2cm 0cm 1.5cm 0cm},clip]{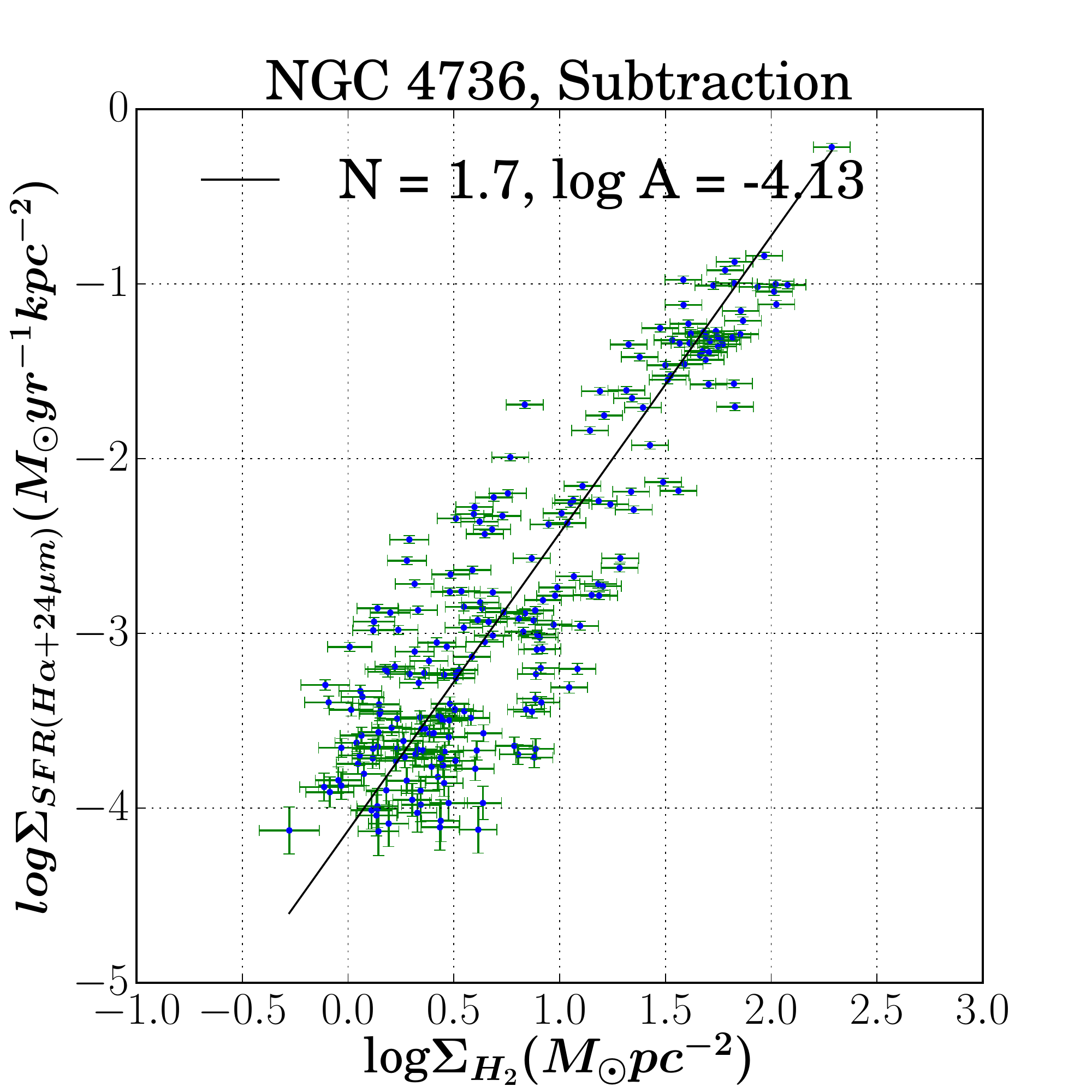}
		\includegraphics[width = 0.33\textwidth,trim={0.2cm 0cm 1.5cm 0cm},clip]{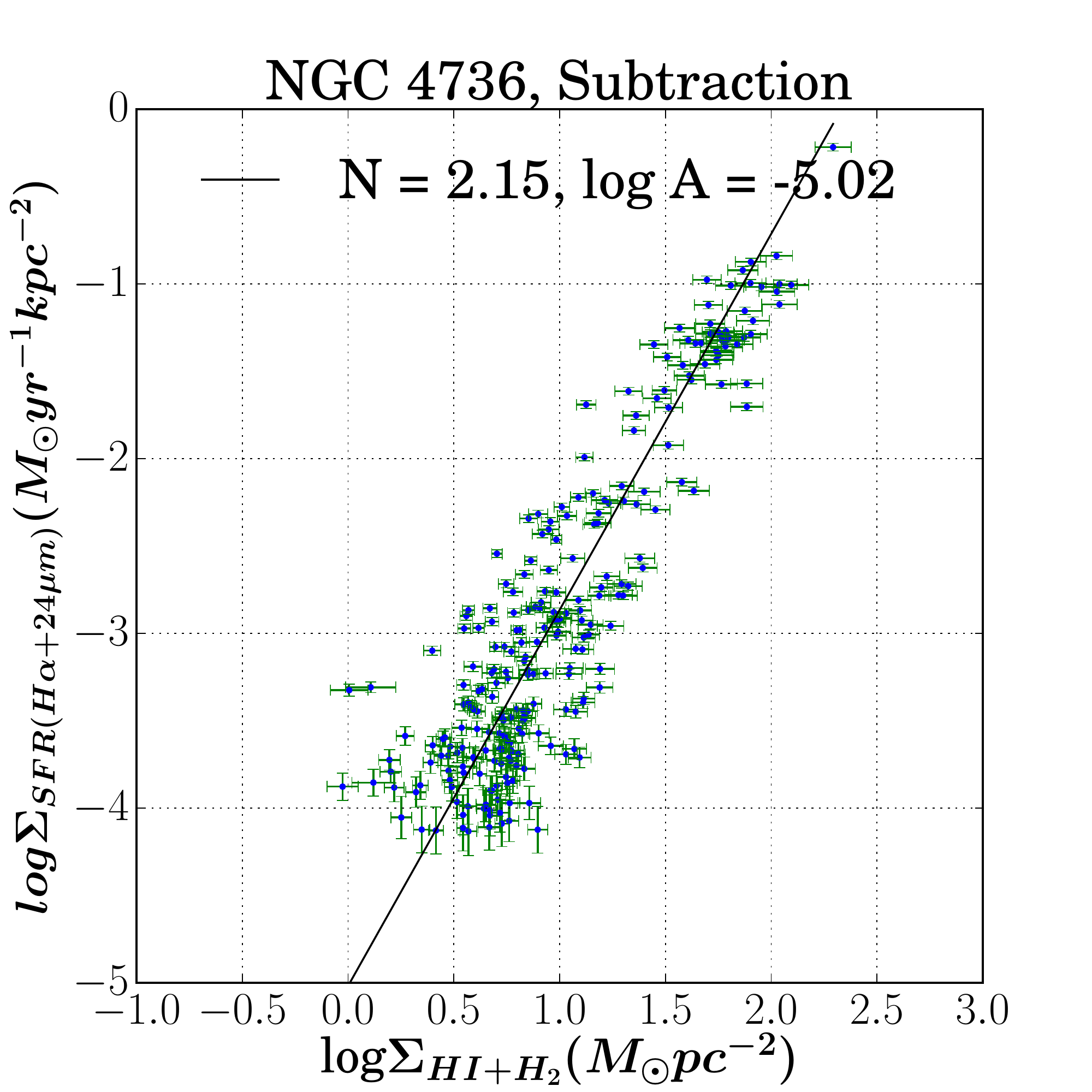}
		\caption{NGC 4736: an aperture size of 15\arcsec is adopted, which corresponds to the physical diameter of $\sim$ 500 pc at a distance of 5.20 Mpc and inclination angle of  41$^{\circ}$.  See caption of Fig. \ref{Figure: NGC 0628} for details.}
		\label{NGC 4736}
	\end{figure*}

\begin{figure*}
		\centering
		\includegraphics[width = 0.33\textwidth,trim={0.2cm 0cm 1.5cm 0cm},clip]{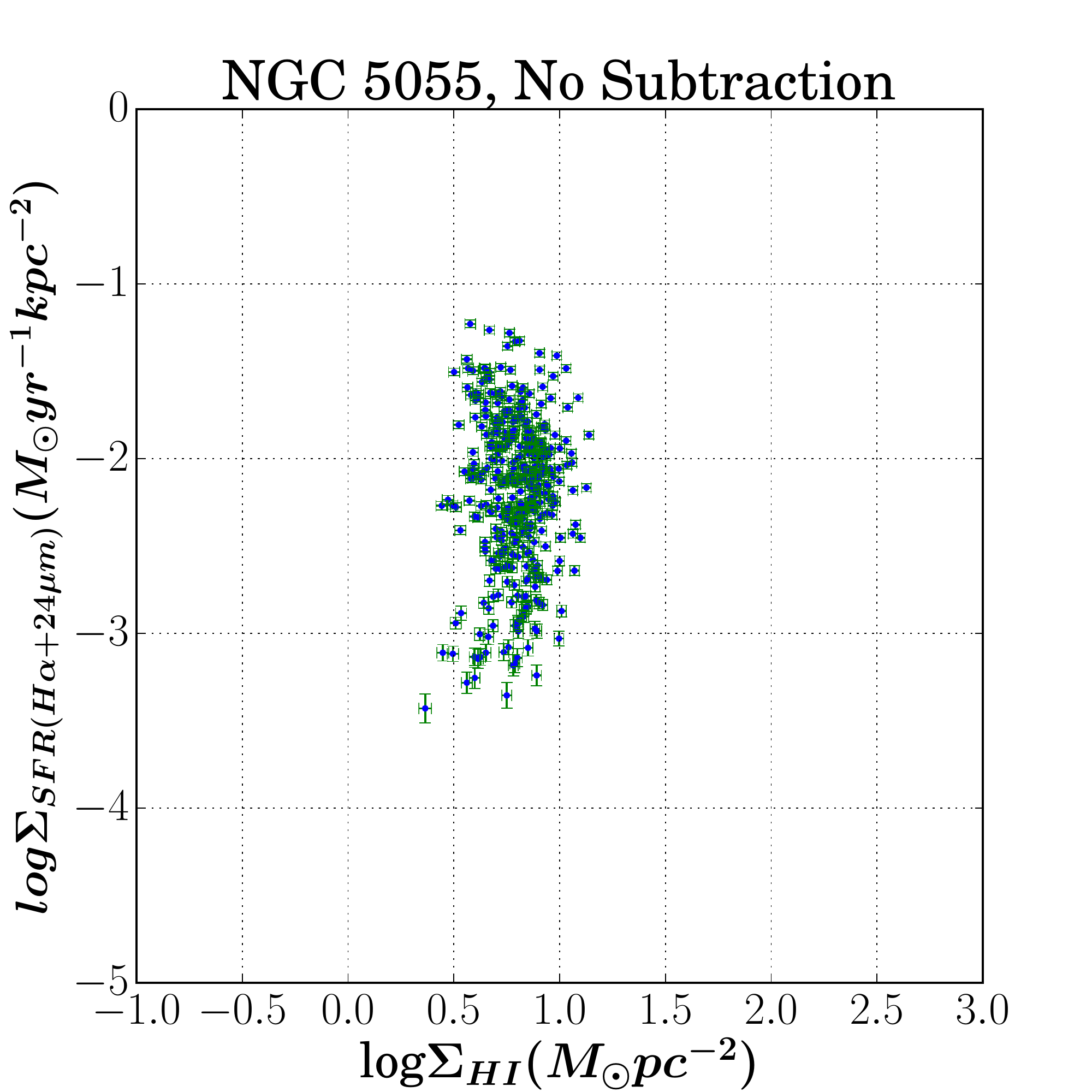}
		\includegraphics[width = 0.33\textwidth,trim={0.2cm 0cm 1.5cm 0cm},clip]{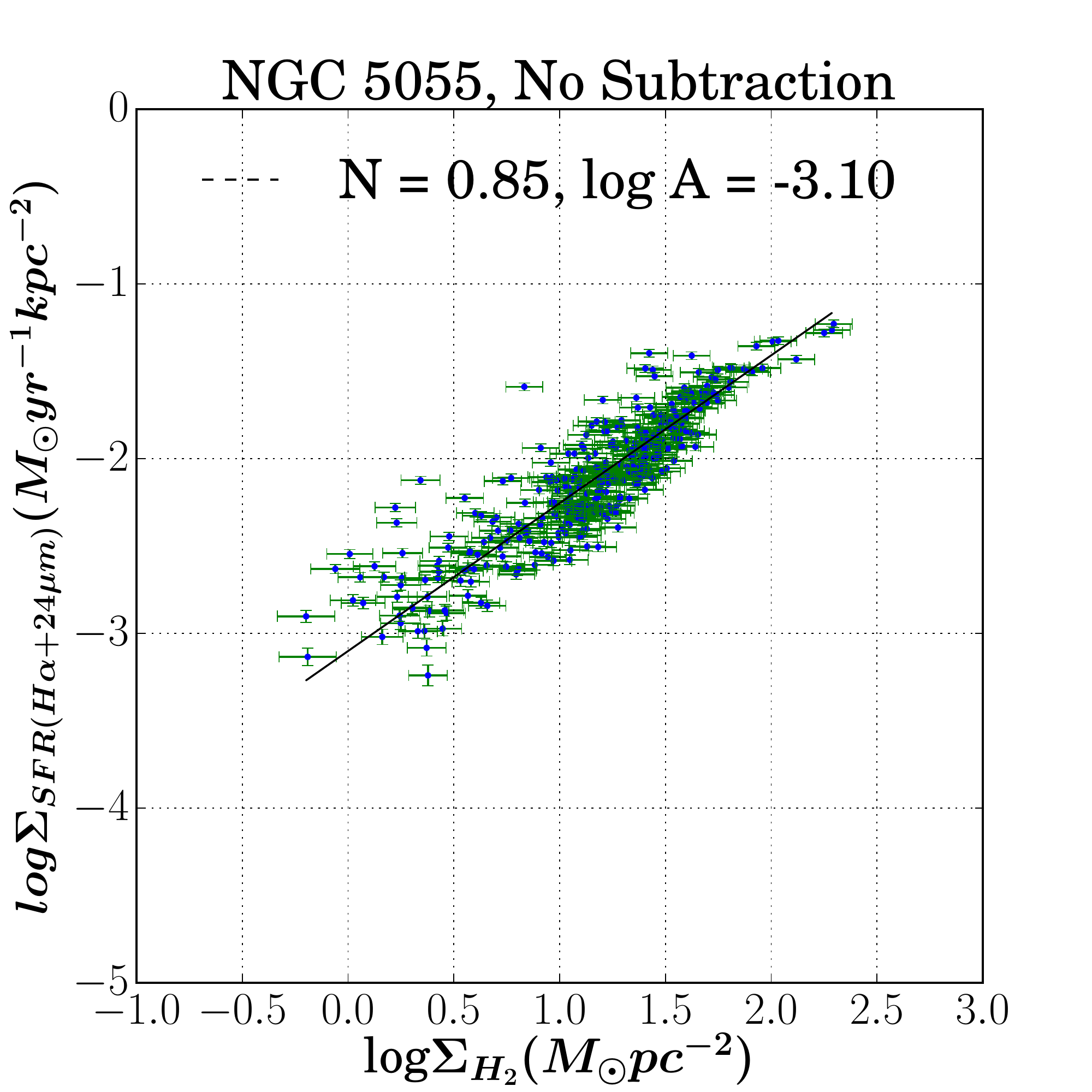}
		\includegraphics[width = 0.33\textwidth,trim={0.2cm 0cm 1.5cm 0cm},clip]{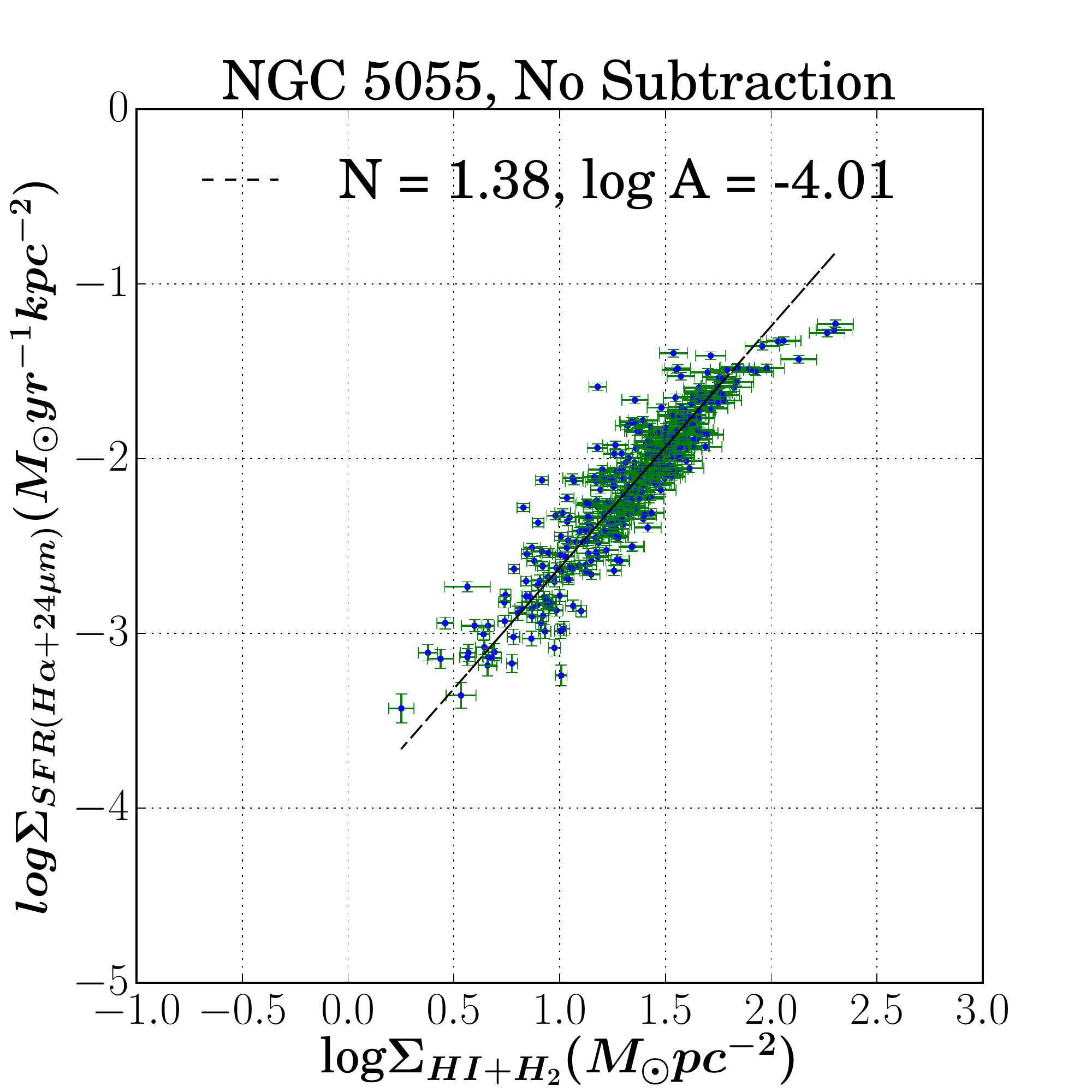}
		\includegraphics[width = 0.33\textwidth,trim={0.2cm 0cm 1.5cm 0cm},clip]{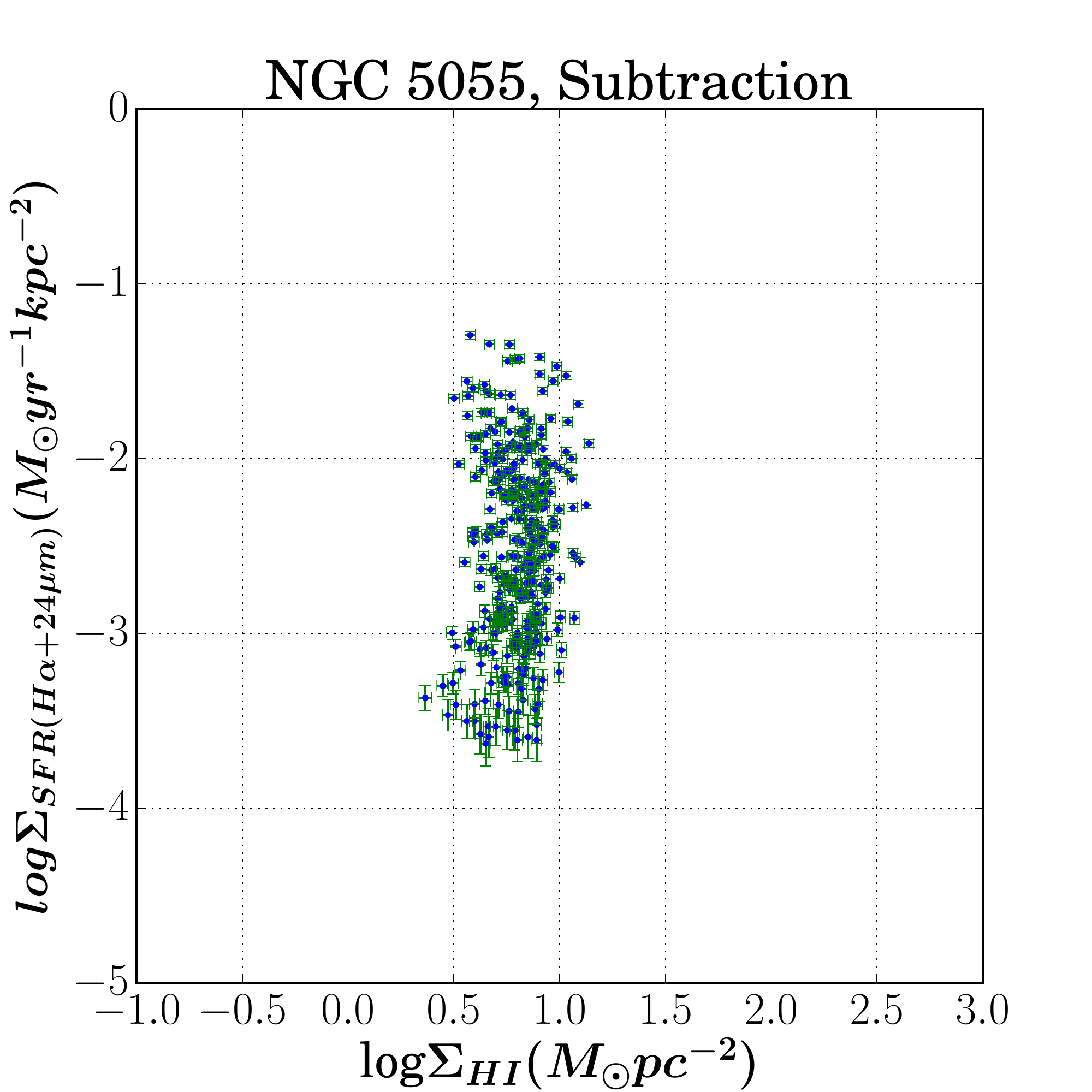}
		\includegraphics[width = 0.33\textwidth,trim={0.2cm 0cm 1.5cm 0cm},clip]{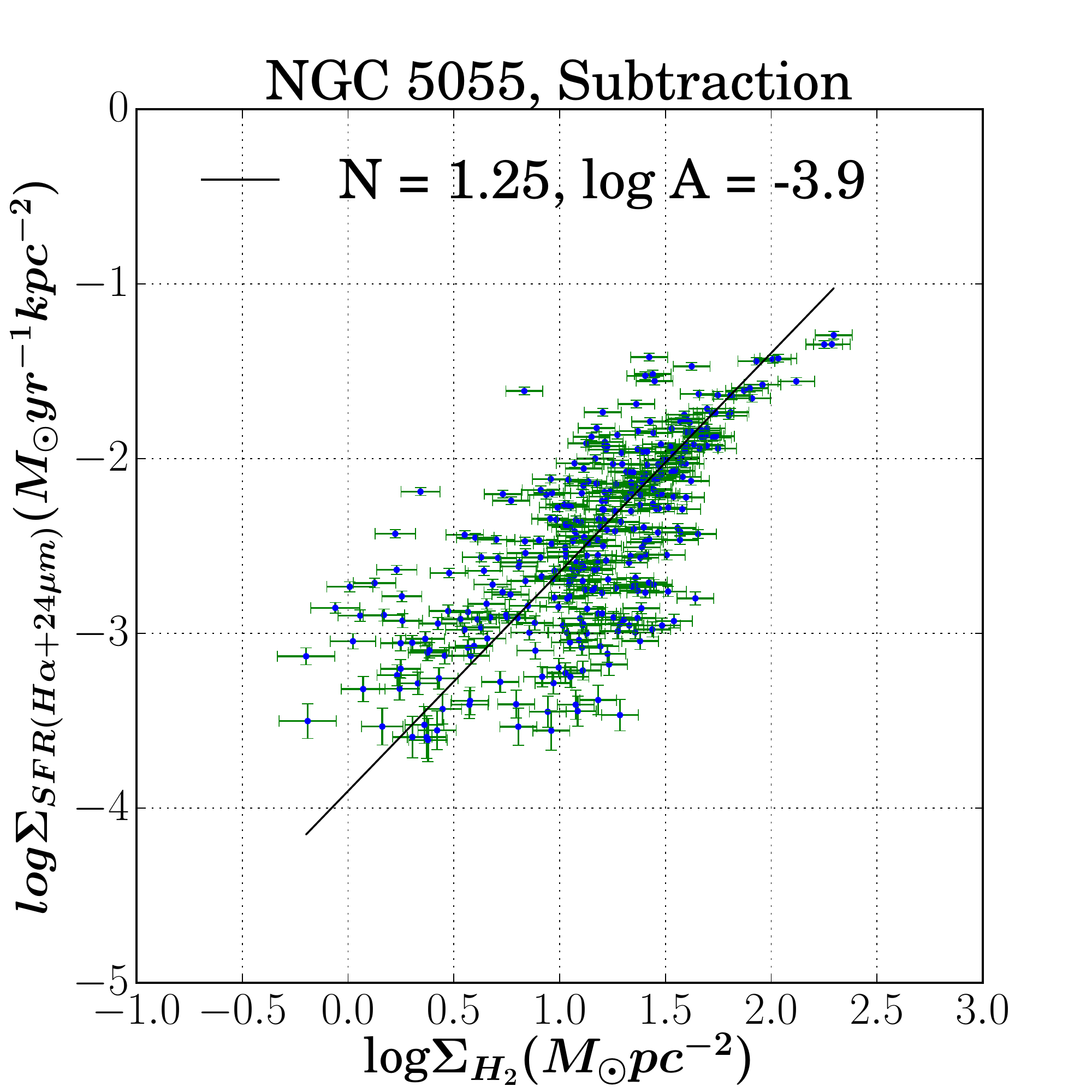}
		\includegraphics[width = 0.33\textwidth,trim={0.2cm 0cm 1.5cm 0cm},clip]{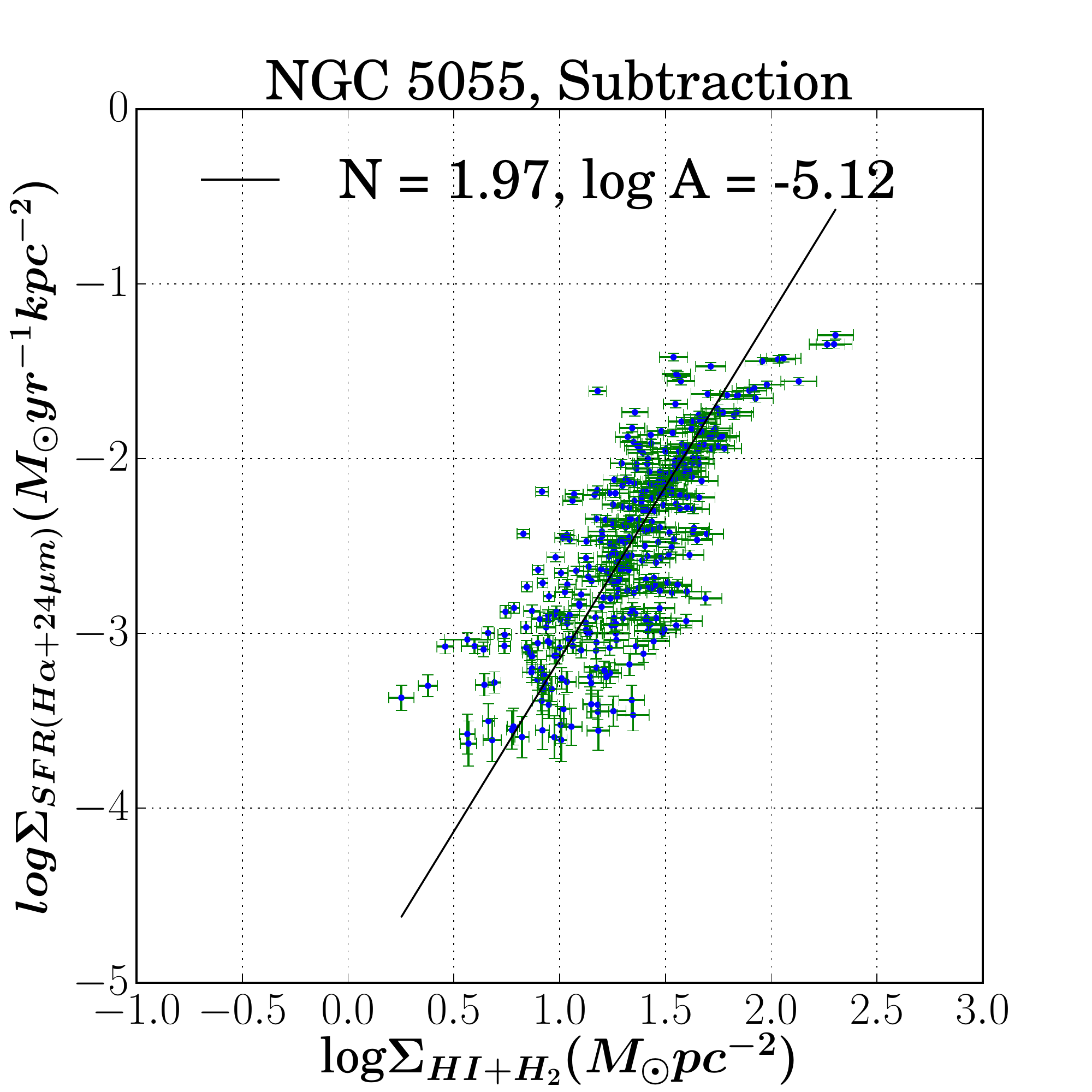}
		\caption{NGC 5055: an aperture size of 12\arcsec is adopted, which corresponds to the physical diameter of $\sim$ 1 kpc at a distance of 7.8 Mpc and inclination angle of  59$^{\circ}$.  See caption of Fig. \ref{Figure: NGC 0628} for details. }
		\label{NGC 5055}
	\end{figure*}

\begin{figure*}
		\centering
		
		\includegraphics[width = 0.33\textwidth,trim={0.2cm 0cm 1.5cm 0cm},clip]{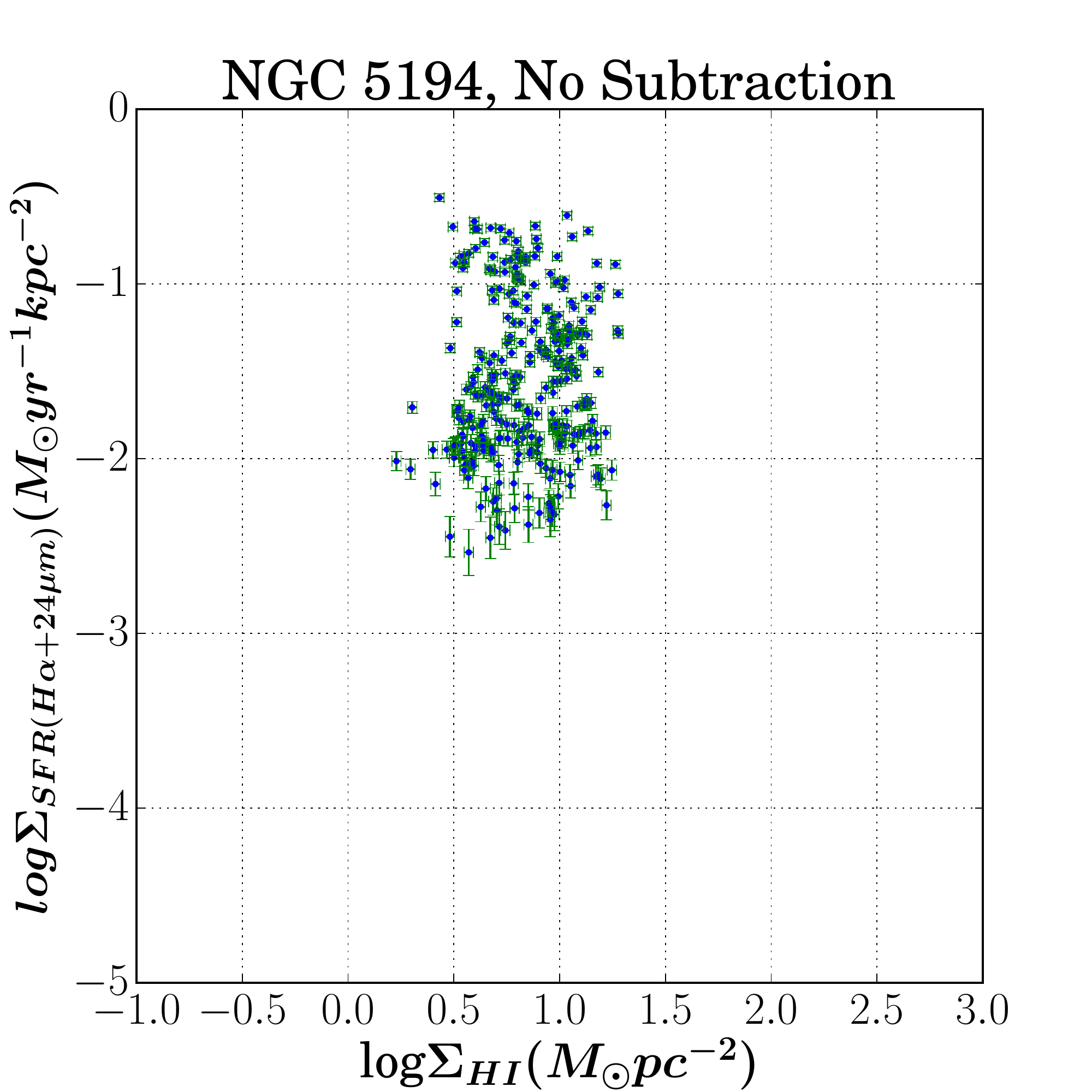}
		\includegraphics[width = 0.33\textwidth,trim={0.2cm 0cm 1.5cm 0cm},clip]{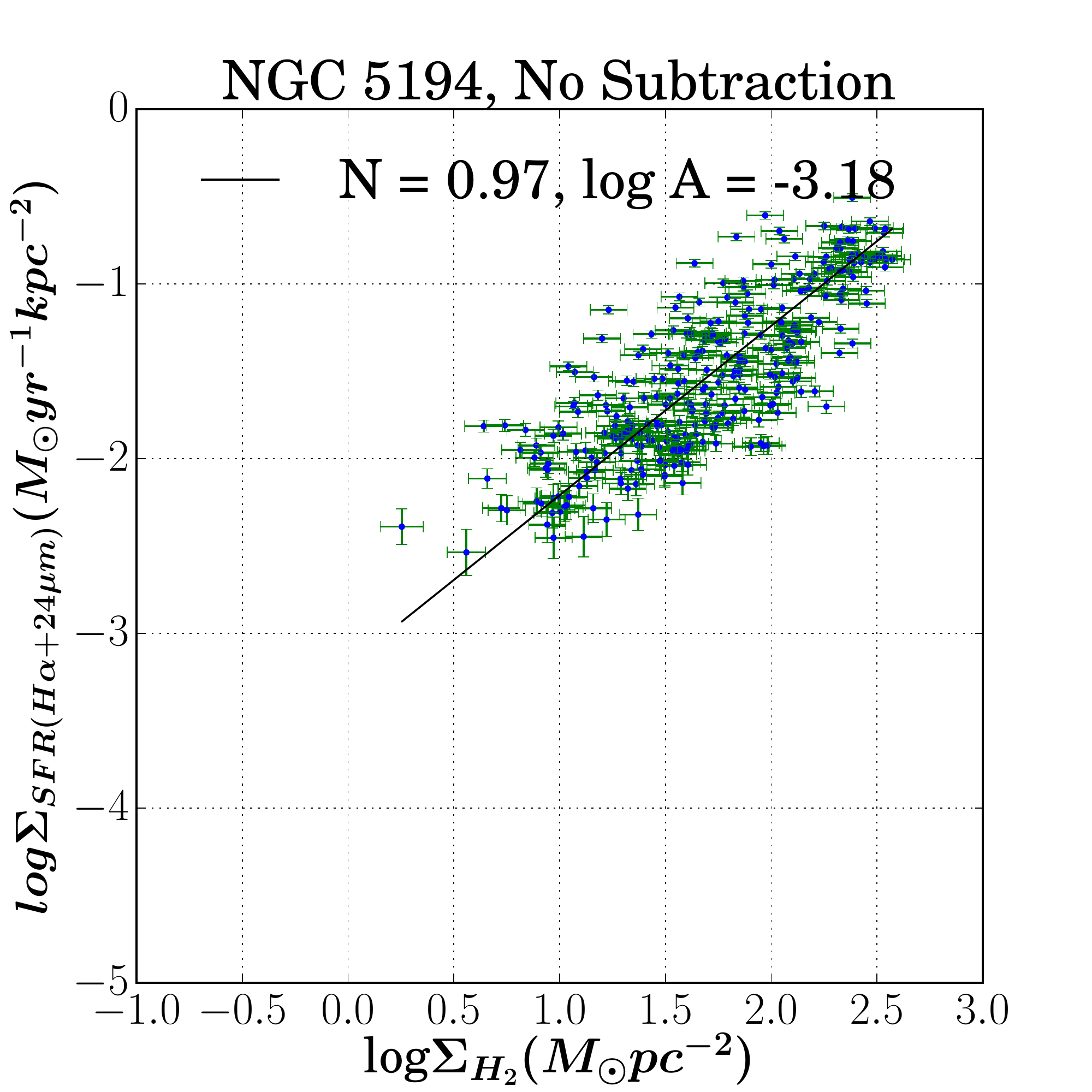}
		\includegraphics[width = 0.33\textwidth,trim={0.2cm 0cm 1.5cm 0cm},clip]{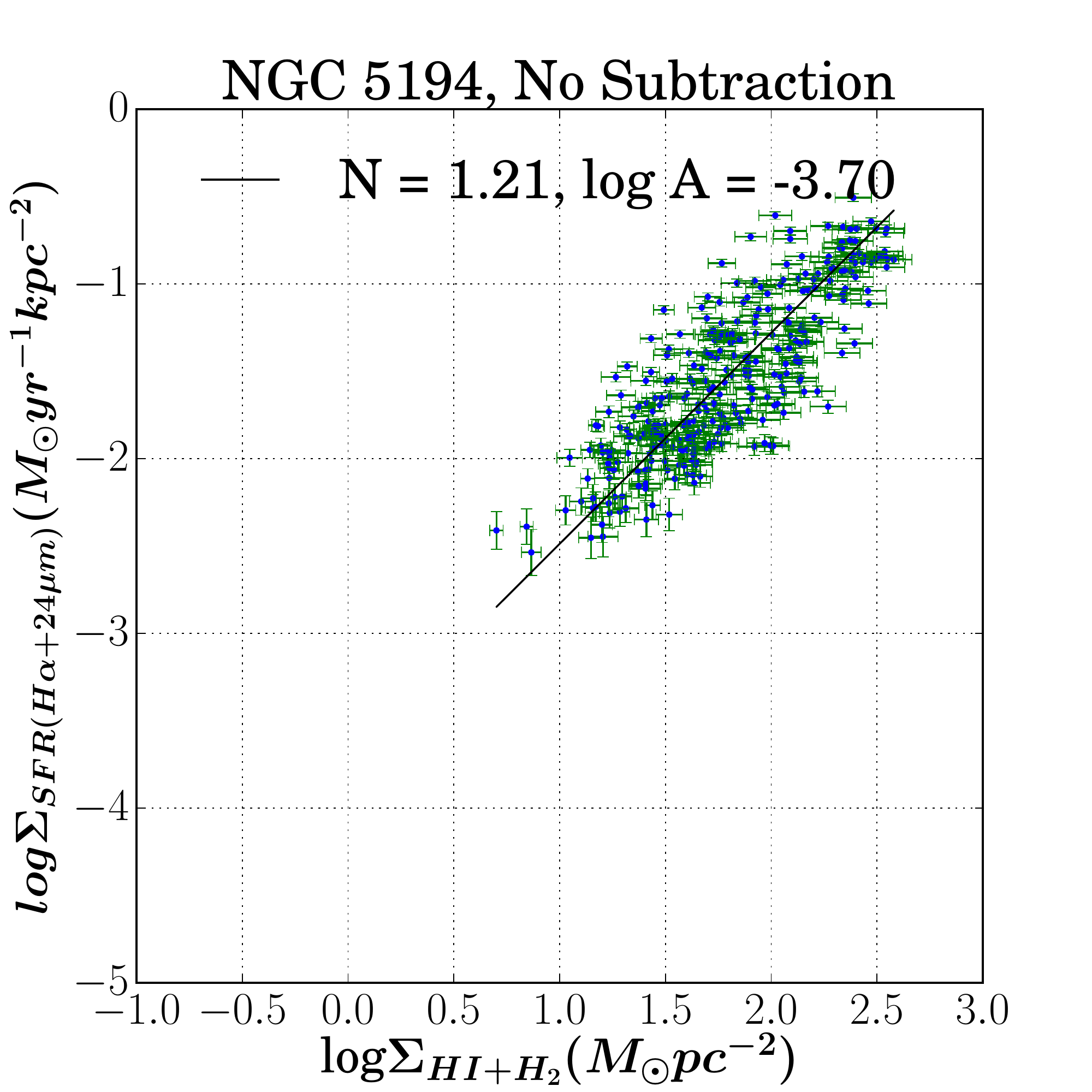}
		\includegraphics[width = 0.33\textwidth,trim={0.2cm 0cm 1.5cm 0cm},clip]{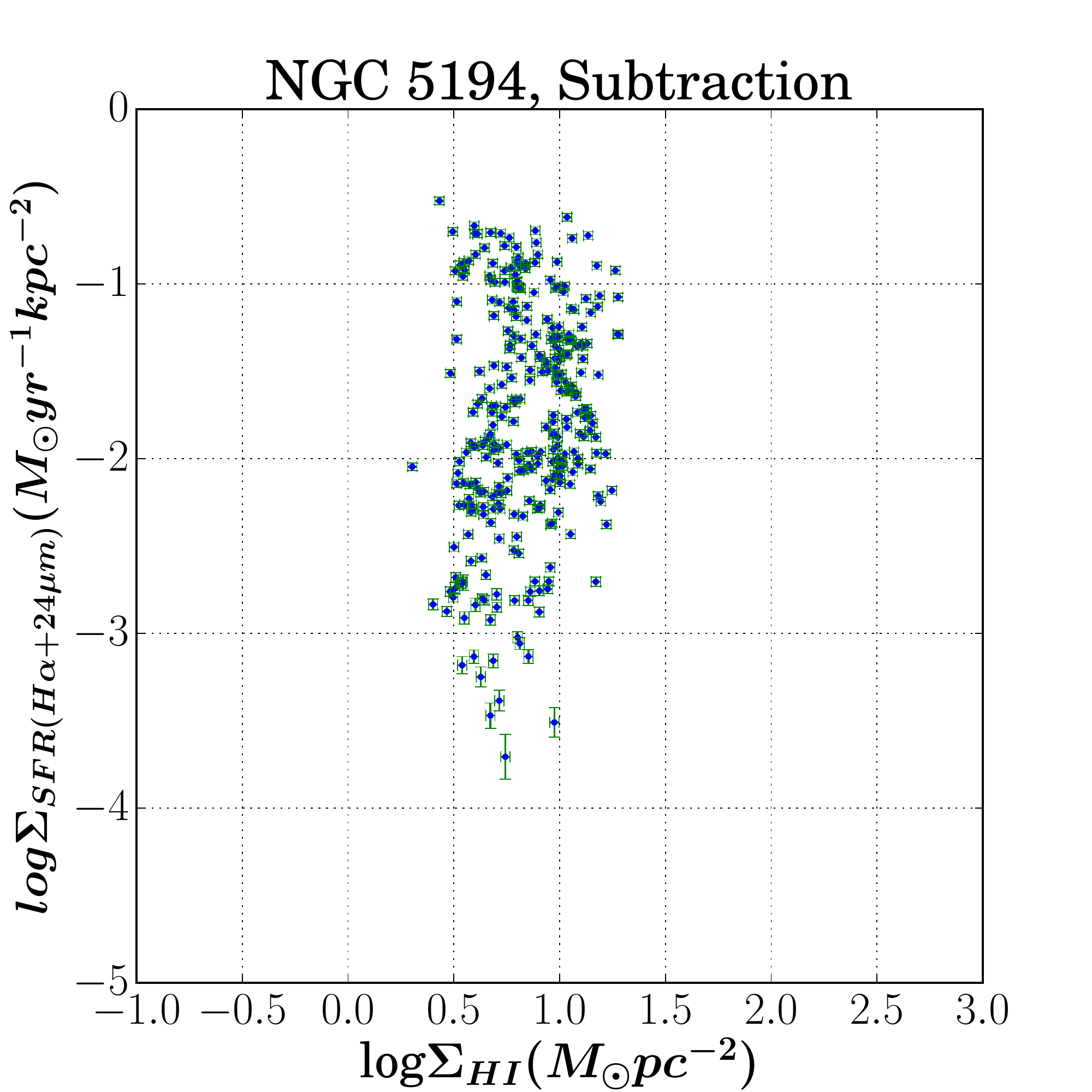}
		\includegraphics[width = 0.33\textwidth,trim={0.2cm 0cm 1.5cm 0cm},clip]{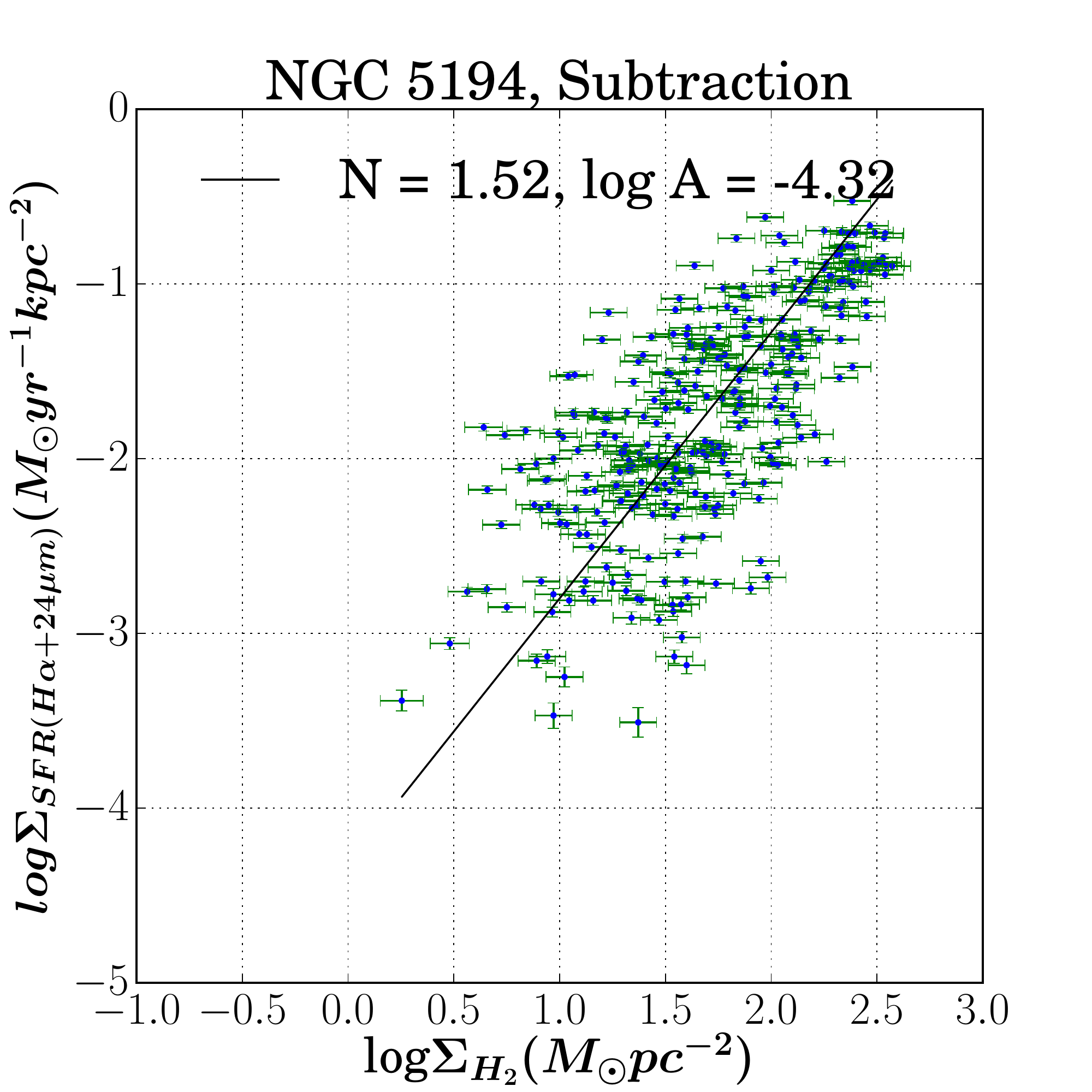}
		\includegraphics[width = 0.33\textwidth,trim={0.2cm 0cm 1.5cm 0cm},clip]{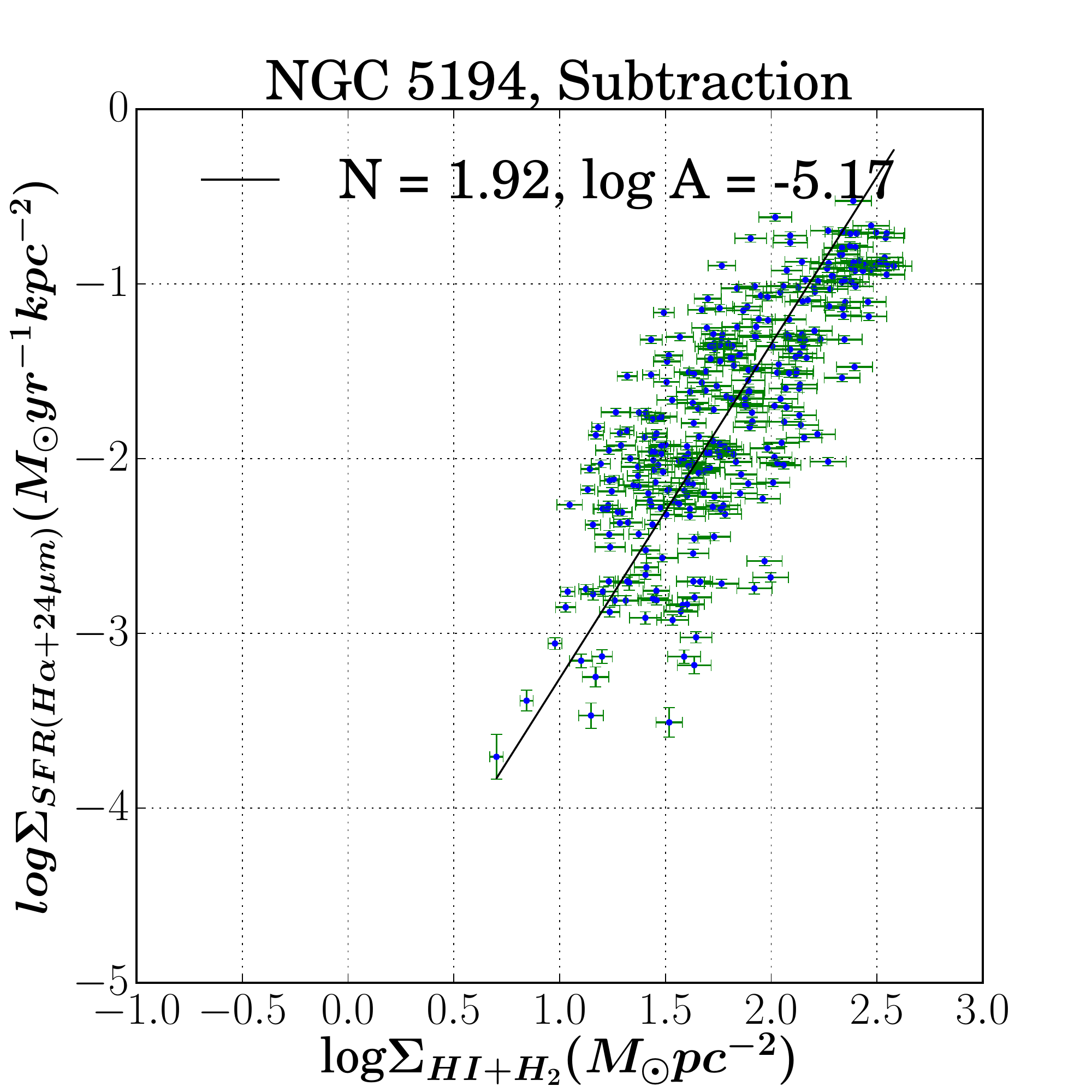}
		
		\caption{NGC 5194: an aperture size of 13\arcsec is adopted, which corresponds to the physical diameter of $\sim$ 520 pc at a distance of 8.2 Mpc and inclination angle of  20$^{\circ}$.  See caption of Fig. \ref{Figure: NGC 0628} for details.}
		\label{NGC 5194}
	\end{figure*}

\begin{figure*}
		\centering
		
		\includegraphics[width = 0.33\textwidth,trim={0.2cm 0cm 1.5cm 0cm},clip]{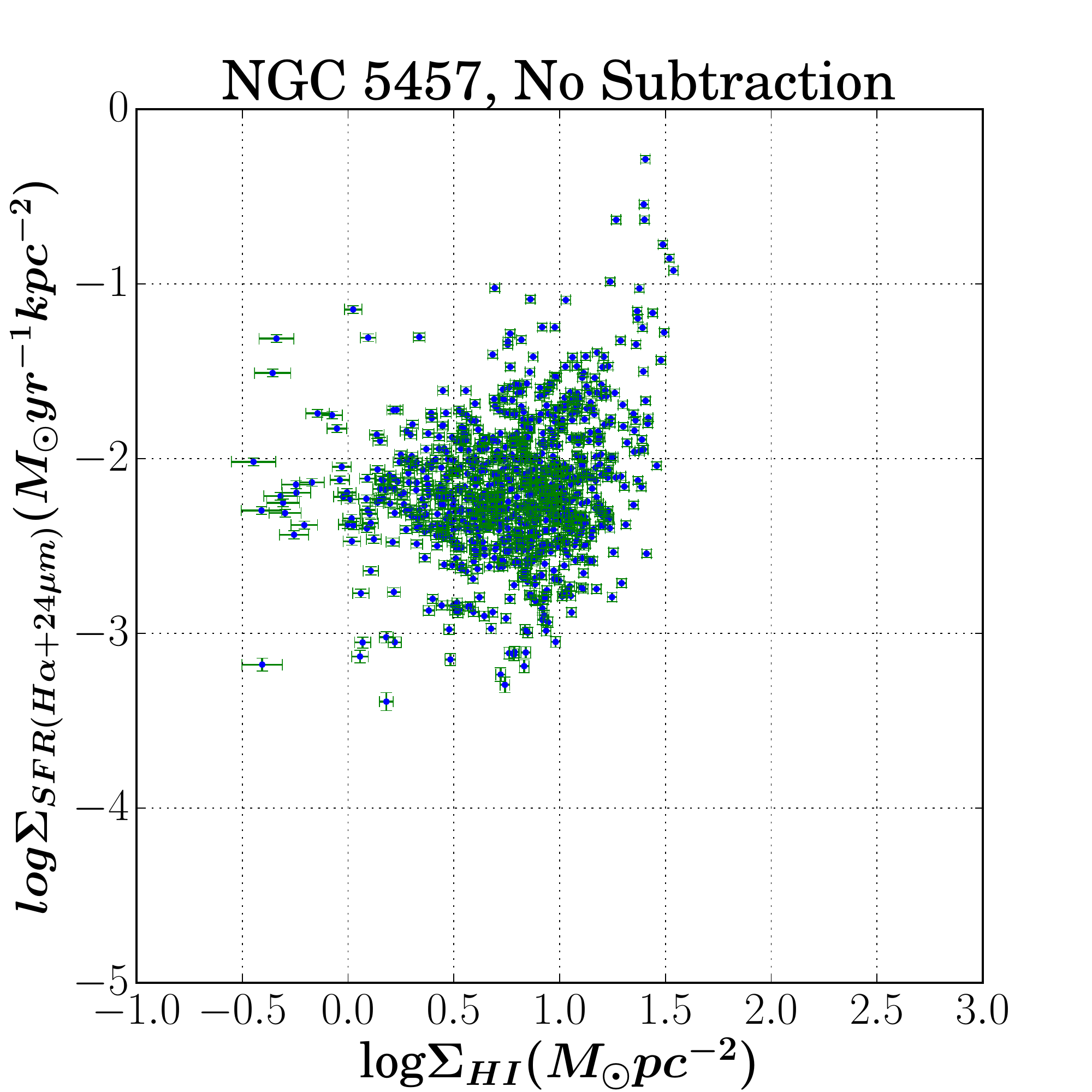}
		\includegraphics[width = 0.33\textwidth,trim={0.2cm 0cm 1.5cm 0cm},clip]{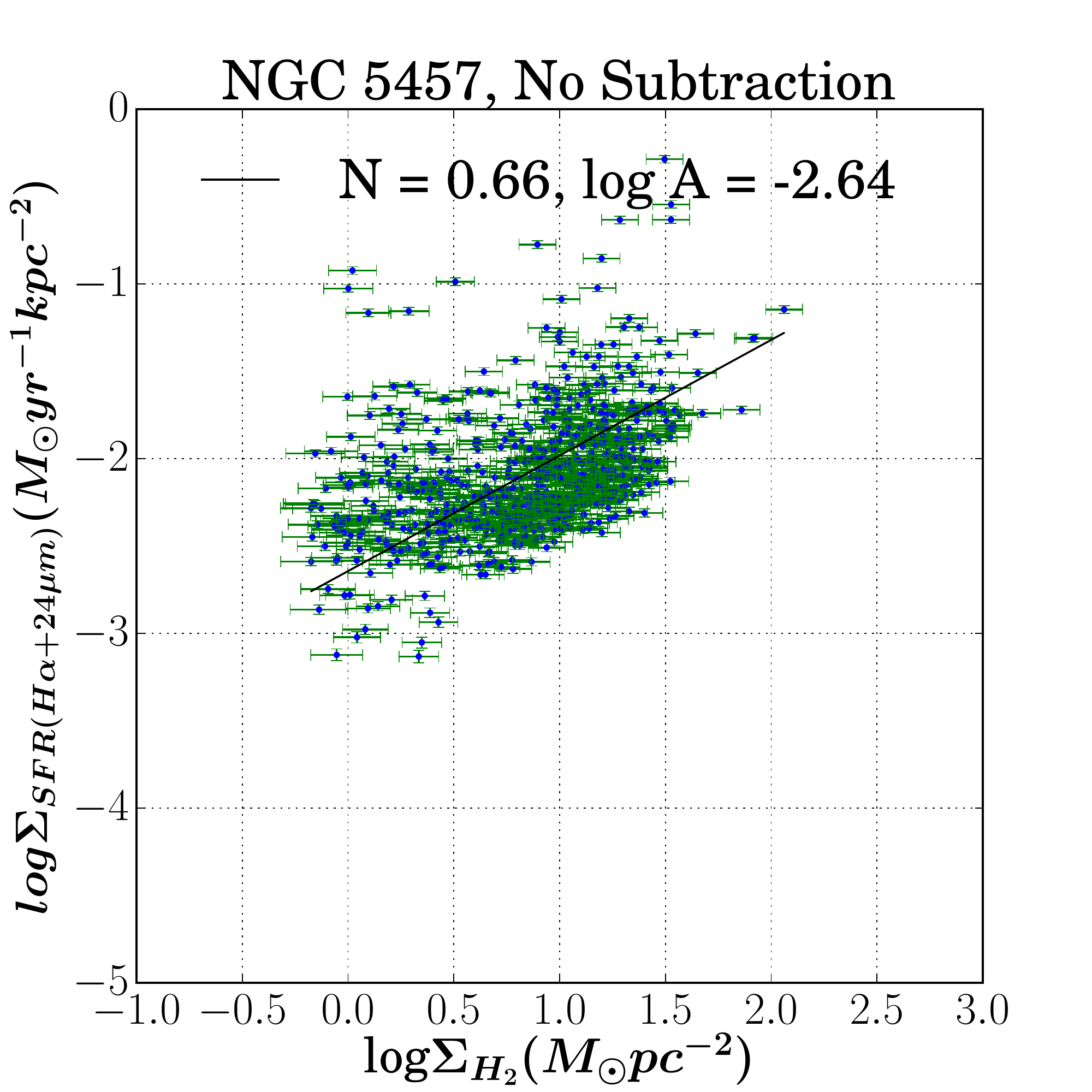}
		\includegraphics[width = 0.33\textwidth,trim={0.2cm 0cm 1.5cm 0cm},clip]{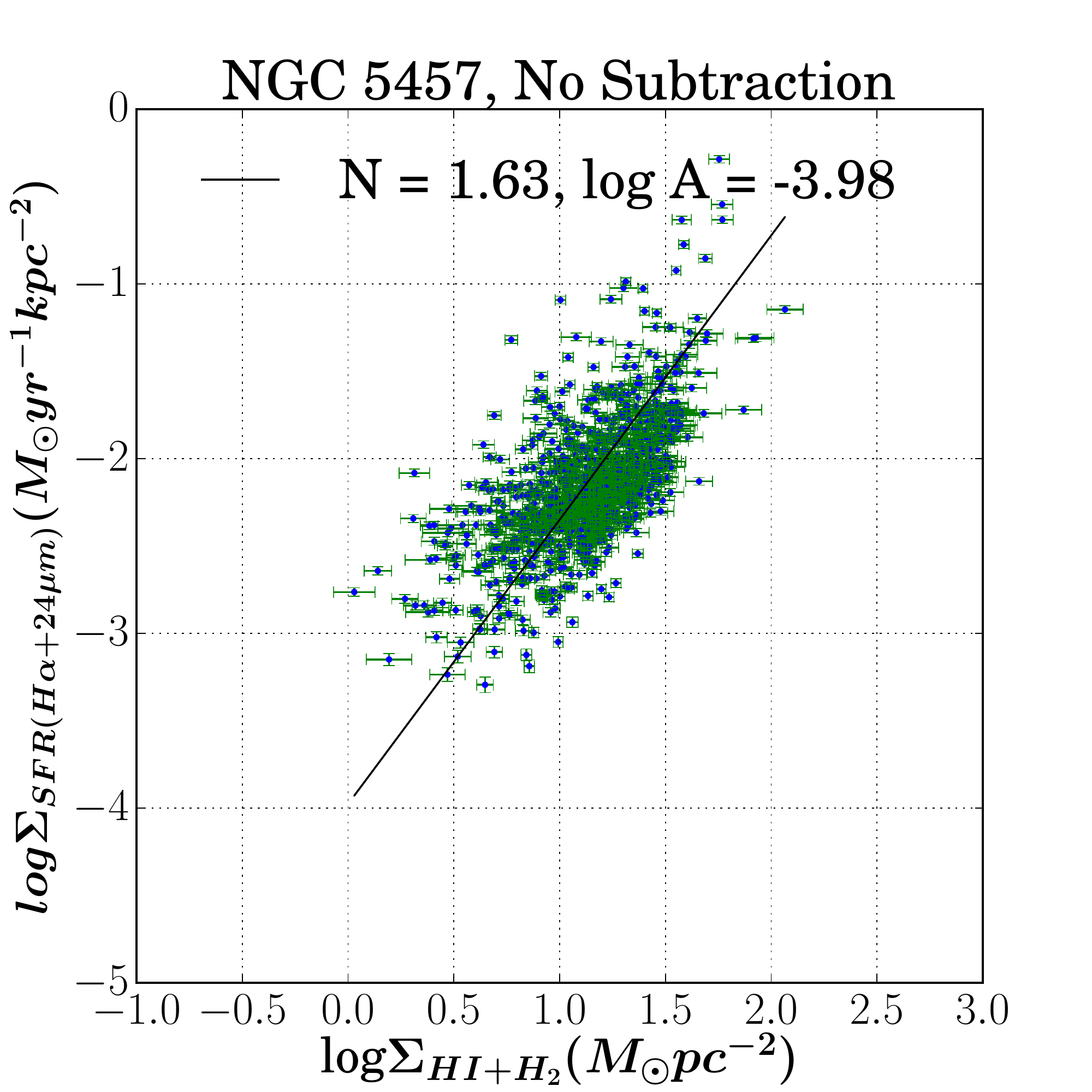}
		
		\includegraphics[width = 0.33\textwidth,trim={0.2cm 0cm 1.5cm 0cm},clip]{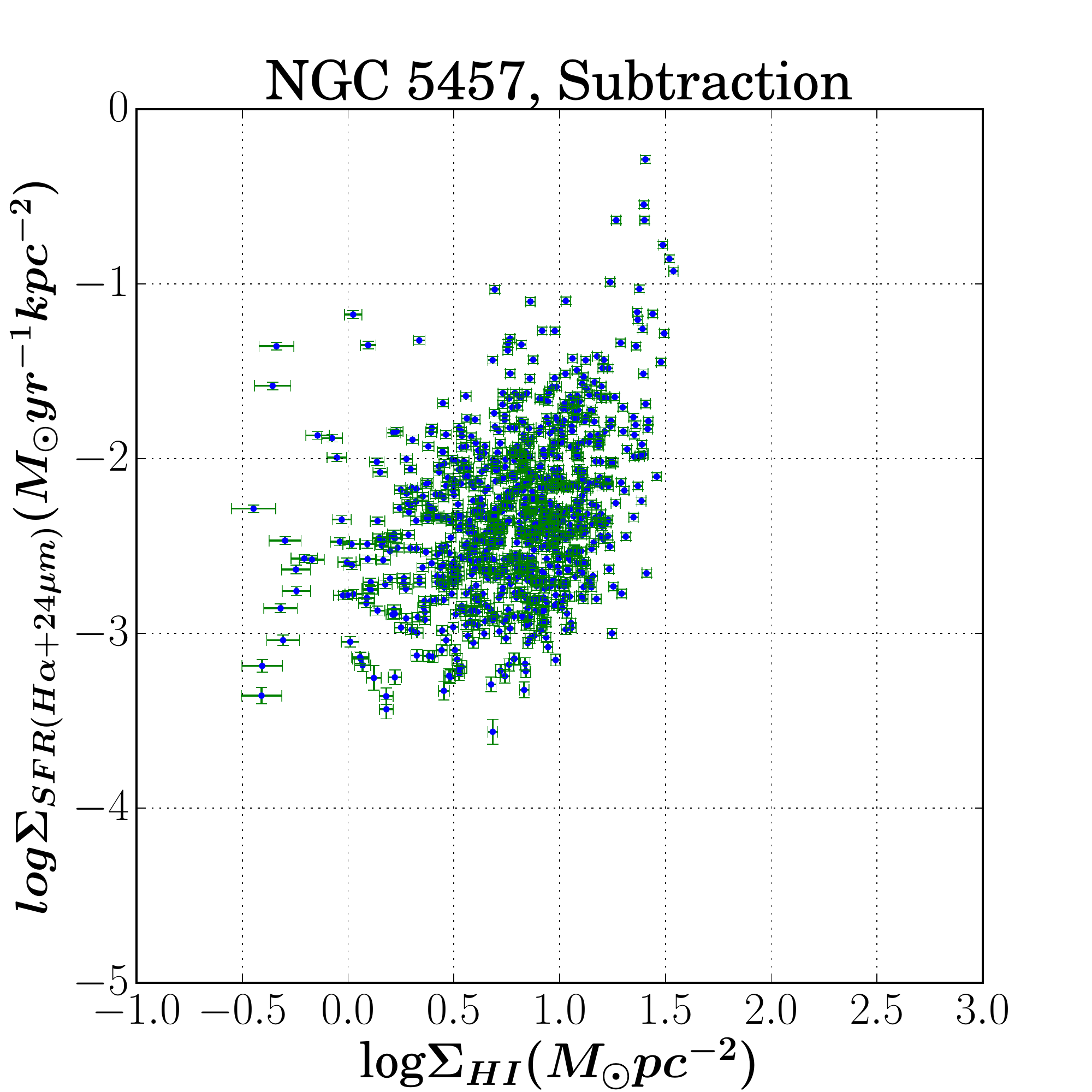}
		\includegraphics[width = 0.33\textwidth,trim={0.2cm 0cm 1.5cm 0cm},clip]{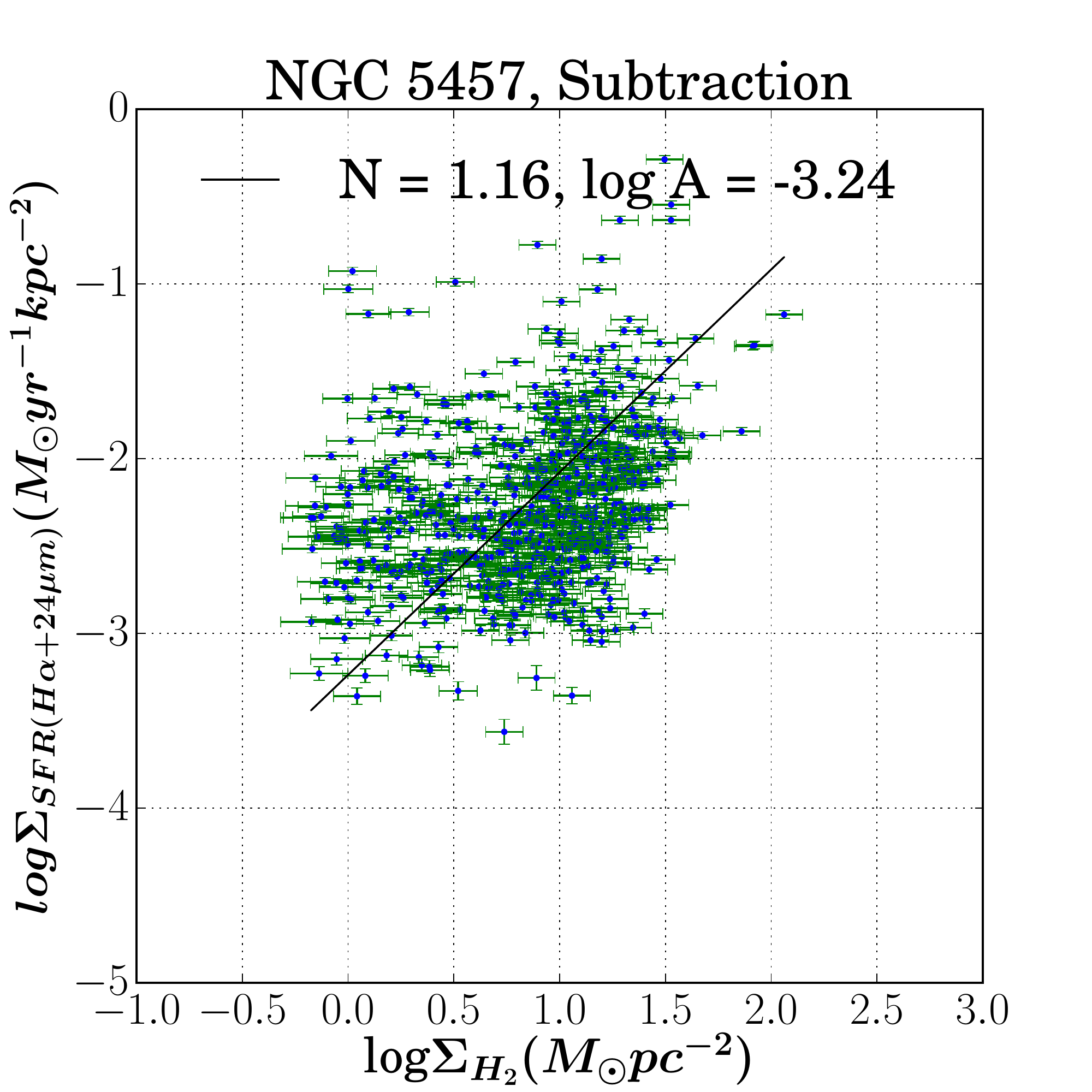}
		\includegraphics[width = 0.33\textwidth,trim={0.2cm 0cm 1.5cm 0cm},clip]{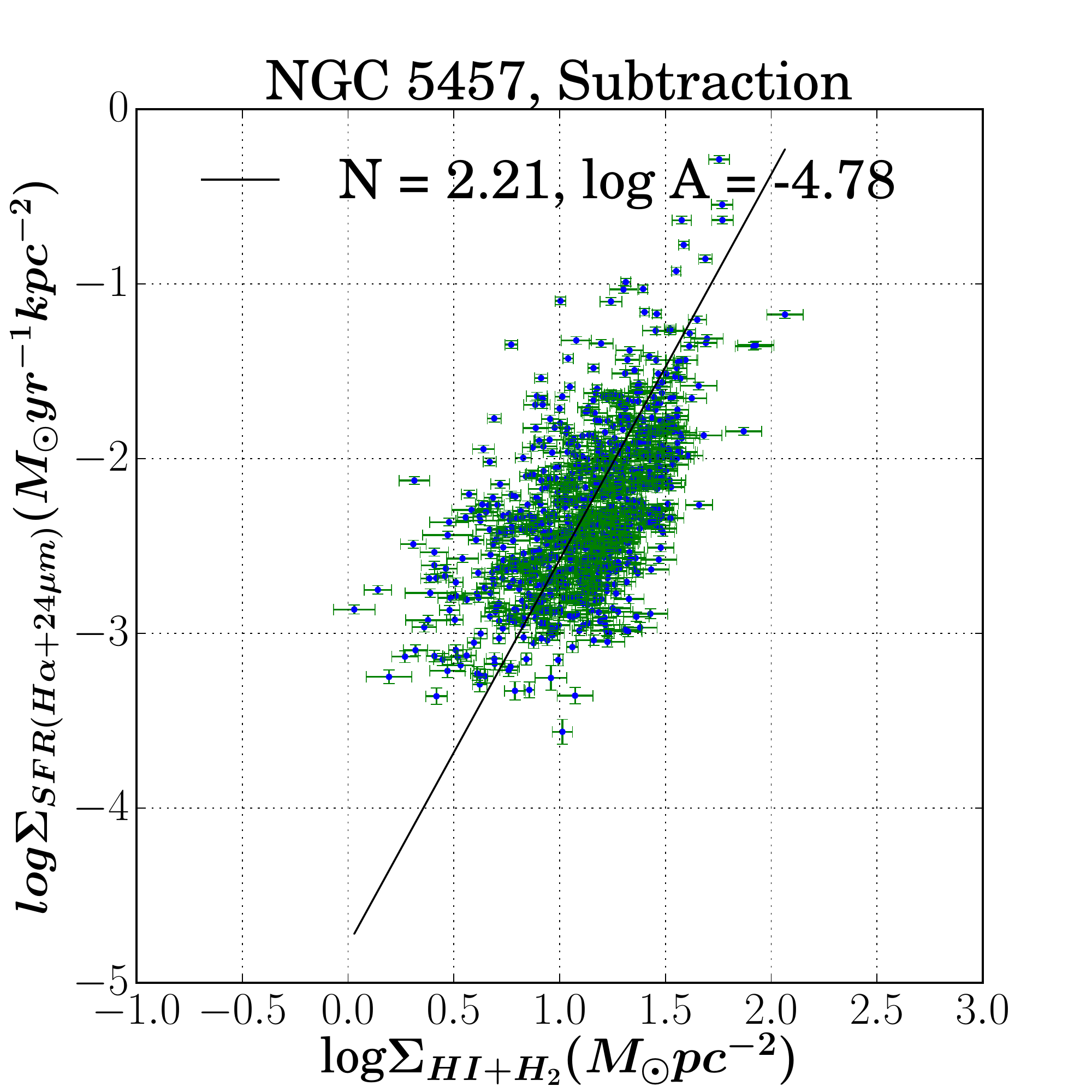}
		\caption{NGC 5457: an aperture size of 14\arcsec is adopted, which corresponds to the physical diameter of $\sim$ 500 pc at a distance of 6.7 Mpc and inclination angle of  18$^{\circ}$.  See caption of Fig. \ref{Figure: NGC 0628} for details.}
		\label{NGC 5457}
	\end{figure*}

\begin{figure*}
		\centering
		\includegraphics[width = 0.33\textwidth,trim={0.2cm 0cm 1.5cm 0cm},clip]{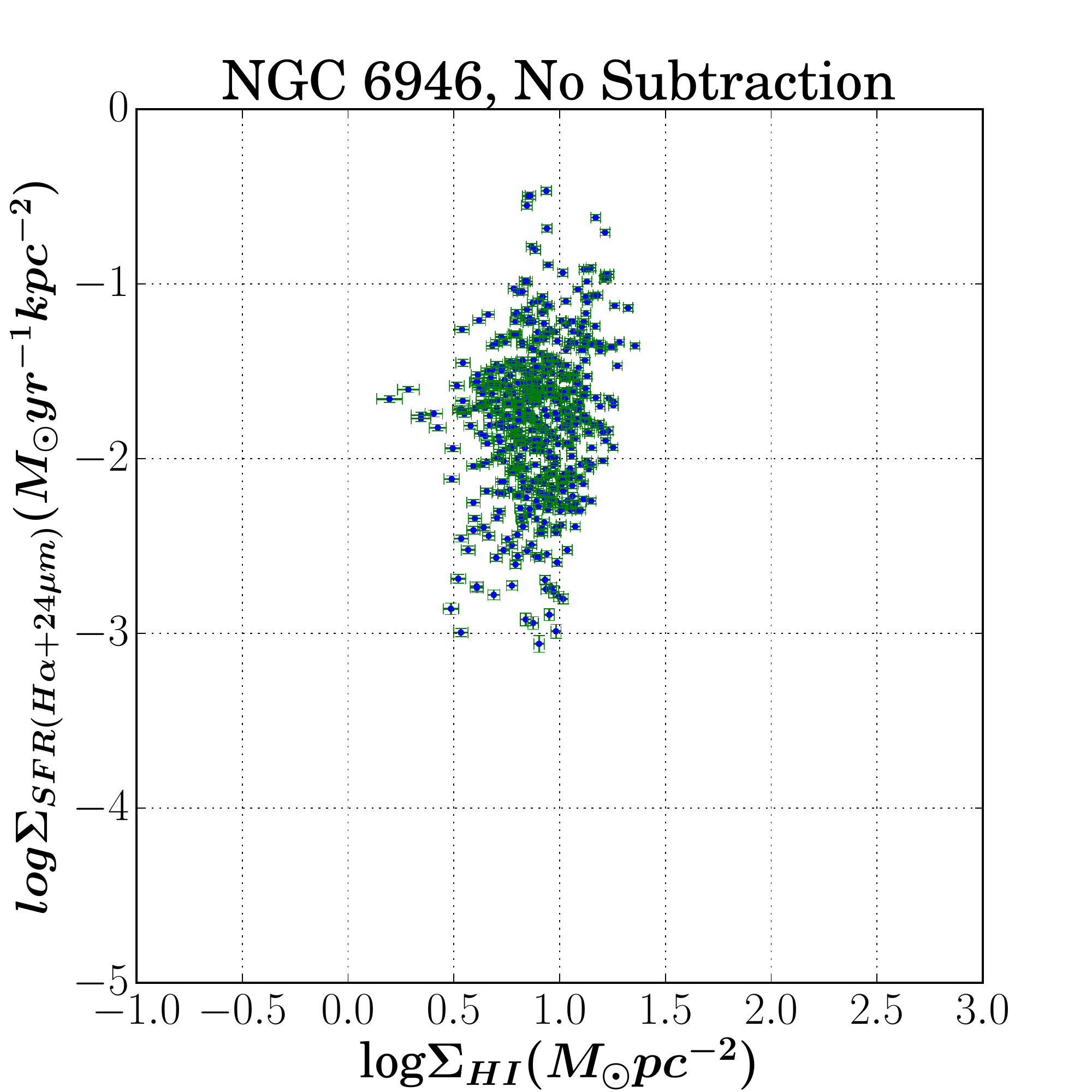}
		\includegraphics[width = 0.33\textwidth,trim={0.2cm 0cm 1.5cm 0cm},clip]{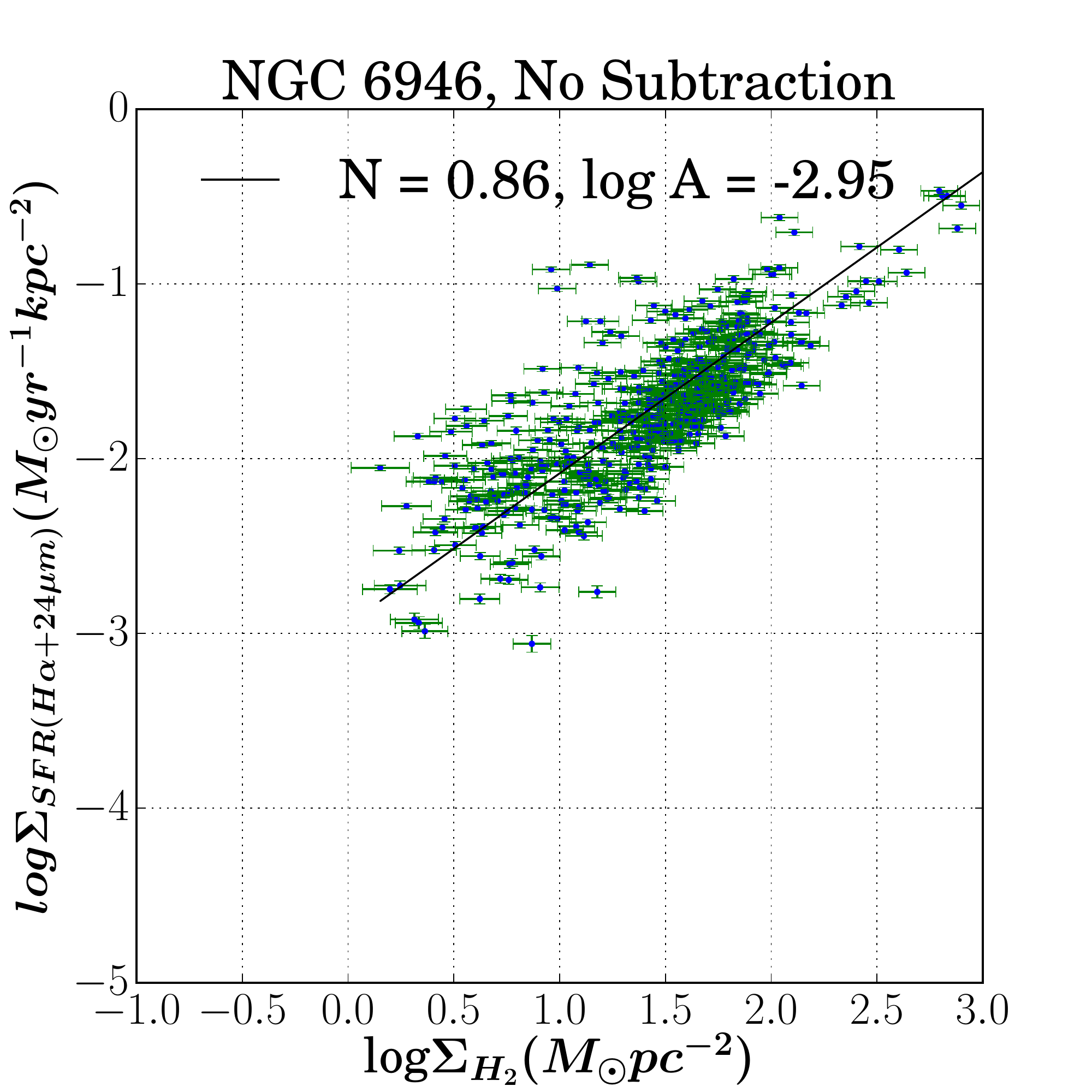}
		\includegraphics[width = 0.33\textwidth,trim={0.2cm 0cm 1.5cm 0cm},clip]{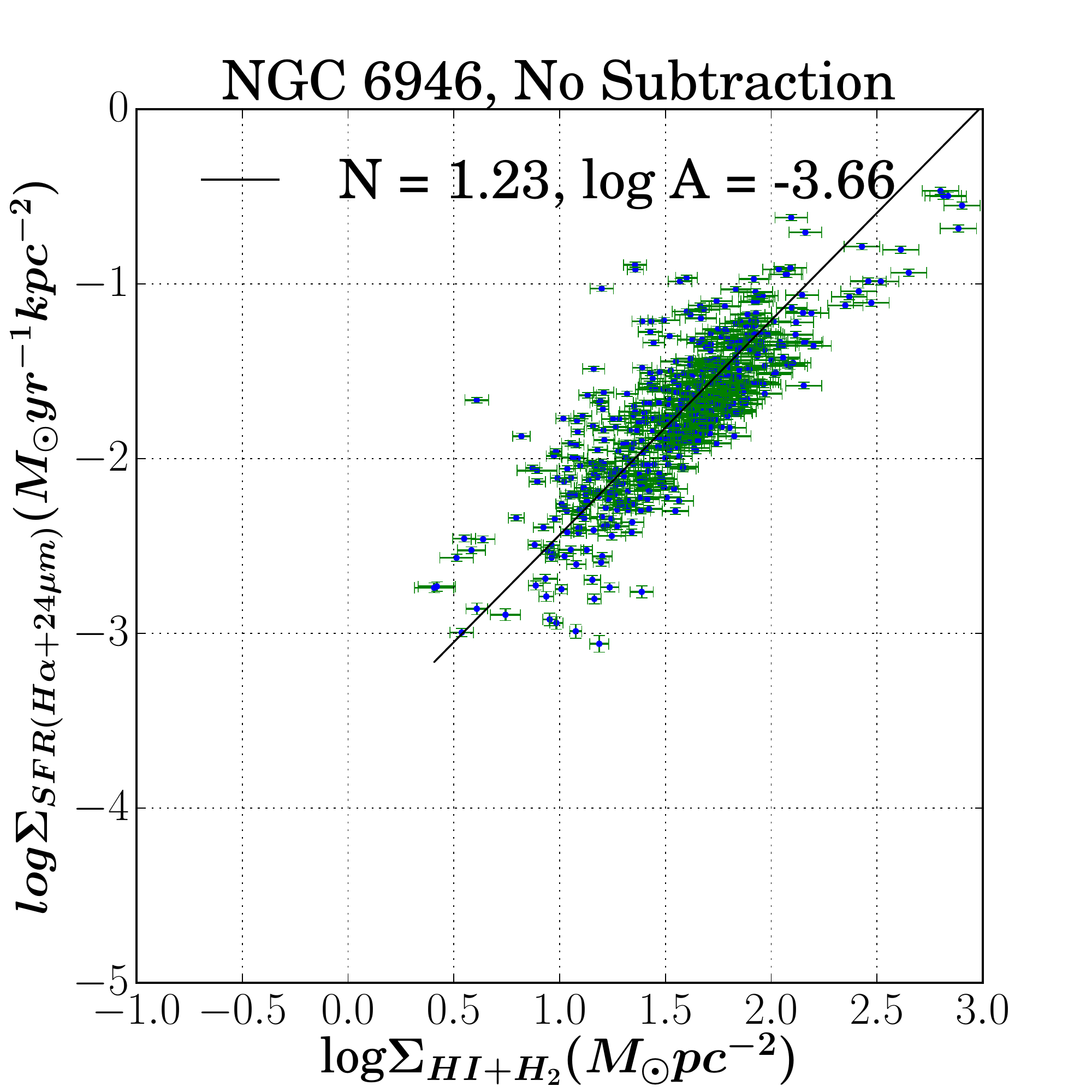}
		\includegraphics[width = 0.33\textwidth,trim={0.2cm 0cm 1.5cm 0cm},clip]{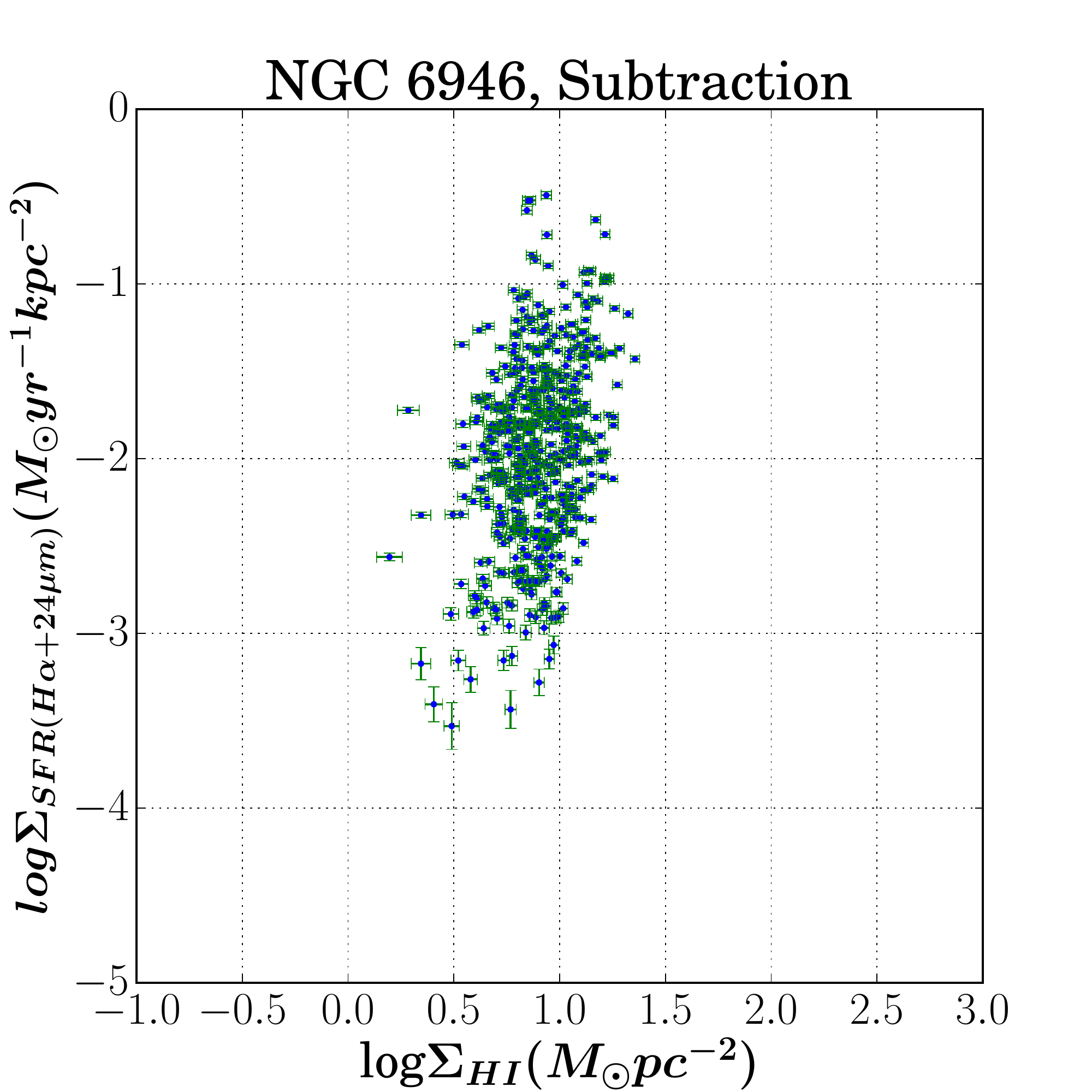}
		\includegraphics[width = 0.33\textwidth,trim={0.2cm 0cm 1.5cm 0cm},clip]{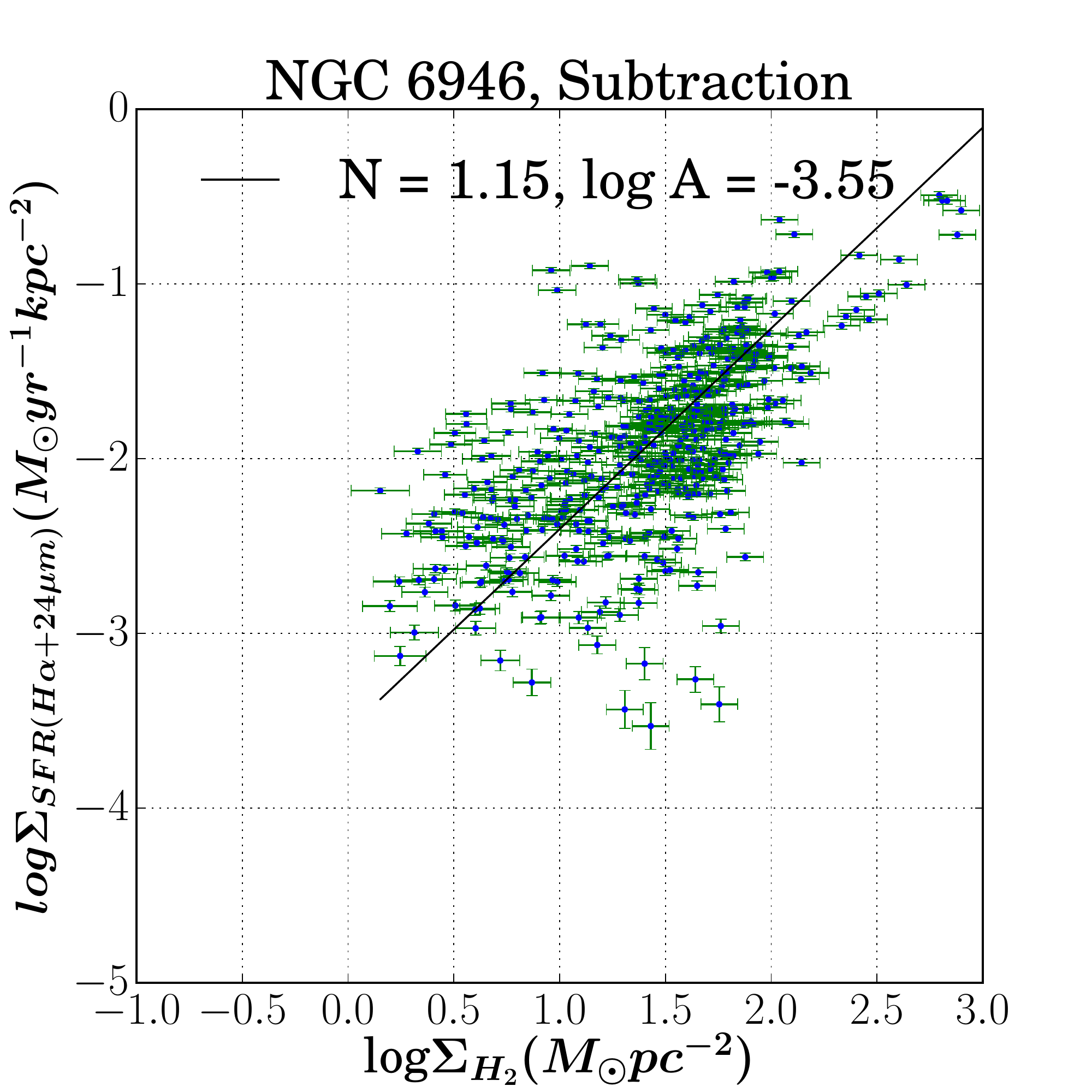}
		\includegraphics[width = 0.33\textwidth,trim={0.2cm 0cm 1.5cm 0cm},clip]{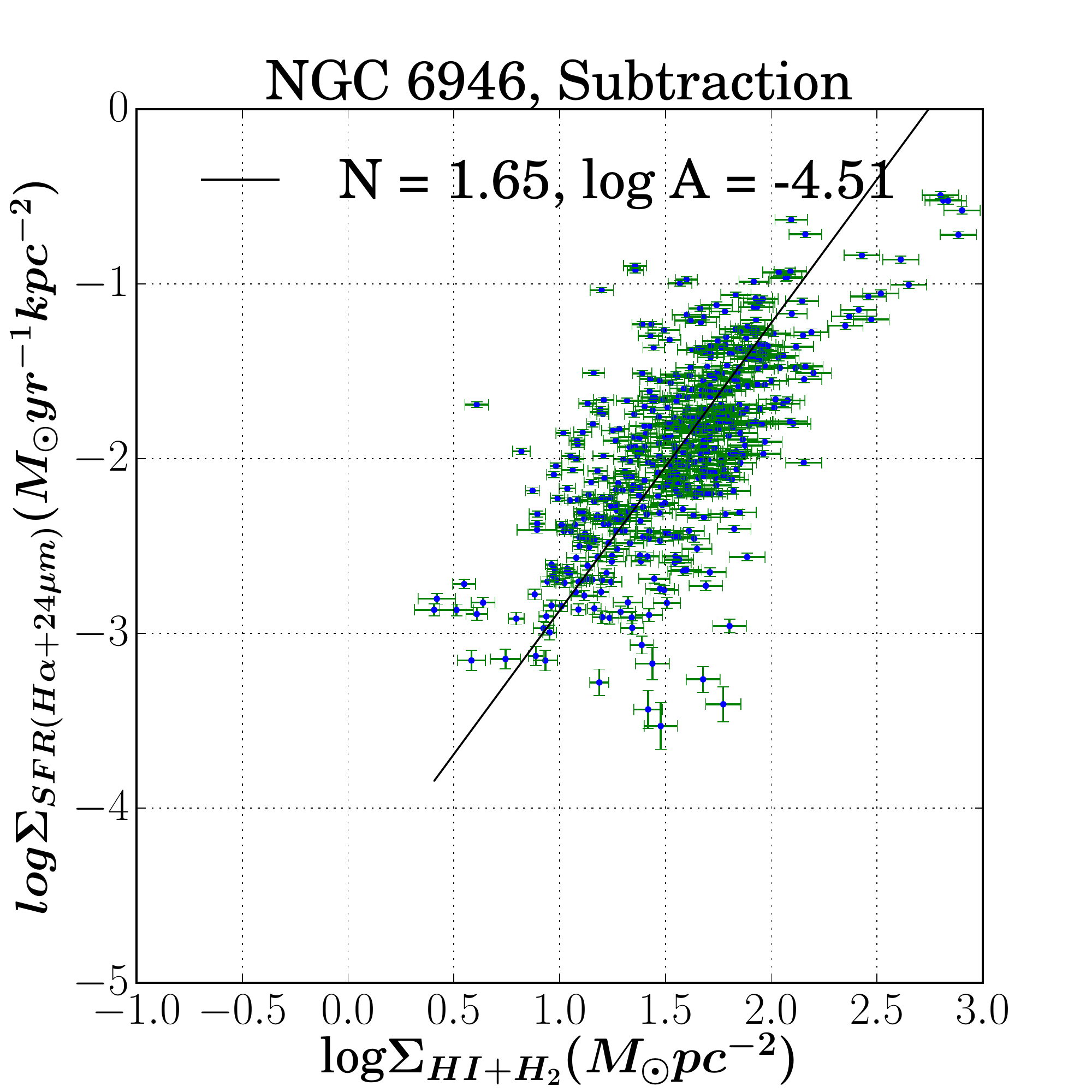}
		\caption{NGC 6946: an aperture size of 12\arcsec is adopted, which corresponds to the physical diameter of $\sim$ 596 pc at a distance of 6.8 Mpc and inclination angle of  33$^{\circ}$.  See caption of Fig. \ref{Figure: NGC 0628} for details.}
		\label{NGC 6946}
	\end{figure*}

\begin{figure*}
	\centering
	\includegraphics[width=0.33\textwidth,trim={0.2cm 0cm 1.5cm 0cm},clip]{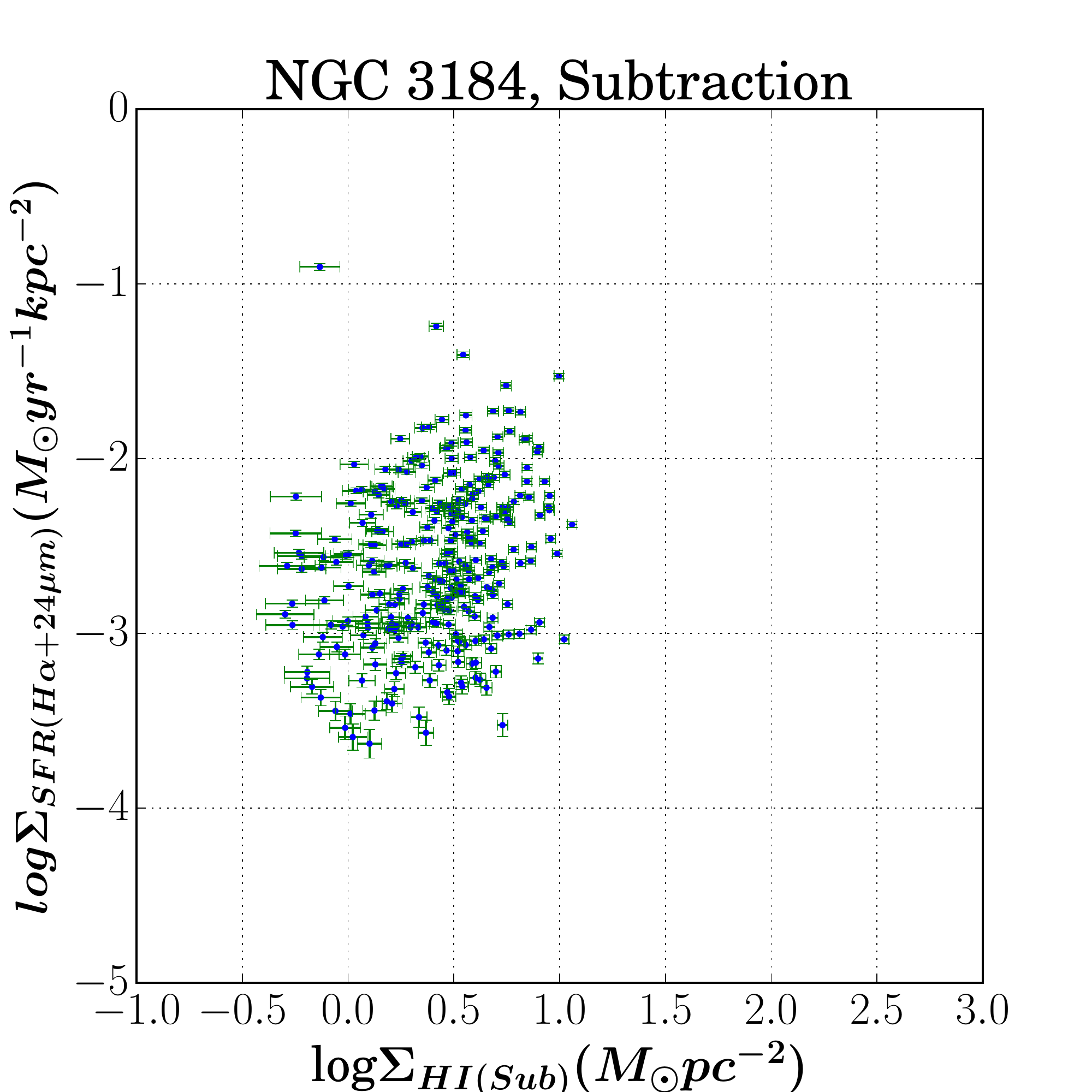}
	\includegraphics[width=0.33\textwidth,trim={0.2cm 0cm 1.5cm 0cm},clip]{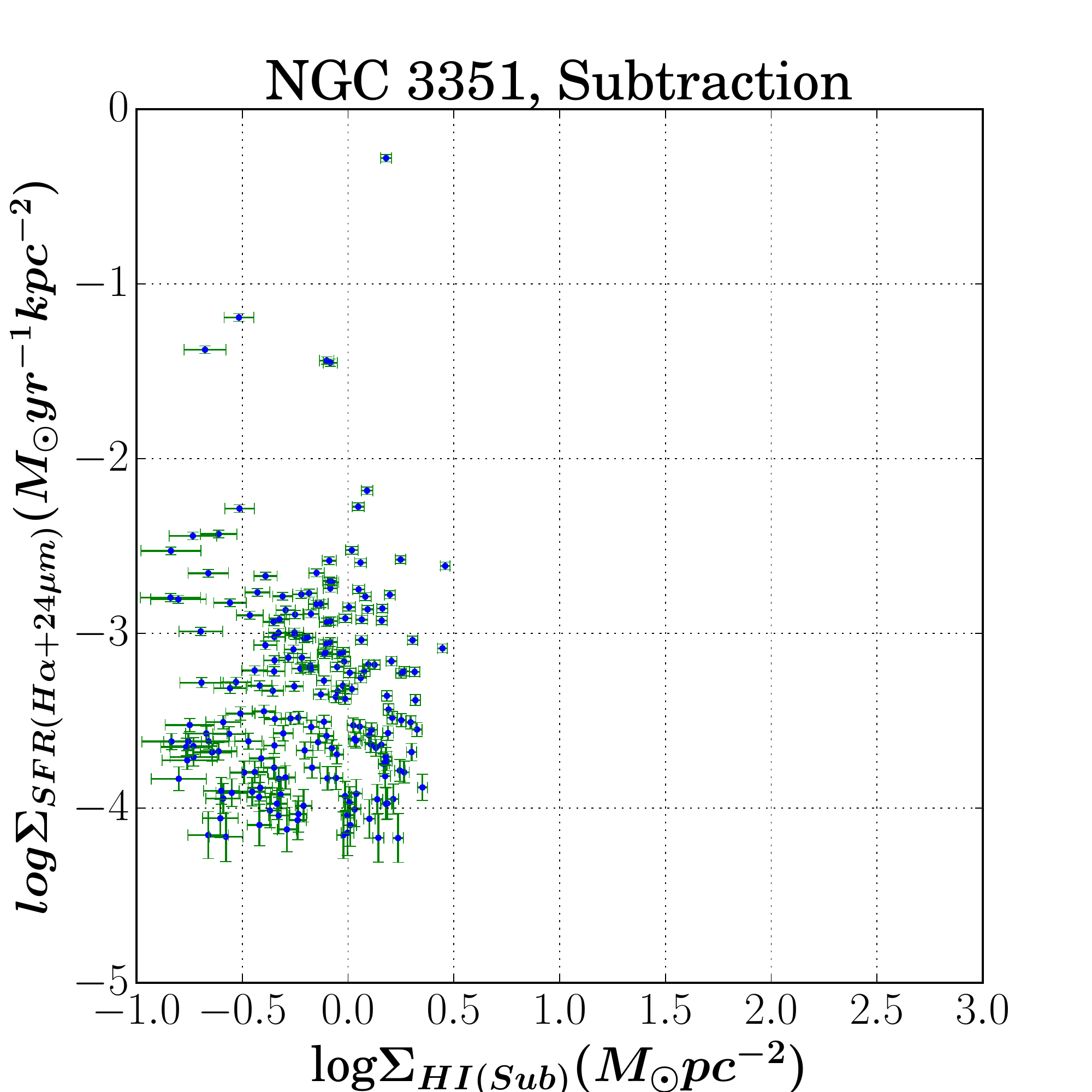}
	\includegraphics[width=0.33\textwidth,trim={0.2cm 0cm 1.5cm 0cm},clip]{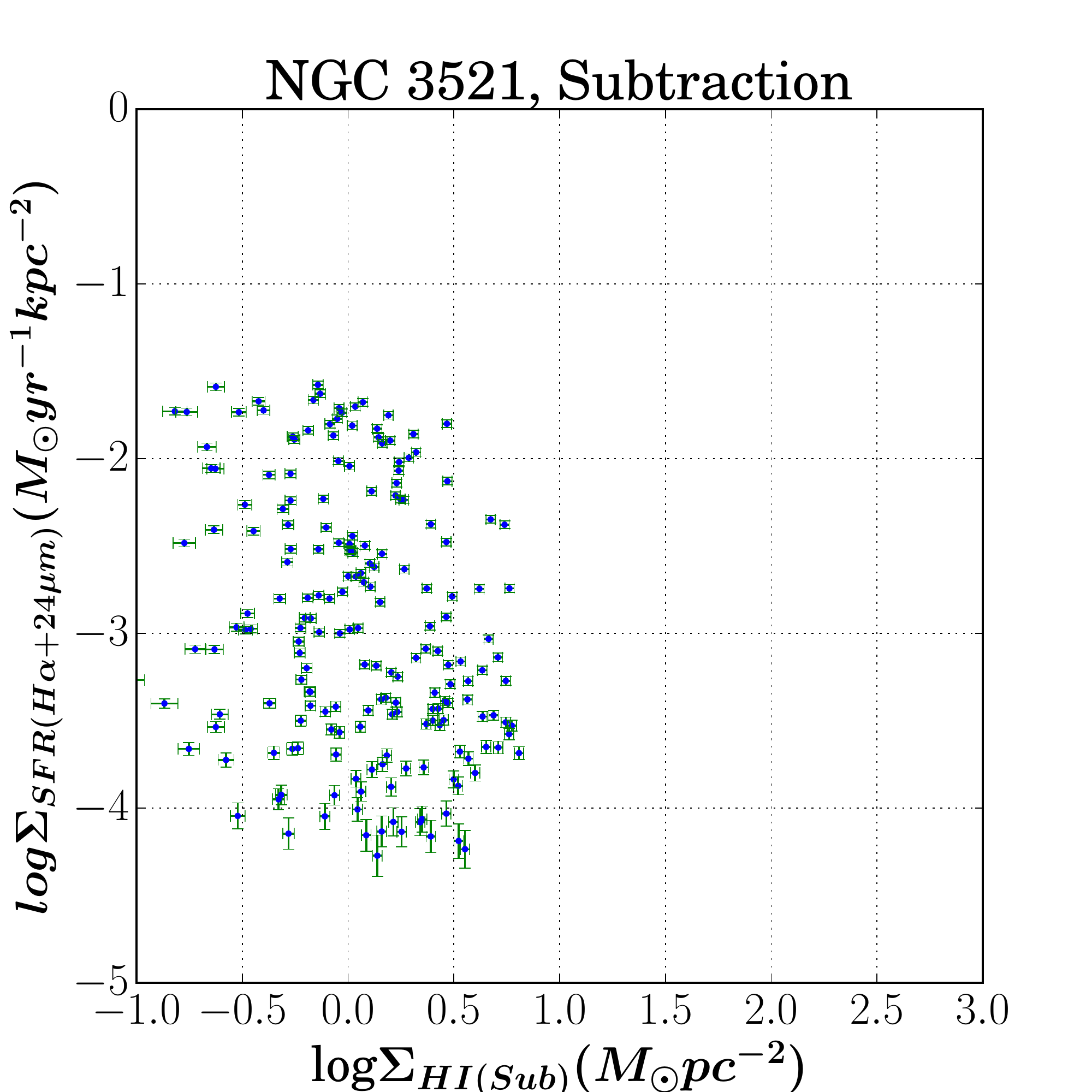}
	
	\includegraphics[width=0.33\textwidth,trim={0.2cm 0cm 1.5cm 0cm},clip]{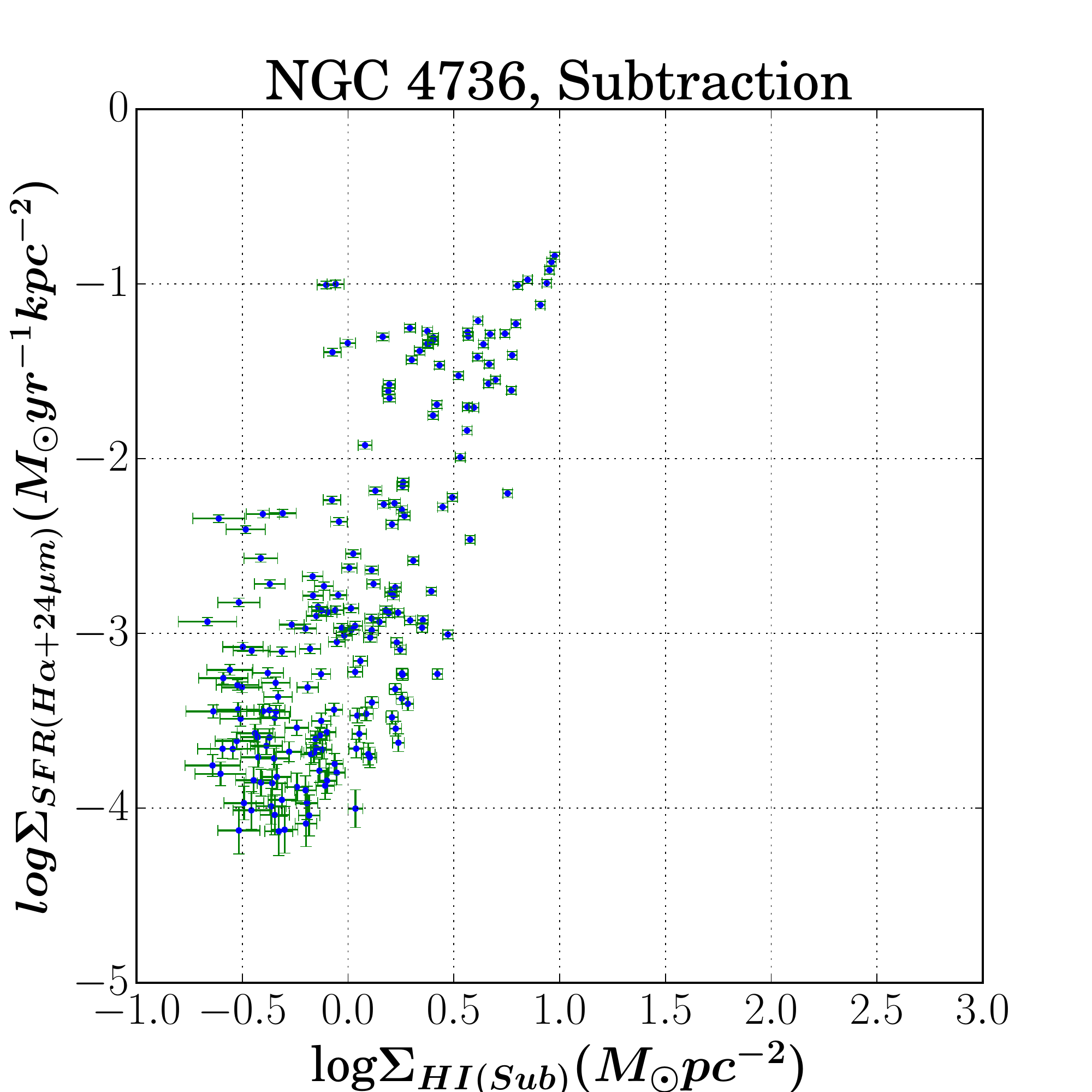}
	\includegraphics[width=0.33\textwidth,trim={0.2cm 0cm 1.5cm 0cm},clip]{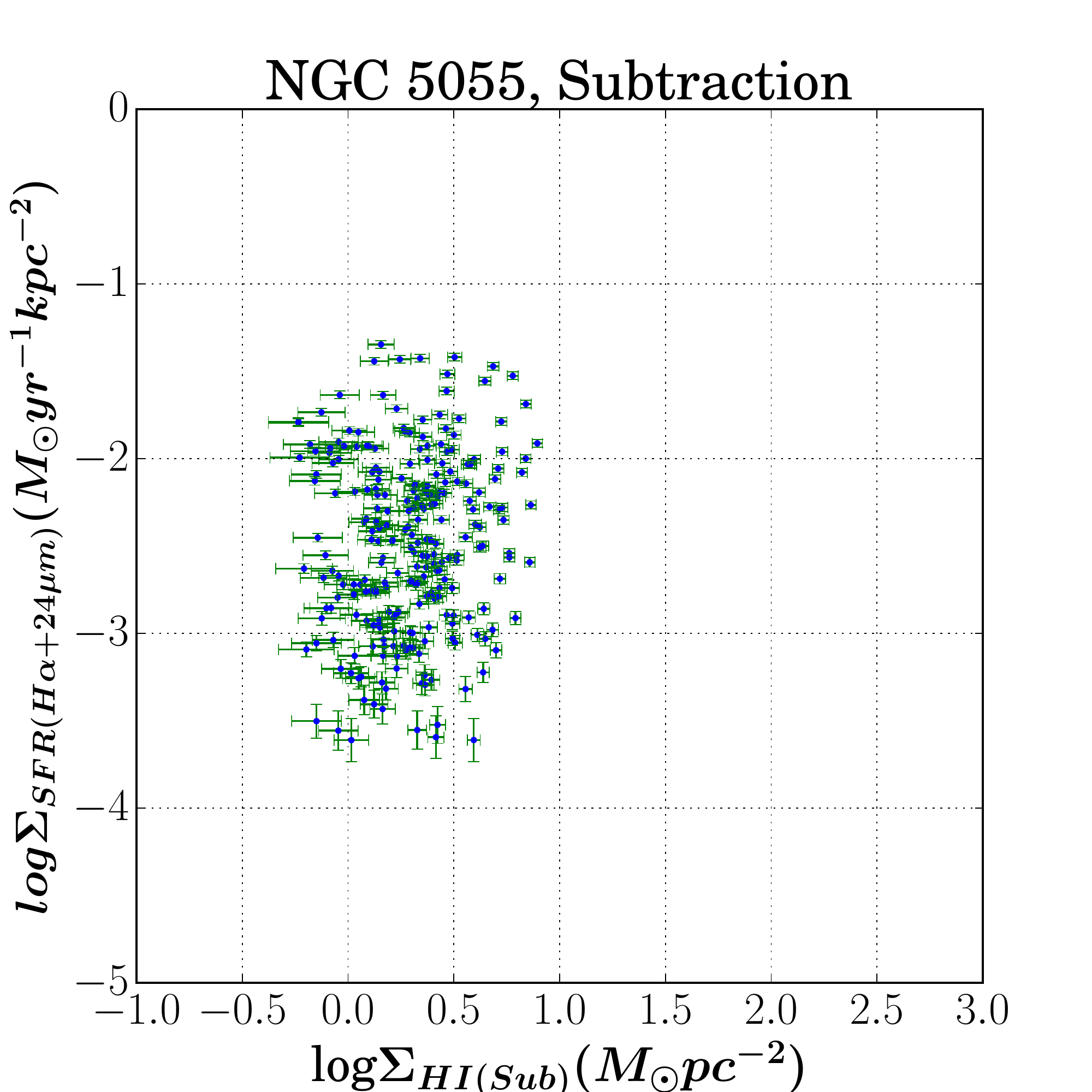}
	\includegraphics[width=0.33\textwidth,trim={0.2cm 0cm 1.5cm 0cm},clip]{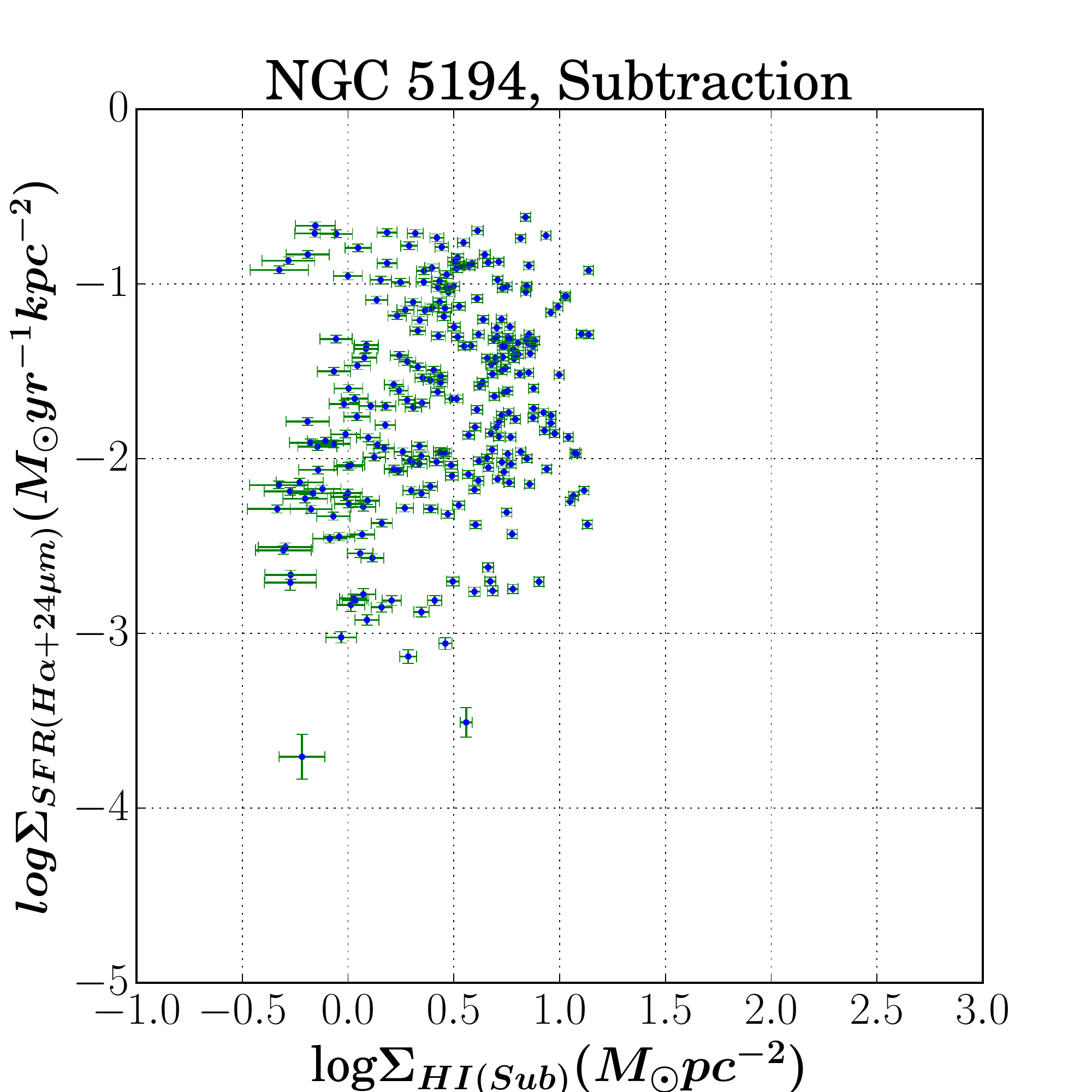}
	
	\includegraphics[width=0.33\textwidth,trim={0.2cm 0cm 1.5cm 0cm},clip]{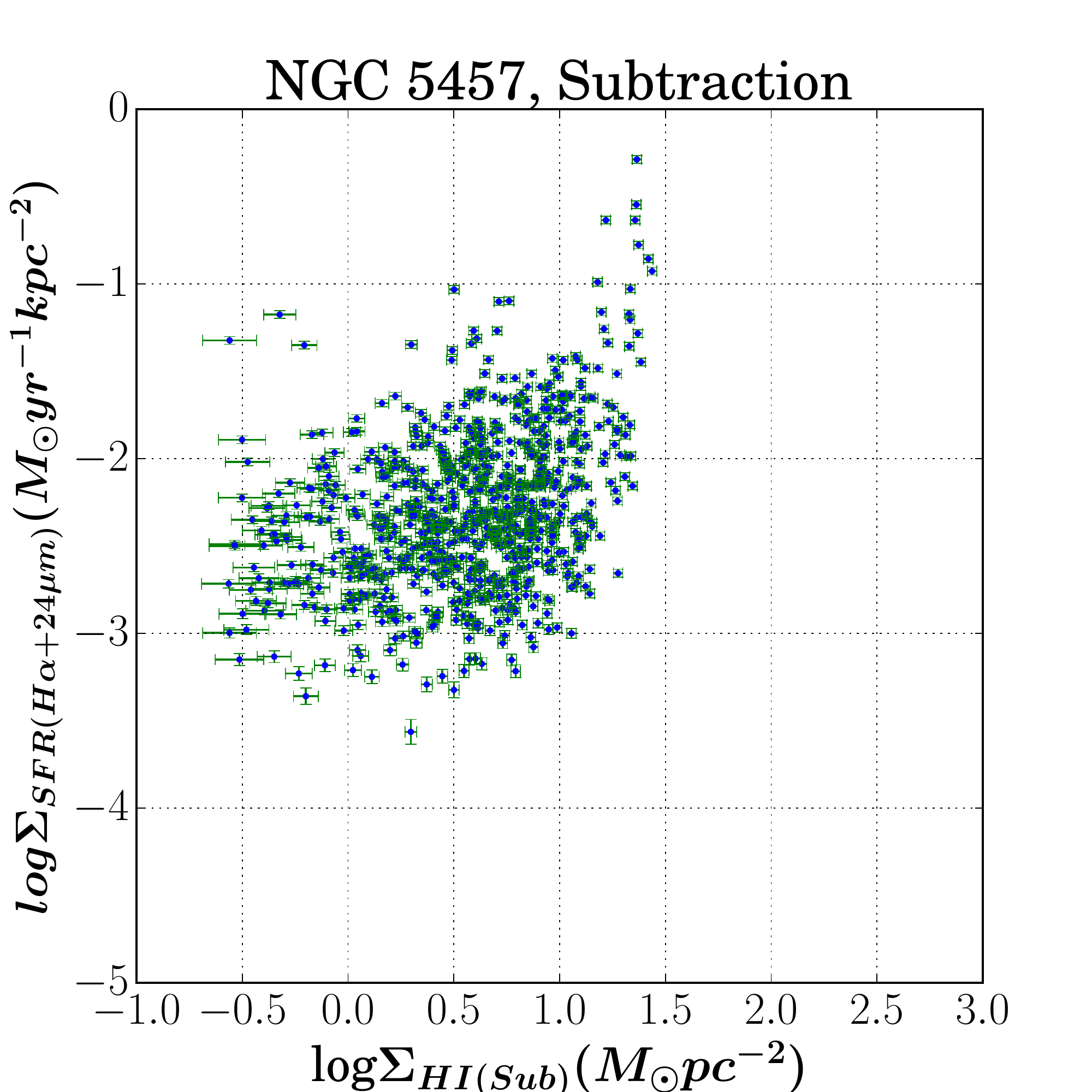}
	\includegraphics[width=0.33\textwidth,trim={0.2cm 0cm 1.5cm 0cm},clip]{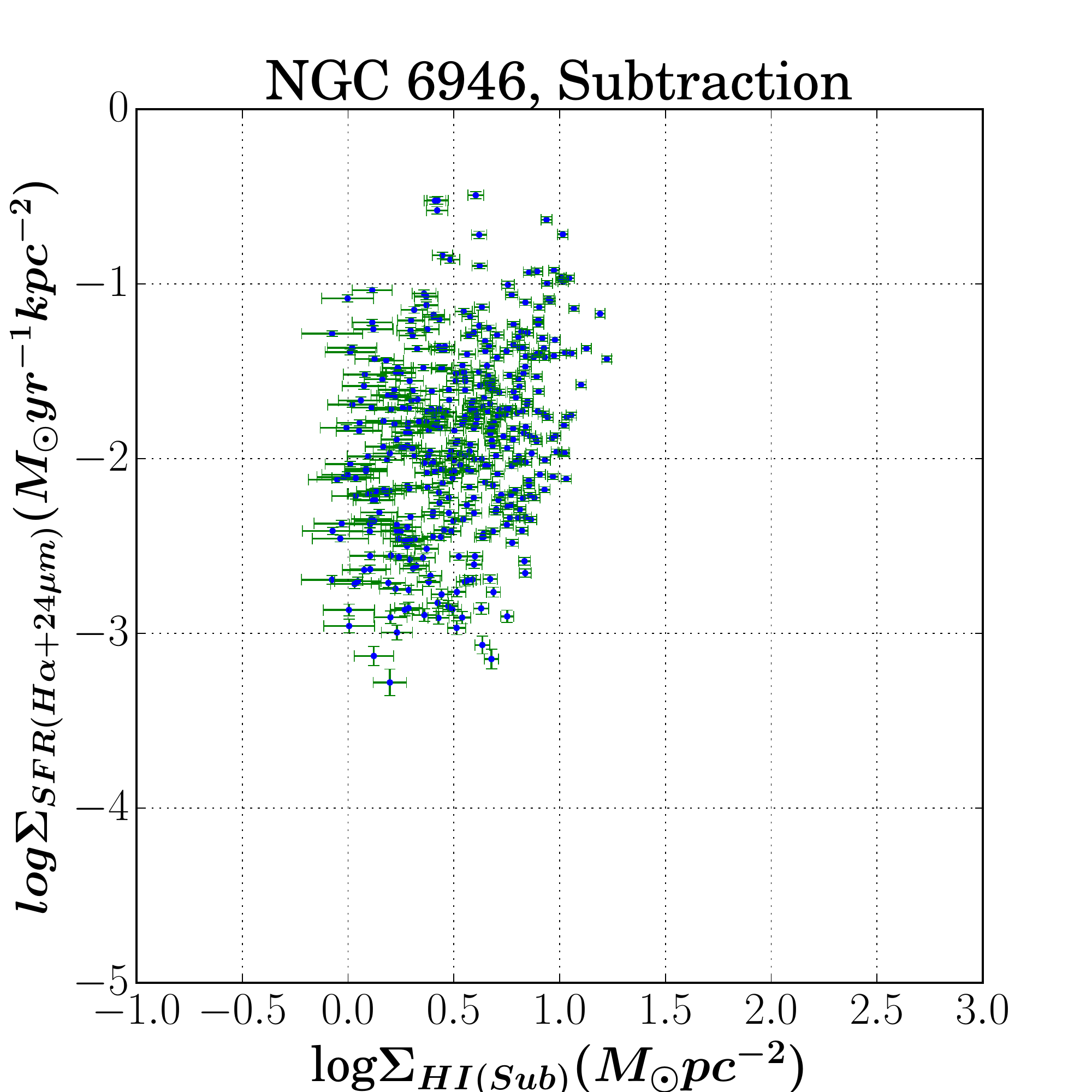}
	\caption{Spatially-resolved atomic gas Schmidt relation ( $\Sigma_{SFR}$ and $\Sigma_{H \textsc{i} (sub)}$). The diffuse background has been subtracted from both SFR tracers and atomic gas.}
	\label{HI sub}
\end{figure*}

\begin{figure*}
	\centering
	\includegraphics[width=0.33\textwidth,trim={0.2cm 0cm 1.5cm 0cm},clip]{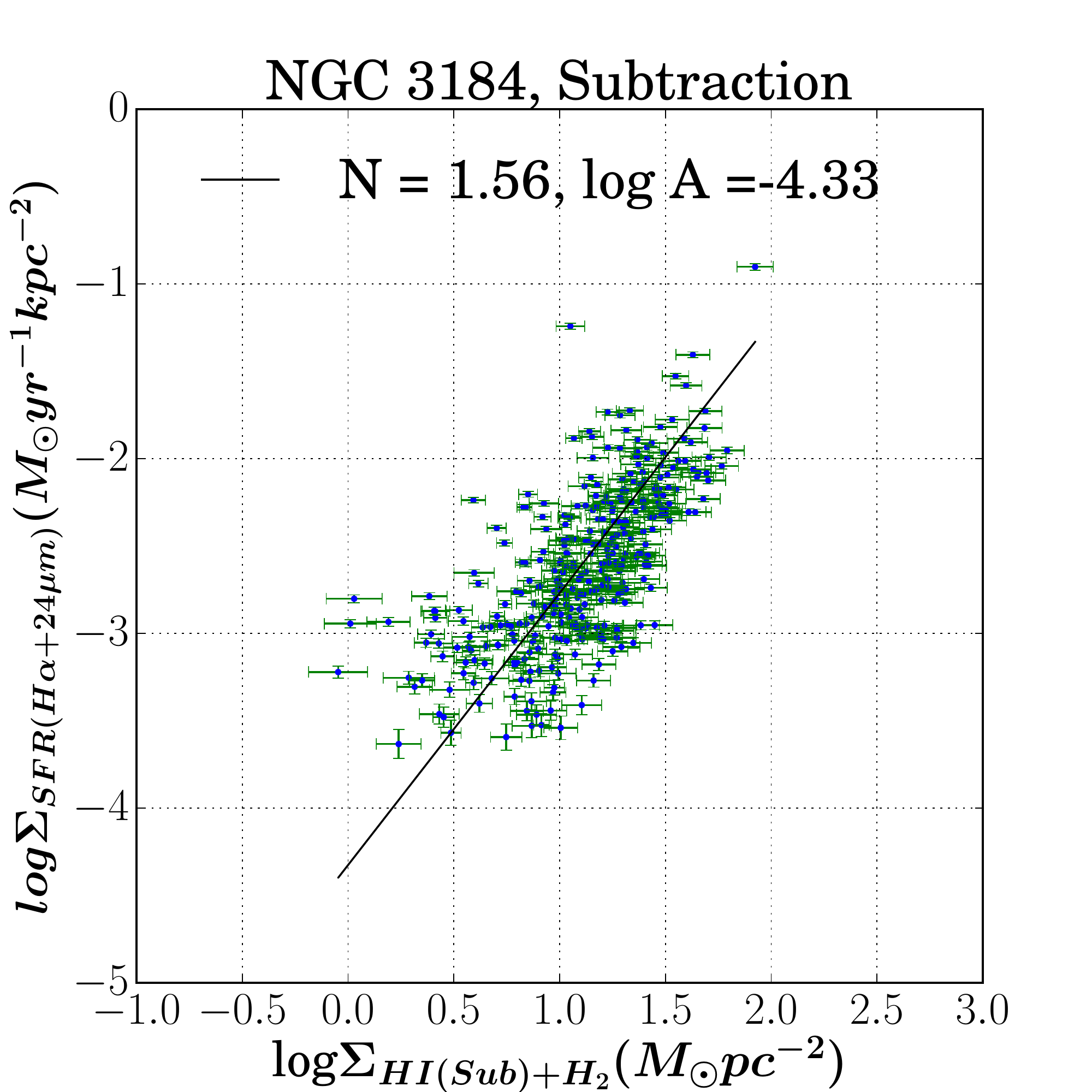}
	\includegraphics[width=0.33\textwidth,trim={0.2cm 0cm 1.5cm 0cm},clip]{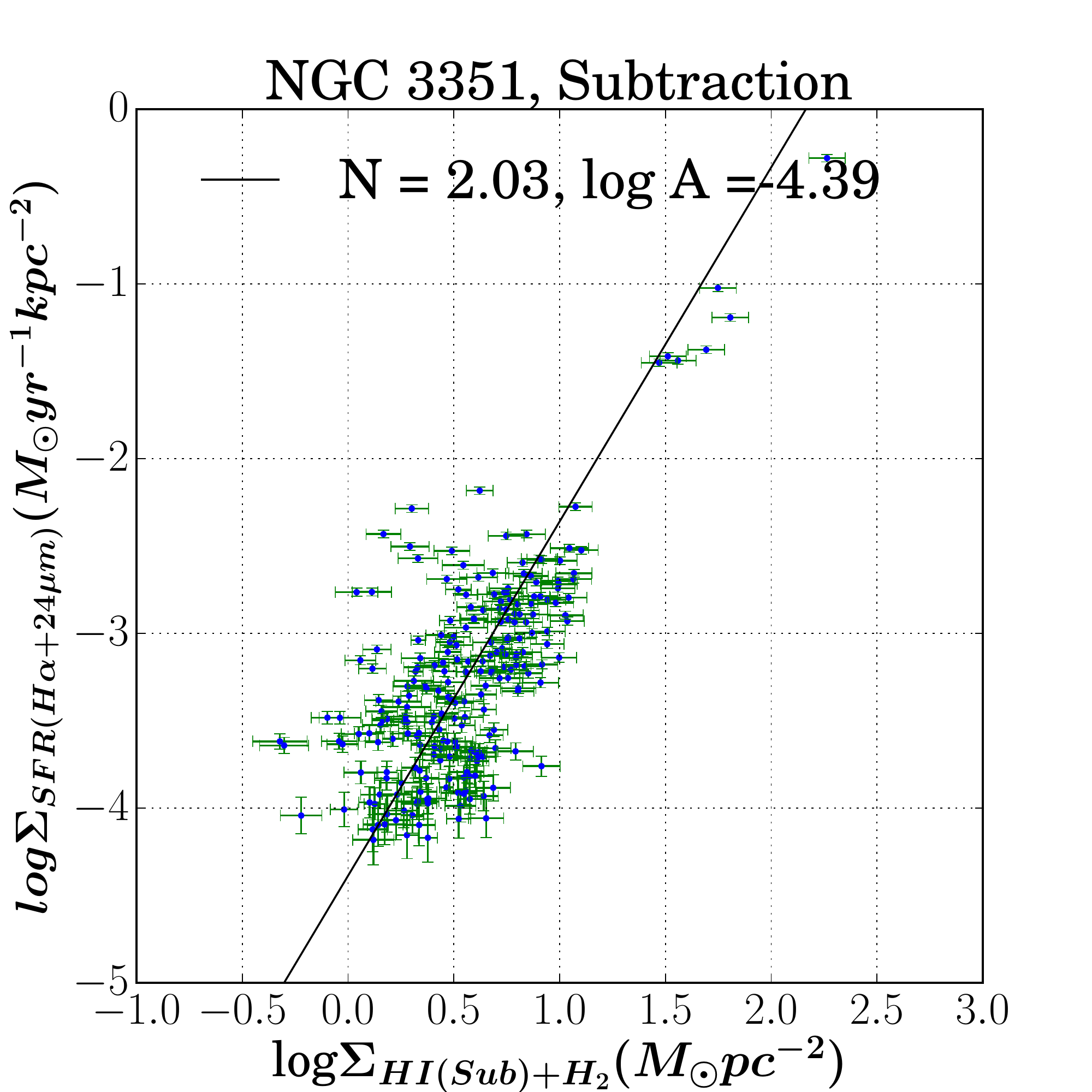}
	\includegraphics[width=0.33\textwidth,trim={0.2cm 0cm 1.5cm 0cm},clip]{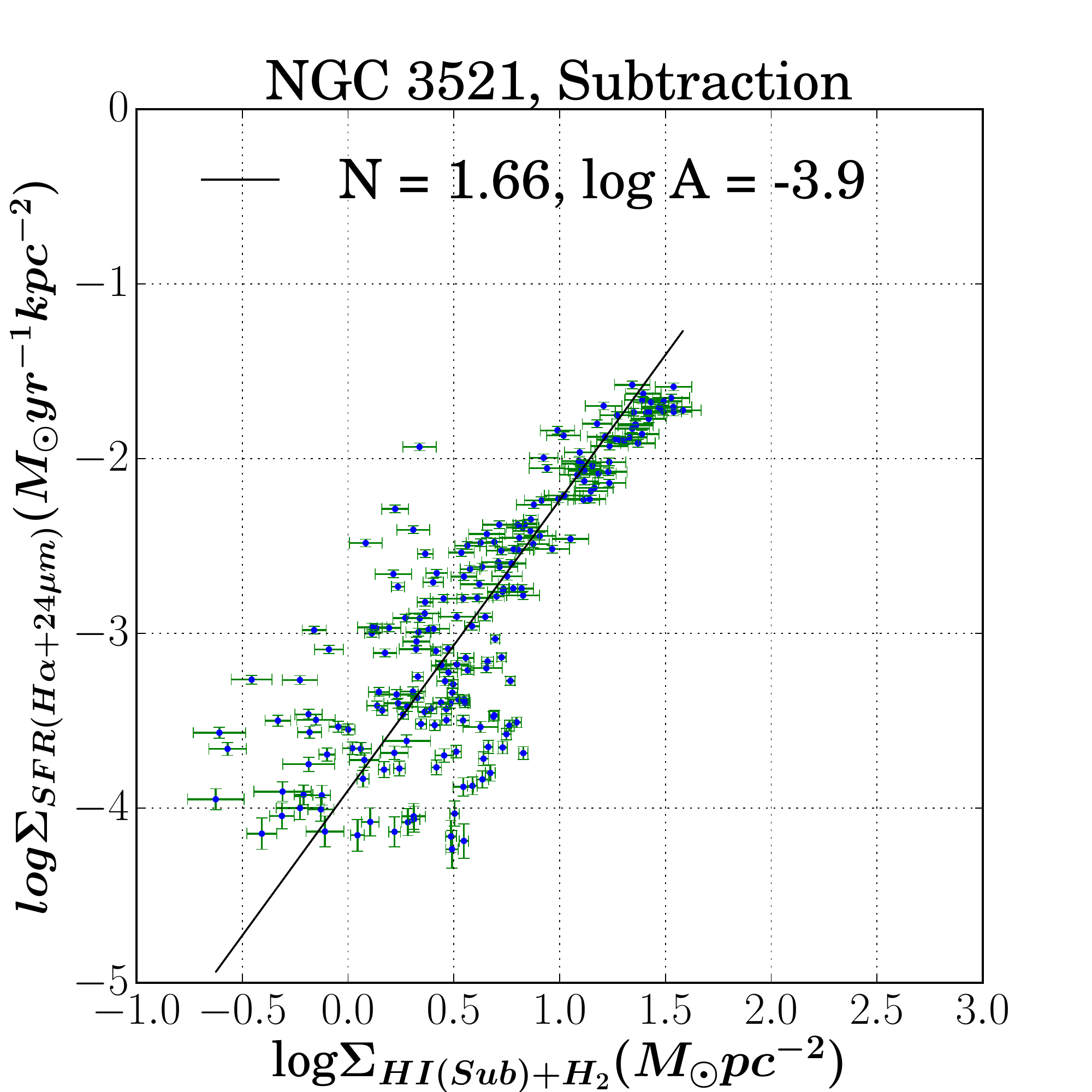}
	
	\includegraphics[width=0.33\textwidth,trim={0.2cm 0cm 1.5cm 0cm},clip]{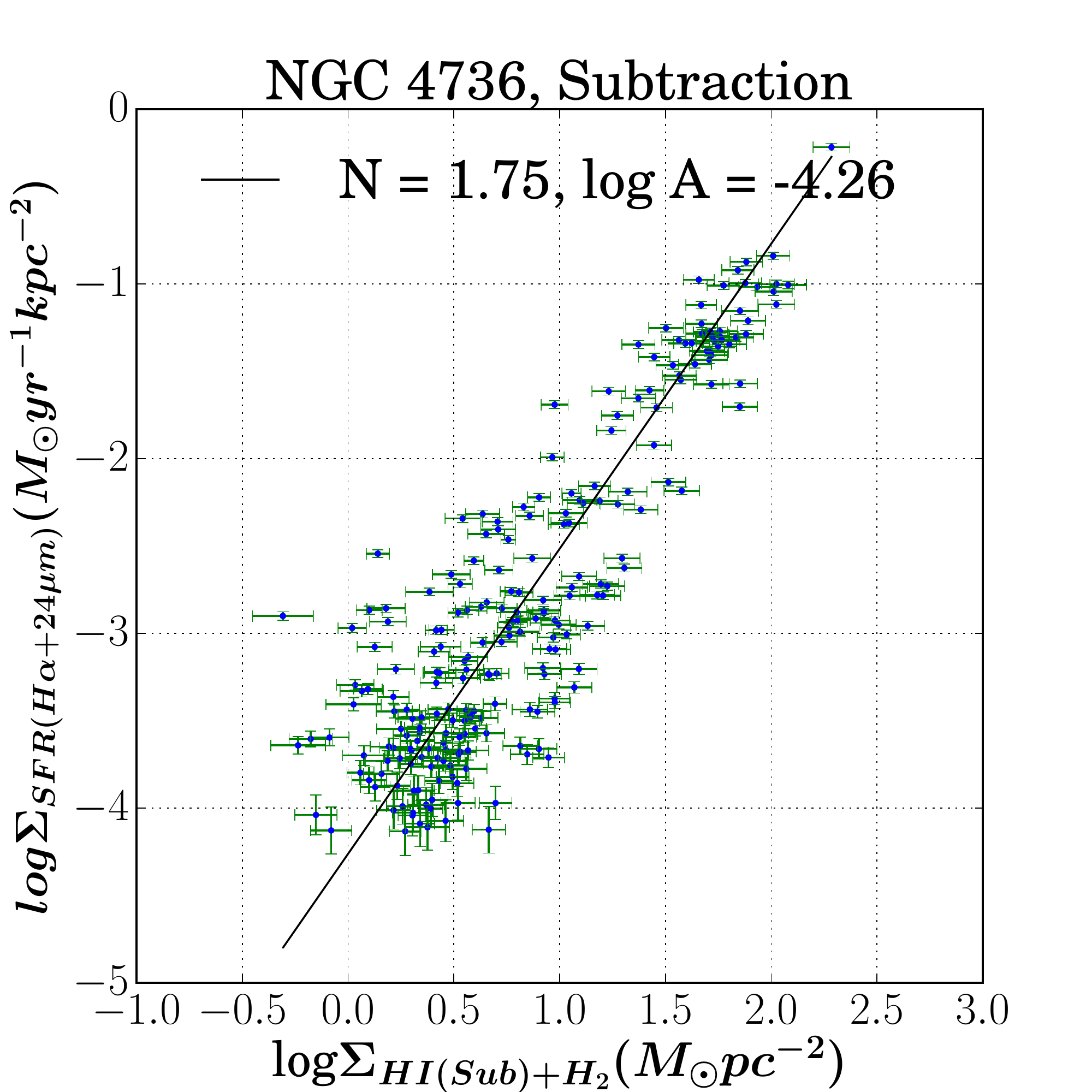}
	\includegraphics[width=0.33\textwidth,trim={0.2cm 0cm 1.5cm 0cm},clip]{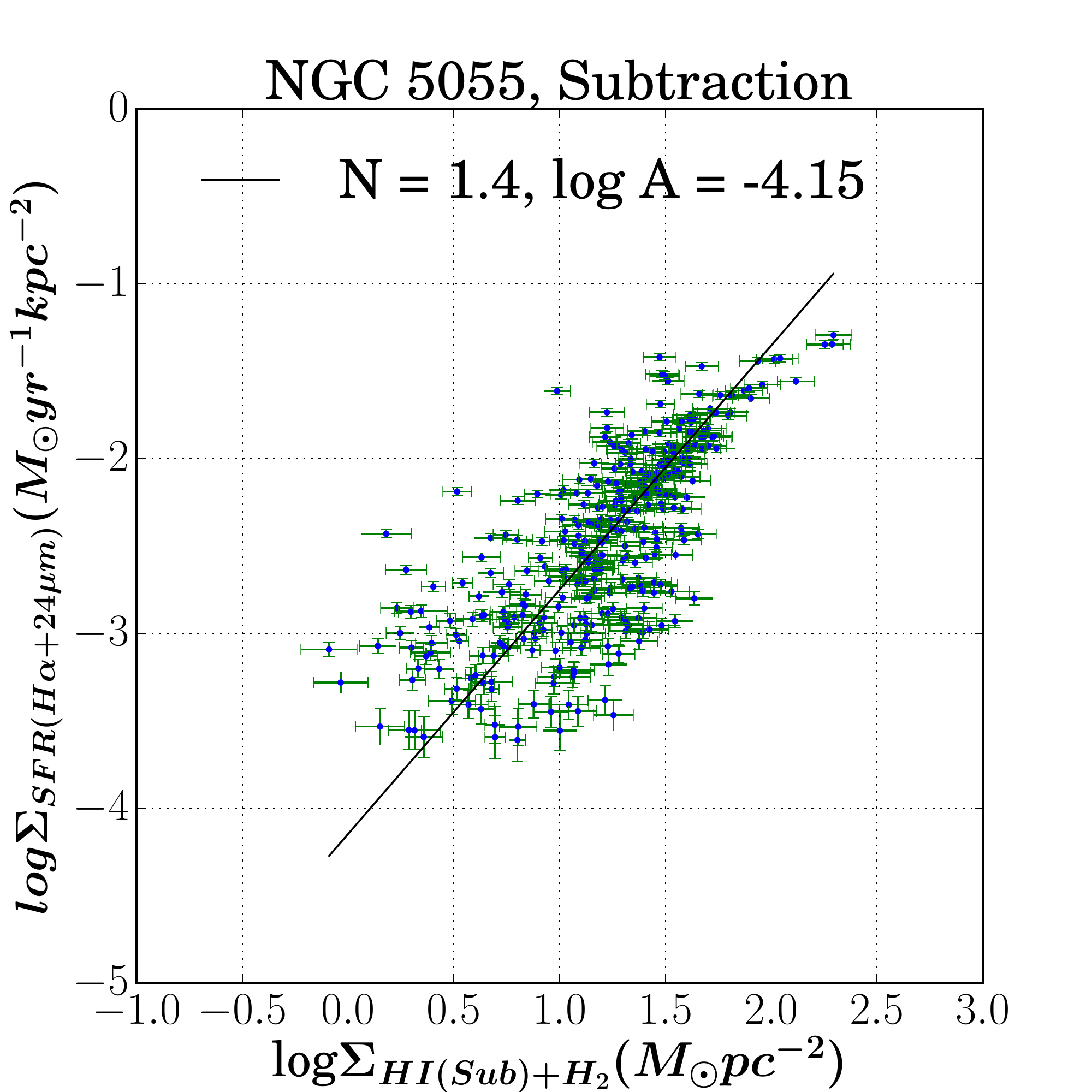}
	\includegraphics[width=0.33\textwidth,trim={0.2cm 0cm 1.5cm 0cm},clip]{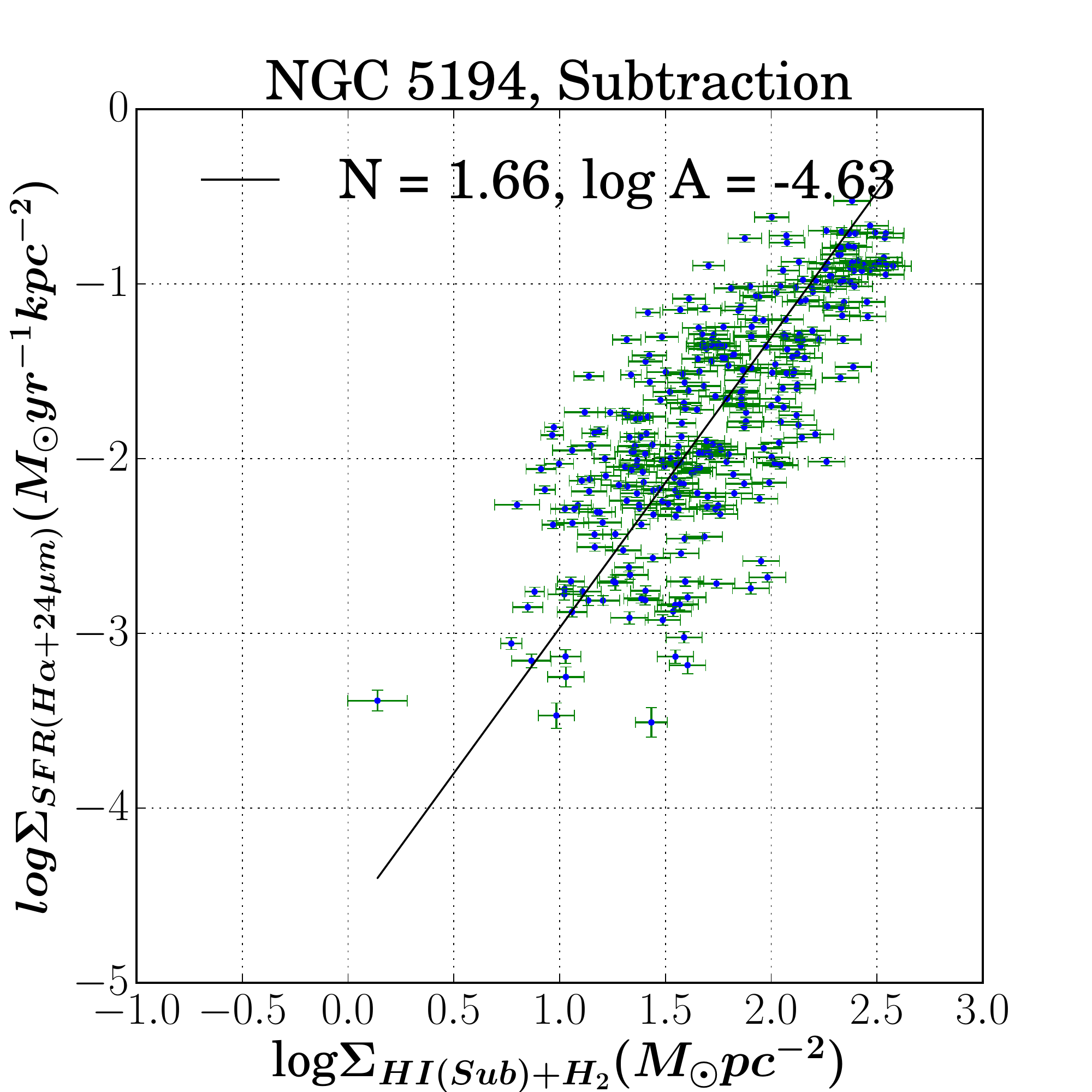}
	
	\includegraphics[width=0.33\textwidth,trim={0.2cm 0cm 1.5cm 0cm},clip]{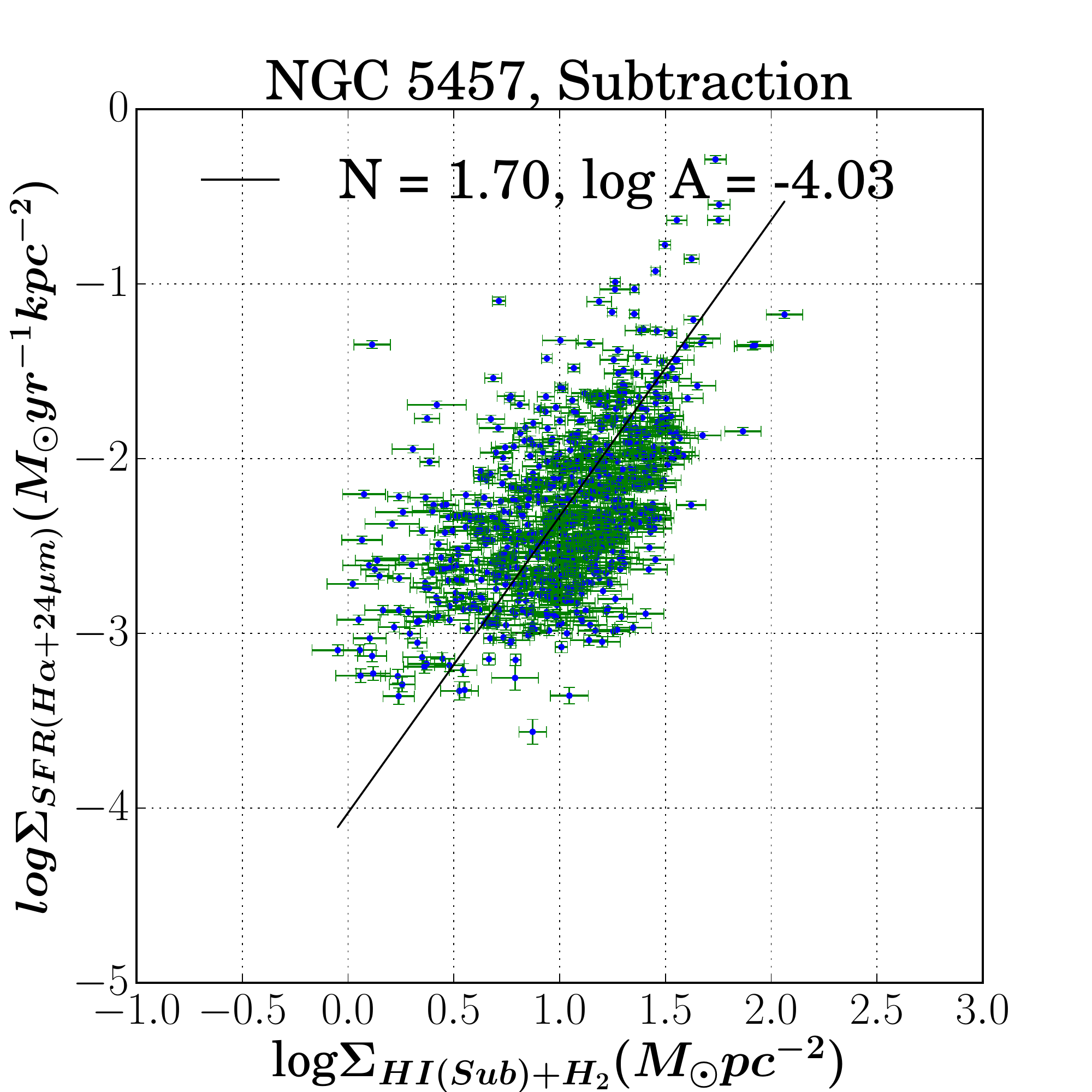}
	\includegraphics[width=0.33\textwidth,trim={0.2cm 0cm 1.5cm 0cm},clip]{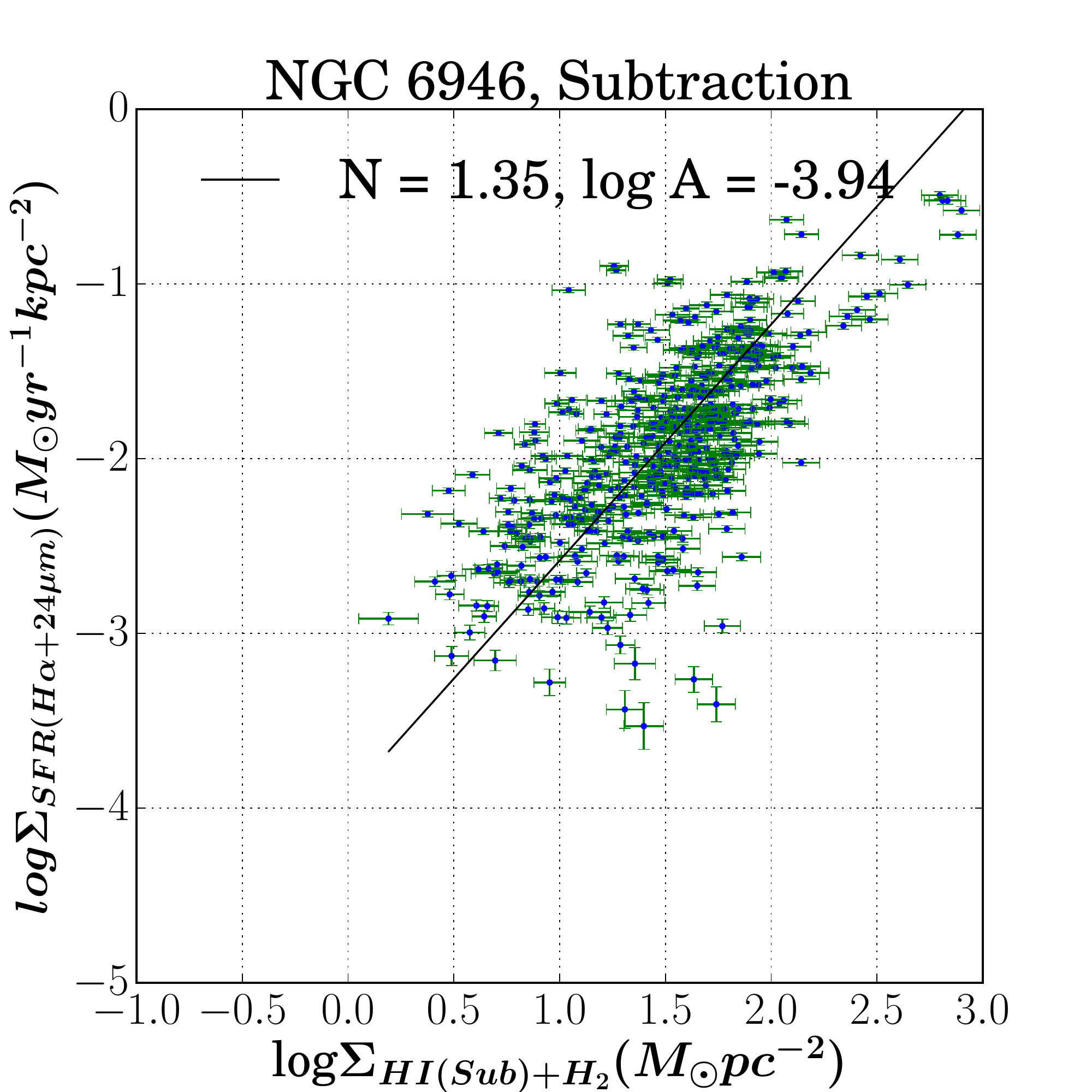}
	\caption{Spatially-resolved total gas Schmidt relation (i.e. $\Sigma_{SFR}$ and $\Sigma_{H \textsc{i} (sub) + H_2}$) for individual galaxies. The diffuse background has been subtracted from both SFR tracers and atomic gas. A Kroupa IMF and a constant X(CO) factor = 2.0$\times$10$^{20}$ cm$^{-2}$ (K km s$^{-1}$)$^{-1}$) have been assumed.}
	\label{total sub}
\end{figure*}

\indent Figs. \ref{NGC 3184}--\ref{NGC 6946} show the effect of inclusion and removal of diffuse background from the SFR tracers for each individual galaxy, except NGC 0628 for which the plots are shown in Fig. \ref{Figure: NGC 0628}. The caption of Fig. \ref{Figure: NGC 0628} holds true for all the eight figures.
 Fig. \ref{HI sub}  shows the Schmidt relation between $\Sigma_{SFR}$ and $\Sigma_{H \textsc{i}}$ after the diffuse background is subtracted from the SFR tracers as well as from the H\,\textsc{i},  individually for all the sample galaxies, except NGC 0628. Fig. \ref{total sub} shows the total gas Schmidt relation after subtraction of diffuse background from SFR tracers and atomic gas for all galaxies except NGC 0628. See  Fig. \ref{Figure: diffuse HI} for the corresponding plots of NGC 0628.
 
	
\begin{figure*}
	\centering
	\includegraphics[width=0.45\textwidth]{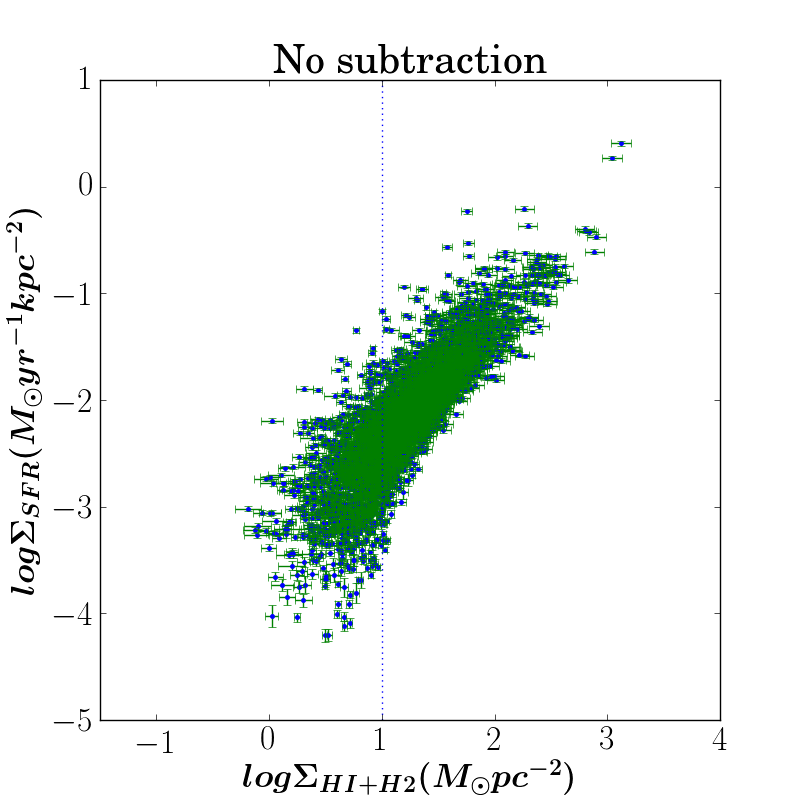}
	\includegraphics[width=0.45\textwidth]{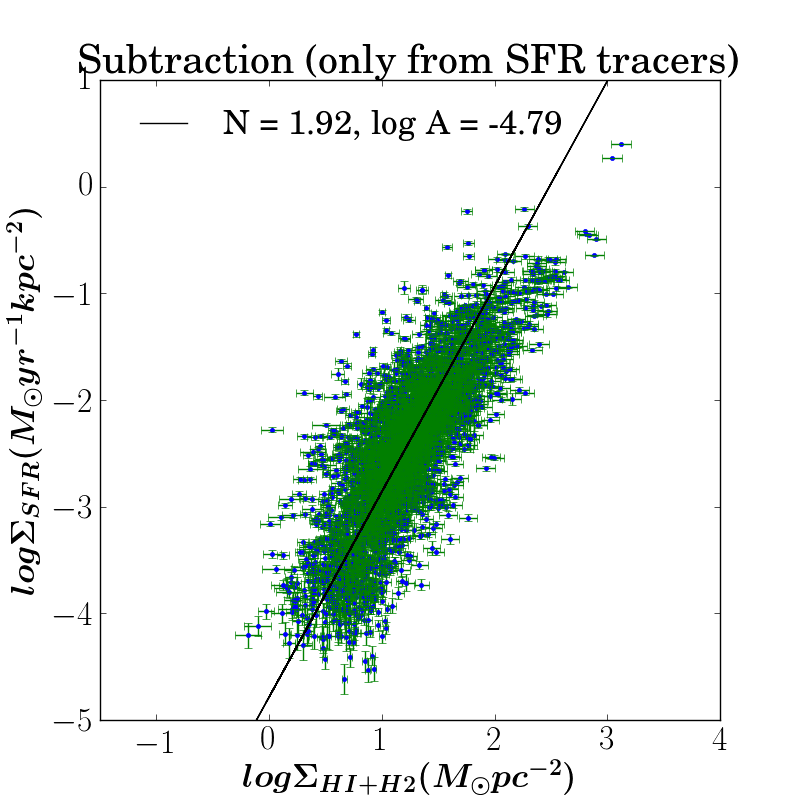}
	\includegraphics[width=0.45\textwidth]{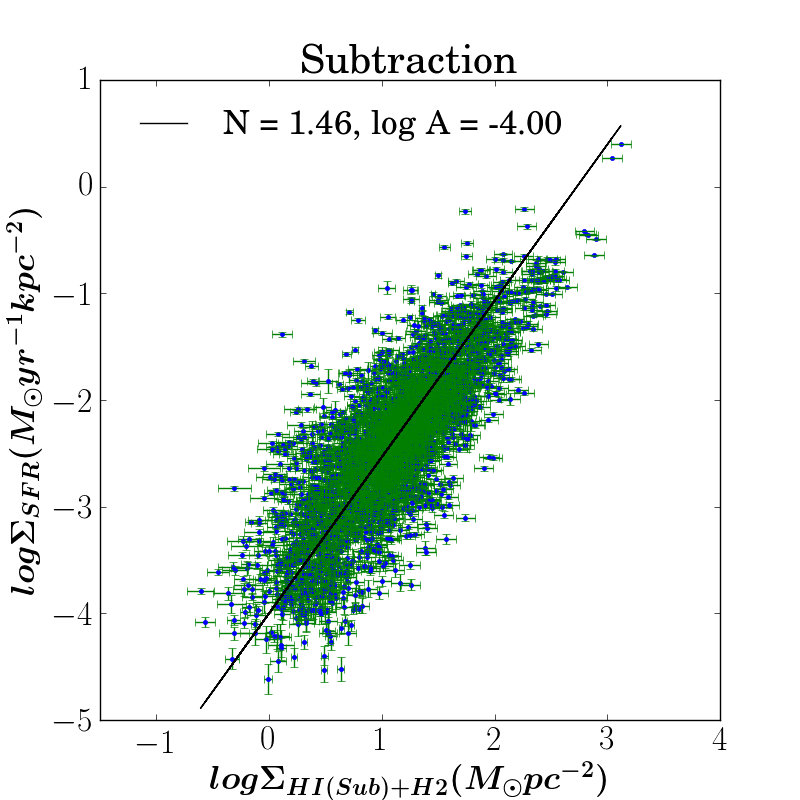}
	\includegraphics[width=0.45\textwidth]{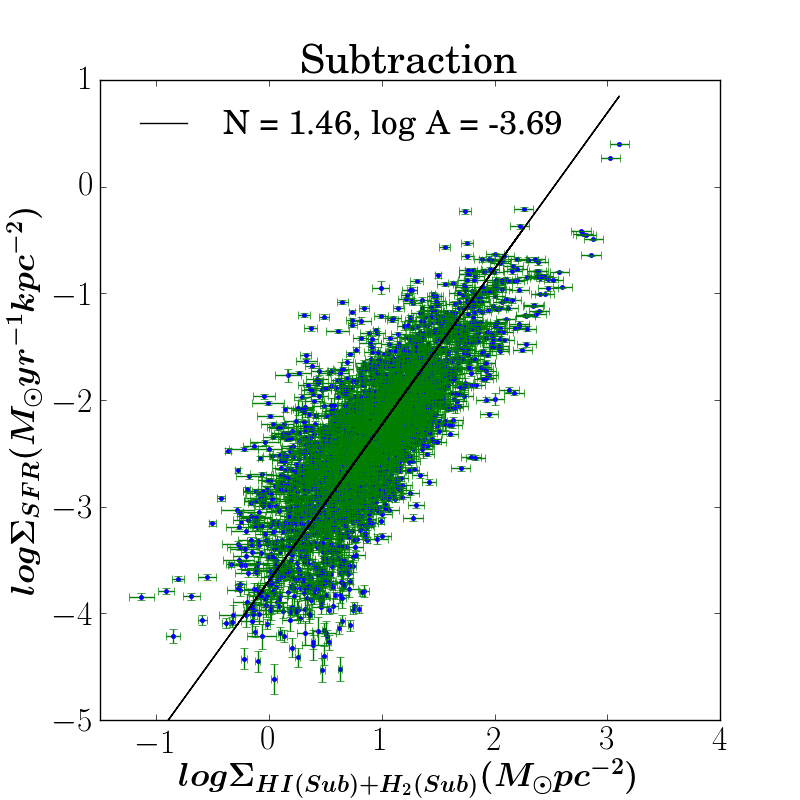}
	\caption{Spatially-resolved total gas Schmidt relation ($\Sigma_{SFR}$ and $\Sigma_{ H \textsc{i} + H_2}$), where SFR is estimated using FUV (equation \ref{sfr_fuv}) rather than using H$\alpha$. In each panel, blue dots with green error bars are the spatially-resolved data, and the solid black line shows the best-fit to the spatially-resolved data for all galaxies. We have assumed a Kroupa IMF and adopted a constant X(CO) factor = 2.0$\times$10$^{20}$ cm$^{-2}$ (K km s$^{-1}$)$^{-1}$. Upper-left panel: No subtraction of diffuse background is done. The vertical dotted line corresponds to 10 M$_{\odot}$ pc$^{-2}$ around which atomic gas surface density saturates. Upper-right panel: The diffuse background is subtracted from the SFR tracers. Lower-left panel: The diffuse background is subtracted from the SFR tracers as well as from the atomic gas. Lower-right panel: The diffuse background is subtracted from the SFR tracers, HI and CO. The trend and best-coefficients of total gas Schmidt relations obtained using FUV agree with those obtained from H$\alpha$ (see Figure \ref{Schmidt HI} and Table \ref{table:KS}).}
	\label{Schmidt HI FUV}
\end{figure*}

\end{appendix}
\end{document}